\documentclass[12pt,letterpaper]{article}

\pdfoutput=1
\usepackage[left=2cm,top=1cm,right=3cm,nohead]{geometry}
\usepackage[dvipsnames]{xcolor}
\usepackage{colortbl}
\usepackage[utf8x]{inputenc} 
\usepackage{color}
\usepackage{amsfonts,amssymb,amsthm,amsmath}
\DeclareFontFamily{U}{mathx}{}
\DeclareFontShape{U}{mathx}{m}{n}{<-> mathx10}{}
\DeclareSymbolFont{mathx}{U}{mathx}{m}{n}
\DeclareMathAccent{\widehat}{0}{mathx}{"70}
\DeclareMathAccent{\widecheck}{0}{mathx}{"71}
\usepackage{epsfig,graphics,latexsym,mathtools,amsbsy,multirow,slashed,mathrsfs,wasysym,textcomp,subfigure,wrapfig,comment,bbold,array,longtable,setspace,pifont, enumitem}
\usepackage{titlesec}
\usepackage{tabularx}
\usepackage{graphicx}
\usepackage{graphbox}
\usepackage{rotate}
\usepackage{accents}
\usepackage[bf]{caption}
\usepackage{braket}
\usepackage{fancyhdr}
\usepackage{subcaption}
\usepackage{tikz}
\usepackage{tikz-3dplot}
\usetikzlibrary{arrows}
\usetikzlibrary{3d}

\usepackage{float}
\usetikzlibrary{mindmap,trees,shadows}
\colorlet{color1}{gray!25}

\usetikzlibrary{positioning}
\usetikzlibrary{intersections} 
\usetikzlibrary{decorations.pathmorphing}
\usetikzlibrary{matrix,decorations.pathreplacing}
\usetikzlibrary{calc,angles,positioning,intersections,quotes,decorations.markings}
\usepackage{tkz-euclide}

\usepackage[toc,page]{appendix}
\usepackage{hhline,geometry,cite,datetime}
\usepackage{bbm} 
\usepackage{xcolor}
\usepackage{diagbox}
\usepackage{slashbox}
\usepackage{dynkin-diagrams}
\numberwithin{equation}{section}

\definecolor{crimson}{RGB}{220, 20, 60}
\def\cola{blue}
\def\colb{applegreen}
\def\colc{RedViolet}
\def\cold{cyan}
\def\colExt{crimson}

\usepackage{subfiles}
\usepackage{simpler-wick}

\usepackage{suffix}
\usepackage{float}

\usepackage[colorlinks=true,linkcolor=Blue,citecolor=Blue,filecolor=Blue,urlcolor=Blue,linktoc=page,pdfstartview=FitV,bookmarksopen=true]{hyperref}
\usepackage[UKenglish]{babel}
\hypersetup{pdftitle={},pdfauthor={},	pdfsubject={}} 

\newlength{\dhatheight}

\theoremstyle{plain}

\newcommand{\cgg}{{\mathfrak{g}}}

\newcommand{\ckk}{{\mathfrak{k}}}

\newcommand{\cpp}{{\mathfrak{p}}}
\newcommand{\cmm}{{\mathfrak m}}
\newcommand{\caa}{{\mathfrak{a}}}
\newcommand{\cnn}{{\mathfrak{n}}}

\newcommand{\dd}{\mathrm{d}}
\newcommand{\ee}{\mathrm{e}}

\DeclareMathOperator{\rank}{rank}
\DeclareMathOperator{\ad}{ad}

\DeclareMathOperator{\conv}{conv}

\def\Ucharge#1{{\text{\tiny$\left(#1\right)$}}}

\WithSuffix\def\splitA1{\mathfrak{sl}_2}
\WithSuffix\def\splitA2{\mathfrak{sl}_3}
\WithSuffix\def\splitA3{\mathfrak{sl}_4}
\WithSuffix\def\splitA4{\mathfrak{sl}_5}
\WithSuffix\def\splitA5{\mathfrak{sl}_6}
\WithSuffix\def\splitA6{\mathfrak{sl}_7}
\WithSuffix\def\splitA7{\mathfrak{sl}_8}
\WithSuffix\def\splitA8{\mathfrak{sl}_9}
\WithSuffix\def\splitAkminus1{\mathfrak{sl}_{k}}
\def\splitAk{\mathfrak{sl}_{k+1}}
\WithSuffix\def\splitD1{\mathfrak{so}(1,1)}
\WithSuffix\def\splitD2{\mathfrak{so}(2,2)}
\WithSuffix\def\splitD3{\mathfrak{so}(3,3)}
\WithSuffix\def\splitD4{\mathfrak{so}(4,4)}
\WithSuffix\def\splitD5{\mathfrak{so}(5,5)}
\WithSuffix\def\splitD6{\mathfrak{so}(6,6)}
\WithSuffix\def\splitD7{\mathfrak{so}(7,7)}
\WithSuffix\def\splitD8{\mathfrak{so}(8,8)}
\def\splitDk{\mathfrak{so}(k,k)}
\WithSuffix\def\splitG2{\mathfrak{g}_{2(2)}}
\WithSuffix\def\splitF4{\mathfrak{f}_{4(4)}}
\WithSuffix\def\splitE2{\mathfrak{e}_{2(2)}}
\WithSuffix\def\splitE3{\mathfrak{e}_{3(3)}}
\WithSuffix\def\splitE4{\mathfrak{e}_{4(4)}}
\WithSuffix\def\splitE5{\mathfrak{e}_{5(5)}}
\WithSuffix\def\splitE6{\mathfrak{e}_{6(6)}}
\WithSuffix\def\splitE7{\mathfrak{e}_{7(7)}}
\WithSuffix\def\splitE8{\mathfrak{e}_{8(8)}}
\WithSuffix\def\splitC1{\mathfrak{sp}_2}
\WithSuffix\def\splitC2{\mathfrak{sp}_4}
\WithSuffix\def\splitC3{\mathfrak{sp}_6}
\WithSuffix\def\splitC4{\mathfrak{sp}_8}
\WithSuffix\def\splitC5{\mathfrak{sp}_{10}}
\WithSuffix\def\splitC6{\mathfrak{sp}_{12}}
\WithSuffix\def\splitC7{\mathfrak{sp}_{14}}
\WithSuffix\def\splitC8{\mathfrak{sp}_{16}}
\def\splitCk{\mathfrak{sp}_{2k}}
\WithSuffix\def\splitB1{\mathfrak{so}(2,1)}
\WithSuffix\def\splitB2{\mathfrak{so}(3,2)}
\WithSuffix\def\splitB3{\mathfrak{so}(4,3)}
\WithSuffix\def\splitB4{\mathfrak{so}(5,4)}
\WithSuffix\def\splitB5{\mathfrak{so}(6,5)}
\WithSuffix\def\splitB6{\mathfrak{so}(7,6)}
\WithSuffix\def\splitB7{\mathfrak{so}(8,7)}
\WithSuffix\def\splitB8{\mathfrak{so}(9,8)}
\def\splitBk{\mathfrak{so}(k+1,k)}

\def\coordinatesTriangularPrism{
\protect\coordinate (O) at (0.95,0.538667);
\protect\coordinate (A) at (0,0);
\protect\coordinate (B) at (1,0);
\protect\coordinate (C) at (0.5,0.866); 
\protect\coordinate (A') at ($(A)+(0.9,0.5)$);
\protect\coordinate (B') at ($(B)+(0.9,0.5)$);
\protect\coordinate (C') at ($(C)+(0.9,0.5)$);}

\def\drawTriangularPrism{
\draw[thick] (A) -- (B) -- (C) -- cycle;
\draw[thick] (B')  -- (C');
\draw[dashed, thick] (C') -- (A') -- (B');
\draw[thick, dashed] (A) -- (A');
\draw[thick] (B) -- (B');
\draw[thick] (C) -- (C');}

\def\coordinatesTriangularPrismExtra{
\coordinate (O) at (0.95,0.538667);
\coordinate (A) at (0,0);
\coordinate (B) at (1,0);
\coordinate (C) at (0.5,0.866); 
\coordinate (A') at ($(A)+(0.6,0.5)$);
\coordinate (B') at ($(B)+(0.6,0.5)$);
\coordinate (C') at ($(C)+(0.6,0.5)$);
\coordinate (AB) at ($0.5*(A)+0.5*(B)$);
\coordinate (BC) at ($0.5*(B)+0.5*(C)$);
\coordinate (CA) at ($0.5*(C)+0.5*(A)$); 
\coordinate (A'B') at ($0.5*(A')+0.5*(B')$);
\coordinate (B'C') at ($0.5*(B')+0.5*(C')$);
\coordinate (C'A') at ($0.5*(C')+0.5*(A')$); 
\coordinate (AA') at ($0.5*(A)+0.5*(A')$);
\coordinate (BB') at ($0.5*(B)+0.5*(B')$);
\coordinate (CC') at ($0.5*(C)+0.5*(C')$);
\coordinate (AB') at ($0.5*(A)+0.5*(B')$);
\coordinate (A'B) at ($0.5*(A')+0.5*(B)$);
\coordinate (BC') at ($0.5*(B)+0.5*(C')$);
\coordinate (B'C) at ($0.5*(B')+0.5*(C)$);
\coordinate (CA') at ($0.5*(C)+0.5*(A')$); 
\coordinate (C'A) at ($0.5*(C')+0.5*(A)$); 
\coordinate (ABC) at ($0.33333*(A)+0.33333*(B)+0.33333*(C)$);
\coordinate (A'B'C') at ($0.33333*(A')+0.33333*(B')+0.33333*(C')$);
\coordinate (Z) at ($0.5*(A'B'C')+0.5*(ABC)$);}

\definecolor{applegreen}{rgb}{0.55, 0.71, 0.0}
\def\cola{blue}
\def\colb{applegreen}
\def\colc{red}
\def\colab{cyan}
\def\colac{magenta}
\def\colbc{yellow}
\def\colabc{white}

\def\isotriangleNoCircleScaled#1{{\tikz[scale=#1]{\draw (-0.707107,-0.801784) -- (0.707107,-0.801784) -- (0,1.06904) -- cycle;}}}

\def\isotriangleStringsScaled#1{{\tikz[scale=#1]{\draw (-0.707107,-0.801784) -- (0.707107,-0.801784) -- (0.707107,0.267261) -- (0,1.06904) -- (-0.707107,0.267261) -- cycle;
}}}

\def\cube{\protect\tikz[scale=0.2]\protect\draw (0,1,0) -- (0,0,0) -- (1,0,0) -- (1,1,0) -- (0,1,0) -- (0,1,-1) -- (1,1,-1) --(1,1,0) (1,0,0) -- (1,0,-1) -- (1,1,-1);}
\def\cubeScaled#1{{\protect\tikz[scale=#1]\protect\draw (0,1,0) -- (0,0,0) -- (1,0,0) -- (1,1,0) -- (0,1,0) -- (0,1,-1) -- (1,1,-1) --(1,1,0) (1,0,0) -- (1,0,-1) -- (1,1,-1);}}

\def\seg{\tikz[scale=1]{\protect\draw[thick] (0,0) -- (0.2,0);\filldraw (0,0) circle (1.1pt);\filldraw (0.2,0) circle (1.1pt);}}

\def\segScaled#1{{\tikz[scale=#1]{\protect\draw[thick] (0,0) -- (0.2,0);\filldraw (0,0) circle (1.1pt);\filldraw (0.2,0) circle (1.1pt);}}}

\def\segconic{\tikz[scale=1.2]{
\protect\draw (0,0) -- (0.2,0);
\filldraw (0,0) circle (.7pt);
\filldraw (0.1,0) circle (.5pt);
\filldraw (0.2,0) circle (.7pt);
}}

\def\squareconic{\protect\tikz[scale=0.3]{
\protect\coordinate (A) at (0,0);
\protect\coordinate (B) at (1,0);
\protect\coordinate (C) at (0,1);
\protect\coordinate (D) at (1,1);
\protect\draw (A) -- (B) -- (D) -- (C) -- cycle;
\filldraw (0.5,0.5) circle (1.5pt);
}}

\def\triangleconic{\protect\tikz[scale=0.23]{
\protect\coordinate (A) at (0,1);
\protect\coordinate (B) at (0.866025,-0.5);
\protect\coordinate (C) at (-0.866025,-0.5);
\foreach \i/\j in {A/B,  B/C,  C/A}{\protect\draw (\i) -- (\j);}
\filldraw (0,0) circle (1.7pt);
}}

\def\coordinatesHexagon{
\protect\coordinate (A) at (1.000, 0.000);
\protect\coordinate (B) at (0.500, 0.866);
\protect\coordinate (C) at (-0.500, 0.866);
\protect\coordinate (D) at (-1.000, 0.000);
\protect\coordinate (E) at (-0.500, -0.866);
\protect\coordinate (F) at (0.500, -0.866);}
\def\drawHexagon{\foreach \i/\j in {A/B,  B/C,  C/D,  D/E,  E/F, F/A}{\protect\draw[thick] (\i) -- (\j);} }

\def\coordinatesSquare{
\protect\coordinate (A) at (1,1);
\protect\coordinate (B) at (-1,1);
\protect\coordinate (C) at (-1,-1);
\protect\coordinate (D) at (1,-1);}

\def\coordinatesRectangle{
\protect\coordinate (A) at (1,0.707);
\protect\coordinate (B) at (-1,0.707);
\protect\coordinate (C) at (-1,-0.707);
\protect\coordinate (D) at (1,-0.707);}

\def\coordinatesTriangle{
\protect\coordinate (A) at (0,1);
\protect\coordinate (B) at (0.866025,-0.5);
\protect\coordinate (C) at (-0.866025,-0.5);}
\def\drawTriangle{\foreach \i/\j in {A/B,  B/C,  C/A}{\protect\draw[thick] (\i) -- (\j);}}

\def\coordinatesOctagon{
\protect\coordinate (A) at (1,-0.5);
\protect\coordinate (B) at (1,0.5);
\protect\coordinate (C) at (0.5,1);
\protect\coordinate (D) at (-0.5,1);
\protect\coordinate (E) at (-1,0.5);
\protect\coordinate (F) at (-1,-0.5);
\protect\coordinate (G) at (-0.5,-1);
\protect\coordinate (H) at (0.5,-1);}

\def\coordinatesTetrahedron{
  \protect\coordinate (A) at (1, 1, 1);
  \protect\coordinate (B) at (-1, -1, 1);
  \protect\coordinate (C) at (-1, 1, -1);
  \protect\coordinate (D) at (1, -1, -1);}

\def\coordinatesOctahedron{
  \protect\coordinate (X)  at (1, 0, 0);
  \protect\coordinate (MX) at (-1, 0, 0);
  \protect\coordinate (Y)  at (0, 1, 0);
  \protect\coordinate (MY) at (0, -1, 0);
  \protect\coordinate (Z)  at (0, 0, 1);
  \protect\coordinate (MZ) at (0, 0, -1);}

\def\octahedronScaled#1{
\tdplotsetmaincoords{100}{70}
\begin{tikzpicture}[scale=#1,tdplot_main_coords]
  \coordinate (X)  at (1, 0, 0);
  \coordinate (MX) at (-1, 0, 0);
  \coordinate (Y)  at (0, 1, 0);
  \coordinate (MY) at (0, -1, 0);
  \coordinate (Z)  at (0, 0, 1);
  \coordinate (MZ) at (0, 0, -1);
  \draw (Y) -- (Z) -- (MX) -- (Y);
  \draw (Y) -- (MZ) -- (MX) -- (MY) -- (Z);
  \draw (MZ) -- (MY);
  \draw[dashed,dash pattern=on 1pt off 1pt,thin] 
  (MY) -- (X) -- (MZ) ;
  \draw[dashed,dash pattern=on 1pt off 1pt,thin] (Z) -- (X) -- (Y) ;
\end{tikzpicture}}

\def\drawSphere{
\begin{scope}[shift={(\shiftx,\shifty)}]
\shade[ball color = gray!40, opacity = 0.2] (0,0) circle(\rad);
\draw[gray,opacity=0.5] (0,0) circle(\rad);
\draw[gray,opacity=0.5] plot[domain=pi:2*pi] ({\rad*cos(\x r)},{\rad*.2*sin(\x r)});
\draw[dashed, gray,opacity=0.5] plot[domain=0:pi] ({\rad*cos(\x r)},{\rad*.2*sin(\x r)});
\draw[gray,opacity=0.5] plot[domain=0.5*pi:1.5*pi] ({\rad*.2*cos(\x r)},{\rad*sin(\x r)});
\draw[dashed, gray,opacity=0.5] plot[domain=-0.5*pi:0.5*pi] ({\rad*.2*cos(\x r)},{\rad*sin(\x r)});
\end{scope}}

\def\labelsOctagonAll{\foreach \pt in {A,B,C,D,E,F,G,H}{ \path  ($1.2*(\pt)$)  node[font=\scriptsize]{$4$};};
\foreach \i/\j in {A/B,C/D,E/F,G/H}{\path  ($1.3*0.5*(\i)+1.3*0.5*(\j)$)  node[font=\scriptsize]{$\infty$};};
\foreach \i/\j in {B/C,D/E,F/G,H/A}{\path  ($1.3*0.5*(\i)+1.3*0.5*(\j)$)  node[font=\scriptsize]{$8$};};}

\def\labelsSquareOneInf{\foreach \pt in {A,B,C,D}{ \path  ($1.2*(\pt)$)  node[font=\scriptsize]{1};};
\foreach \i/\j in {A/B,B/C,C/D,D/A}{\path  ($1.3*0.5*(\i)+1.3*0.5*(\j)$)  node[font=\scriptsize]{$\infty$};};}
\def\labelsSquareOneTwo{\foreach \pt in {A,B,C,D}{ \path  ($1.2*(\pt)$)  node[font=\scriptsize]{1};};
\foreach \i/\j in {A/B,B/C,C/D,D/A}{\path  ($1.3*0.5*(\i)+1.3*0.5*(\j)$)  node[font=\scriptsize]{$2$};};}

\def\labelsRectangleOne{\foreach \pt in {A,B,C,D}{ \path  ($1.2*(\pt)$)  node[font=\scriptsize]{$1$};};
\foreach \i/\j in {A/B,C/D}{\path  ($1.3*0.5*(\i)+1.3*0.5*(\j)$)  node[font=\scriptsize]{$\infty$};};
\foreach \i/\j in {B/C,D/A}{\path  ($1.3*0.5*(\i)+1.3*0.5*(\j)$)  node[font=\scriptsize]{$2$};};}

\def\labelsTriangleOneInf{\foreach \pt in {A,B,C}{ \path  ($1.2*(\pt)$)  node[font=\scriptsize]{$1$};};
\foreach \i/\j in {A/B,B/C,C/A}{\path  ($1.3*0.5*(\i)+1.3*0.5*(\j)$)  node[font=\scriptsize]{$\infty$};};}

\def\labelsTriangleOneTwo{\foreach \pt in {A,B,C}{ \path  ($1.2*(\pt)$)  node[font=\scriptsize]{$1$};};
\foreach \i/\j in {A/B,B/C,C/A}{\path  ($1.3*0.5*(\i)+1.3*0.5*(\j)$)  node[font=\scriptsize]{$2$};};}

\def\labelsHexagonOneTwo{
\foreach \pt in {A,B,C,D,E,F}{ 
\path  ($1.2*(\pt)$)  node[font=\scriptsize]{1};
};
\foreach \i/\j in {A/B,B/C,C/D,D/E,E/F,F/A}{ \path  ($1.3*0.5*(\i)+1.3*0.5*(\j)$)  node[font=\scriptsize]{2};};
}

\def\labelsHexagonThreeInf{
\foreach \pt in {A,B,C,D,E,F}{ 
\path  ($1.2*(\pt)$)  node[font=\scriptsize]{$3$};
};
\foreach \i/\j in {A/B,B/C,C/D,D/E,E/F,F/A}{ \path  ($1.3*0.5*(\i)+1.3*0.5*(\j)$)  node[font=\scriptsize]{$\infty$};};}

\def\IIB{\protect\seg}
\def\IIA{\protect\seg}

\def\coordinatesCubeExtra{
\coordinate (AB) at ($0.5*(A)+0.5*(B)$);
\coordinate (BC) at ($0.5*(B)+0.5*(C)$);
\coordinate (CD) at ($0.5*(C)+0.5*(D)$);
\coordinate (DA) at ($0.5*(D)+0.5*(A)$);
\coordinate (A'B') at ($0.5*(A')+0.5*(B')$);
\coordinate (B'C') at ($0.5*(B')+0.5*(C')$);
\coordinate (C'D') at ($0.5*(C')+0.5*(D')$);
\coordinate (D'A') at ($0.5*(D')+0.5*(A')$);
\coordinate (AA') at ($0.5*(A)+0.5*(A')$);
\coordinate (BB') at ($0.5*(B)+0.5*(B')$);
\coordinate (CC') at ($0.5*(C)+0.5*(C')$);
\coordinate (DD') at ($0.5*(D)+0.5*(D')$);
\coordinate (AC) at ($0.5*(A)+0.5*(C)$);
\coordinate (AB') at ($0.5*(A)+0.5*(B')$);
\coordinate (DC') at ($0.5*(D)+0.5*(C')$);
\coordinate (A'C') at ($0.5*(A')+0.5*(C')$);}

\def\triprism{\protect\tikz[scale=0.3]{
\protect\coordinate (A) at (0,0);
\protect\coordinate (B) at (1,0);
\protect\coordinate (C) at (0.5,0.866); 
\protect\coordinate (A') at ($(A)+(0.9,0.5)$);
\protect\coordinate (B') at ($(B)+(0.9,0.5)$);
\protect\coordinate (C') at ($(C)+(0.9,0.5)$);
\protect\draw (A) -- (B) -- (C) -- cycle;
\protect\draw (B')  -- (C');
\protect\draw (B) -- (B');
\protect\draw (C) -- (C');}}

\def\rectangle{\sqsubset \!\!\! \sqsupset}

\def\Ffour{\text{F4}}
\def\Efour{\text{E4}}
\def\Efive{\text{E5}}
\def\Esix{\text{E6}}
\def\Eseven{\text{E7}}
\def\Eeight{\text{E8}}

\def\FfourAlt{\text{oc}}

\def\EfourAlt{\text{tetro}}
\def\EfiveAlt{\text{demi}}
\def\EsixAlt{V_{27}}
\def\EsevenAlt{V_{56}}
\def\EeightAlt{V_{240}}

\def\cuboctahedron{\tikz[scale=0.2]
{\coordinate (A) at (1.000, 0.000);
\coordinate (B) at (0.500, 0.866);
\coordinate (C) at (-0.500, 0.866);
\coordinate (D) at (-1.000, 0.000);
\coordinate (E) at (-0.500, -0.866);
\coordinate (F) at (0.500, -0.866);
\coordinate (G) at (0.000, -0.577);
\coordinate (H) at (-0.500, 0.289);
\coordinate (I) at (0.500, 0.289);
\coordinate (J) at (0.000, 0.490);
\coordinate (K) at (-0.425, -0.246);
\coordinate (L) at (0.425, -0.246);
\foreach \i/\j in {A/B,  B/C,  C/D,  D/E,  E/F, F/A, G/H, H/I, I/G,  I/A, I/B, H/C, H/D, G/E, G/F}{\draw (\i) -- (\j);}}}

\def\rectangleScaled#1{\tikz[scale=#1]{
\coordinate (A) at (1,0.707);
\coordinate (B) at (-1,0.707);
\coordinate (C) at (-1,-0.707);
\coordinate (D) at (1,-0.707);
\foreach \i/\j in {A/B,B/C,C/D,D/A}{\draw[thick] (\i) -- (\j);};}}

\def\recstringScaled#1{\tikz[scale=#1]{
\coordinate (A) at (1,0.707);
\coordinate (B) at (-1,0.707);
\coordinate (C) at (-1,-0.707);
\coordinate (D) at (1,-0.707);
\foreach \i/\j in {A/B,C/D}{\draw[thick] (\i) -- ($(\i)+(\j)$) -- (\j);\path  ($(\i)+(\j)$);};
\foreach \i/\j in {B/C,D/A}{\draw[thick] (\i) -- (\j);\path  ($0.5*(\i)+0.5*(\j)$);};}}

\def\cuboctahedronScaled#1{\tikz[scale=#1]{\coordinate (A) at (1.000, 0.000);
\coordinate (B) at (0.500, 0.866);
\coordinate (C) at (-0.500, 0.866);
\coordinate (D) at (-1.000, 0.000);
\coordinate (E) at (-0.500, -0.866);
\coordinate (F) at (0.500, -0.866);
\coordinate (G) at (0.000, -0.577);
\coordinate (H) at (-0.500, 0.289);
\coordinate (I) at (0.500, 0.289);
\coordinate (J) at (0.000, 0.490);
\coordinate (K) at (-0.425, -0.246);
\coordinate (L) at (0.425, -0.246);
\foreach \i/\j in {A/B,  B/C,  C/D,  D/E,  E/F, F/A, G/H, H/I, I/G,  I/A, I/B, H/C, H/D, G/E, G/F}{\draw (\i) -- (\j);}}}

\def\rhombicdodecScaled#1{
\tdplotsetmaincoords{70}{77}
\tdplotsetrotatedcoords{-2}{15}{5}
\tikz[tdplot_main_coords,tdplot_rotated_coords, scale=#1]
{\coordinate (A) at (-.5,-.5,-.5);
\coordinate (B) at (.5,-.5,-.5);
\coordinate (C) at (.5,.5,-.5);
\coordinate (D) at (-.5,.5,-.5);
\coordinate (A') at (-.5,-.5,.5);
\coordinate (B') at (.5,-.5,.5);
\coordinate (C') at (.5,.5,.5);
\coordinate (D') at (-.5,.5,.5);
\coordinate (ABCD) at ($0.5*(A)+0.5*(B)+0.5*(C)+0.5*(D)$);
\coordinate (A'B'C'D') at ($0.5*(A')+0.5*(B')+0.5*(C')+0.5*(D')$);
\coordinate (ABB'A') at ($0.5*(A)+0.5*(B)+0.5*(B')+0.5*(A')$);
\coordinate (ADD'A') at ($0.5*(A)+0.5*(D)+0.5*(D')+0.5*(A')$);
\coordinate (CDD'C') at ($0.5*(C)+0.5*(D)+0.5*(D')+0.5*(C')$);
\coordinate (CBB'C') at ($0.5*(C)+0.5*(B)+0.5*(B')+0.5*(C')$);
\foreach \i/\j in {CBB'C'/B',CBB'C'/C',CBB'C'/B,CBB'C'/C,
A'B'C'D'/B',A'B'C'D'/C',A'B'C'D'/A',A'B'C'D'/D',ABCD/B,ABCD/C,ABB'A'/B,ABB'A'/B',ABB'A'/A',CDD'C'/C,CDD'C'/C',CDD'C'/D'}{
\draw[black] (\i) -- (\j);};
\foreach \i/\j in {ABCD/A,ABCD/D,ABB'A'/A,CDD'C'/D,ADD'A'/A,ADD'A'/D,ADD'A'/D',ADD'A'/A'}{\draw[black,dash pattern=on 1pt off 1pt,thin] (\i) -- (\j);};
}}

\def\triprism{\protect\tikz[scale=0.3]{
\protect\coordinate (A) at (0,0);
\protect\coordinate (B) at (1,0);
\protect\coordinate (C) at (0.5,0.866); 
\protect\coordinate (A') at ($(A)+(0.9,0.5)$);
\protect\coordinate (B') at ($(B)+(0.9,0.5)$);
\protect\coordinate (C') at ($(C)+(0.9,0.5)$);
\protect\draw (A) -- (B) -- (C) -- cycle;
\protect\draw (B')  -- (C');
\protect\draw (B) -- (B');
\protect\draw (C) -- (C');}}

\def\triprismScaled#1{\tikz[scale=#1]{
\protect\coordinate (A) at (0,0);
\protect\coordinate (B) at (1,0);
\protect\coordinate (C) at (0.5,0.866); 
\protect\coordinate (A') at ($(A)+(0.6,0.5)$);
\protect\coordinate (B') at ($(B)+(0.6,0.5)$);
\protect\coordinate (C') at ($(C)+(0.6,0.5)$);
\protect\draw (A) -- (B) -- (C) -- cycle;
\protect\draw (B')  -- (C');
\protect\draw (B) -- (B');
\protect\draw (C) -- (C');}}

\def\drawCube{\draw[thick] (A) -- (B) -- (C) -- (D) -- cycle;
\draw[thick] (B) -- (B');
\draw[thick] (C) -- (C');
\draw[thick] (D) -- (D');
\draw[dashed, thick] (A) -- (A');
\draw[dashed, thick] (A') -- (B');
\draw[dashed, thick] (A') -- (D');
\draw[thick] (B') -- (C');
\draw[thick] (C') -- (D');}

\def\coordinatesCube{
\def\xoff{0.5}
\def\yoff{0.5}
\coordinate (A) at ($(0,0)-(0.75,0.75)$);
\coordinate (B) at ($(1,0)-(0.75,0.75)$);
\coordinate (C) at ($(1,1)-(0.75,0.75)$);
\coordinate (D) at ($(0,1)-(0.75,0.75)$);
\coordinate (A') at ($(A)+(\xoff,\yoff)$);
\coordinate (B') at ($(B)+(\xoff,\yoff)$);
\coordinate (C') at ($(C)+(\xoff,\yoff)$);
\coordinate (D') at ($(D)+(\xoff,\yoff)$);}

\def\coordinatesCuboctahedronTilted{
\coordinate (A) at (0.992, -0.087);
\coordinate (B) at (0.584, 0.805);
\coordinate (C) at (-0.409, 0.892);
\coordinate (D) at (-0.992, 0.087);
\coordinate (E) at (-0.584, -0.805);
\coordinate (F) at (0.409, -0.892);
\coordinate (G) at (-0.108, -0.438);
\coordinate (H) at (-0.517, 0.454);
\coordinate (I) at (0.476, 0.367);
\coordinate (J) at (0.099, 0.352);
\coordinate (K) at (-0.397, -0.332);
\coordinate (L) at (0.447, -0.406);}

\def\drawCuboctahedron{\foreach \i/\j in {
  A/B,  B/C,  C/D,  D/E,  E/F, F/A, G/H, H/I, I/G,  I/A, I/B, H/C, H/D, G/E, G/F}{\draw[thick] (\i) -- (\j);}
\foreach \i/\j in {J/K,K/L,L/J,J/B,J/C,K/D,K/E,L/F,L/A}{\draw[dashed, thick] (\i) -- (\j);}}

\newcommand{\dext}{d}
\newcommand{\dt}{k}
\newcommand{\dimAbelian}{p}
\newcommand{\nvertex}{n_{\emptyset}}
\newcommand{\pmax}{p_{\text{max}}}

\begin{document}

\begin{titlepage}

\vspace*{-2cm} 
\begin{flushright}
{\tt \phantom{xxx}  IFT-UAM/CSIC-26-14} \qquad \qquad \\
{\tt \phantom{xxx}  Imperial-TP-2026-DW-2} \qquad \qquad
\end{flushright}

\vspace*{0.8cm} 
\begin{center}

{\Huge EFTs with Symmetric Moduli Spaces: \\
\vspace{0.3cm}
the Landscape and the Swampland}\\

 \vspace*{1.5cm}
{\bf Stephanie Baines},$^{1}$ {\bf Veronica Collazuol},$^{2}$ {\bf Bernardo Fraiman},$^{2,4}$\\ 
\vspace{0.2cm}
{\bf Mariana Gra\~na}$^{3}$   and {\bf Daniel Waldram}$^{1}$\\

\begin{flushleft}
\begin{small}
 \vspace*{0.4cm} 
$^1$ {\it Abdus Salam Centre for Theoretical Physics, Imperial College London,
London, SW7 2AZ, UK}\\[2mm]

$^2$ {\it Instituto de Física Teórica UAM/CSIC, 
Cantoblanco, 28049 Madrid, Spain
}\\[2mm]

$^3$ {\it Institut de Physique Th\'eorique, Universit\'e Paris Saclay, CEA, CNRS, 
Orme des Merisiers, 91191 Gif-sur-Yvette CEDEX, France}\\[2mm]

$^4$ {\it Max-Planck-Institut f\"ur Physik, Boltzmannstrasse 8, 85748 Garching bei M\"unchen, Germany}\\[2mm]
\end{small}
\end{flushleft}

\vspace{1cm}
\small{\bf Abstract} \\[3mm]\end{center}

\noindent

The Swampland Distance Conjecture (SDC) states that, for any infinite-distance limit in the moduli space of a quantum gravity effective field theory (EFT), there should exist an infinite tower of states that become exponentially light. According to the Emergent String Conjecture, such a tower should consist either of tensionless strings or of Kaluza–Klein modes, each with a mass-decay rate that depends in a precise way on the dimension of the effective field theory.

In this paper, we use the results obtained in \cite{Baines:2025upi} on the SDC for symmetric moduli spaces and how these rates are encoded in the weight polytope of the corresponding particle-state representations to determine the symmetric space EFTs and representations that have these decay rates. Remarkably, assuming that the particle states transform in an irreducible representation, the list of possible polytopes and moduli spaces is finite.
Different EFTs are related by embedding one moduli space in another or by taking a decompactification limit.  
Requiring compatibility of the particle representations under such branching, we find that, while most of the theories can be obtained from an EFT based on $E_{8(8)}$, there remain three in our list that appear to be impossible to get from M- or string-theory compactifications. Using the same embedding procedure, we also identify the string and brane representations that should be present in the spectrum.

\vspace{0.3cm}

\end{titlepage}

\newpage
\tableofcontents 
\newpage

\section{Introduction}\label{sec:intro}

Within the Swampland program~\cite{Vafa:2005ui} (reviewed in depth in~\cite{Brennan:2017rbf,Palti:2019pca,Agmon:2022thq,vanBeest:2021lhn,Grana:2021zvf}), it has become clear that only very particular Effective Field Theories (EFTs) are compatible with principles from quantum gravity. Indeed, guided both from string compactifications and from heuristic arguments, it seems that there are very specific universal features that arise, and one of the richest corners of the theory where to look for these appears to be at infinite distance regions of moduli spaces. Here, the presence of towers of states gives rise to very particular behaviours that are not captured purely by EFT considerations. Among these are the presence of infinitely many states becoming massless~\cite{Ooguri:2006in} and a suppression of the Planck scale~\cite{Dvali:2007hz,Dvali:2007wp,Dvali:2009ks,Dvali:2010vm}, which call for a new effective description of the theory.
In particular, the {\it Swampland Distance Conjecture} (SDC) \cite{Ooguri:2006in}\footnote{The original motivation for this conjecture comes purely from top-down considerations. For more recent bottom-up arguments, see e.g. \cite{Hamada:2021yxy,Stout:2021ubb,Stout:2022phm,Calderon-Infante:2023ler}.} quantifies the way in which the tower of states becomes massless as one moves along a geodesic approaching the boundary of moduli space. It states that if $d(x,x_0)$ is the minimum geodesic distance between two points in the moduli space, then as $x$ approaches the boundary, the mass of the tower depends exponentially on $d(x,x_0)$, that is $m \sim e^{- \alpha d}$, with $\alpha$ an $\mathcal{O}(1)$ constant. Furthermore, the {\it Emergent String Conjecture} (ESC) states that the only two leading towers that arise at the boundaries of moduli spaces are either KK modes or oscillator states of a critical, tensionless string~\cite{Lee:2018urn,Lee:2019wij}. It is further conjectured that the mass decay rate $\alpha$ is bounded by
\cite{Etheredge:2022opl}\footnote{Previous work on this subject can be found in \cite{Gendler:2020dfp,Bedroya:2019snp,Andriot:2020lea,Lanza:2020qmt}. } 
\begin{equation*}
\label{eqn:sharpeneddc}
	\alpha \geq \frac{1}{\sqrt{\dext-2}} \, ,
\end{equation*}
in a theory with $\dext$ external dimensions. The SDC together with this bound go under the name of {\it Sharpened Distance Conjecture}. The value of $\alpha$ along a specific limiting geodesic is theory-dependent, but there are some universal values that appear in simple string compactifications for the two possible kinds of towers\footnote{For the case of KK modes in warped compactifications, see \cite{Etheredge:2023odp,Raucci:2026fzp}.}, that are respectively
\cite{Etheredge:2022opl}
\begin{equation} \label{eqn:alphakkstring}
    \alpha_{\text{KK};n} = \sqrt{\frac{1}{\dext-2}+\frac1n} \, , \quad \alpha_{\text{St}} = \frac{1}{\sqrt{\dext-2}} \, ,
\end{equation}
where $n$ is the number of dimensions that decompactify. In the following, we will refer to \eqref{eqn:alphakkstring} as `{\it ESC rates}'.

All these conjectures, and the repercussions that they can have on the EFTs, have driven a considerable amount of work in the community~\cite{Heidenreich:2015nta,Blumenhagen:2017cxt,Grimm:2018ohb,Heidenreich:2018kpg,Blumenhagen:2018nts,Lee:2018spm,Ooguri:2018wrx,Corvilain:2018lgw,Grimm:2018cpv,Buratti:2018xjt,Lust:2019zwm,Joshi:2019nzi,Erkinger:2019umg,Marchesano:2019ifh,Font:2019cxq,Baume:2020dqd,Perlmutter:2020buo,Klaewer:2020lfg,Lanza:2021udy,Lee:2021qkx,Lee:2021usk,Rudelius:2022gbz,Baume:2023msm,Rudelius:2023mjy,Alvarez-Garcia:2023gdd,Alvarez-Garcia:2023qqj,Castellano:2023jjt,Castellano:2023stg,Ooguri:2024ofs,Calderon-Infante:2024oed,Ashmore:2015joa,Aoufia:2024awo,Friedrich:2025gvs,Baume:2016psm,Klaewer:2016kiy,Gendler:2020dfp,Lanza:2020qmt}. 

In a previous paper~\cite{Baines:2025upi}, we showed that, under fairly mild assumptions, EFTs with locally symmetric moduli spaces satisfy the SDC, including the fact that the tower of states necessarily transform in a representation $\rho$ of the local isometry group $G$ of $\mathcal{M}$. In this paper we refine this analysis to answer the following question
\begin{quote}\samepage
    \textit{Which locally symmetric spaces $\mathcal{M}$ and representations $\rho$ are compatible with the Emergent String Conjecture rates?}
\end{quote}
The group $G$ is reductive and for simplicity we  assume that $\rho$ is irreducible, although the analysis can in principle be extended to any finite dimensional representation. There are then two classes of solutions: one where $G$ includes an Abelian factor which we refer to as ``perturbative'' in analogy to the case of the weak coupling limit of theories with a dilaton, and one where $G$ is semi-simple which we refer to as ``global'' in analogy to theories with their full U-duality group. The ESC then restricts the perturbative moduli spaces to lie in simple families and the moduli spaces to a \emph{finite set}, each with a minimal allowed representation. These are listed in Tables~\ref{tab:1irrep_integern_scaled_perturbative},~\ref{tab:1irrep_integern_scaled_global} and~\ref{tab:1irrep_integern_extra}. 

The key observation we use is that for any infinite distance limit the behaviour of the exponentially-decaying masses of the various towers can be summarised in a convex hull formulation~\cite{Etheredge:2023odp,Etheredge:2023zjk,Etheredge:2024tok,Calderon-Infante:2023ler,Etheredge:2023usk,Grieco:2025bjy}. For each state, with mass $m(\phi^i)$ in Planck units as a function of the canonically-normalised moduli $\phi^i$, one can define a vector whose components read 
\begin{equation*} \label{eqn:zetas}
    \alpha_i = - \frac{\partial \log m}{\partial \phi^i} \, ,
\end{equation*}
where the derivative is taken holding the $\dext$-dimensional Planck mass fixed. If $\vec{\phi}$ is a unit-norm vector tangent to a boundary geodesic, the exponential rate of the tower is then given as
\begin{equation*}
    \alpha = \vec{\alpha} \cdot \vec{\phi} \, ,
\end{equation*}
with $\cdot$ the scalar product with respect to the moduli space metric. Several geodesics can asymptote to the same boundary point, and it is natural to choose a canonical representative, which in the locally symmetric space case corresponds to fixing the unit-norm vector $\vec{\phi}$ to lie in the Cartan subalgebra of the Lie algebra $\cgg$ of $G$. Within this subspace, one can then draw one vector $\vec{\alpha}$ for each of the towers in the tangent space at a given point, and the full set of vectors for a given theory define a {\it convex hull}, which describes the exponential rate of the leading tower along each limit defined by $\vec{\phi}$. As such, from \eqref{eqn:sharpeneddc} it should contain a ball of radius $\frac{1}{\sqrt{\dext-2}}$. Using this perspective, Etheredge et al.~\cite{Etheredge:2024tok} analysed how the ESC constrain the allowed building blocks of the convex hull in a given duality frame, and then, under some natural assumptions, how these blocks can fit together to make a global polytope. This led, in particular, to a finite list of two-dimensional polytopes in $d\geq6$. For locally symmetric spaces, as we will see, this limited number of polytopes is directly related to finiteness in the classification of reductive Lie algebras of a given rank, and comes without the need for assumptions about how the duality frame building blocks can fit together. 

This same procedure can also be applied by considering the asymptotic fall-off of the tensions of higher-dimensional objects \cite{Etheredge:2024amg,Etheredge:2025ahf}, and can be quantified in the {\it Brane Distance Conjecture} (BDC) \cite{Etheredge:2024amg}. By consistency of the Sharpened Distance Conjecture with dimensional reduction of higher-dimensional objects, the exponential rate of decay of the mass/tension of the combination of particles ($p=1$) with strings ($p=2$) and all other extended objects of (external) spatial dimension up to $\pmax$, should satisfy the bound 
\begin{equation} \label{eqn:BDC}
    \alpha_{\pmax} \geq \frac{1}{\sqrt{\dext-\pmax-1}} \, , 
\end{equation}
where the saturation cannot hold for $1 < p_\text{max} < d-3$, because it would imply the emergence of weakly coupled objects of dimension higher than strings (or their duals).

Our analysis builds on that we presented in~\cite{Baines:2025upi} as follows. All locally symmetric spaces are of the form $\mathcal{M}=\Gamma \backslash G / K$, where $G$ is a non-compact, centerless, reductive Lie group, $K$ its maximal compact subgroup and $\Gamma$ a discrete subgroup of $G$, which in string compactifications is the duality group. The fundamental group of $\mathcal{M}$ is equal to $\Gamma$ and so states of the theory have charges transforming in some representation of $\Gamma$, as has already been argued for in the supersymmetric setting~\cite{Delgado:2024skw}. This means the states lie in a lattice that is mapped to itself under the action of $\Gamma$. Strictly, the states must lie in an orbit of $\Gamma$ within the lattice, and different sets of states can lie in different orbits within different rescaled lattices. In such a setting, we showed in \cite{Baines:2025upi} that the SDC is always satisfied, independently of the representation of the charges of the theory (of course, with the exception of the trivial representation), and under a mild assumption about the number of orbits of states. A crucial step was that the super-rigidity theorems of Margulis~\cite{Margulis-rigidity} imply that the representations of $\Gamma$ lift to representations of $G$. The latter can be decomposed under the Cartan subalgebra $\caa\subset\cgg$ into weight spaces and the $\vec{\alpha}$ convex hull becomes simply the convex hull of the weights. Our goal is then to understand how the ESC constrains the weight polytope. For simplicity we consider irreducible representations but the analysis can be extended to the general case. 

The result is that we find a very limited set of possibilities, as listed in Tables~\ref{tab:1irrep_integern_scaled_perturbative},~\ref{tab:1irrep_integern_scaled_global} and~\ref{tab:1irrep_integern_extra}. Only a finite number of $\cgg$ are allowed in the global polytope case, each with a particular ``minimal'' representation, although there are scaling symmetries such that there are an infinite number of additional solutions for each minimal theory, either with larger representations or more directions decompactifying. What is particularly striking is that the minimal theories have a maximal decompactified dimension of eleven, although no direct input from string theory was used. The analysis goes in two steps. The irreducibility condition implies that all the vertices of the polytope are at equal distance and so necessarily correspond to decompactifying the same number of dimensions (or emergent string limits in the rank one case). One can then find all the polytopes of this form consistent with the ESC (and a smaller set than those allowed by the general building blocks found in~\cite{Etheredge:2024tok} because of the irreducibility constraint). The second step is to find which of these arise as symmetric space EFTs. What again is very striking is that all of them do. In fact, the constraints of the ESC, most notably the integrality of the number of decompactifying dimensions and the external space-time dimension $d$ require that the polytope is related to a root algebra. 

We also show that there are well defined notions of getting one symmetric space EFT as a descendent of another, either by freezing some of the moduli, or by decompactifying. These correspond to a totally geodesic submanifold or a blow-up of a point on the boundary. The allowed EFTs then come in family trees of descendents from a given seed EFT. Many of the minimal theories descend from the standard compactification of M-theory on $T^8$, with only three additional seed theories appearing listed in Section \ref{sec:stringinter}.  

As a by-product of the analysis, we see that in all of the allowed examples, in the emergent string direction there is always a weight in some representation of $G$ whose distance from the origin is twice that of the string oscillator $\alpha_{\text{St}}$ vector. From string theory such a behaviour is expected, and this allows us in several cases to construct the representation and convex hull of the strings, where the distance to the origin measures the decay rate of the tension.  In fact, the same group-theoretic techniques can be extended to the BDC, which requires to study the combined convex hull coming from both particles and higher-dimensional objects, which also arrange in representations of $G$. By considering decompactification limits we show how one can read off the convex hulls of brane states in the descendent theory from that of the particle convex hull of the parent. 

This paper is structured as follows. In Section~\ref{sec:math}, we briefly review the group-theoretic description of locally symmetric spaces, and describe their geodesics and boundaries. In Section~\ref{sec:weight-polytopes}, we first summarise the analysis of~\cite{Baines:2025upi} showing that the exponential decay rates of the particle states are defined by the convex hull of the weights of a given representation of $G$. Restricting to an irreducible representation we further give a formal derivation of the distances of a Weyl polytope faces in terms of algebraic quantities, and review how the faces are represented using certain ``$\lambda$-extended'' Dynkin diagrams. In Section~\ref{sec:SDC} we then present the set of symmetric space EFTs that are compatible with the ESC. We first derive the allowed polytopes and then classify when these are Weyl polytopes for a Lie algebra $\cgg$ with highest weight representation $\lambda$. In Section~\ref{sec:tree} we show how the allowed EFTs are related by either freezing some moduli or by taking decompactification limits. In Section~\ref{sec:bdc} we discuss the relation of our analysis to the Brane Distance Conjecture, focussing primarily on string states. We summarize our results and discuss open questions in Section~\ref{Conclusions}. Some further details on polytopes and descendent theories are given in the Appendices. 

\section{Locally symmetric spaces and their boundaries}\label{sec:math}

In this section, we give a summary of the mathematical formalism used to describe symmetric spaces and their boundary that we introduced in~\cite{Baines:2025upi}, to which we refer for a comprehensive explanation, explicit examples and a list of relevant citations. 

The type of $\dext$-dimensional EFTs in which we are interested have an action of the form 
\begin{equation} \label{eqn:dimredaction}
    S=\frac{M_{P}^{\dext-2}}{2} \int d^{\dext}x \sqrt{-g^E} \left(R^E - G_{ij}(\phi) \partial \phi^i \partial\phi^j\right) + \dots \, ,
\end{equation}
where $M_P$ is the $\dext$-dimensional Planck scale, $g^E$ is the space-time metric in the Einstein frame, and the moduli $\phi^i$ can be viewed as coordinates on a moduli space $\mathcal{M}$ with metric $G_{ij}(\phi)$. We will restrict the moduli space to be a locally symmetric space of non-compact type, meaning it has constant Riemann curvature. Group theoretically it implies one can write
\begin{equation}
\label{eq:M-product}
    \mathcal{M} = \Gamma\backslash G / K ,
\end{equation}
where $G=\mathbb{R}^{\dimAbelian}\times G_1\times \dots\times G_s$ is a reductive group with Abelian factor $\mathbb{R}^{\dimAbelian}$ and non-compact simple factors $G_a$, $K=K_1\times\dots\times K_s$ is its maximally compact subgroup and $\Gamma$ is a discrete group. Furthermore, each $G_a$ is centreless\footnote{This implies $G$ is \emph{algebraic}, meaning it can be defined by a set of polynomial equations in $GL(k,\mathbb{R})$, except for the rank one examples where the Lie algebra contains $\mathfrak{so}(n,1)$, $\mathfrak{su}(n,1)$, $\mathfrak{sp}(n,1)$ or $\mathfrak{f}_{-4(20)}$ factor. In these cases one can still take $\mathcal{M}$ as the same form $\Gamma\backslash G/K$ with $G$ algebraic but now $G$ has a centre that acts trivially on $\mathcal{M}$.} (that is the adjoint form of the Lie group) and $G$ is the local isometry group of $\mathcal{M}$. The group structure uniquely defines the metric on $\mathcal{M}$ up to some positive constants $\kappa_a$ so that
\begin{equation}
\label{eq:modspace-metric}
    G_{ij}(\phi) \dd \phi^i \dd \phi^j = \dd s_0^2 + \sum_a \kappa_a
    \dd \tilde{s}_a^2 \, ,
\end{equation}
where $\dd s_0^2$ is the flat metric on $\mathbb{R}^{\dimAbelian}$ and each $\dd\tilde{s}_a^2$ is (locally) an Einstein metric on $G_a/K_a$, normalised so that $\tilde{R}^{(a)}_{ij}=-\frac12\tilde{g}^{(a)}_{ij}$ where $\tilde{R}^{(a)}_{ij}$ and $\tilde{g}^{(a)}_{ij}$ are Ricci tensors and metrics respectively. In the following we will denote this metric by the bracket $\langle \cdot,\cdot\rangle$. 

The key point is that, under the further assumption that $\Gamma$ is \emph{arithmetic}, then geodesics and boundaries of $\mathcal{M}$ can be systematically described by rational parabolic subgroups $P(\mathbb{Q})\subset G$, as we briefly explain below.\footnote{An arithmetic group is defined as $\Gamma=G\cap GL(k,\mathbb{Z})$ for any algebraic group $G\subset GL(k,\mathbb{R})$ defined via polynomial equations with rational coefficients.} Furthermore the condition that $\Gamma$ is arithmetic follows directly\footnote{Strictly speaking this implication does not apply to the cases which have rank-one factors of $\mathfrak{so}(n,1)$ and $\mathfrak{su}(n,1)$ in the Lie algebra of $G$.} from the physical ``compactifiability'' criterion on $\mathcal{M}$ introduced in~\cite{Delgado:2024skw}\footnote{See also \cite{Grimm:2025lip}, in which compactifiability is related to the proposal that moduli spaces admit tame isometric embeddings into Euclidean space.} and so is natural from the point of view of quantum gravity. 

To describe how this works, one first notes that using the Killing form for each $G_a$ factor one can decompose the Lie algebra $\cgg$ of $G$ as
\begin{equation}
\label{eqn:cartandecomposition}
    \cgg =   \cpp \oplus \ckk \, ,
\end{equation}
where $\ckk$ is the Lie algebra of $K$ and $\cpp$ is the orthogonal complement. One then has 
\begin{align}\label{lem:1}
    [\ckk,\ckk]\subseteq \ckk,\qquad 
    [\ckk,\cpp]\subseteq \cpp,\qquad 
    [\cpp,\cpp]\subseteq \ckk \, .
\end{align}
Furthermore, by definition, the tangent space at any point $x\in\mathcal{M}$ satisfies $T_x\mathcal{S}\simeq\cpp$. A general geodesic passing through $x=g\cdot o$, where $o\in\mathcal{M}$ is the point stabilised by $K\subset G$, can be written as  
\begin{equation}  \label{eqn:genericgeodesic}
    \gamma (t)= g \, e^{tX} \cdot o \, ,
\end{equation}
where $X\in\cpp$. Furthermore, if we fix the normalization such that $\left< X,X\right>=1$ then $t$ is an affine parameter giving the geodesic distance away from $x$. For the SDC, we would like to consider geodesics that go to infinite distance, that is, all the way to the boundary. This naturally defines the boundary $\partial\mathcal{M}$ as the ``geodesic compactification'' as follows. 
Given two points $x_1,x_2\in\mathcal{M}$ let $d(x_1,x_2)$ be the Riemannian distance function, defined as the minimum geodesic distance between the two points. One defines an ``eventually distance minimizing'' (EDM) geodesic as one where there exists some $t_0$ such that for all $t_1,t_2\geq t_0$, $d(\gamma(t_1),\gamma(t_2))=|t_1-t_2|$. This just means that the geodesic defines the shortest path between points along itself, thereby excluding geodesics that form closed cycles or have an ergodic motion as they go to the boundary. One defines an equivalence class of EDM geodesics as 
\begin{equation}
    \gamma_1(t) \sim \gamma_2(t) \quad \text{iff} \quad  \limsup_{t\to \infty} d(\gamma_1(t),\gamma_2(t))< \infty \, , 
\end{equation} 
that is, the distance between them is bounded from above as $t\to\infty$. The boundary of $\mathcal{M}$ is then given by set of equivalence classes
\begin{align}
    \partial\mathcal{M} = \{ \text{EDM geodesics in $\mathcal{M}$} \} / \sim \, 
    .
\end{align}

To describe the relation to parabolic subgroups, we first note that since $G$ is reductive its Lie algebra $\cgg$ admits a real Cartan subalgebra $\caa$ and a decomposition into root spaces. The former is a maximal subset $\caa\subset\cgg$ of elements $X\in\cgg$ such that $\ad_X$ can be simultaneously diagonalised over the reals. One defines the real rank of both $G$ and $\mathcal{M}$ as $r=\rank{\mathcal{M}} = \rank_\mathbb{R} \cgg=\dim\caa$. The set of (restricted) roots $\beta\in \caa^*$ are defined in the usual way via the root spaces
\begin{equation}
    \cgg_\beta = \left\{ X \in \cgg : [H,X] = \beta(H) X , \forall H\in \caa \right\} \neq 0 \, , 
\end{equation}
for $\beta\neq 0$. The corresponding restricted root system $\Phi$ has a Weyl group $W\subset K$ and one can fix a positive Weyl chamber $\caa^+$, its closure $\overline{\caa^+}$ and sets of positive and negative roots $\Phi^+$ and $\Phi^-$. Note that in general the real rank of $\cgg$ can be smaller than the complex rank (where one diagonalises the Cartan subalgebra over the complex numbers). One can again introduce a set of simple roots $\{r_i\}$, but unlike the complex case, one can have $\dim \cgg_{r_i} >1$ and also sometimes $\cgg_{2r_i}\neq0$. Finally, recall that for each complex simple Lie algebra there is always a real simple Lie algebra $\cgg$, known as the ``split form'' such that $\rank_\mathbb{R}\cgg = \rank_\mathbb{C} \cgg$ and the restricted and conventional complex root systems agree. 

As reviewed in~\cite{Baines:2025upi}, the Lie algebra of every parabolic subgroup $P\subset G$ contains a natural subalgebra $\caa_P$ of a real Cartan subalgebra $\caa\subset\cgg$ (called the split component), with a corresponding positive Weyl chamber $\caa_P^+$. If $\Gamma$ is arithmetic, that is, is a subgroup $\Gamma\subset GL(k,\mathbb{Z})$ defined by a set of polynomial equations with rational coefficients\footnote{Note that in general there are infinitely many non-isomorphic arithmetic subgroups of a given algebraic group $G$, and hence there is no ``natural'' way of assigning one canonical $\Gamma$ to $G$. In many string theory examples for which $G$ splits over $\mathbb{R}$, though, the duality groups turn out to be the so-called {\it Chevalley arithmetic subgroups}.}, and $\mathcal{M}$ is noncompact but has finite volume, then points on the boundary $\partial\mathcal{M}$ are associated to proper \emph{rational} parabolic subgroups $P(\mathbb{Q})\subset GL(k,\mathbb{Q})$ of $G$, in the following way. Let $P$ be the real locus of $P(\mathbb{Q})$, defined as the corresponding algebraic group over $\mathbb{R}$ rather than $\mathbb{Q}$, and write $\caa_{P(\mathbb{Q})}$ for the corresponding real Abelian subalgebra, and $\caa_{P(\mathbb{Q}),1}^+$ for the space of unit length elements of $\caa_{P(\mathbb{Q})}^+$. One finds that each pair $(P(\mathbb{Q}),H)$ of proper rational parabolic subgroup $P(\mathbb{Q})\subset G$ and $H\in\caa_{P(\mathbb{Q}),1}^+$ defines a point in $\partial\mathcal{M}$, such that the corresponding equivalence class of geodesics has a canonical representative 
\begin{align}
\label{eqn:canonicalgeodesic}
    \gamma_H(t) = \ee^{tH}\cdot o \, . 
\end{align}
The boundary is then given by the disjoint union 
\begin{align}
\label{eq:boundaryM}
    \partial\mathcal{M} = \Gamma \mathbin{\big\backslash} \coprod_{P(\mathbb{Q})} \caa^+_{P(\mathbb{Q}),1} \, ,
\end{align}
where different pairs $(P(\mathbb{Q}),H)$ define the same boundary point if they are related by an element of the discrete group and so we have to quotient by the action of $\Gamma$. (The topology on $\partial\mathcal{M}$ comes from $\caa^+_{P(\mathbb{Q}),1}\subset\overline{\caa^+_{P'(\mathbb{Q}),1}}$ if $P'(\mathbb{Q})\subset P(\mathbb{Q})$.) An important result is that there are only a finite number of such equivalence classes and so the boundary has a finite number of disconnected pieces or ``cusps''. 

\section{Representations, weight spaces and convex hulls}
\label{sec:weight-polytopes}

In~\cite{Baines:2025upi}, we showed that, under some fairly simple physical assumptions, the SDC holds for essentially any locally symmetric moduli space $\mathcal{M}$. Specifically, we required that $\mathcal{M}$ was ``compactifiable'' in the sense of~\cite{Delgado:2024skw} and a condition on the spectrum of states that we called ``semi-completeness'' that we will discuss below. The compactifiability condition implied that $\mathcal{M}$ had finite volume and hence that $\Gamma$ was algebraic, allowing us to use a number of classic mathematical results. In particular, we showed that the exponential decay of the masses of states in the theory was determined by the weight spaces of some representation of the group $G$. Although in~\cite{Baines:2025upi}, and in most of our discussion below, the focus is on the case of particle states, the same arguments can be applied to the Brane Distance Conjecture, as we discuss in some more detail in Section~\ref{sec:bdc}. 

In this section we review these ideas and then focus on the particular case of a single irreducible representation defined by some highest weight. In this case the rate at which states become light is controlled by what is known as the Weyl polytope. Requiring that the towers become massless according to the ESC rates then constrains the size of the polytope, along with the allowed groups and representations. 

\subsection{General analysis}

The first step in showing that SDC holds was to note that $\Gamma$ is the fundamental group of $\mathcal{M}$. While the spectrum of states, be it particles or branes, should be unchanged when moving around a closed loop in $\mathcal{M}$, we can have monodromies, where individual states can map into each other. Thus in general we expect that the set of states transforms under some representation of the fundamental group $\rho:\Gamma\to GL(N,\mathbb{R})$. If $V\simeq\mathbb{R}^{N}$ is the vector space on which the representation acts, the set of states must form a subset $\Pi\subset V$ that is invariant under the action of $\rho$. One can then apply the ``super-rigidity'' theorem of Margulis \cite{Margulis-rigidity} that if $\Gamma$ is algebraic and $\mathcal{M}$ has finite volume then any representation of $\Gamma$ extends to a representation of $G$ and a given orbit of states under the action of $\Gamma$ hence lies on a lattice $L\subset V$ (and we have $\Pi \subseteq L \subset V$). That the physical states do not fill the whole lattice is very familiar from string and M-theory examples. Consider, for example, the string on a two-torus so that  $\mathcal{M}=\mathbb{R}\times SO(2,2;\mathbf{Z})\backslash SO(2,2)/SO(2)\times SO(2)$. There is a four-dimensional lattice of states labelled by winding and momentum charges $q=(n_1,n_2,w^1,w^2)\in L \simeq \mathbf{Z}^4$. However, the orbit of the physical half-supersymmetric BPS states $\Pi\subset L$ contains only those states for which $\eta(q,q)=n_1w^1+n_2w^2=0$ and it is this subset of states that includes those that become massless at the boundary of the moduli space. 

In fact the situation can be slightly more general. The duality groups under which the states and moduli transform need not be quite the same (see e.g.~\cite{Cribiori:2024qsv}, footnote~1), the classic example being S duality where the moduli transform under $PSL(2,\mathbb{Z})=SL(2,\mathbb{Z})/\mathbb{Z}_2$ and the F- and D-strings transform under $SL(2,\mathbb{Z})$. In general the representation $\rho$ under which the states transform will be a faithful representation for some central extension $\tilde{\Gamma}$ of the moduli group $\Gamma$. Via the super-rigidity theorems, it will in turn be a faithful representation of the corresponding central extension $\tilde{G}$ of $G$ (with the same Lie algebra). Recall that $G$ was centreless. Depending on the representation, the group $\tilde{G}$ can then be any central extension of $G$, all the way up to the largest one which is the simply-connected form. One can still write the symmetric space in the form $\tilde{\Gamma}\backslash\tilde{G}/\tilde{K}$, since the maximal compact subgroup $\tilde{K}$ includes the centre. However $\tilde{G}$ is no longer the local isometry group. In the following, we will typically refer to representations of $G$ for simplicity. However, given the symmetric space takes the same form, the analysis goes through the same way for any extension $\tilde{G}$. 

Since $G$ is reductive its representations are well understood and can be characterised using a decomposition into weight spaces. Suppose we are approaching a point on $\partial\mathcal{M}$ labelled by $(P(\mathbb{Q}),H)$ along the canonical geodesic~\eqref{eqn:canonicalgeodesic}. We can then decompose $V$ into weight spaces under the action of the subalgebra $\caa_{P(\mathbb{Q})}$, that is  
\begin{equation}
    V= \bigoplus_{w \in W} V_{w} \, ,
\end{equation}
where $W$ is the set of weights\footnote{For simplicity, we will use the metric to view weights as living in the dual space to the weight space, namely $w \in \caa_P$.}. Assuming that one state in a given orbit is massless, we showed that, as one approaches the boundary along the canonical geodesic the mass of a state $q\in\Pi\subseteq L$ has the form 
\begin{align}
\label{eq:leading mass}
    m^2(t,H,q) 
       = \sum_{w\in W} \ee^{-2t\langle w,H\rangle} |q_w|^2 + \dots \, , 
\end{align}
where $q_w$ is the projection of $q$ onto the subspace $V_w$, $|\cdot|^2$ is a $K$-invariant positive-definite metric on $V$ and the dots represent sub-leading corrections, in the sense that they do not affect the decay rate of the fastest decaying states as one approaches the boundary. Indeed, the leading tower of massless states corresponds to those in the space 
\begin{align}
\label{eq:V*}
    V_* \equiv \sum_{\substack{w\in W \\ \langle w,H\rangle =\alpha(H)}} V_w \, , 
\end{align}
where the fastest decay rate as a function of $H$ is given by 
\begin{align} \label{eq:alpha-w*}
    \alpha(H) \equiv \max_{w\in W} \langle w,H\rangle > 0 \, . 
\end{align}
Crucially, since $P(\mathbb{Q})$ is rational, there is always an infinite number of lattice points in $V_*$ and hence potentially  an infinite number of fastest decaying states in $\Pi$. 

It is relatively easy to show that the weight spaces in $V_*$ always lie on the boundary of the convex hull $C$ of the $w\in W$ defined by 
\begin{equation}
    C = \Bigl\{ {\textstyle \sum_{w\in E} \mu_w w : \mu_w \geq 0 , \sum_{w\in E}\mu_w=1 } \Bigr\} \, . 
\end{equation}
In fact, if $E\subset C$ is the ``vertex subset'' of those weights that form the vertices of $C$, then $V_*$ always includes $V_w$ for some $w\in E$. This leads to a natural ``semi-complete'' spectrum assumption: if there are an infinite number of  states  in $V_w$ for each $w\in E$ then the SDC holds\footnote{In general, these states can lie in different orbits and so, in principle, can be in different rescaled lattices $L$ in $V$. However, if $q$ can be interpreted as a quantised charge then the states will lie in the same lattice.}. An equivalent way, and perhaps more physical way, of stating the assumption is 
\begin{equation*}
    \text{semi-completeness} \quad \Leftrightarrow \quad
    \text{the SDC tower has the fastest possible decay}\, . 
\end{equation*}
By this we mean the following. Consider all the possible charge vectors $q\in V$, physical or not. For any given $H$ there will be a set of vectors that give the fastest possible decay, namely those in $V_*$. There will of course be other vectors that give a decay and so also satisfy the SDC, but generically the rate will be slower. The semi-completeness assumption is that we choose the subset of physical states $\Pi\subset L$ such that, for any $H$, there is always an infinite number of physical states in $V_*$. In this sense, the physical states always saturate the bound on the fastest allowed decay among all possible charge vectors $q\in V$. (This is exactly the case, for example, for the half-BPS states in compactifications of M-theory or type II string theory on tori.) Under this assumption, we proved in~\cite{Baines:2025upi} that the SDC holds and furthermore
\begin{equation*}
    \text{local extrema of $\alpha(H)$} \quad \Leftrightarrow \quad
    \text{extremal distance points to $C$}\, ,
\end{equation*}
reducing the problem of finding extrema to a purely group theoretic one. 

It is important to note that thus far we have fixed a particular rational parabolic group $P(\mathbb{Q})$ representing one of the equivalence classes in~\eqref{eq:boundaryM}. The weight space and convex hull are defined relative to the subalgebra $\mathfrak{a}_{P(\mathbb{Q})}\subseteq \mathfrak{a}$  and give the mass decay rates for the set of boundary geodesics that are stabilized by $P(\mathbb{Q})$. (In general, this will be a projection of the full weight space based on the Cartan subalgebra $\mathfrak{a}$.) A different equivalence class will have a different $\mathfrak{a}_{P(\mathbb{Q})}$ and in principle a different convex hull and set of decay rates. However, the possible  $P(\mathbb{Q})$ will always include ``minimal parabolics'', that is cases where $\mathfrak{a}_{P(\mathbb{Q})}=\mathfrak{a}$, so that the weight space and convex hull are those of the full group $G$. Thus there is always at least one cusp on the boundary $\partial\mathcal{M}$ that has the same convex hull as that of the full group $G$. Furthermore, boundaries of closure of the positive Weyl chamber $\overline{\mathfrak{a}^+}$ in this case correspond to non-minimal rational parabolic subgroup. Hence, analysing the rates of a minimal parabolic cusp will cover the possible rates at all cusps. In other words:
\begin{quote}
    \textit{For any given $P(\mathbb{Q})$ the possible $\alpha(H)$ are equal to or a subset of those defined via the convex hull of $\rho$ as a $G$-representation}
\end{quote}

A second important point to note is that the boundary points are defined for $H$ in the positive Weyl chamber $\mathfrak{a}_{P(\mathbb{Q}),1}^+$, whereas we think of the convex hull as being defined over the whole of $\mathfrak{a}_{P(\mathbb{Q})}$ or simply $\mathfrak{a}$ when $P(\mathbb{Q})$ is minimal. In general a connected Weyl chamber can either correspond to a different part of the same cusp or to an equivalent description of the same cusp if the chamber is related to $\mathfrak{a}_{P(\mathbb{Q}),1}^+$ by an element of the duality group $\Gamma$. In the latter case the different chambers correspond to different duality frames. 

Following~\cite{Calderon-Infante:2020dhm}, one finds that in string compactifications the local extrema and global minima of $\alpha(H)$ are KK or tensionless-string exponential rates respectively. Our objective in this paper is to study the set of extrema for general symmetric spaces and representation spaces $V$, checking when they have the ESC rates form
\begin{equation} \label{eqn:defalphas}
    \text{local extremum of $\alpha(H)$} \stackrel{?}{\in} \{ \alpha_{\text{KK};n} , \alpha_{\text{St}} \}\, .
\end{equation} 
From the comments above, this problem can be analysed by considering the convex hull of the weights of $\rho$ as a representation of $G$. 
 
\subsection{Highest weights and Weyl polytopes}
\label{sec:Weylpolytopes}

Let us now restrict to the most basic example where $G$ is semi-simple and the representation $\rho$ is irreducible\footnote{For string theoretic examples on tori this assumption is enough in order to characterise particle towers. As we will see in the following, in order to deal with the Brane Distance Conjecture \cite{Etheredge:2024amg} one has to consider reducible representations with multiple highest weights.}, defined by a single highest weight $\lambda$. In this case (see for example~\cite{hall2003lie}) the convex hull $C$ of the weights is just the convex hull of the Weyl orbit of $\lambda$, which is known as the Weyl polytope $C(\lambda)$. 

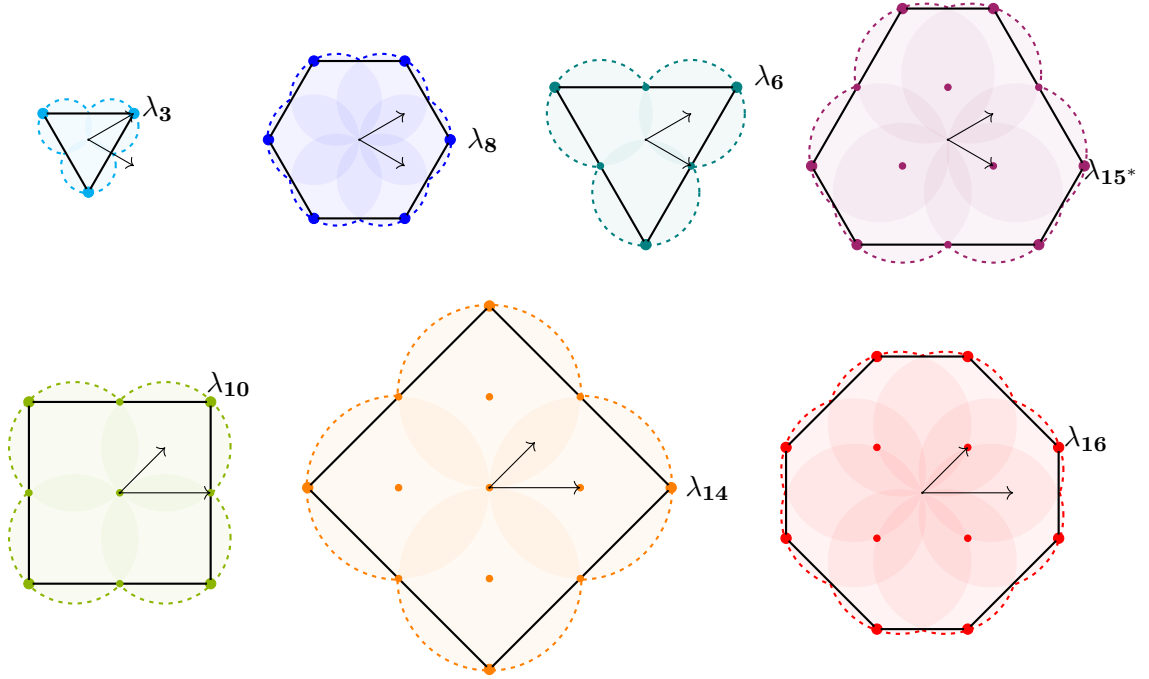
\begin{figure}[t]
\centering
\def\colB{orange}
\def\colone{blue}
\def\coltwo{black}
\def\sizesmall{0.4}
\def\opaBubbles{0.05}
\def\opaInsideHull{0.0}
\def\radone{0.08}
\def\radtwo{0.06}
\def\fontLabels{\small}
\def\scale{0.85}
\def\radInternal{0.05}
\def\colA{cyan}
\begin{tikzpicture}[scale=\scale]\coordinate (AA) at (-0.353553, -0.204124); \coordinate (AB) at (0., -0.816497); \coordinate (AC) at (0.353553, -0.204124); \coordinate (AD) at (0.707107, 0.408248); \coordinate (AE) at (0., 0.408248); \coordinate (AF) at (-0.707107, 0.408248);
\foreach \P/\r/\c in {AB/0.408248/\colA,AD/0.408248/\colA,AF/0.408248/\colA}{\draw[dotted,ultra thick,\c] ($0.5*(\P)$) circle (\r);};
\foreach \P/\r/\c in {AB/0.408248/\colA,AD/0.408248/\colA,AF/0.408248/\colA}{\fill[white] ($0.5*(\P)$) circle (\r);};\foreach \P/\r/\c in {AB/0.408248/\colA,AD/0.408248/\colA,AF/0.408248/\colA}{\fill[\c,opacity=\opaBubbles] ($0.5*(\P)$) circle (\r);
\filldraw[\c] (\P) circle(\radone);};\foreach \P/\Q/\R/\cP/\cQ/\cR in {AA/AF/AB/\coltwo/\colone/\colone,AC/AB/AD/\coltwo/\colone/\colone,AE/AF/AD/\coltwo/\colone/\colone}{
\fill[\cQ, opacity=\opaInsideHull](\P) -- (\Q) -- (0,0) -- cycle;
\fill[\cR, opacity=\opaInsideHull](\P) -- (\R) -- (0,0) -- cycle;
\draw[\cP, thick] (\Q) -- (\R);};\foreach \P/\n/\c/\rad/\dist/\name in {AD/$1$/\colone/\radone/\sqrt{\frac{2}{3}}/$\quad\lambda_{\mathbf{3}}$}{
\path ($1.2*(\P)$) node[font=\fontLabels]{\name};
};\foreach \P/\c in {(0.707107, 0.408248)/\colA,(0., -0.816497)/\colA,(-0.707107, 0.408248)/\colA}{\filldraw[\c] \P circle(\radInternal);};
\filldraw[black] (0,0) circle(0.01);
\draw[->] (0,0)-- (0.707107, 0.408248)node[color=black,font=\fontLabels,
xshift=0.2cm, yshift=0.2cm]{};
\draw[->] (0,0)-- (0.707107, -0.408248)node[color=black,font=\footnotesize,
xshift=0.2cm, yshift=0.2cm]{};\foreach \P in {(-2.82843, 0.), (-1.41421, -2.44949), (-1.41421, 2.44949), (1.41421, -2.44949), (1.41421, 2.44949), (2.82843, 0.)}{
\filldraw \P circle(0);};\end{tikzpicture}
\hspace{-4em}
\def\colA{blue}
\begin{tikzpicture}[scale=\scale]
\coordinate (AA) at (-1.06066, -0.612372); \coordinate (AB) at (-0.707107, -1.22474); \coordinate (AC) at (-0.353553, -0.612372); \coordinate (AD) at (0., -1.22474); \coordinate (AE) at (0.353553, -0.612372); \coordinate (AF) at (0.707107, -1.22474); \coordinate (AG) at (1.06066, -0.612372); \coordinate (AH) at (0.707107, 0.); \coordinate (AI) at (1.41421, 0.); \coordinate (AJ) at (1.06066, 0.612372); \coordinate (AK) at (0.353553, 0.612372); \coordinate (AL) at (0.707107, 1.22474); \coordinate (AM) at (0., 1.22474); \coordinate (AN) at (-0.707107, 1.22474); \coordinate (AO) at (-0.353553, 0.612372); \coordinate (AP) at (-1.06066, 0.612372); \coordinate (AQ) at (-0.707107, 0.); \coordinate (AR) at (-1.41421, 0.); \coordinate (AS) at (0., 0.); \coordinate (AT) at (0., 0.); \coordinate (AU) at (0., 0.);
\foreach \P/\r/\c in {AB/0.707107/\colA,AF/0.707107/\colA,AI/0.707107/\colA,AL/0.707107/\colA,AN/0.707107/\colA,AR/0.707107/\colA}{\draw[dotted,ultra thick,\c] ($0.5*(\P)$) circle (\r);};
\foreach \P/\r/\c in {AB/0.707107/\colA,AF/0.707107/\colA,AI/0.707107/\colA,AL/0.707107/\colA,AN/0.707107/\colA,AR/0.707107/\colA}{\fill[white] ($0.5*(\P)$) circle (\r);};\foreach \P/\r/\c in {AB/0.707107/\colA,AF/0.707107/\colA,AI/0.707107/\colA,AL/0.707107/\colA,AN/0.707107/\colA,AR/0.707107/\colA}{\fill[\c,opacity=\opaBubbles] ($0.5*(\P)$) circle (\r);
\filldraw[\c] (\P) circle(\radone);};\foreach \P/\Q/\R/\cP/\cQ/\cR in {AA/AR/AB/\coltwo/\colone/\colone,AD/AB/AF/\coltwo/\colone/\colone,AG/AF/AI/\coltwo/\colone/\colone,AJ/AL/AI/\coltwo/\colone/\colone,AM/AN/AL/\coltwo/\colone/\colone,AP/AR/AN/\coltwo/\colone/\colone}{
\fill[\cQ, opacity=\opaInsideHull](\P) -- (\Q) -- (0,0) -- cycle;
\fill[\cR, opacity=\opaInsideHull](\P) -- (\R) -- (0,0) -- cycle;
\draw[\cP, thick] (\Q) -- (\R);};\foreach \P/\n/\c/\rad/\dist/\name in {AI/$1$/\colone/\radone/\sqrt{2}/$\quad\lambda_{\mathbf{8}}$}{
\path ($1.2*(\P)$) node[font=\fontLabels]{\name};
};\foreach \P/\c in {(1.41421, 0.)/\colA,(0.707107, -1.22474)/\colA,(0.707107, 1.22474)/\colA,(-0.707107, -1.22474)/\colA,(-0.707107, 1.22474)/\colA,(-1.41421, 0.)/\colA}{\filldraw[\c] \P circle(\radInternal);};
\filldraw[black] (0,0) circle(0.01);
\draw[->] (0,0)-- (0.707107, 0.408248)node[color=black,font=\fontLabels,
xshift=0.2cm, yshift=0.2cm]{};
\draw[->] (0,0)-- (0.707107, -0.408248)node[color=black,font=\footnotesize,
xshift=0.2cm, yshift=0.2cm]{};\foreach \P in {(-2.82843, 0.), (-1.41421, -2.44949), (-1.41421, 2.44949), (1.41421, -2.44949), (1.41421, 2.44949), (2.82843, 0.)}{
\filldraw \P circle(0);};
\end{tikzpicture}
\hspace{-3.5em}
\def\colA{teal}
\begin{tikzpicture}[scale=\scale]\coordinate (AA) at (-0.707107, -0.408248); \coordinate (AB) at (0., -1.63299); \coordinate (AC) at (0.707107, -0.408248); \coordinate (AD) at (1.41421, 0.816497); \coordinate (AE) at (0., 0.816497); \coordinate (AF) at (-1.41421, 0.816497);
\foreach \P/\r/\c in {AB/0.816497/\colA,AD/0.816497/\colA,AF/0.816497/\colA}{\draw[dotted,ultra thick,\c] ($0.5*(\P)$) circle (\r);};
\foreach \P/\r/\c in {AB/0.816497/\colA,AD/0.816497/\colA,AF/0.816497/\colA}{\fill[white] ($0.5*(\P)$) circle (\r);};\foreach \P/\r/\c in {AB/0.816497/\colA,AD/0.816497/\colA,AF/0.816497/\colA}{\fill[\c,opacity=\opaBubbles] ($0.5*(\P)$) circle (\r);
\filldraw[\c] (\P) circle(\radone);};\foreach \P/\Q/\R/\cP/\cQ/\cR in {AA/AF/AB/\coltwo/\colone/\colone,AC/AB/AD/\coltwo/\colone/\colone,AE/AF/AD/\coltwo/\colone/\colone}{
\fill[\cQ, opacity=\opaInsideHull](\P) -- (\Q) -- (0,0) -- cycle;
\fill[\cR, opacity=\opaInsideHull](\P) -- (\R) -- (0,0) -- cycle;
\draw[\cP, thick] (\Q) -- (\R);};\foreach \P/\n/\c/\rad/\dist/\name in {AD/$1$/\colone/\radone/2 \sqrt{\frac{2}{3}}/$\quad \lambda_{\mathbf{6}}$}{
\path ($1.2*(\P)$) node[font=\fontLabels]{\name};
};\foreach \P/\c in {(1.41421, 0.816497)/\colA,(0.707107, -0.408248)/\colA,(0., -1.63299)/\colA,(0., 0.816497)/\colA,(-0.707107, -0.408248)/\colA,(-1.41421, 0.816497)/\colA}{\filldraw[\c] \P circle(\radInternal);};
\filldraw[black] (0,0) circle(0.01);
\draw[->] (0,0)-- (0.707107, 0.408248)node[color=black,font=\fontLabels,
xshift=0.2cm, yshift=0.2cm]{};
\draw[->] (0,0)-- (0.707107, -0.408248)node[color=black,font=\footnotesize,
xshift=0.2cm, yshift=0.2cm]{};\foreach \P in {(-2.82843, 0.), (-1.41421, -2.44949), (-1.41421, 2.44949), (1.41421, -2.44949), (1.41421, 2.44949), (2.82843, 0.)}{
\filldraw \P circle(0);};
\end{tikzpicture}
\hspace{-3em}
\def\colA{RedViolet}
\begin{tikzpicture}[scale=\scale]\coordinate (AA) at (-2.12132, -0.408248); \coordinate (AB) at (-1.76777, -1.02062); \coordinate (AC) at (-1.41421, -1.63299); \coordinate (AD) at (-0.353553, -1.02062); \coordinate (AE) at (0., -1.63299); \coordinate (AF) at (0., -0.408248); \coordinate (AG) at (0.353553, -1.02062); \coordinate (AH) at (1.41421, -1.63299); \coordinate (AI) at (1.76777, -1.02062); \coordinate (AJ) at (2.12132, -0.408248); \coordinate (AK) at (1.06066, 0.204124); \coordinate (AL) at (0.353553, 0.204124); \coordinate (AM) at (1.41421, 0.816497); \coordinate (AN) at (0.707107, 0.816497); \coordinate (AO) at (0.707107, 2.04124); \coordinate (AP) at (0., 2.04124); \coordinate (AQ) at (-0.707107, 2.04124); \coordinate (AR) at (-0.707107, 0.816497); \coordinate (AS) at (-0.353553, 0.204124); \coordinate (AT) at (-1.41421, 0.816497); \coordinate (AU) at (-1.06066, 0.204124);
\foreach \P/\r/\c in {AA/1.08012/\colA,AC/1.08012/\colA,AH/1.08012/\colA,AJ/1.08012/\colA,AO/1.08012/\colA,AQ/1.08012/\colA}{\draw[dotted,ultra thick,\c] ($0.5*(\P)$) circle (\r);};
\foreach \P/\r/\c in {AA/1.08012/\colA,AC/1.08012/\colA,AH/1.08012/\colA,AJ/1.08012/\colA,AO/1.08012/\colA,AQ/1.08012/\colA}{\fill[white] ($0.5*(\P)$) circle (\r);};\foreach \P/\r/\c in {AA/1.08012/\colA,AC/1.08012/\colA,AH/1.08012/\colA,AJ/1.08012/\colA,AO/1.08012/\colA,AQ/1.08012/\colA}{\fill[\c,opacity=\opaBubbles] ($0.5*(\P)$) circle (\r);
\filldraw[\c] (\P) circle(\radone);};\foreach \P/\Q/\R/\cP/\cQ/\cR in {AB/AA/AC/\coltwo/\colone/\colone,AE/AC/AH/\coltwo/\colone/\colone,AI/AH/AJ/\coltwo/\colone/\colone,AM/AO/AJ/\coltwo/\colone/\colone,AP/AQ/AO/\coltwo/\colone/\colone,AT/AA/AQ/\coltwo/\colone/\colone}{
\fill[\cQ, opacity=\opaInsideHull](\P) -- (\Q) -- (0,0) -- cycle;
\fill[\cR, opacity=\opaInsideHull](\P) -- (\R) -- (0,0) -- cycle;
\draw[\cP, thick] (\Q) -- (\R);};\foreach \P/\n/\c/\rad/\dist/\name in {AJ/$1$/\colone/\radone/\sqrt{\frac{14}{3}}/$\lambda_{\mathbf{15}^*}$}{
\path ($1.2*(\P)$) node[font=\fontLabels]{\name};
};\foreach \P/\c in {(2.12132, -0.408248)/\colA,(1.41421, -1.63299)/\colA,(1.41421, 0.816497)/\colA,(0.707107, -0.408248)/\colA,(0.707107, 2.04124)/\colA,(0., -1.63299)/\colA,(0., 0.816497)/\colA,(-0.707107, -0.408248)/\colA,(-0.707107, 2.04124)/\colA,(-1.41421, -1.63299)/\colA,(-1.41421, 0.816497)/\colA,(-2.12132, -0.408248)/\colA}{\filldraw[\c] \P circle(\radInternal);};
\filldraw[black] (0,0) circle(0.01);
\draw[->] (0,0)-- (0.707107, 0.408248)node[color=black,font=\fontLabels,
xshift=0.2cm, yshift=0.2cm]{};
\draw[->] (0,0)-- (0.707107, -0.408248)node[color=black,font=\footnotesize,
xshift=0.2cm, yshift=0.2cm]{};\foreach \P in {(-2.82843, 0.), (-1.41421, -2.44949), (-1.41421, 2.44949), (1.41421, -2.44949), (1.41421, 2.44949), (2.82843, 0.)}{
\filldraw \P circle(0);};
\end{tikzpicture}

\def\colA{applegreen}
\begin{tikzpicture}[scale=\scale]\coordinate (A) at (-1.41421, -1.41421); \coordinate (B) at (0., -1.41421); \coordinate (C) at (1.41421, -1.41421); \coordinate (D) at (1.41421, 0.); \coordinate (E) at (1.41421, 1.41421); \coordinate (F) at (0., 1.41421); \coordinate (G) at (-1.41421, 1.41421); \coordinate (H) at (-1.41421, 0.);
\foreach \P/\r/\c in {A/1./\colA,C/1./\colA,E/1./\colA,G/1./\colA}{\draw[dotted,ultra thick,\c] ($0.5*(\P)$) circle (\r);};
\foreach \P/\r/\c in {A/1./\colA,C/1./\colA,E/1./\colA,G/1./\colA}{\fill[white] ($0.5*(\P)$) circle (\r);};\foreach \P/\r/\c in {A/1./\colA,C/1./\colA,E/1./\colA,G/1./\colA}{\fill[\c,opacity=\opaBubbles] ($0.5*(\P)$) circle (\r);
\filldraw[\c] (\P) circle(\radone);};\foreach \P/\Q/\R/\cP/\cQ/\cR in {B/C/A/\coltwo/\colone/\colone,D/E/C/\coltwo/\colone/\colone,F/E/G/\coltwo/\colone/\colone,H/G/A/\coltwo/\colone/\colone}{
\fill[\cQ, opacity=\opaInsideHull](\P) -- (\Q) -- (0,0) -- cycle;
\fill[\cR, opacity=\opaInsideHull](\P) -- (\R) -- (0,0) -- cycle;
\draw[\cP, thick] (\Q) -- (\R);};\foreach \P/\n/\c/\rad/\dist/\name in {E/$1$/\colone/\radone/2/\lambda_{\mathbf{10}}}{
\path ($1.2*(\P)$) node[font=\fontLabels]{$\name$};
};\foreach \P/\c in {(-1.41421, -1.41421)/\colA,(-1.41421, 0.)/\colA,(-1.41421, 1.41421)/\colA,(0., -1.41421)/\colA,(0., 0.)/\colA,(0., 1.41421)/\colA,(1.41421, -1.41421)/\colA,(1.41421, 0.)/\colA,(1.41421, 1.41421)/\colA}{\filldraw[\c] \P circle(\radInternal);};
\filldraw[black] (0,0) circle(0.01);
\draw[->] (0,0)-- (0.707107, 0.707107)node[color=black,font=\fontLabels,
xshift=0.2cm, yshift=0.2cm]{};
\draw[->] (0,0)-- (1.41421, 0.)node[color=black,font=\footnotesize,
xshift=0.2cm, yshift=0.2cm]{};\foreach \P in {(2.82843, 0.), (0., 2.82843), (0., -2.82843), (-2.82843, 0.)}{
\filldraw \P circle(0);};\end{tikzpicture}\def\colA{orange}\begin{tikzpicture}[scale=\scale]\coordinate (A) at (-1.41421, -1.41421); \coordinate (B) at (0., -2.82843); \coordinate (C) at (1.41421, -1.41421); \coordinate (D) at (2.82843, 0.); \coordinate (E) at (1.41421, 1.41421); \coordinate (F) at (0., 2.82843); \coordinate (G) at (-1.41421, 1.41421); \coordinate (H) at (-2.82843, 0.);
\foreach \P/\r/\c in {B/1.41421/\colA,D/1.41421/\colA,F/1.41421/\colA,H/1.41421/\colA}{\draw[dotted,ultra thick,\c] ($0.5*(\P)$) circle (\r);};
\foreach \P/\r/\c in {B/1.41421/\colA,D/1.41421/\colA,F/1.41421/\colA,H/1.41421/\colA}{\fill[white] ($0.5*(\P)$) circle (\r);};\foreach \P/\r/\c in {B/1.41421/\colA,D/1.41421/\colA,F/1.41421/\colA,H/1.41421/\colA}{\fill[\c,opacity=\opaBubbles] ($0.5*(\P)$) circle (\r);
\filldraw[\c] (\P) circle(\radone);};\foreach \P/\Q/\R/\cP/\cQ/\cR in {A/B/H/\coltwo/\colone/\colone,C/D/B/\coltwo/\colone/\colone,E/D/F/\coltwo/\colone/\colone,G/F/H/\coltwo/\colone/\colone}{
\fill[\cQ, opacity=\opaInsideHull](\P) -- (\Q) -- (0,0) -- cycle;
\fill[\cR, opacity=\opaInsideHull](\P) -- (\R) -- (0,0) -- cycle;
\draw[\cP, thick] (\Q) -- (\R);};\foreach \P/\n/\c/\rad/\dist/\name in {D/$1$/\colone/\radone/2 \sqrt{2}/\lambda_{\mathbf{14}}}{
\path ($1.2*(\P)$) node[font=\fontLabels]{$\name$};
};\foreach \P/\c in {(-2.82843, 0.)/\colA,(-1.41421, -1.41421)/\colA,(-1.41421, 0.)/\colA,(0., -2.82843)/\colA,(-1.41421, 1.41421)/\colA,(0., -1.41421)/\colA,(0., 0.)/\colA,(0., 1.41421)/\colA,(1.41421, -1.41421)/\colA,(0., 2.82843)/\colA,(1.41421, 0.)/\colA,(1.41421, 1.41421)/\colA,(2.82843, 0.)/\colA}{\filldraw[\c] \P circle(\radInternal);};
\filldraw[black] (0,0) circle(0.01);
\draw[->] (0,0)-- (0.707107, 0.707107)node[color=black,font=\fontLabels,
xshift=0.2cm, yshift=0.2cm]{};
\draw[->] (0,0)-- (1.41421, 0.)node[color=black,font=\footnotesize,
xshift=0.2cm, yshift=0.2cm]{};\foreach \P in {(2.82843, 0.), (0., 2.82843), (0., -2.82843), (-2.82843, 0.)}{
\filldraw \P circle(0);};\end{tikzpicture}\def\colA{red}\begin{tikzpicture}[scale=\scale]\coordinate (A) at (-2.12132, -0.707107); \coordinate (B) at (-1.41421, -1.41421); \coordinate (C) at (-0.707107, -2.12132); \coordinate (D) at (0., -2.12132); \coordinate (E) at (0.707107, -2.12132); \coordinate (F) at (1.41421, -1.41421); \coordinate (G) at (2.12132, -0.707107); \coordinate (H) at (2.12132, 0.); \coordinate (I) at (2.12132, 0.707107); \coordinate (J) at (1.41421, 1.41421); \coordinate (K) at (0.707107, 2.12132); \coordinate (L) at (0., 2.12132); \coordinate (M) at (-0.707107, 2.12132); \coordinate (N) at (-1.41421, 1.41421); \coordinate (O) at (-2.12132, 0.707107); \coordinate (P) at (-2.12132, 0.);
\foreach \P/\r/\c in {A/1.11803/\colA,C/1.11803/\colA,E/1.11803/\colA,G/1.11803/\colA,I/1.11803/\colA,K/1.11803/\colA,M/1.11803/\colA,O/1.11803/\colA}{\draw[dotted,ultra thick,\c] ($0.5*(\P)$) circle (\r);};
\foreach \P/\r/\c in {A/1.11803/\colA,C/1.11803/\colA,E/1.11803/\colA,G/1.11803/\colA,I/1.11803/\colA,K/1.11803/\colA,M/1.11803/\colA,O/1.11803/\colA}{\fill[white] ($0.5*(\P)$) circle (\r);};\foreach \P/\r/\c in {A/1.11803/\colA,C/1.11803/\colA,E/1.11803/\colA,G/1.11803/\colA,I/1.11803/\colA,K/1.11803/\colA,M/1.11803/\colA,O/1.11803/\colA}{\fill[\c,opacity=\opaBubbles] ($0.5*(\P)$) circle (\r);
\filldraw[\c] (\P) circle(\radone);};\foreach \P/\Q/\R/\cP/\cQ/\cR in {B/C/A/\coltwo/\colone/\colone,D/E/C/\coltwo/\colone/\colone,F/G/E/\coltwo/\colone/\colone,H/I/G/\coltwo/\colone/\colone,J/I/K/\coltwo/\colone/\colone,L/K/M/\coltwo/\colone/\colone,N/M/O/\coltwo/\colone/\colone,P/O/A/\coltwo/\colone/\colone}{
\fill[\cQ, opacity=\opaInsideHull](\P) -- (\Q) -- (0,0) -- cycle;
\fill[\cR, opacity=\opaInsideHull](\P) -- (\R) -- (0,0) -- cycle;
\draw[\cP, thick] (\Q) -- (\R);};\foreach \P/\n/\c/\rad/\dist/\name in {I/$1$/\colone/\radone/\sqrt{5}/\lambda_{\mathbf{16}}}{
\path ($1.2*(\P)$) node[font=\fontLabels]{$\name$};
};\foreach \P/\c in {(-2.12132, -0.707107)/\colA,(-2.12132, 0.707107)/\colA,(-0.707107, -2.12132)/\colA,(-0.707107, -0.707107)/\colA,(-0.707107, 0.707107)/\colA,(0.707107, -2.12132)/\colA,(-0.707107, 2.12132)/\colA,(0.707107, -0.707107)/\colA,(0.707107, 0.707107)/\colA,(0.707107, 2.12132)/\colA,(2.12132, -0.707107)/\colA,(2.12132, 0.707107)/\colA}{\filldraw[\c] \P circle(\radInternal);};
\filldraw[black] (0,0) circle(0.01);
\draw[->] (0,0)-- (0.707107, 0.707107)node[color=black,font=\fontLabels,
xshift=0.2cm, yshift=0.2cm]{};
\draw[->] (0,0)-- (1.41421, 0.)node[color=black,font=\footnotesize,
xshift=0.2cm, yshift=0.2cm]{};\foreach \P in {(2.82843, 0.), (0., 2.82843), (0., -2.82843), (-2.82843, 0.)}{
\filldraw \P circle(0);};\end{tikzpicture}
\caption{Convex hulls with respect to the fundamental weight basis for the $\mathbf{3}$, $\mathbf{8}$, $\mathbf{6}$, $\mathbf{15}^*$ irreducible representations of $\mathfrak{sl}(3,\mathbb{R})$ and for the $\mathbf{10}$, $\mathbf{14}$ and $\mathbf{16}$ irreducible
representations of $\mathfrak{sp}(4,\mathbb{R})$. Weights are shown as dots, with the largest ones being those in the Weyl orbit of the highest weight $\lambda$. The dashed line indicates the $\alpha$-hull. One sees that local extrema of $\alpha(H)$ occur at edges of the convex hull where pairs of bubbles of the $\alpha$-hull meet.}
\label{fig:SU3_irreps}
\end{figure} 

Examples of Weyl polytopes for $\mathfrak{sl}(3,\mathbb{R})$ and $\mathfrak{sp}(4,\mathbb{R})$ are shown in Figure~\ref{fig:SU3_irreps}. Plotting the ``$\alpha$-hull'', that is the vector $\vec{\alpha}=\alpha(H)H$ as a function of $H$, it is also clear that local extrema of $\alpha(H)$ indeed lie on the convex hull of the $\lambda$ representation. Since the Weyl polytope is generated by Weyl orbit of the highest weight $\lambda$, for a given group, $\alpha(H)$ and hence the extrema, depend only on $\lambda$ and the overall normalisation of the metric $\langle\cdot,\cdot\rangle$ on the real Cartan subalgebra, or equivalently the length of the simple roots $\{r_i\}$. 

One can construct the extrema directly from the Lie algebra data as follows. As noted, the vertices of the convex hull $C(\lambda)$ correspond to the Weyl images of the highest weight $\lambda$. Vertices are connected through faces, whose dimension ranges from $1$ to $r-1$ (with $r$ the rank of the group). 
Due to the Weyl symmetry,  we can restrict the analysis to the faces containing the highest weight. Each face is generated by choosing a particular subset $I\in \Delta$ of $k_I$ simple roots. The $k_I$-dimensional face $F_I$ then contains weights of the form
\begin{equation} \label{eqn:fI}
    \lambda - \sum_{r_i \in I} \gamma^i r_i \, ,
\end{equation}
for non-negative real coefficients $\gamma^i$ that are bounded depending on the shape of the face itself. In particular, the convex hull of the face $F_I$ is the convex hull of the orbit of $\lambda$ under the subgroup of the Weyl group generated by reflections in the planes orthogonal to the fundamental roots in $I$. Not all subsets of roots define such faces. In particular, one needs a subset that is connected to the highest weight $\lambda$ in the following sense~\cite{Satake60,Dhillon17}: take the Dynkin diagram and add a point representing $\lambda$ and draw a line\footnote{This $\lambda$-extended diagram is just a special case of a marked Coxeter diagram used to described uniform polytopes.} between a simple root $r_i$ and $\lambda$ if $\langle r_i,\lambda\rangle\neq0$, then each subset $I$ of roots  that together with the $\lambda$ node form a connected graph of the extended diagram define a face $F_I$. We will give a concrete example below.  

\begin{figure}[t]
	\centering
\def\cola{RedViolet}
\def\colb{blue}
\def\colc{applegreen}
\begin{tikzpicture}[scale=4,>=stealth]{
\coordinatesHexagon
\coordinate (O) at (0,0);
\coordinate (oI) at ($(A)!0.5!(B)$);
\draw[cyan, line width=2.7pt,opacity=0.5] (A) -- (B) node[midway,above right=-2pt, color=cyan] {$F_I$};
\draw[<->, shorten >=2pt, shorten <=2pt, \colc,thick,dashed] (O) -- ($(A)!0.5!(B)$) node[midway, above,color=\colc] {$d_I$};
\draw[->,\cola, very thick] (O) -- (B) node[midway, above left=-2pt, color=\cola] {$\lambda$};
\draw[black, dashed] (O) -- (A) node[right, color=black] {};
\foreach \pt in {B}{ 
\path (\pt) node[circle, fill=\cola, inner sep=2pt]{};};
\foreach \pt in {O,A}{ \path (\pt) node[circle, fill=black, inner sep=2pt]{};};
\tkzMarkRightAngle[size=0.1,thick](B,oI,O);
}\end{tikzpicture}
\caption{Representation of the face $F_I$ at a distance $d_I$ from the centre of the weight diagram.}
		    \label{fig:rhoI}
	
\end{figure}
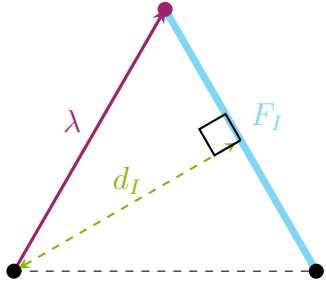

In order to determine what the local extrema of the function $\alpha(H)$ are we first note that the face $F_I$ is, by definition, orthogonal to each fundamental weight in the complement of $I$ that is $w_i\in\Delta-I$. The metric (that is the Gram matrix) on the space spanned by $w_i\in\Delta-I$ is given by 
\begin{align}
    (W^\perp_I)^{ij}= \langle w^i, w^j \rangle \, \qquad \text{for $w^i, w^j \in \Delta-I$} \, . 
\end{align}
Since the face $F_I$ includes the highest weight $\lambda$, the perpendicular (and hence minimal) distance $d_I$ to $F_I$ is given by 
\begin{align} \label{eqn:constraintsW}
    d_I^{\, \,2} = \sum_{i,j\in \Delta-I} 
       \langle w^i, \lambda \rangle  \, (W^\perp_I)^{-1}_{ij} \, 
       \langle w^j, \lambda \rangle \, . 
\end{align}
This is shown pictorially in Figure~\ref{fig:rhoI}. 
The local extrema of $\alpha(H)$ are thus the values of $d_I$ for each face $F_I$. 

Geometrically it is clear that the global minimum of $\alpha(H)$ is the minimum distance from the origin to the set of faces of dimension $k_I=r-1$. In this case the face is defined by a choice of $r-1$ simple roots, that is $I=\Delta-r_i$ for some root $r_i$, such that $I$ is connected to $\lambda$. Equivalently, the face is orthogonal to the fundamental weight $w^i$. From ~\eqref{eqn:constraintsW} the distance to the face is then given by 
\begin{equation}
\label{eqn:alphai}
   d_{\Delta-r_i}= \frac{\langle w^i, \lambda \rangle}{\sqrt{\langle w^i, w^i \rangle}}  \, .
\end{equation}
Even if $\Delta-r_i$ is not connected to $\lambda$ and hence does not define a face, the Weyl symmetry of $C(\lambda)$ implies that $d_{\Delta-r_i}$ is still the minimum distance to some lower-dimensional face. Thus, for a module $V$ with highest weight $\lambda = \sum_{i=1}^r \lambda_i w^i$, one has 
\begin{equation}
   \min_{H \in \mathfrak{a}_1} \alpha(H) = \min_{i=1,...,r} d_{\Delta-r_i} = \min_{i=1,...,r} \sum_{j=1}^r \frac{\langle w^i, w^j \rangle}{\sqrt{\langle w^i, w^i \rangle}} \, \lambda_j =  \min_{i=1,...,r} \sum_{j=1}^r \frac{W^{ij}}{\sqrt{W^{ii}}} \, \lambda_j \, ,   
\end{equation}
where $W^{ij}=\langle w^i, w^j \rangle$ is the Gram matrix of the full set of fundamental weights. 

As a final remark, note that if $\rho$ corresponds to the representation of particles, then each weight corresponds to the $\alpha$-vector for the mass of the relative tower in Planck units. On the other hand, if $\rho$ is the representation under which higher-dimensional objects transform, then the same weights would describe the corresponding tension in Planck units. 

\subsubsection*{Example: $E_{4(4)} \cong SL(5,\mathbb{R}) $}

Let us show how this machinery reproduces the known results for M-theory on $T^4$, for example for the particles and strings representations\footnote{The former corresponding to the four momenta and winding numbers of M2-branes wrapping the six independent two-cycles of the $T^4$, and the latter to windings of M2-branes wrapping one-cycles of the $T^4$ and M5-branes wrapping the whole $T^4$.}, respectively the ${\bf 10}^*$ and $\bf 5$ of $E_{4(4)}$. The corresponding highest weights (where $\mathfrak{sl}(5,\mathbb{R})_\mathbb{C}\simeq \mathfrak{su}(5)_\mathbb{C} = A_4$) are\footnote{The other fundamental weights in our notation are $w^2= \left( 0, \, 0, \, 0, \, -1 \right)$ and $ w^4 = \left( \frac{2}{3}, \, -\frac{1}{3}, \, -\frac{1}{3}, \, -\frac{1}{3} \right)$ and we are using the same conventions as in~\cite{Baines:2025upi}.}
\begin{align}
    \lambda_{\bf 10^*} = w^3 = \left( \tfrac{1}{3}, \, \tfrac{1}{3}, \, -\tfrac{2}{3}, \, -\tfrac{2}{3} \right) \, , \quad
    \lambda_{\bf 5}=w^1 &= \left( -\tfrac{1}{3}, \, -\tfrac{1}{3}, \, -\tfrac{1}{3}, \, -\tfrac{1}{3} \right) \, ,  
\end{align}
and the weights are normalized as 
\begin{equation}
\label{eqn:bil}
    \langle w^i, \, w^j \rangle = W^{ij} = \frac{1}{5} \begin{pmatrix}
        4 & 3 & 2 & 1 \\
        3 & 6 & 4 & 2 \\
        2 & 4 & 6 & 3 \\
        1 & 2 & 3 & 4 \\
    \end{pmatrix} \, .  
\end{equation}

For the particle convex hull, since $\langle w^i,r_j\rangle=\delta^i_j$, the ``$\lambda$-extended'' Dynkin diagram has the form 
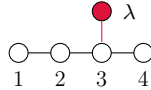
\begin{equation*}
    \tikzset{/Dynkin diagram,root radius=.125cm,edge length=.55cm}
    \begin{dynkinDiagram}[label,mark=o]{A}{4}
    \node[circle,inner sep=0pt,minimum size=2.5mm,fill={\colExt},draw,label=right:{\scriptsize $\lambda$}] (A) at ($(root 3) + (0,0.55)$) {};
    \draw[\colExt] (A) -- (root 3); 
    \end{dynkinDiagram}
\end{equation*}
The convex hull $C$ is a four-dimensional ``tetroctahedric''. The faces and distances are given by 
\begin{align*}
\tikzset{/Dynkin diagram,root radius=.125cm,edge length=.55cm}
\begin{dynkinDiagram}[edge/.style={white,thick}]{A}{o***}
    \node[circle,inner sep=0pt,minimum size=2.5mm,fill={\colExt},draw] (A) at ($(root 3) + (0,0.55)$) {};
    \draw[\colExt] (A) -- (root 3); 
    \draw (root 2) -- (root 3) -- (root 4);
\end{dynkinDiagram} &&
I &= \{ 2,3,4\} \, , & 
d_I &= \sqrt{\frac{1}{5}} \, , && \text{octahedron} \, , \\*[5pt]
\tikzset{/Dynkin diagram,root radius=.125cm,edge length=.55cm}
\begin{dynkinDiagram}[edge/.style={white,thick}]{A}{***o}
    \node[circle,inner sep=0pt,minimum size=2.5mm,fill={\colExt},draw] (A) at ($(root 3) + (0,0.55)$) {};
    \draw[\colExt] (A) -- (root 3); 
    \draw (root 1) -- (root 2) -- (root 3);
\end{dynkinDiagram} &&I &= \{ 1,2,3\} \, , & 
d_I &= \sqrt{\frac{1}{5}+\frac{1}{4}} \, , && \text{tetrahedron} \, , \\*[5pt]
\tikzset{/Dynkin diagram,root radius=.125cm,edge length=.55cm}
\begin{dynkinDiagram}[edge/.style={white,thick}]{A}{oo**}
    \node[circle,inner sep=0pt,minimum size=2.5mm,fill={\colExt},draw] (A) at ($(root 3) + (0,0.55)$) {};
    \draw[\colExt] (A) -- (root 3); 
    \draw (root 3) -- (root 4);
\end{dynkinDiagram} &&
I &= \{3,4\} \, , & 
d_I &= \sqrt{\frac{1}{5}+\frac{1}{3}} \, , && \text{triangle} \, , \\*[5pt]
\tikzset{/Dynkin diagram,root radius=.125cm,edge length=.55cm}
\begin{dynkinDiagram}[edge/.style={white,thick}]{A}{o**o}
    \node[circle,inner sep=0pt,minimum size=2.5mm,fill={\colExt},draw] (A) at ($(root 3) + (0,0.55)$) {};
    \draw[\colExt] (A) -- (root 3); 
    \draw (root 2) -- (root 3);
\end{dynkinDiagram} &&
I &= \{ 2,3\} \, , & 
d_I &= \sqrt{\frac{1}{5}+\frac{1}{3}} \, , && \text{triangle} \, , \\*[5pt]
\tikzset{/Dynkin diagram,root radius=.125cm,edge length=.55cm}
\begin{dynkinDiagram}[edge/.style={white,thick}]{A}{oo*o}
    \node[circle,inner sep=0pt,minimum size=2.5mm,fill={\colExt},draw] (A) at ($(root 3) + (0,0.55)$) {};
    \draw[\colExt] (A) -- (root 3); 
\end{dynkinDiagram} &&
I &= \{3\} \, , & 
d_I &= \sqrt{\frac{1}{5}+\frac{1}{2}} \, , && \text{edge} \, , \\*[5pt]
\tikzset{/Dynkin diagram,root radius=.125cm,edge length=.55cm}
\begin{dynkinDiagram}[edge/.style={white,thick}]{A}{oooo}
    \node[circle,inner sep=0pt,minimum size=2.5mm,fill={\colExt},draw] (A) at ($(root 3) + (0,0.55)$) {};
\end{dynkinDiagram} &&
I &= \emptyset \, , & 
d_I &= \sqrt{\frac{1}{5}+1} \, , && \text{vertex} \, .  
\end{align*}
We see that the string tower rate is a global minimum and comes from the octahedral face defined by $I=\{2,3,4\}$ with 
\begin{equation}
    d_{\{2,3,4\}} = \alpha_{\text{St}} = \sqrt{\frac15} \, , 
\end{equation}
while the other faces give rates
\begin{equation} \label{alphan10}
\alpha_{\text{KK};n} = \sqrt{\frac{1}{5}+ \frac{1}{n}} \ , \, \qquad n=1,2,3,4 \, ,
\end{equation} 
and as expected are associated to decompactifications of $n$ directions. 

If alternatively we look at the convex hull of the string representation $\lambda_{\mathbf{5}}=w^1$, $C$ is a ``hypertetrahedron'' or 4-simplex, and the faces are given by 
\begin{align*}
\tikzset{/Dynkin diagram,root radius=.125cm,edge length=.55cm}
\begin{dynkinDiagram}[edge/.style={white,thick}]{A}{***o}
    \node[circle,inner sep=0pt,minimum size=2.5mm,fill={\colExt},draw] (A) at ($(root 1) + (0,0.55)$) {};
    \draw[\colExt] (A) -- (root 1); 
    \draw (root 1) -- (root 2) -- (root 3); 
\end{dynkinDiagram} &&
I &= \{ 1,2,3\} \, , & 
d_I &= \sqrt{-\frac{1}{5}+\frac{1}{4}} \, , && \text{tetrahedron} \ , \\*[5pt]
\tikzset{/Dynkin diagram,root radius=.125cm,edge length=.55cm}
\begin{dynkinDiagram}[edge/.style={white,thick}]{A}{**oo}
    \node[circle,inner sep=0pt,minimum size=2.5mm,fill={\colExt},draw] (A) at ($(root 1) + (0,0.55)$) {};
    \draw[\colExt] (A) -- (root 1); 
    \draw (root 1) -- (root 2); 
\end{dynkinDiagram} &&
I &= \{ 1,2\} \, , & 
d_I &= \sqrt{-\frac{1}{5}+\frac{1}{3}} \, , && \text{triangle} \, , \\*[5pt]
\tikzset{/Dynkin diagram,root radius=.125cm,edge length=.55cm}
\begin{dynkinDiagram}[edge/.style={white,thick}]{A}{*ooo}
    \node[circle,inner sep=0pt,minimum size=2.5mm,fill={\colExt},draw] (A) at ($(root 1) + (0,0.55)$) {};
    \draw[\colExt] (A) -- (root 1); 
\end{dynkinDiagram} &&
I &= \{1\} \, , & 
d_I &= \sqrt{-\frac{1}{5}+\frac{1}{2}} \, , && \text{edge} \, , \\*[5pt]
\tikzset{/Dynkin diagram,root radius=.125cm,edge length=.55cm}
\begin{dynkinDiagram}[edge/.style={white,thick}]{A}{oooo}
    \node[circle,inner sep=0pt,minimum size=2.5mm,fill={\colExt},draw] (A) at ($(root 1) + (0,0.55)$) {};
\end{dynkinDiagram} &&
I &= \emptyset \, , & 
d_I &= \sqrt{-\frac{1}{5}+1} \, , && \text{vertex} \, .
\end{align*}
We see that the distances fit into a sequence
\begin{equation}
    d_I = \sqrt{ - \frac{1}{5} + \frac{1}{n_I}}\, , \qquad n_I=1,2,3,4\, , 
\end{equation}
with the global minimum is given by 
\begin{equation}
    d_{\{1,2,3\}} = \frac{1}{2\sqrt5} \, ,
\end{equation}
which violates the bound $\alpha \geq \frac{1}{\sqrt{\dext-2}}=\frac{1}{\sqrt{5}}$. This means that a theory with $E_{4(4)}$ moduli space and particles with charges in the $\mathbf{5}$ representation is inconsistent if we assume the sharpened version of the SDC. On the other hand, string charges in this representation are allowed and in particular agree with the BDC~\eqref{eqn:BDC}, as we will see in the Section \ref{sec:bdc}. 

\subsection{Perturbative hulls}

In the previous subsection, we assumed the isometry $G$ of the symmetric space is semi-simple. However, the general analysis applies to any reductive group. To see how what changes when one includes an Abelian factor, consider a general reductive isometry group $G=G'\times \mathbb{R}^\dimAbelian$ where $G'$ is semi-simple, of rank $r'$. States in a given irreducible representation of $G$  must have the same definite charge $q=(q_1,\dots,q_\dimAbelian)$ under $\mathbb{R}^p$, that is, transform under some particular one-dimensional subgroup $\mathbb{R}\subset\mathbb{R}^\dimAbelian$. Without loss of generality, we can thus consider duality groups of the form $G=G'\times\mathbb{R}$. The weights split into a component along $\mathbb{R}$ and those in the Cartan subalgebra of $G'$, so that an irreducible representation is defined by $\lambda=(\lambda',q)$, where $\lambda'$ is the highest weight of the $G'$ representation and $q$ is the $\mathbb{R}$ charge. In the asymptotic mass formula~\eqref{eq:leading mass} the $\mathbb{R}$-contribution appears as an overall prefactor 
\begin{align}
    m^2(t,H',q) 
       = \ee^{-2qh_0t}\sum_{w'\in W'} \ee^{-2t\langle w',H'\rangle} |q_{w'}|^2 + \dots \, , 
\end{align}
where the sum is now over the weights of the $G'$ representation and we have split $H=(h_0,H')$ where $H'\in\mathfrak{a}'^+$, the positive Weyl chamber of the real Cartan of $G'$. Assuming we have normalised the metric on $\mathbb{R}$ to one, to get the correct affine parameter, we then have 
\begin{equation}
    h_0^2 + \langle H', H' \rangle = 1 \, . 
\end{equation}

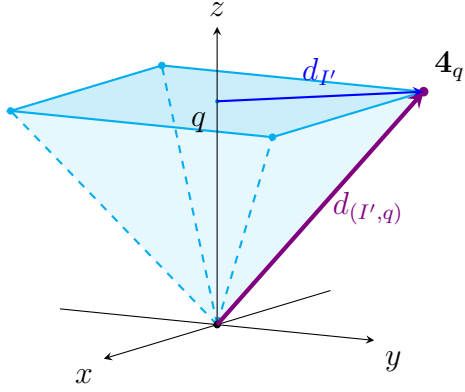
\begin{figure}[t]
\centering
\tdplotsetmaincoords{80}{120} 
\begin{tikzpicture}[tdplot_main_coords, scale=2,>=stealth]

\def\col{cyan}

\coordinate (A) at (1,1,1.5);
\coordinate (B) at (-1,1,1.5);
\coordinate (C) at (-1,-1,1.5);
\coordinate (D) at (1,-1,1.5);
\coordinate (O) at (0,0,0);

\fill[\col, opacity=0.1] (A) -- (B) -- (O) -- cycle;
\fill[\col, opacity=0.1] (D) -- (A) -- (O) -- cycle;
\fill[\col, opacity=0.2] (D) -- (A) -- (B) -- (C) -- cycle;
\draw[thick, \col] (A) -- (B) -- (C) -- (D) -- cycle;
\draw[dashed, thick, \col] (O) -- (A);
\draw[dashed, thick, \col] (O) -- (B);
\draw[dashed, thick, \col] (O) -- (C);
\draw[dashed, thick, \col] (O) -- (D);
\filldraw[black] (O) circle (0.6pt);
\foreach \p in {A,C,D}
    \filldraw[\col] (\p) circle (0.6pt);
\foreach \p in {B}
    \filldraw[violet] (\p) circle (0.8pt);
\filldraw[\col] (0,0,1.5) circle (0.3pt);    

\node at (B) [above right] {$\mathbf{4}_{q}$};
\node at (0,0,1.5) [below left] {$q$};
\draw[->] (-1.5,0,0) -- (1.5,0,0) node[anchor=north east]{$x$};
\draw[->] (0,-1.2,0) -- (0,1.2,0) node[anchor=north west]{$y$};
\draw[->] (0,0,0) -- (0,0,2) node[anchor=south]{$z$};
\draw[->,thick,blue] (0,0,1.5) -- (B) node[pos=0.5, anchor=south] {$d_{I'}$};
\draw[->,ultra thick,violet] (0,0,0) -- (B) node[pos=0.5, anchor=west] {$d_{(I',q)}$};
\end{tikzpicture}
\caption{Perturbative hull for $SO(2,2) \times \mathbb{R}$ for the $\mathbf{4}$ representation of $SO(2,2)$ with $ \mathbb{R}$ weight $q$ forming the base of a cone parallel to the $x - y$ plane, at height $q$ on the $z$ axis).}
\label{fig:conichull}
\end{figure}

As already mentioned above, the highest-weight representation $\lambda=(\lambda',q)$ has an associated convex hull $C(\lambda)$. However, it is no longer centred on the origin but, because of the $q$ charge, sits in a $(r-1)$-dimensional plane at a distance $q$ from the origin, forming the base of a cone, as shown in Figure \ref{fig:conichull} for the representation $(\mathbf{4},q)$ of $SO(2,2) \times \mathbb{R}$. The extremal values of $\alpha(H)$ still come from extremal distances to the convex hull $C(\lambda)$. The faces $F_{(I',q)}$ correspond to choosing  subsets of simple roots $I'\subset\Delta'$ in $G'$ connected to $\lambda'$, as described above. (As simple roots in the Cartan of $G$, they have the form $r_i=(r'_i,0)$.) We denote the minimum distance to the face as $d_{(I',q)}$, and they are given by 
\begin{align}\label{face123}
        d_{(I',q)}^2=d^2_{I'}+q^2\, , 
\end{align}
where $d_{I'}$ is the corresponding distance to the convex hull $C_{\lambda'}$ in the $G'$ weight space. One key new point is that we should view that whole convex hull $C(\lambda)$ as one of the faces, since there is now also a non-zero extremal distance corresponding to the point at the centre of $C(\lambda)$, where $H'=0$. This corresponds to keeping all the roots in $\Delta'$, and the distance is given by $d_{(\Delta',q)}=q$. By construction this is also the global minimum
\begin{equation} 
    \min_{H \in \mathfrak{a}_1} \alpha(H) = d_{(\Delta',q)} = q \, . 
\end{equation}

\begin{figure}[t]
\def\colone{blue}\def\coltwo{applegreen}\def\colthree{cyan}\def\colinf{red}\def\colhull{gray}
\centering

\subfigure{\begin{tikzpicture}[scale=2]
\def\opaedges{1}
\def\opafill{0.2}
\def\opafilltwo{0.03}
\coordinatesCuboctahedronTilted
\def\opaedges{1}\def\opafill{0.1}\def\opafillback{0.05}
\def\sizebig{1.2pt}\def\sizesmall{1.0pt}
\coordinate(O) at (0,0,0); 
\foreach \i/\j/\k in {A/B/I,C/D/H,E/F/G,G/H/I}{
\fill[\colthree,opacity=\opafill](\i) -- (\j) -- (\k);
};
\foreach \i/\j/\k in {J/K/L,B/C/J,F/A/L,D/E/K}{
\fill[\colthree,opacity=\opafillback](\i) -- (\j) -- (\k);
};
\foreach \i/\j/\k/\l in {B/C/H/I,H/D/E/G,G/F/A/I}{
\fill[\colinf,opacity=\opafill](\i) -- (\j) -- (\k) -- (\l);
};
\foreach \i/\j/\k/\l in {B/J/L/A,E/F/L/K,C/D/K/J}{\fill[\colinf,opacity=\opafillback](\i) -- (\j) -- (\k) -- (\l);
};
\foreach \i/\j in {A/B,B/C,C/D,D/E,E/F,F/A, G/H, H/I, I/G,  I/A, I/B, H/C, H/D, G/E, G/F}{\draw[\coltwo, thick] (\i) -- (\j);
};
\foreach \i/\j in {J/K,K/L,L/J,J/B,J/C,K/D,K/E,L/F,L/A}{
\draw[\coltwo, thick,dashed] (\i) -- (\j);
};
\foreach \i in  {A,B,C,D,E,F,G,H,I,J,K,L}{
\path ($(\i)$) node[circle, fill=\colone, inner sep=\sizebig]{{}};};
\end{tikzpicture}}\quad\quad
\subfigure{\begin{tikzpicture}[scale=2]
\coordinatesCuboctahedronTilted
\def\sizebig{1.2pt}\def\sizesmall{1.0pt}
\coordinate(O) at (0,0,0); 

\def\opaedges{0.2}\def\opafill{0.04}\def\opafillback{0.01}\def\opanodes{0.2}
\foreach \i/\j/\k in {A/B/I,C/D/H,E/F/G,G/H/I}{
\fill[\colthree,opacity=\opafill](\i) -- (\j) -- (\k);};
\foreach \i/\j/\k in {J/K/L,B/C/J,F/A/L,D/E/K}{
\fill[\colthree,opacity=\opafillback](\i) -- (\j) -- (\k);};
\foreach \i/\j/\k/\l in {B/C/H/I,H/D/E/G,G/F/A/I}{
\fill[\colinf,opacity=\opafill](\i) -- (\j) -- (\k) -- (\l);};
\foreach \i/\j/\k/\l in {B/J/L/A,E/F/L/K,C/D/K/J}{\fill[\colinf,opacity=\opafillback](\i) -- (\j) -- (\k) -- (\l);};
\foreach \i/\j in {A/B,B/C,C/D,D/E,E/F,F/A, G/H, H/I, I/G,  I/A, I/B, H/C, H/D, G/E, G/F}{\draw[\coltwo, thick,opacity=\opaedges] (\i) -- (\j);};
\foreach \i/\j in {J/K,K/L,L/J,J/B,J/C,K/D,K/E,L/F,L/A}{
\draw[\coltwo, thick,dashed,opacity=\opaedges] (\i) -- (\j);};
\foreach \i in  {A,B,C,D,E,F,G,H,I,J,K,L}{
\path ($(\i)$) node[circle, fill=\colone, inner sep=\sizebig,opacity=\opanodes]{{}};};

\def\opaedges{1}\def\opafill{0.2}\def\opafillback{0.05}

\fill[\colhull,opacity=\opafill] (A) -- (B) -- (I) -- cycle;
\draw[dashed, thick] (O) -- (A);
\draw[dashed, thick] (O) -- (B);
\draw[dashed, thick] (O) -- (I);
\draw[thick] (A) -- (B) -- (I) -- cycle;
\foreach \i/\j/\k in {A/B/I}{
\fill[\colthree,opacity=\opafill](\i) -- (\j) -- (\k);
\path  ($0.333*(\i)+0.333*(\j)+0.333*(\k)$)  
node[circle, fill=\colthree, inner sep=\sizesmall]{{}};};

\foreach \i/\j/\k in {A/B/O,A/I/O,I/B/O}{
\fill[\colthree,opacity=\opafillback](\i) -- (\j) -- (\k);};

\foreach \i/\j in {A/B,I/A,I/B}{\draw[\coltwo, thick] (\i) -- (\j);\path  ($0.5*(\i)+0.5*(\j)$) node[circle, fill=\coltwo, inner sep=\sizesmall]{{}};};
\foreach \i in  {A,B,C,D,E,F,G,H,I,J,K,L}{\path ($(\i)$) node[]{{}};};
\foreach \i in  {A,B,I}{\path ($(\i)$) node[circle, fill=\colone, inner sep=\sizebig]{{}};};
\end{tikzpicture}}\quad\quad
\subfigure{\begin{tikzpicture}[scale=2]
\coordinatesCuboctahedronTilted
\def\sizebig{1.2pt}\def\sizesmall{1.0pt}
\coordinate(O) at (0,0,0); 

\def\opaedges{0.2}\def\opafill{0.04}\def\opafillback{0.01}\def\opanodes{0.2}
\foreach \i/\j/\k in {A/B/I,C/D/H,E/F/G,G/H/I}{
\fill[\colthree,opacity=\opafill](\i) -- (\j) -- (\k);};
\foreach \i/\j/\k in {J/K/L,B/C/J,F/A/L,D/E/K}{
\fill[\colthree,opacity=\opafillback](\i) -- (\j) -- (\k);};
\foreach \i/\j/\k/\l in {B/C/H/I,H/D/E/G,G/F/A/I}{
\fill[\colinf,opacity=\opafill](\i) -- (\j) -- (\k) -- (\l);};
\foreach \i/\j/\k/\l in {B/J/L/A,E/F/L/K,C/D/K/J}{\fill[\colinf,opacity=\opafillback](\i) -- (\j) -- (\k) -- (\l);};
\foreach \i/\j in {A/B,B/C,C/D,D/E,E/F,F/A, G/H, H/I, I/G,  I/A, I/B, H/C, H/D, G/E, G/F}{\draw[\coltwo, thick,opacity=\opaedges] (\i) -- (\j);};
\foreach \i/\j in {J/K,K/L,L/J,J/B,J/C,K/D,K/E,L/F,L/A}{
\draw[\coltwo, thick,dashed,opacity=\opaedges] (\i) -- (\j);};
\foreach \i in  {A,B,C,D,E,F,G,H,I,J,K,L}{
\path ($(\i)$) node[circle, fill=\colone, inner sep=\sizebig,opacity=\opanodes]{{}};};

\def\opaedges{1}\def\opafill{0.2}\def\opafillback{0.05}

\fill[\colhull,opacity=\opafill] (B) -- (C) -- (H) -- (I) -- cycle;
\draw[dashed, thick] (O) -- (B);
\draw[dashed, thick] (O) -- (C);
\draw[dashed, thick] (O) -- (H);
\draw[dashed, thick] (O) -- (I);
\draw[thick] (B) -- (C) -- (H) -- (I) -- cycle;
\foreach \i/\j/\k/\l in {B/C/H/I}{
\fill[\colinf,opacity=\opafill](\i) -- (\j) -- (\k) -- (\l);
\path ($0.25*(\i)+0.25*(\j)+0.25*(\k)+0.25*(\l)$) node[circle, fill=\colinf, inner sep=\sizesmall]{{}};};

\foreach \i/\j/\k in {B/C/O,H/I/O, H/C/O,B/I/O}{
\fill[\colinf,opacity=\opafillback](\i) -- (\j) -- (\k);};

\foreach \i/\j in {B/C,H/I, H/C,B/I}{\draw[\coltwo, thick] (\i) -- (\j);\path  ($0.5*(\i)+0.5*(\j)$) node[circle, fill=\coltwo, inner sep=\sizesmall]{{}};};
\foreach \i in  {A,B,C,D,E,F,G,H,I,J,K,L}{\path ($(\i)$) node[]{{}};};
\foreach \i in  {B,C,H,I}{\path ($(\i)$) node[circle, fill=\colone, inner sep=\sizebig]{{}};};
\end{tikzpicture}}
\caption{$SL(4,\mathbb{R})$, $SL(3,\mathbb{R}) \times\mathbb{R}$ and $SO(2,2)\times \mathbb{R}$ polytopes of the ${\bf 12}$ representation of $SL(4,\mathbb{R})$.} 
\label{fig:perturbative_cuboctahedron}
\end{figure}
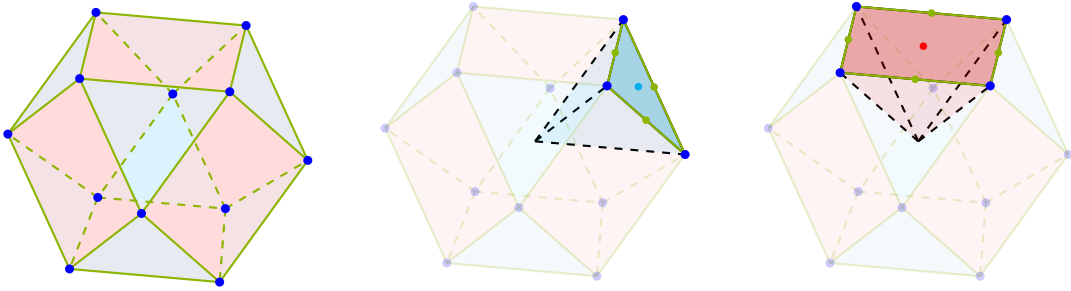

We call such a polytope a {\it perturbative hull}, borrowing the notation from string theory, since such a representation of a reductive group only describes particles in a ``small coupling'' region of the moduli space in some duality frame, if we think of the $\mathbb{R}$ factor as being the analogue of the dilaton. If the duality group also acts non-trivially on $\mathbb{R}$, for example by $\mathbb{Z}_2$, then we would also need to include a set of weights at $-q$. In the $\mathbb{Z}_2$ case the convex hull would become a prism and would again be centred on the origin. In general there might also be states with trivial charge. The perturbative convex hulls then form single faces of the full (or {\it global}) convex hull, each corresponding to a different duality frame. An example of this is shown in Figure~\ref{fig:perturbative_cuboctahedron}, where the global convex hull is the cuboctahedron of the $\mathbf{12}$ representation of $SL(4,\mathbb{R})$. The individual triangular and square faces are the perturbative hulls for the reductive subalgebras
$SL(3,\mathbb{R}) \times \mathbb{R}\subset SL(4,\mathbb{R})$ and $SO(2,2) \times \mathbb{R} \subset SL(4,\mathbb{R})$. In this sense, we can always identify a face with a ``sub-convex hull''. Let $\mathfrak{g}_I\subset\mathfrak{g}$ be the reductive Lie algebra defined by the subset of simple roots $I\subset\Delta$ defining the face $F_I$ and which contains the full Cartan subalgebra $\mathfrak{a}$ of $\mathfrak{g}$, that $\lambda$ is a weight of $\mathfrak{g}_I$. The sub-convex hull defined by the face $F_I$ is then simply the perturbative hull of the representation of $\mathfrak{g}_I$ with highest weight $\lambda$.   

\section{Allowed EFTs: general results}\label{sec:SDC}

In this section, we focus on EFTs in $\dext\geq3$ space-time dimensions with locally symmetric moduli spaces $\mathcal{M}=\Gamma\backslash G/K$ satisfying the compactifiability condition, which implies the duality group $\Gamma$ is arithmetic. As we have seen, the particle states have to transform in some representation $\rho$ of $G$, and the SDC holds under some mild assumptions about the spectrum of states. The question we now wish to ask is: 
\begin{quote}
    \textit{which locally symmetric spaces and representations $\rho$ are compatible with the Emergent String Conjecture rates?}
\end{quote}
For simplicity we will require that $\rho$ is irreducible. As we will see, this leads to an elegant classification of the allowed EFTs. 

We will actually address this problem in two separate steps. We first analyse what convex hulls $C$ are compatible with the ESC rates, without making any reference to symmetric spaces or requiring $C$ to be the convex hull of the weights of some representation. The only condition we will impose is that all the vertices of $C$ are at the same distance from the origin. This is the statement that they each correspond to decompactifying the same number of dimensions $\nvertex$, and geometrically means they lie on a sphere, and are known as ``spherical polytopes''. For symmetric spaces it is a necessary condition on the Weyl polytope for $\rho$ to be irreducible. Finding the allowed $C$ follows closely the analysis in~\cite{Etheredge:2024tok}, except that, as we will see, the spherical condition leads to a complete and fairly simple classification without the need for further assumptions. Requiring the ESC rates are realised imposes conditions not only on the distance to the vertices but also the perpendicular distance to the faces of $C$, since these correspond to decompactifying multiple dimensions, or an emergent string. 
As a result we show that the consistent convex hulls have facets (that is codimension-one faces) that are either regular simplices or orthoplexes (for the precise definitions, see Appendix \ref{app:regandcross}). This means they are examples of ``semi-regular'' polytopes, a class that has been completely classified, starting with the work of Gosset~\cite{Gosset1900} in 1900, with the final proof that the list was complete given by Blind and Blind in 1991~\cite{Blind1991}. The spherical polytopes compatible with the ESC rates then form a subclass within the list of semi-regular polytopes, each with a unique space-time dimension for the EFT. 

The second step is to see which algebras $\cgg$ and irreducible representations can realise the allowed polytopes. For simplicity we assume that the simple factors in $\cgg$ are split.\footnote{We leave a comprehensive study of the non-split forms of the groups for future work. The same analysis applies and since for each Lie group they are a finite number, they will not make the list considerably longer.} In the case where $G$ has an Abelian factor, we follow the discussion above on perturbative hulls to reduce the problem to one of the form $G=G'\times\mathbb{R}$ where $G'$ is semi-simple. The list is given in Table~\ref{tab:1irrep_integern_scaled_perturbative}, \ref{tab:1irrep_integern_scaled_global} and \ref{tab:1irrep_integern_extra} (the latter shows theories that are allowed -- and are not rescaled versions of the ones in Table \ref{tab:1irrep_integern_scaled_global} -- by relaxing the requirement that there should be a limit where only one direction decompactifies). This characterises the massless scalar sector of the EFTs, up to rescalings of the highest weight accompanied by a change in the normalization of the root system of the duality algebra. Remarkably we see that every allowed spherical polytope is realised by at least one symmetric-space EFT. In fact one can view the ESC rate conditions as imposing certain integrality constraints on the lengths and angles of the polytope that are directly related to those of root systems. 

We postpone the physical interpretation of these results to Section \ref{sec:stringinter}.

\subsection{Polytopes and ESC rates: shapes and sizes} 
\label{sec:wpshapes}

In line with the nomenclature above, we are interested in two types of polytope: ``global'' ones of dimension $r$, where the origin lies at the centre of the polytope (corresponding to $G$ being semi-simple) and ``perturbative'' ones of dimension $r-1$ where the polytope lies in a plane at distance $q$ from the origin (corresponding to reductive groups with a single $\mathbb{R}$ factor). In addition, we will assume that $C$ is spherical, which, as we have already mentioned, is a necessary condition for the weight polytope to be that of an irreducible representation. 

For simplicity, we start by assuming that the vertices of the polytope correspond to the decompactification of a single dimension (or if $r=1$ to either a KK tower or an emergent string). We will relax this requirement later in this section. Let $d_0$ be the distance to a vertex, then the ESC rate constraint implies\footnote{We use the notation $\nvertex$ since for a Weyl polytope the highest weight vertex corresponds to choosing the subset of simple roots defining the face $I\subset\Delta$ to be empty.}
\begin{align}
\label{eq:1d-face}
   d_0^2 = \frac{1}{\dext-2} + \frac{1}{\nvertex} \, , \quad \nvertex \in \begin{cases}
       \{ 1, \infty \} & r=1 \, , \\
       \{1\} & r\geq 2 \, ,
   \end{cases} 
\end{align}
corresponding to either decompactifying one dimension ($\nvertex=1$) or the appearance of an emergent string ($\nvertex=\infty$). 

One-dimensional faces connect two vertices $v_1$ and $v_2$. Moving in directions $H_1,H_2\in\mathfrak{a}$ in moduli space parallel to $v_1$ and $v_2$, by assumption, each leads to a KK tower with $\nvertex=1$ (since $r>1$). As can be seen from Figure~\ref{fig:rhoI}, the extremal distance to the face is the midpoint of the interval between $v_1$ and $v_2$. If the independent moduli $H_1$ and $H_2$ are radii of two different decompactifying circles then the extremal point corresponds to both circles decompactifying at the same rate giving a KK spectrum with $n=2$. Alternatively, the extremal point may correspond to a dilaton, in which case it gives an emergent string and $n=\infty$. Thus, the square of the distances to the vertices and midpoint are given by 
\begin{align}
   d_0^2 &= \frac{1}{\dext-2} + 1 , & 
   d_1^2 &= \frac{1}{\dext-2} + \frac{1}{n} \, , \quad n \in \begin{cases}
       \{ 2, \infty \} & r=2 \, , \\
       \{2\} &  r \geq 3 \, .
   \end{cases} \label{norm_rho_n1}
\end{align}
Note that this is equivalent to the ``taxonomy rule'' of Etheredge et al.~\cite{Etheredge:2024tok}, namely 
\begin{equation}
\label{eq:taxonomy}
    \langle v_a,v_b\rangle = \frac{1}{d-2} + \frac{1}{\nvertex}\delta_{ab} \, , 
\end{equation}
in the special case where $\nvertex=1$. It also implies that the distance between the two vertices, which is independent of $d$, is given by
\begin{equation}
\label{eq:edges}
    l = 2\sqrt{1- \frac{1}{n}} \in \begin{cases}
       \{ \sqrt{2},2 \} & r=2 \, , \\
       \{ \sqrt{2} \} &  r\geq 3 \,  \\
   \end{cases}  \, .
\end{equation}
Thus we have two possible types of one-dimensional face
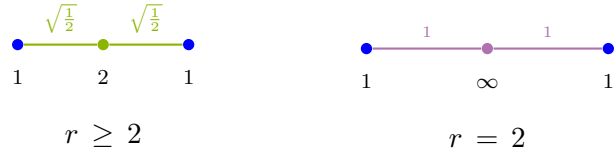
\begin{figure}[H]
\centering
\def\colone{blue}
\def\coltwo{applegreen}
\def\colthree{cyan}
\def\colinf{Orchid}
\subfigure{\begin{tikzpicture}[scale=1.6]
\def\sizebig{1.5pt}\def\sizesmall{1.0pt}
\node[circle, fill=\colone, inner sep=\sizebig, label=below:{\scriptsize$1$}] (A) at (0,0) {};
\node[circle, fill=\colone, inner sep=\sizebig, label=below:{\scriptsize$1$}] (B) at (1.414,0) {};
\node[circle, fill=\coltwo, inner sep=\sizebig, label=below:{\scriptsize$2$}] (C) at (0.707,0) {};
\draw[\coltwo, thick] (A) -- node [above,midway] {\tiny$\sqrt{\frac12}$} (C);
\draw[\coltwo, thick] (C) -- node [above,midway] {\tiny$\sqrt{\frac12}$} (B);
\node[below=5pt of current bounding box.south, text width=2cm, align=center] 
            {\small $r \geq 2$};
\end{tikzpicture}}
\qquad \qquad 
\subfigure{
\begin{tikzpicture}[scale=1.6]
\def\sizebig{1.5pt}\def\sizesmall{1.0pt}
\node[circle, fill=\colone, inner sep=\sizebig, label=below:{\scriptsize$1$}] (A) at (0,0) {};
\node[circle, fill=\colone, inner sep=\sizebig, label=below:{\scriptsize$1$}] (B) at (2,0) {};
\node[circle, fill=\colinf, inner sep=\sizebig, label=below:{\scriptsize$\infty$}] (C) at (1,0) {};
\draw[\colinf, thick] (A) -- node [above,midway] {\tiny$1$} (C);
\draw[\colinf, thick] (C) -- node [above,midway] {\tiny$1$} (B);
\node[below=5pt of current bounding box.south, text width=2cm, align=center] 
            {\small $r = 2$};
\end{tikzpicture}}
\end{figure}
\noindent
where the numbers below list the number of decompactifying dimensions $n$ at the vertices and for the one-dimensional face, and those above list the length of each segment.  

For two-dimensional faces, we have $r\geq3$ and repeating the argument as above we have the minimal squared-distances to vertices, edges and the face are given by 
\begin{equation}
\begin{aligned}
   d_0^2 &= \frac{1}{\dext-2} + 1 , \\
   d_1^2 &= \frac{1}{\dext-2} + \frac12 , \\
   d_2^2 &= \frac{1}{\dext-2} + \frac{1}{n} \, , \quad n \in \begin{cases}
       \{ 3, \infty \} & r=3 \, , \\
       \{ 3 \} &  r \geq 4 \, . 
   \end{cases} 
\end{aligned}
\end{equation}
Eliminating the dependence on $d$, one finds that the circumradius $R$ of the face is given by  
\begin{equation}
\begin{aligned}    
R &= \sqrt{d_0^2 - d_2^2}=\sqrt{1-\frac{1}{n}} \in \begin{cases}
       \{ \sqrt{\tfrac23},\,1 \} & r=3 \, , \\
       \{ \sqrt{\tfrac23} \} & r\geq 4 \,  \\
   \end{cases}  \, , 
\end{aligned}
\end{equation}
and we also have the constraint on the edges~\eqref{eq:edges}, that $l=1/\sqrt{2}$. The only two-dimensional face polytopes that are compatible with these conditions are the simplex (solid triangle) and the orthoplex (solid square):
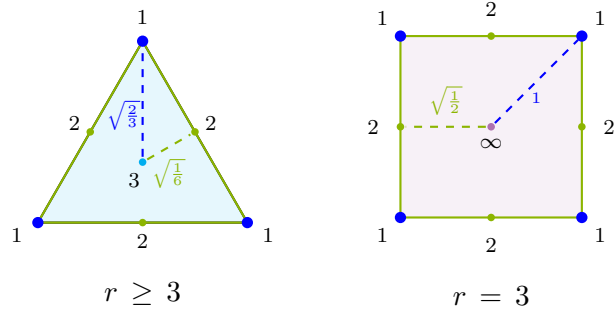
\begin{figure}[H]\centering
\def\colone{blue}
\def\coltwo{applegreen}
\def\colthree{cyan}
\def\colinf{Orchid}
\subfigure{\begin{tikzpicture}[scale=1.6]{
\coordinatesTriangle\drawTriangle\labelsTriangleOneTwo
\def\sizebig{1.5pt}\def\sizesmall{1.0pt}
\def\opafill{0.1}\def\opafillback{0.05}\def\sizebig{1.5pt}\def\sizesmall{1.0pt}
\foreach \i/\j/\k in {A/B/C}{
\fill[\colthree,opacity=\opafill](\i) -- (\j) -- (\k);
\node[circle, fill=\colthree, inner sep=\sizesmall, label={[label distance=-6pt] below left:{\scriptsize$3$}}] (O) at ($0.33*(\i)+0.33*(\j)+0.33*(\k)$) {};};
\foreach \i/\j in {A/B,B/C,C/A}{
\draw[\coltwo, thick] (\i) -- (\j);\path  ($0.5*(\i)+0.5*(\j)$) node[circle, fill=\coltwo, inner sep=\sizesmall]{};};
\foreach \i in {A,B,C}{\path ($(\i)$) node[circle, fill=\colone, inner sep=\sizebig]{};};
\draw[shorten >=2pt, shorten <=2pt, \colone,thick,dashed] (0,0) -- (A) node[midway,yshift=-5pt,left,color=\colone,inner sep=0.5pt]{\tiny$\sqrt{\frac23}$}; 
\draw[shorten >=2pt, shorten <=2pt, \coltwo,thick,dashed] (0,0) -- ($(A)!0.5!(B)$) node[midway, below,color=\coltwo]{\tiny$\sqrt{\frac16}$}; 
}
\node[below=5pt of current bounding box.south, text width=2cm, align=center] 
            {\small $r \geq 3$};
\end{tikzpicture}}
\quad\quad
\subfigure{\begin{tikzpicture}[scale=1.2]{
\coordinatesSquare
\labelsSquareOneTwo
\def\opafill{0.1}\def\opafillback{0.05}
\def\sizebig{1.5pt}\def\sizesmall{1.0pt}
\foreach \i/\j/\k/\l in {A/B/C/D}{\fill[\colinf,opacity=\opafill](\i) -- (\j) -- (\k) -- (\l);
\node[circle, fill=\colinf, inner sep=\sizesmall, label={[label distance=-6pt] below:{\scriptsize$\infty$}}] (O) at ($0.25*(\i)+0.25*(\j)+0.25*(\k)+0.25*(\l)$) {};};
\foreach \i/\j in {A/B,B/C,C/D,D/A}{
\draw[\coltwo, thick] (\i) -- (\j);\path  ($0.5*(\i)+0.5*(\j)$) node[circle, fill=\coltwo, inner sep=\sizesmall]{};};
\foreach \i in {A,B,C,D}{\path ($(\i)$) node[circle, fill=\colone, inner sep=\sizebig]{};};
\draw[shorten >=2pt, shorten <=2pt, \colone,thick,dashed] (0,0) -- (A) node[midway, below,color=\colone] {\tiny$1$}; 
\draw[shorten >=2pt, shorten <=2pt, \coltwo,thick,dashed] (0,0) -- ($(B)!0.5!(C)$) node[midway, above,color=\coltwo] {\tiny$\sqrt{\frac12}$}; 
}
\node[below=5pt of current bounding box.south, text width=2cm, align=center] 
            {\small $r = 3$};
\end{tikzpicture}} 
\end{figure}
\noindent
where again the numbers $1$, $2$, $3$, $\infty$ denote the value of $n$ on each face. In the language of~\cite{Etheredge:2024tok}, the triangle is a ``frame polytope'' in the ``Planckian phase'', while the right-angle triangle formed in the figure on the right by two vertices, one side and the centre of the square form a frame polytope in the ``stringy phase'', since the centre point corresponds to an emergent string. Our spherical polytope condition means that these are the only two such frame polytopes and hence that the triangle and the square are the only two two-dimensional faces that can be built of such frame polytopes. 

This pattern continues so that three-dimensional faces must be either a simplex (that is a tetrahedron) describing a decompactification limit with $n=4$, or an orthoplex (that is an octohedron) describing an emergent string limit:
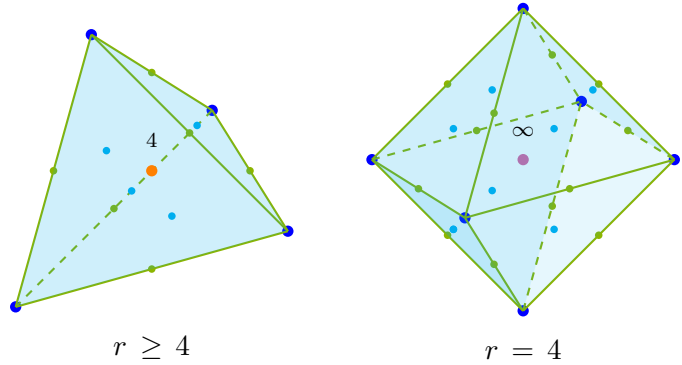
\begin{figure}[H]\centering
\def\colone{blue}
\def\coltwo{applegreen}
\def\colthree{cyan}
\def\colfour{orange}
\def\colinf{Orchid}
\subfigure{\begin{tikzpicture}[scale=1.3]{
\coordinatesTetrahedron
\def\opaedges{1}\def\opafill{0.1}
\def\sizebig{1.5pt}\def\sizesmall{1.0pt}
\foreach \i in {A,B,C,D}{\path ($(\i)$) node[circle, fill=\colone, inner sep=\sizebig]{};};
\foreach \i/\j in {A/C,B/D,B/C,C/D,A/D}{
\draw[\coltwo, thick] (\i) -- (\j);\path  ($0.5*(\i)+0.5*(\j)$) node[circle, fill=\coltwo, inner sep=\sizesmall]{};};
\foreach \i/\j in {A/B}{
\draw[\coltwo, thick,dashed] (\i) -- (\j);\path  ($0.5*(\i)+0.5*(\j)$) node[circle, fill=\coltwo, inner sep=\sizesmall]{};};
\foreach \i/\j/\k in {A/B/C,B/C/D,A/B/D,A/C/D}{
\fill[\colthree,opacity=\opafill](\i) -- (\j) -- (\k);
\path  ($0.333*(\i)+0.333*(\j)+0.333*(\k)$)  node[circle, fill=\colthree, inner sep=\sizesmall]{};};
\node[below=5pt of current bounding box.south, text width=2cm, align=center] 
            {\small $r \ge 4$};
\foreach \i in  {O}{\path ($(\i)$) node[circle, fill=\colfour, inner sep=\sizebig, label=above:{\scriptsize$4$}]{};};
}\end{tikzpicture}}\quad\quad
\subfigure{
\begin{tikzpicture}[scale=2]{
\coordinatesOctahedron
\def\opaedges{1}\def\opafill{0.1}
\def\sizebig{1.5pt}\def\sizesmall{1.0pt}
\foreach \i in  {X,Y,Z,MX,MY,MZ}{\path ($(\i)$) node[circle, fill=\colone, inner sep=\sizebig]{};};
\foreach \i/\j in {X/Y,X/Z,X/MY,MX/MY,MX/Y,MX/Z,Y/Z,MY/Z}{
\draw[\coltwo, thick] (\i) -- (\j);\path  ($0.5*(\i)+0.5*(\j)$) node[circle, fill=\coltwo, inner sep=\sizesmall]{};};
\foreach \i/\j in {X/MZ,MX/MZ,MY/MZ,Y/MZ}{
\draw[\coltwo, thick,dashed] (\i) -- (\j);\path  ($0.5*(\i)+0.5*(\j)$) node[circle, fill=\coltwo, inner sep=\sizesmall]{};};
\foreach \i/\j/\k in {X/Y/Z,MX/MY/Z,X/Y/MZ,MX/MY/Z,MX/Y/Z,X/MY/Z,MX/Y/MZ,MX/MY/MZ}{
\fill[\colthree,opacity=\opafill](\i) -- (\j) -- (\k);
\path  ($0.333*(\i)+0.333*(\j)+0.333*(\k)$)  node[circle, fill=\colthree, inner sep=\sizesmall]{};};
\node[circle, fill=\colinf, inner sep=\sizebig, label=above:{\scriptsize$\infty$}] (O) at (0,0,0) {};
}
\node[below=5pt of current bounding box.south, text width=2cm, align=center] 
            {\small $r = 4$};
\end{tikzpicture}}
\end{figure}

Summarising this analysis, we see that for each rank the $(n-1)$-dimensional faces of the particle polytope $C$  must satisfy that the minimal squared distance to the face is given by
\begin{equation}
\label{eq:dn2}
    d_{n-1}^2 = \begin{cases} 
        \frac{1}{d-2} + \frac{1}{n} \,, &  \text{if $n<r-1$} \, , \\
        \frac{1}{d-2} + \frac{1}{n} \ \text{or} \ \frac{1}{d-2} & \text{if $n=r-1$} \, ,         \end{cases} 
\end{equation}
corresponding to decompactifying $n$ dimensions or, in the second of the $n=r-1$ cases, an emergent string limit. As we show in Appendix~\ref{app:regandcross} this means that:
\begin{itemize}
\item The facets (that is the co-dimension-1 faces) must either simplices or orthoplexes.
\item This implies that co-dimension-2 faces are all of the same type (simplices), but may be inequivalent, meaning that they are boundaries of different types of codimension-1 faces. (We will see an example of this below, namely the triangular prism, for which the edges are either connecting a triangle and a square, or two squares.)
\item It also implies that higher co-dimension faces are all of the same type (simplices) and equivalent.
\item The full polytope does not need to be either a simplex or an orthoplex, but all its lower-dimensional faces have to satisfy the above.
\end{itemize}
The regular $(n-1)$-dimensional simplices $\Delta_{n-1}$ correspond to decompactification limits and the $(r-1)$-dimensional orthoplexes $\beta_{r-1}$ correspond to emergent string limits. 

Crucially these properties allow us to derive all the allowed polytopes. Consider  first the perturbative case which is particularly simple. Here $G$ includes a single Abelian factor $\mathbb{R}$ and the polytope is centred on a plane a distance $q$ from the origin. The polytope has dimension $r-1$ and so must be either a simplex or an orthoplex with the value of $q$ chosen appropriately to satisfy~\eqref{eq:dn2}. These are summarised in Table~\ref{tab:perturb_polytopes}. Note that the one-dimensional simplex and orthoplex are both simply an interval. However we distinguish them by whether the centre point corresponds to decompactification (simplex) or an emergent string (orthoplex). 
\begin{table}[ht!]
\centering
\renewcommand{\arraystretch}{1.5}
\begin{tabular}
{|>{$}c<{$}|>{$}c<{$}|>{$}c<{$}|>{$}c<{$}|} \hline
\text{polytope, $C$} & q & \dext & n \\
\hline\hline
\beta_{r-1} & \sqrt{\frac{1}{d-2}} & d & 1,\dots,r-1,\infty \\
\hline
\multirow{2}{*}{$\protect\segScaled{1.6}$} & \sqrt{\frac{1}{d-2}} & d & 1,\infty \\
\cline{2-4}
& \sqrt{\frac{1}{d-2}+\frac12} & d & 1,2 \\
\hline
\Delta_{r-1} & \sqrt{\frac{1}{d-2}+\frac{1}{r}} & d & 1,\dots,r \\
\hline
\end{tabular}
\caption{Perturbative polytopes consistent with ESC rates and with $\nvertex=1$.}
\label{tab:perturb_polytopes}
\end{table}

For the global polytopes, where $C$ is of dimension $r$, the condition that the facets of $C$ are regular polytopes (in this case simplices or orthoplexes) puts us, by definition, in the realm of what are known as ``semi-regular'' polytopes. These have been completely classified~\cite{Gosset1900,Blind1991}. Restricting to those built from simplices and orthoplexes and $r>2$ one has
\begin{itemize}
    \item two infinite families of regular polytopes: simplices $\Delta_r$ and orthoplexes $\beta_r$; 
    \item seven additional exceptional cases with $r=3$: cube, icosahedron, cuboctahedron, rhombicuboctahedron, snub cuboctahedron, square antiprism and triangular prism;
    \item three additional exceptional cases with $r=4$: octacube (or 24-cell), tetraplex (or 600-cell) and tetroctahedric;
    \item four additional exceptional cases which together with the triangular prism and the tetroctahedric form the family of ``Gosset polytopes'': demipenteract, $\EsixAlt$, $\EsevenAlt$ and $\EeightAlt$ with $r=5,6,7,8$ respectively.
\end{itemize}
Since $r>2$ in all cases the edges of the polytope have squared-length $l^2=2$, while the circumradius must satisfy
\begin{equation}
\label{eq:circumrad}
    R^2 = \frac{1}{d-2}+1 \, , 
\end{equation}
and hence $R^2$ must be rational. The rationality condition excludes the icosahedron, rhombicuboctahedron, snub cuboctahedron, square antiprism and tetroplex. The simplices and orthoplexes have $R^2=r/(r+1)$ and $R^2=1$ respectively and so cannot satisfy~\eqref{eq:circumrad}. This leaves a finite list:
\begin{equation}
\begin{aligned}
    r=3: & && \text{triangular prism ($d=8$)}\,, \text{cube ($d=4$)}\,, \text{cuboctahedron ($d=4$)}\,, \\
    r=4: & && \text{tetrohedric ($d=7$)}\,, \text{octacube ($d=3$)}\,,  \\
    r=5: & && \text{demipenteract ($d=6$)}\,, \\
    r=6: & && \EsixAlt\ (d=5)\,, \\
    r=7: & && \EsevenAlt\ (d=4)\,, \\
    r=8: & && \EeightAlt\ (d=3)\, .
\end{aligned}
\end{equation}
In particular, the three allowed polyhedra for $r=3$ are shown below
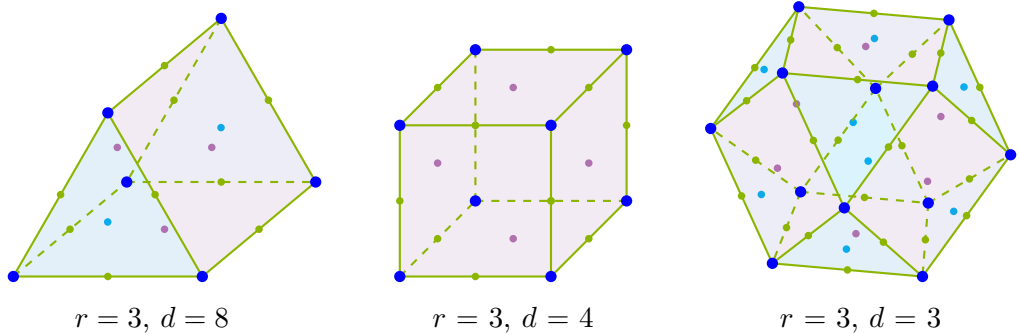
\begin{figure}[H]\centering
\def\colone{blue}
\def\coltwo{applegreen}
\def\colthree{cyan}
\def\colinf{Orchid}
\subfigure{\begin{tikzpicture}[scale=2.5]
\coordinatesTriangularPrismExtra
\def\opaedges{1}\def\opafill{0.1}\def\opafillback{0.05}
\def\sizebig{1.5pt}\def\sizesmall{1.0pt}
\foreach \i/\j/\k in {A/B/C}{
\fill[\colthree,opacity=\opafill](\i) -- (\j) -- (\k);
\path  ($0.333*(\i)+0.333*(\j)+0.333*(\k)$)  
node[circle, fill=\colthree, inner sep=\sizesmall]{{}};};
\foreach \i/\j/\k in {A'/B'/C'}{
\fill[\colthree,opacity=\opafillback](\i) -- (\j) -- (\k);
\path  ($0.333*(\i)+0.333*(\j)+0.333*(\k)$)  
node[circle, fill=\colthree, inner sep=\sizesmall]{{}};};
\foreach \i/\j/\k/\l in {B/B'/C'/C}{
\fill[\colinf,opacity=\opafill](\i) -- (\j) -- (\k) -- (\l);
\path ($0.25*(\i)+0.25*(\j)+0.25*(\k)+0.25*(\l)$) node[circle, fill=\colinf, inner sep=\sizesmall]{{}};};
\foreach \i/\j/\k/\l in {A/A'/B'/B,A/A'/C'/C}{\fill[\colinf,opacity=\opafillback](\i) -- (\j) -- (\k) -- (\l);\path ($0.25*(\i)+0.25*(\j)+0.25*(\k)+0.25*(\l)$) node[circle, fill=\colinf, inner sep=\sizesmall]{{}};};
\foreach \i/\j in {B/B',C/C',A/B,B/C,C/A,B'/C'}{\draw[\coltwo, thick] (\i) -- (\j);\path  ($0.5*(\i)+0.5*(\j)$) node[circle, fill=\coltwo, inner sep=\sizesmall]{{}};};
\foreach \i/\j in {A/A',A'/B',C'/A'}{
\draw[\coltwo, thick,dashed] (\i) -- (\j);\path  ($0.5*(\i)+0.5*(\j)$) node[circle, fill=\coltwo, inner sep=\sizesmall]{{}};};
\foreach \i in  {A,B,C,A',B',C'}{\path ($(\i)$) node[circle, fill=\colone, inner sep=\sizebig]{{}};};
\node[below=5pt of current bounding box.south, text width=2.5cm] 
            {\small $\,  r = 3$, $\dext=8$};
\end{tikzpicture}} \qquad 
\subfigure{\begin{tikzpicture}[scale=2]
\def\opafill{0.1}\def\opafillback{0.05}\def\sizebig{1.5pt}\def\sizesmall{1.0pt}
\coordinatesCube
\coordinate (Z) at (0,0);\def\dlabels{0.3}\def\mult{1.3}
\foreach \i/\j/\k/\l in {A/B/C/D,C/D/D'/C',B/C/C'/B'}{\fill[\colinf,opacity=\opafill](\i) -- (\j) -- (\k) -- (\l);\path ($0.25*(\i)+0.25*(\j)+0.25*(\k)+0.25*(\l)$) node[circle, fill=\colinf, inner sep=\sizesmall]{{}};};
\foreach \i/\j/\k/\l in {A'/B'/C'/D',A/B/B'/A',A/D/D'/A'}{\fill[\colinf,opacity=\opafillback](\i) -- (\j) -- (\k) -- (\l);\path ($0.25*(\i)+0.25*(\j)+0.25*(\k)+0.25*(\l)$) node[circle, fill=\colinf, inner sep=\sizesmall]{{}};};
\foreach \i/\j in {A/B,B/C,C/D,D/A,B/B',C/C',D/D',C'/D',B'/C'}{\draw[\coltwo, thick] (\i) -- (\j);\path  ($0.5*(\i)+0.5*(\j)$) node[circle, fill=\coltwo, inner sep=\sizesmall]{{}};};
\foreach \i/\j in {A'/B',D'/A',A/A'}{\draw[\coltwo, thick,dashed] (\i) -- (\j);\path  ($0.5*(\i)+0.5*(\j)$) node[circle, fill=\coltwo, inner sep=\sizesmall]{{}};};
\foreach \i in {A,B,C,D,A',B',C',D'}{\path ($(\i)$) node[circle, fill=\colone, inner sep=\sizebig]{{}};};
\node[below=5pt of current bounding box.south, text width=2.5cm, align=center] 
            {\small $r =3$, $\dext=4$};
\end{tikzpicture}} \qquad
\subfigure{\begin{tikzpicture}[scale=2]
\coordinatesCuboctahedronTilted
\def\opaedges{1}\def\opafill{0.1}\def\opafillback{0.05}
\def\sizebig{1.5pt}\def\sizesmall{1.0pt}
\foreach \i/\j/\k in {A/B/I,C/D/H,E/F/G,G/H/I}{
\fill[\colthree,opacity=\opafill](\i) -- (\j) -- (\k);
\path  ($0.333*(\i)+0.333*(\j)+0.333*(\k)$)  
node[circle, fill=\colthree, inner sep=\sizesmall]{{}};};
\foreach \i/\j/\k in {J/K/L,B/C/J,F/A/L,D/E/K}{
\fill[\colthree,opacity=\opafillback](\i) -- (\j) -- (\k);
\path  ($0.333*(\i)+0.333*(\j)+0.333*(\k)$)  
node[circle, fill=\colthree, inner sep=\sizesmall]{{}};};
\foreach \i/\j/\k/\l in {B/C/H/I,H/D/E/G,G/F/A/I}{
\fill[\colinf,opacity=\opafill](\i) -- (\j) -- (\k) -- (\l);
\path ($0.25*(\i)+0.25*(\j)+0.25*(\k)+0.25*(\l)$) node[circle, fill=\colinf, inner sep=\sizesmall]{{}};};
\foreach \i/\j/\k/\l in {B/J/L/A,E/F/L/K,C/D/K/J}{\fill[\colinf,opacity=\opafillback](\i) -- (\j) -- (\k) -- (\l);\path ($0.25*(\i)+0.25*(\j)+0.25*(\k)+0.25*(\l)$) node[circle, fill=\colinf, inner sep=\sizesmall]{{}};};
\foreach \i/\j in {A/B,B/C,C/D,D/E,E/F,F/A, G/H, H/I, I/G,  I/A, I/B, H/C, H/D, G/E, G/F}{\draw[\coltwo, thick] (\i) -- (\j);\path  ($0.5*(\i)+0.5*(\j)$) node[circle, fill=\coltwo, inner sep=\sizesmall]{{}};};
\foreach \i/\j in {J/K,K/L,L/J,J/B,J/C,K/D,K/E,L/F,L/A}{
\draw[\coltwo, thick,dashed] (\i) -- (\j);\path  ($0.5*(\i)+0.5*(\j)$) node[circle, fill=\coltwo, inner sep=\sizesmall]{{}};};
\foreach \i in  {A,B,C,D,E,F,G,H,I,J,K,L}{\path ($(\i)$) node[circle, fill=\colone, inner sep=\sizebig]{{}};};
\node[below=5pt of current bounding box.south, text width=2.5cm, align=center] 
            {\small $r = 3$, $\dext=3$};
\end{tikzpicture}}
\end{figure}
\noindent 
Here blue points correspond to vertices with $n=1$, green points have $n=2$, light blue points at the centre of triangular facets have $n=3$ and purple points at the centre of square facets have $n=\infty$. A further example with $r=4$ is the convex hull formed by the weights of the $\mathbf{10}$ representation of $\mathfrak{sl}(5,\mathbb{R})$, which characterises M-theory on $T^4$ discussed in the previous section and in~\cite{Baines:2025upi}. This is the “tetroctahedric” (also known as rectified 5-cell), whose three-dimensional faces are of both types -- simplices (tetrahedra) and orthoplexes (octahedra). 

For $r=1,2$ that allowed polytopes are as follows. For $r=1$ the vertices have either $\nvertex=1$ or $\nvertex=\infty$ and so the only possible polytopes are 
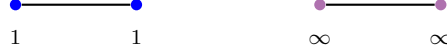
\begin{figure}[H]
\centering
\def\colone{blue}
\def\coltwo{applegreen}
\def\colthree{cyan}
\def\colinf{Orchid}
\subfigure{\begin{tikzpicture}[scale=1.6]
\def\sizebig{1.5pt}\def\sizesmall{1.0pt}
\node[circle, fill=\colone, inner sep=\sizebig, label=below:{\scriptsize$1$}] (A) at (0,0) {};
\node[circle, fill=\colone, inner sep=\sizebig, label=below:{\scriptsize$1$}] (B) at (1,0) {};
\draw[thick] (A) -- (B);
\end{tikzpicture}}
\qquad \qquad 
\subfigure{
\begin{tikzpicture}[scale=1.6]
\def\sizebig{1.5pt}\def\sizesmall{1.0pt}
\node[circle, fill=\colinf, inner sep=\sizebig, label=below:{\scriptsize$\infty$}] (A) at (0,0) {};
\node[circle, fill=\colinf, inner sep=\sizebig, label=below:{\scriptsize$\infty$}] (B) at (1,0) {};
\draw[thick] (A) -- (B);
\end{tikzpicture}}
\caption*{$r=1$}
\end{figure}
\noindent
where the lengths between the vertices are $2\sqrt{\frac{1}{d-2}+1}$ and $2\sqrt{\frac{1}{d-2}}$ respectively. 

For $r=2$ the convex hull is a polygon with sides either of length $\sqrt{2}$ or $2$. Put more precisely, the allowed two-dimensional convex hulls are :
\begin{itemize}
\item regular polygons whose edges all have $l = \sqrt{2}$; 
\item regular polygons whose edges all have $l=2$;
\item semi-regular polygons with alternating edges of $l=2$ and $l=\sqrt{2}$.
\end{itemize}
Satisfying the conditions~\eqref{norm_rho_n1} with $n\in\{2,\infty\}$ and $d$ integer, implies that $\cos\theta$ is rational, where $\theta$ is the angle subtended by an edge. For a regular polygon with $N>2$ sides, Niven's theorem~\cite{MR80123} implies that only $N\in\{3,4,6\}$ are allowed. For $N=3$ the only solution with $d\in {\mathbb Z}$ is a triangle with $l_i=2$, possible only for $\dext = 5 $ 
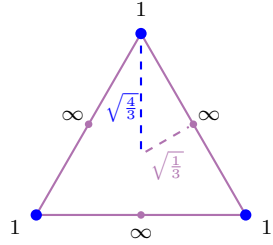
\begin{figure}[H]
\centering

\def\colone{blue}
\def\coltwo{applegreen}
\def\colthree{cyan}
\def\colinf{Orchid}

\begin{tikzpicture}[scale=1.6]
{
\coordinatesTriangle
\labelsTriangleOneInf
\def\sizebig{1.5pt}\def\sizesmall{1.0pt}
\foreach \i/\j in {A/B,B/C,C/A}{\draw[\colinf, thick] (\i) -- (\j);\path  ($0.5*(\i)+0.5*(\j)$) node[circle,fill=\colinf, inner sep=\sizesmall]{};};
\foreach \i in {A,B,C}{\path ($(\i)$) node[circle, fill=\colone, inner sep=\sizebig]{};};
\draw[shorten >=2pt, shorten <=2pt, \colone,thick,dashed] (0,0) -- (A) node[midway,yshift=-5pt,left,color=\colone,inner sep=0.5pt]{\tiny$\sqrt{\frac43}$}; 
\draw[shorten >=2pt, shorten <=2pt, \colinf,thick,dashed] (0,0) -- ($(A)!0.5!(B)$) node[midway, below,color=\colinf]{\tiny$\sqrt{\frac13}$}; 
}
\end{tikzpicture}
\caption*{$r=2$, $\dext=5$}
\end{figure}
\noindent $N=4$ and $6$ are only possible in $\dext = 3$, and correspond to
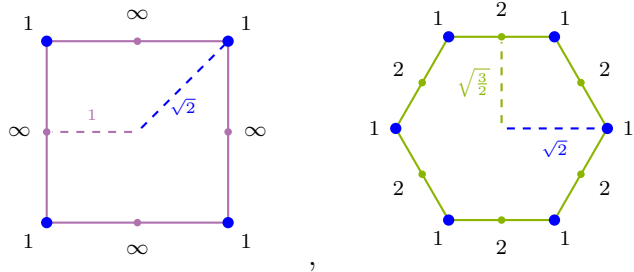
\begin{figure}[H]
\def\colone{blue}
\def\coltwo{applegreen}
\def\colthree{cyan}
\def\colinf{Orchid}
\centering
\subfigure{
\begin{tikzpicture}[scale=1.2,>=stealth]
{\coordinatesSquare
\labelsSquareOneInf
\def\sizebig{1.5pt}\def\sizesmall{1.0pt}
\foreach \i/\j in {A/B,B/C,C/D,D/A}{
\draw[\colinf, thick] (\i) -- (\j);\path  ($0.5*(\i)+0.5*(\j)$) node[circle, fill=\colinf, inner sep=\sizesmall]{};};
\foreach \i in {A,B,C,D}{\path ($(\i)$) node[circle, fill=\colone, inner sep=\sizebig]{};};
\draw[shorten >=2pt, shorten <=2pt, \colone,thick,dashed] (0,0) -- (A) node[midway, below,color=\colone] {\tiny$\sqrt{2}$}; 
\draw[shorten >=2pt, shorten <=2pt, \colinf,thick,dashed] (0,0) -- ($(B)!0.5!(C)$) node[midway, above,color=\colinf] {\tiny$1$}; 
}
\end{tikzpicture}}\quad,\quad\subfigure{
\begin{tikzpicture}[scale=1.4]
{\coordinatesHexagon
\labelsHexagonOneTwo
\def\sizebig{1.5pt}\def\sizesmall{1.0pt}
\foreach \i/\j in {A/B,B/C,C/D,D/E,E/F,F/A}{
\draw[\coltwo, thick] (\i) -- (\j);\path  ($0.5*(\i)+0.5*(\j)$) node[circle, fill=\coltwo, inner sep=\sizesmall]{};};
\foreach \i in {A,B,C,D,E,F}{\path ($(\i)$) node[circle, fill=\colone, inner sep=\sizebig]{};};
\draw[shorten >=2pt, shorten <=2pt, \colone,thick,dashed] (0,0) -- (A) node[midway, below,color=\colone] {\tiny$\sqrt{2}$}; 
\draw[shorten >=2pt, shorten <=2pt, \coltwo,thick,dashed] (0,0) -- ($(B)!0.5!(C)$) node[midway, left,color=\coltwo] {\tiny$\sqrt{\frac32}$};
}\end{tikzpicture}} 
\caption*{$r=2$, $\dext=3$}
\end{figure}
\noindent 
For the semi-regular polytopes, it can be proven that the only possibility with edges of alternating lengths as in \eqref{eq:edges}, and whose vertices are at distance given by \eqref{norm_rho_n1} with $\nvertex=1$, is the rectangle for $\dext = 4$
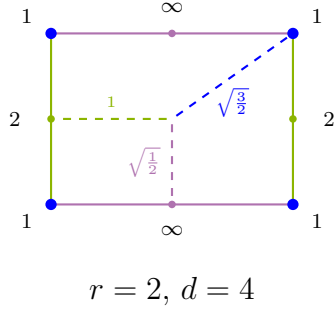
\begin{figure}[H]
\def\colone{blue}
\def\coltwo{applegreen}
\def\colthree{cyan}
\def\colinf{Orchid}
\centering
\begin{tikzpicture}[scale=1.6]{
\coordinatesRectangle
\labelsRectangleOne
\def\sizebig{1.5pt}\def\sizesmall{1.0pt}
\foreach \i/\j in {A/B,C/D}{
\draw[\colinf, thick] (\i) -- (\j);\path  ($0.5*(\i)+0.5*(\j)$) node[circle, fill=\colinf, inner sep=\sizesmall]{};};
\foreach \i/\j in {B/C,D/A}{
\draw[\coltwo, thick] (\i) -- (\j);\path  ($0.5*(\i)+0.5*(\j)$) node[circle, fill=\coltwo, inner sep=\sizesmall]{};};
\foreach \i in {A,B,C,D}{\path ($(\i)$) node[circle, fill=\colone, inner sep=\sizebig]{};};
\draw[shorten >=2pt, shorten <=2pt, \colone,thick,dashed] (0,0) -- (A) node[midway, below,color=\colone] {\tiny$\sqrt{\frac32}$}; 
\draw[shorten >=2pt, shorten <=2pt, \coltwo,thick,dashed] (0,0) -- ($(B)!0.5!(C)$) node[midway, above,color=\coltwo] {\tiny$1$}; 
\draw[shorten >=2pt, shorten <=2pt, \colinf,thick,dashed] (0,0) -- ($(C)!0.5!(D)$) node[midway, left,color=\colinf] {\tiny$\sqrt{\frac12}$}; 
}
\end{tikzpicture}
\caption*{$r = 2$, $\dext=4$}
\end{figure}

In summary, the allowed global polytopes with $\nvertex=1$ are those of Table~\ref{tab:global_polytopes}. For details of the geometry of these different semi-regular polytopes see Appendix~\ref{app:regandcross}. 
\begin{table}[ht!]
\centering
\renewcommand{\arraystretch}{1.1}
\begin{tabular}{|>{$}c<{$}|>{$}c<{$}|>{$}c<{$}|>{$}c<{$}|}
\hline
\text{polytope, $C$} & r & \dext & n \\
\hline\hline
\multirow{2}{*}{$\protect\segScaled{1.6}$} & \multirow{2}{*}{1} & d & \infty \\
&& d & 1 \\
\hline
  \text{\Large$\triangle$} & \multirow{4}{*}{2} & 5 & 1,\infty \\
  \text{\Large$\rectangle$} && 4 & 1,2,\infty \\
  \text{\Large$\square$} && 3 & 1,\infty \\
  \text{\LARGE\hexagon} && 3 & 1,2 \\
  \hline
  \triprismScaled{0.4} & \multirow{3}{*}{3} & 8 & 1,2,3,\infty \\
  \cubeScaled{0.4} && 4 & 1,2,\infty \\
  \cuboctahedronScaled{0.35} && 3 & 1,2,3,\infty \\
  \hline
  \text{tetroctahedric} & \multirow{2}{*}{4} & 7 & 1,\dots,4,\infty \\
  \text{octacube} && 3 & 1,2,3,\infty \\
  \hline
  \text{demipenteract} & 5 & 6 & 1,\dots,5,\infty \\
  \hline
  \EsixAlt & 6 & 5 & 1,\dots,6,\infty \\
  \hline 
  \EsevenAlt & 7 & 5 & 1,\dots,7,\infty \\
  \hline 
  \EeightAlt & 8 & 5 & 1,\dots,8,\infty \\
  \hline 
\end{tabular}
\caption{Global polytopes consistent with ESC rates and with $\nvertex=1$.}
\label{tab:global_polytopes}
\end{table}

Recall that to obtain these allowed polytopes we assumed that their vertices correspond to KK towers associated to decompactifying one dimension, i.e.~we assumed $\nvertex=1$. Relaxing this assumption so that each vertex decompactifies $\nvertex>1$ dimensions\footnote{A situation where one is not allowed to decompactify only one dimension is for example a $T^2$ orbifold that fixes the complex structure but not the volume.} is straightforward. First one notes that the same argument that implied a $(p-1)$-dimensional face corresponds to decompactifying $p$ dimensions (or potentially an emergent string if $p=r-1$) now implies that a $(p-1)$-dimensional face corresponds to decompactifying $p\nvertex$ dimensions (or potentially an emergent string if $p=r-1$). Thus we can write
\begin{equation}
\label{eq:dn2-prime}
    d_{p-1}^2 = \begin{cases} 
        \frac{1}{d-2} + \frac{1}{p\nvertex} \,, &  \text{if $p<r-1$} \, , \\
        \frac{1}{d-2} + \frac{1}{p\nvertex} \ \text{or} \ \frac{1}{d-2} & \text{if $p=r-1$} \, . 
        \end{cases} 
\end{equation}
We see the $d_{I}^2$ are proportional to the ones in~\eqref{eq:dn2} since we can write
\begin{equation}
\label{eq:distance_zz}
    d_{p-1}^{\, 2} = \frac{d_{p-1}^{\prime\, 2}}{\nvertex} \, , 
    \qquad \text{where} \quad 
    d_{p-1}^{\prime\, 2} = \frac{1}{\dext'-2}+\frac{1}{p} \, , 
    \quad \dext' = \frac{d-2}{\nvertex} + 2 \, .  
\end{equation}
Thus changing the external dimension $\dext \to d' \in \frac{\mathbb{N}_+}{\nvertex}$, the classification follows through as above (with all decompactification dimensions scaled by $\nvertex$) except that one needs to allow for appropriate non-integer values of $d'$. Carrying out this analysis in detail, there are only two additional polytopes which cannot be obtained by rescaling one with $\nvertex=1$: a regular hexagon with $\nvertex=3$ and a semi-regular octagon with $\nvertex=4$, both for $r=2$ in $d=3$. These have the form
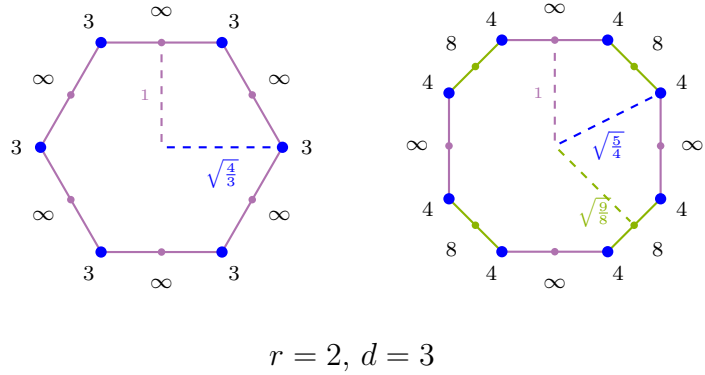
\begin{figure}[H]
\def\colone{blue}
\def\coltwo{applegreen}
\def\colthree{cyan}
\def\colinf{Orchid}
\centering
\subfigure{
\begin{tikzpicture}[scale=1.6]
\coordinatesHexagon
\labelsHexagonThreeInf
\def\sizebig{1.5pt}\def\sizesmall{1.0pt}
\foreach \i/\j in {A/B,B/C,C/D,D/E,E/F,F/A}{
\draw[\colinf, thick] (\i) -- (\j);
\path ($0.5*(\i)+0.5*(\j)$) node[circle, fill=\colinf, inner sep=\sizesmall]{};
};
\foreach \i in {A,B,C,D,E,F}{
\path ($(\i)$) node[circle, fill=\colone, inner sep=\sizebig]{};};
\draw[shorten >=2pt, shorten <=2pt, \colone,thick,dashed](0,0) -- (A) node[midway, below,color=\colone]{\tiny$\sqrt{\frac43}$}; 
\draw[shorten >=2pt, shorten <=2pt, \colinf,thick,dashed](0,0) -- ($(B)!0.5!(C)$) node[midway, left,color=\colinf]{\tiny$1$};
\end{tikzpicture}}
\quad\quad\subfigure{\begin{tikzpicture}[scale=1.4]{
\coordinatesOctagon
\labelsOctagonAll
\def\sizebig{1.5pt}\def\sizesmall{1.0pt}
\foreach \i/\j in {A/B,C/D,E/F,G/H}{
\draw[\colinf, thick] (\i) -- (\j);
\path ($0.5*(\i)+0.5*(\j)$) node[circle, fill=\colinf, inner sep=\sizesmall]{};
};
\foreach \i/\j in {B/C,D/E,F/G,H/A}{
\draw[\coltwo, thick] (\i) -- (\j);
\path ($0.5*(\i)+0.5*(\j)$) node[circle, fill=\coltwo, inner sep=\sizesmall]{};
};
\foreach \i in {A,B,C,D,E,F,G,H}{
\path ($(\i)$) node[circle, fill=\colone, inner sep=\sizebig]{};};
\draw[shorten >=2pt, shorten <=2pt, \colone,thick,dashed](0,0) -- (B) node[midway, below,color=\colone]{\tiny$\sqrt{\frac54}$}; 
\draw[shorten >=2pt, shorten <=2pt, \coltwo,thick,dashed](0,0) -- ($(H)!0.5!(A)$) node[midway, below,color=\coltwo]{\tiny$\sqrt{\frac98}$}; 
\draw[shorten >=2pt, shorten <=2pt, \colinf,thick,dashed](0,0) -- ($(C)!0.5!(D)$) node[midway, left,color=\colinf]{\tiny$1$};
}
\end{tikzpicture}}\label{fig:rank2polytopes_additional}
\caption*{$r=2$, $\dext=3$}
\end{figure}
\noindent
and listed in Table~\ref{tab:extra_polytopes}.
\begin{table}[ht!]
\centering
\renewcommand{\arraystretch}{1.1}
\begin{tabular}{|>{$}c<{$}|>{$}c<{$}|>{$}c<{$}|>{$}c<{$}|}
\hline
\text{polytope, $C$} & r & \dext & n \\
\hline\hline
  \text{\Large$\hexagon$} & \multirow{2}{*}{2} & 3 & 3,\infty \\
  \text{\Large$\octagon$} && 3 & 4,8,\infty \\
  \hline
\end{tabular}
\caption{Additional global polytopes consistent with ESC rates and with minimal $\nvertex$.}
\label{tab:extra_polytopes}
\end{table}

It is important to note that the analysis in this section relied only on the assumption that $C$ was spherical. For convenience, we also imposed that $\nvertex$, the number of dimensions decompactified by moving in the direction of a vertex of $C$ was minimal: either $\nvertex=1$ for the polytopes in Tables~\ref{tab:perturb_polytopes} and~\ref{tab:global_polytopes} and $\nvertex=3$ and $\nvertex=4$ for the polytopes in Tables~\ref{tab:extra_polytopes}. However, as we discuss in the next Section, each polytope actually describes an infinite family of solutions with $\nvertex'=z\nvertex^{\text{min}}$ and $d'=z(d-2)+2$ for $z\in\mathbb{N}$. 

\subsection{Polytopes and ESC rates: groups and irreps} \label{sec:classification}

Having understood the constraints on the geometry of the polytopes that satisfy the Sharpened Distance Conjecture, we now turn to classifying which Lie algebras $\cgg$ and irreps $\rho$ with highest weight $\lambda$ realise the possible polytope as the convex hull of the weights $C(\lambda)$. Recall that the invariant metric $\langle \cdot,\cdot\rangle$ is fixed only up to scale, and so this is an additional input. Thus, we want to find a list of 
\begin{equation} \label{eq:theory}
    ( \mathfrak{g}, \lambda, \langle \cdot,\cdot\rangle) \, 
\end{equation}
for each of the polytopes in Tables~\ref{tab:perturb_polytopes},~\ref{tab:global_polytopes} and~\ref{tab:extra_polytopes}. For simplicity we will restrict to split forms of $\cgg$. However, extending to include non-split forms should be straightforward. What is remarkable is that we will find that all the allowed polytopes have realisations in terms of EFTs with locally symmetric moduli spaces. 

It is worth making a couple of comments first. By restricting to irreducible representations, we impose that all the vertices of the polytope are at the same distance and hence encode the same kind of decay rates (all decompactifying the same number of dimensions $\nvertex$). As we have seen this strongly restricts the allowed polytopes. Allowing for reducible representations clearly gives considerably more freedom and cases that appear physically, for example, some vertices corresponding to decompactifying one dimension and some to decompactifying two dimensions. That said, our analysis does still apply to certain reducible representations. In particular, there could be additional particles forming another representation with highest weight $\lambda'$ if its convex hull $C_{\lambda'}$ lies strictly inside $C(\lambda)$. We will not concern ourselves with the classification of these subleading particle representation here. The second point to note is that while there must be weights corresponding to vertices of $C(\lambda)$ there is no need for there to be weights at the extremal points on a given face $F_I$. In particular, in general points characterised by $n= \infty$ can, but do not need to, be weights in the particle representation. As we discuss in Section~\ref{sec:bdc}, the $n=\infty$ points should however correspond to weight in the representation that describes string states. 

Before turning to the full analysis it is worth noting that there is an obvious symmetry of the set $( \mathfrak{g}, \lambda, \langle \cdot,\cdot\rangle)$ that preserves the geometry of $C(\lambda)$, namely
\begin{align}
\label{eq:rescalingroots}
    \langle \cdot,\cdot\rangle' = \frac{1}{\gamma^2} \langle \cdot,\cdot\rangle \, , \quad 
    \lambda' = \gamma \lambda \, ,
\end{align}
such that $\lambda'$ is also an element of the weight lattice\footnote{Note that if $\cgg$ has more than one simple factor there is a more general symmetry where each factor has an independent scaling.}. By scaling $\lambda$ and the metric in opposite directions we preserve the length of the highest weight $|\lambda|^2=\langle \lambda,\lambda\rangle$ and hence the geometry of $C(\lambda)$. Thus in our analysis we can only expect to classify $( \mathfrak{g}, \lambda, \langle \cdot,\cdot\rangle)$ up to this symmetry and for any given allowed polytope, there is actually an infinite family of solutions (all at the same $d$). However there will always be a minimal representation $\lambda_{\rm min}$ such that there is no $\gamma<1$ such that $\gamma\lambda_{\rm min}$ lies in the weight lattice. The metric $\langle\cdot,\cdot\rangle_{\rm min}$ and hence the length of the longest roots $\langle r_{\rm long},r_{\rm long}\rangle_{\rm min}$ will be maximal in this case. Highest weights $\gamma \lambda_{\rm min}$ and metrics $\langle\cdot,\cdot\rangle= (1/\gamma^2)\langle\cdot,\cdot\rangle_{\rm min}$ with $\gamma\in\mathbb{N}$ then describe the full family of solutions. 

\begin{figure}[th]
	\centering
\def\cola{RedViolet}
\def\colb{blue}
\def\cold{teal}
\def\colc{black}
\def\radinner{0.75}
\def\fontLabels{\small}
\begin{tikzpicture}[scale=2.2,>=stealth]{
\coordinatesHexagon
\drawHexagon
\coordinate (O) at (0,0);
\foreach \x/\y/\z/\w/\r in {
B/\cola/1/$\lambda_{\mathbf{7}}$/1,
A/\cola/1/{}/1,
F/\cola/1/{}/1,
C/\cola/1/{}/1,
D/\cola/1/{}/1,
E/\cola/1/{}/1,
O/\cola/0/{}/\radinner}
{
\draw[dotted,thick,color=\y] ($0.5*(\x)$) circle (\z*0.5);
\draw[dotted,thick,color=\y,fill=\y,opacity=0.05] ($0.5*(\x)$) circle (\z*0.5);
\path (\x) node[circle, fill=\y, inner sep=\r*1.5pt]{};
\node[above right, xshift=0.1cm, yshift=0.1cm] at (\x) {\w};
};
\draw[->] (0,0)-- (0., 1.73)node[color=black,font=\fontLabels,
xshift=0.2cm, yshift=0.2cm]{};
\draw[->] (0,0)-- (0.5,0.866)node[color=black,font=\footnotesize,
xshift=0.2cm, yshift=0.2cm]{};
}
\end{tikzpicture}
\begin{tikzpicture}[scale=2.2,>=stealth]{
\coordinatesHexagon
\drawHexagon
\coordinate (O) at (0,0);
\coordinate (OA) at ($(O)!0.5!(A)$);
\coordinate (OB) at ($(O)!0.5!(B)$);
\coordinate (OC) at ($(O)!0.5!(C)$);
\coordinate (OD) at ($(O)!0.5!(D)$);
\coordinate (OE) at ($(O)!0.5!(E)$);
\coordinate (OF) at ($(O)!0.5!(F)$);
\coordinate (AB) at ($(A)!0.5!(B)$);
\coordinate (AF) at ($(A)!0.5!(F)$);
\coordinate (EF) at ($(E)!0.5!(F)$);
\coordinate (DE) at ($(E)!0.5!(D)$);
\coordinate (CD) at ($(C)!0.5!(D)$);
\coordinate (BC) at ($(B)!0.5!(C)$);
\foreach \x/\y/\z/\w/\r in {
B/\cola/1/$\lambda_{\mathbf{27}}$/1,
A/\cola/1/{}/1,
F/\cola/1/{}/1,
C/\cola/1/{}/1,
D/\cola/1/{}/1,
E/\cola/1/{}/1,
AB/\cold/0.866/{}/\radinner,
AF/\cold/0.866/{}/\radinner,
EF/\cold/0.866/{}/\radinner,
DE/\cold/0.866/{}/\radinner,
CD/\cold/0.866/{}/\radinner,
BC/\cold/0.866/{}/\radinner,
OA/\cold/0/{}/\radinner,
OB/\cold/0/{}/\radinner,
OC/\cold/0/{}/\radinner,
OD/\cold/0/{}/\radinner,
OE/\cold/0/{}/\radinner,
OF/\cold/0/{}/\radinner,
O/\cola/0/{}/\radinner}
{
\draw[dotted,thick,color=\y] ($0.5*(\x)$) circle (\z*0.5);
\draw[dotted,thick,color=\y,fill=\y,opacity=0.05] ($0.5*(\x)$) circle (\z*0.5);
\path (\x) node[circle, fill=\y, inner sep=\r*1.5pt]{};
\node[above right, xshift=0.1cm, yshift=0.1cm] at (\x) {\w};
};
\draw[->] (0,0)-- (0., 0.866)node[color=black,font=\fontLabels,
xshift=0.2cm, yshift=0.2cm]{};
\draw[->] (0,0)-- (0.25,0.433)node[color=black,font=\footnotesize,
xshift=0.2cm, yshift=0.2cm]{};
}
\end{tikzpicture}
\begin{tikzpicture}[scale=2.2,>=stealth]{
\coordinatesHexagon
\drawHexagon
\coordinate (O) at (0,0);
\coordinate (OAA) at ($(O)!0.33!(A)$);
\coordinate (OOA) at ($(O)!0.67!(A)$);
\coordinate (OBB) at ($(O)!0.33!(B)$);
\coordinate (OOB) at ($(O)!0.67!(B)$);
\coordinate (OCC) at ($(O)!0.33!(C)$);
\coordinate (OOC) at ($(O)!0.67!(C)$);
\coordinate (ODD) at ($(O)!0.33!(D)$);
\coordinate (OOD) at ($(O)!0.67!(D)$);
\coordinate (OEE) at ($(O)!0.33!(E)$);
\coordinate (OOE) at ($(O)!0.67!(E)$);
\coordinate (OFF) at ($(O)!0.33!(F)$);
\coordinate (OOF) at ($(O)!0.67!(F)$);
\coordinate (AAB) at ($(A)!0.33!(B)$);
\coordinate (ABB) at ($(A)!0.67!(B)$);
\coordinate (AAF) at ($(A)!0.33!(F)$);
\coordinate (AFF) at ($(A)!0.67!(F)$);
\coordinate (EEF) at ($(E)!0.33!(F)$);
\coordinate (EFF) at ($(E)!0.67!(F)$);
\coordinate (DDE) at ($(D)!0.33!(E)$);
\coordinate (DEE) at ($(D)!0.67!(E)$);
\coordinate (CCD) at ($(C)!0.33!(D)$);
\coordinate (CDD) at ($(C)!0.67!(D)$);
\coordinate (BBC) at ($(B)!0.33!(C)$);
\coordinate (BCC) at ($(B)!0.67!(C)$);
\coordinate (OAB) at ($0.67*(A)!0.5!(B)$);
\coordinate (OBC) at($0.67*(B)!0.5!(C)$);
\coordinate (OCD) at($0.67*(C)!0.5!(D)$);
\coordinate (ODE) at($0.67*(D)!0.5!(E)$);
\coordinate (OEF) at($0.67*(E)!0.5!(F)$);
\coordinate (OFA) at($0.67*(F)!0.5!(A)$);
\foreach \x/\y/\z/\w/\r in {
B/\cola/1/$\lambda_{\mathbf{77}}$/1,
A/\cola/1/{}/1,
F/\cola/1/{}/1,
C/\cola/1/{}/1,
D/\cola/1/{}/1,
E/\cola/1/{}/1,
ABB/\colb/0.882/{}/\radinner,
AAB/\colb/0.882/{}/\radinner,
AAF/\colb/0.882/{}/\radinner,
AFF/\colb/0.882/{}/\radinner,
EEF/\colb/0.882/{}/\radinner,
EFF/\colb/0.882/{}/\radinner,
DDE/\colb/0.882/{}/\radinner,
DEE/\colb/0.882/{}/\radinner,
CCD/\colb/0.882/{}/\radinner,
CDD/\colb/0.882/{}/\radinner,
BBC/\colb/0.882/{}/\radinner,
BCC/\colb/0.882/{}/\radinner,
OOA/\colb/0/{}/\radinner,
OOB/\colb/0/{}/\radinner,
OOC/\colb/0/{}/\radinner,
OOD/\colb/0/{}/\radinner,
OOE/\colb/0/{}/\radinner,
OOF/\colb/0/{}/\radinner,
OAA/\colb/0/{}/\radinner,
OBB/\colb/0/{}/\radinner,
OCC/\colb/0/{}/\radinner,
ODD/\colb/0/{}/\radinner,
OEE/\colb/0/{}/\radinner,
OFF/\colb/0/{}/\radinner,
OAB/\colb/0/{}/\radinner,
OBC/\colb/0/{}/\radinner,
OCD/\colb/0/{}/\radinner,
ODE/\colb/0/{}/\radinner,
OEF/\colb/0/{}/\radinner,
OFA/\colb/0/{}/\radinner,
O/\cola/0/{}/\radinner}
{\draw[dotted,thick,color=\y] ($0.5*(\x)$) circle (\z*0.5);
\draw[dotted,thick,color=\y,fill=\y,opacity=0.05] ($0.5*(\x)$) circle (\z*0.5);
\path (\x) node[circle, fill=\y, inner sep=\r*1.5pt]{};
\node[above right, xshift=0.1cm, yshift=0.1cm] at (\x) {\w};
};
\draw[->] (0,0)-- (0., 0.577)node[color=black,font=\fontLabels,
xshift=0.2cm, yshift=0.2cm]{};
\draw[->] (0,0)-- (0.167,0.289)node[color=black,font=\footnotesize,
xshift=0.2cm, yshift=0.2cm]{};
}
\end{tikzpicture}
\caption{
Convex hulls with respect to the fundamental weights basis for the $\mathbf{7}$, $\mathbf{27}$ and $\mathbf{77}$ irreducible representations of $\splitG2$ where the metric $\langle\cdot,\cdot\rangle$ in each case has been chosen so that length squared of the long roots are $6$, $\frac32$ and $1$ respectively, so that the three polytopes are identical. The minimal representation $\lambda_{\rm min}$ is the $\mathbf{7}$, which is also the only representation without additional weights on the edges.} 
\label{fig:nface}
\end{figure}
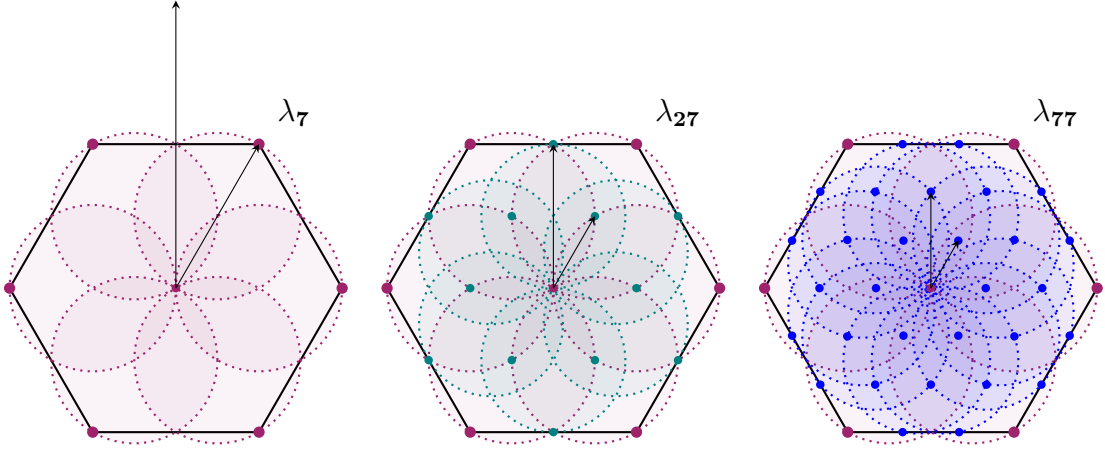

This situation is represented pictorially for $\splitG2$ in Figure~\ref{fig:nface}. Notice that the rescaling \eqref{eq:rescalingroots} increases the number of weights in the convex hull, both adding weights on the faces and inside the polytope\footnote{Moreover, the new weights may come with higher multiplicities.}. The exponential rates of the weights that do not lie on vertices are always subleading with respect to the ones that do, except when $H$ lies along the centroid of the face, in which case all the rates on the face become equal. 

One way to find the possible $( \mathfrak{g}, \lambda, \langle \cdot,\cdot\rangle)$ is to use the description of the faces of $C(\lambda)$ in terms of $\lambda$-extended Dynkin diagrams discussed in section~\ref{sec:Weylpolytopes}. The key point, discussed in Appendix~\ref{app:regandcross}, is that regular simplices and orthoplexes correspond to convex hull of only certain irreps, namely, using the $\lambda$-extended Dynkin diagram notation, those given by
\begin{align}
    \label{eq:simplex-dynkin}
    \text{simplex $\Delta_k$:} & \qquad \mathfrak{sl}(k+1,\mathbb{R})  \qquad
    \tikzset{/Dynkin diagram,root radius=.125cm,edge length=.55cm}
    \begin{dynkinDiagram}{A}{oo.oo}
    \node[circle,inner sep=0pt,minimum size=2.5mm,fill={\colExt},draw] (A) at ($(root 1) + (0,0.55)$) {}; 
    \draw[\colExt] (A) -- (root 1);
    \end{dynkinDiagram} \\*[7pt]
    \label{eq:orthoplex-dynkin}
    \text{orthoplex $\beta_k$:} & \qquad 
    \begin{cases}
    \mathfrak{so}(k,k, \mathbb{R})) & \qquad
    \tikzset{/Dynkin diagram,root radius=.125cm,edge length=.55cm}
    \begin{dynkinDiagram}{D}{oo.ooo}
    \node[circle,inner sep=0pt,minimum size=2.5mm,fill={\colExt},draw] (A) at ($(root 1) + (0,0.55)$) {}; 
    \draw[\colExt] (A) -- (root 1);
    \end{dynkinDiagram} \\*[3pt]
    \mathfrak{so}(k,k+1, \mathbb{R}) & \qquad
    \tikzset{/Dynkin diagram,root radius=.125cm,edge length=.55cm}
    \begin{dynkinDiagram}{B}{oo.ooo}
    \node[circle,inner sep=0pt,minimum size=2.5mm,fill={\colExt},draw] (A) at ($(root 1) + (0,0.55)$) {}; 
    \draw[\colExt] (A) -- (root 1);
    \end{dynkinDiagram} \\*[3pt]
    \mathfrak{sp}(2k,\mathbb{R}) & \qquad
    \tikzset{/Dynkin diagram,root radius=.125cm,edge length=.55cm}
    \begin{dynkinDiagram}{C}{oo.ooo}
    \node[circle,inner sep=0pt,minimum size=2.5mm,fill={\colExt},draw] (A) at ($(root 1) + (0,0.55)$) {}; 
    \draw[\colExt] (A) -- (root 1);
    \end{dynkinDiagram} 
    \end{cases}
\end{align}

In each case, we have $\lambda=\gamma\lambda_{\rm min}$ where $\lambda_{\rm min}$ is the fundamental vector representation of the Lie algebra. It is easy to see that any subset of roots connected to $\lambda$ in each case gives the $\lambda$-extended simplex diagram of lower rank -- that is, as required, all the faces of a simplex are simplices and all the faces of an orthoplex are also simplices. Note that for the low-rank $\splitDk$ examples the diagrams degenerate to\footnote{\label{foot:so22} In the second case the highest weight and the metric must be such that $\langle\lambda,r_i\rangle=\langle\lambda,r_j\rangle$ for the two roots so that the face is a square rather than a rectangle.}
\begin{equation}
\begin{aligned}
    \splitD3 &\simeq \mathfrak{sl}(4,\mathbb{R}) && \quad
    \tikzset{/Dynkin diagram,root radius=.125cm,edge length=.55cm}
    \begin{dynkinDiagram}{D}{ooo}
    \node[circle,inner sep=0pt,minimum size=2.5mm,fill={\colExt},draw] (A) at ($(root 2) + (0,0.55)$) {}; 
    \draw[\colExt] (A) -- (root 2);
    \end{dynkinDiagram} \\*[3pt]
    \splitD2 &\simeq \mathfrak{sl}(2,\mathbb{R}) \oplus \mathfrak{sl}(2,\mathbb{R}) && \quad
    \tikzset{/Dynkin diagram,root radius=.125cm,edge length=.55cm}
    \begin{dynkinDiagram}[edge/.style={\colExt}]{D}{ooo}
    \node[circle,inner sep=0pt,minimum size=2.5mm,fill={\colExt},draw] (A) at (root 2) {}; 
    \end{dynkinDiagram}
\end{aligned}   
\end{equation}

For the perturbative polytopes given in Table~\ref{tab:perturb_polytopes}, where the convex hull is either $\Delta_{r-1}$ or $\beta_{r-1}$, the possible pairs $(\mathfrak{g},\lambda)$ are then simply any of those in~\eqref{eq:simplex-dynkin} or~\eqref{eq:orthoplex-dynkin} with $k=r-1$ and a suitably chosen $\mathbb{R}$-charge $q$. These are listed in Table~\ref{tab:1irrep_integern_scaled_perturbative} together with $\mathfrak{g}$, $\rho_{\rm min}$ and $\dext$, giving, in particular, the theories that would correspond to the biggest possible normalization for $|r_{\text{long}}|$.

\begin{table}[ht]\renewcommand{\arraystretch}{1.2}
\centering
\begin{tabular}
{|@{\,}>{$}c<{$}@{\,}|@{\,}>{$}c<{$}@{\,}|>{$}c<{$}|@{\,}>{$}c<{$}@{\,}|
@{\,}>{$}c<{$}@{\,}|@{\,}>{$}c<{$}@{\,}|@{\,}>{$}c<{$}@{\,}|@{}>{$}c<{$}@{}|} \hline
\mathfrak{g} 
& \left|\tfrac{r_{\text{long}}}{\sqrt{2}}\right| \times  \lambda & \dext & n 
    & D_{\text{max}} 
& \rho_{\text{min}} & |r_{\text{long}}|^2_{\text{max}} & \text{ polytope } \\ \hline\hline

 \splitDk \oplus \mathbb{R}
&
\left(1,0_{k-1}\right) \Ucharge{\sqrt{\frac{1}{\dext-2}}} &
\multirow{3}{*}{$\dext$} & \multirow{3}{*}{$1,...,k,\infty$}&
\multirow{3}{*}{$\dext+k$}  
& \mathbf{2k} & 2 & \multirow{3}{*}{$\beta_{k}$}\\
  \splitBk \oplus \mathbb{R} & \left(1,0_{k-1}\right)\Ucharge{\sqrt{\frac{1}{\dext-2}}} &&&
  &\mathbf{2k+1} & 2
 & \\  
 \splitCk \oplus \mathbb{R}  
 &\left(\sqrt{2},0_{k-1}\right)\Ucharge{\sqrt{\frac{1}{\dext-2}}} &

&&
& \mathbf{2k} & 4
&\\\hline
\splitA1 \oplus \mathbb{R} &\left(\sqrt{2}\right)\Ucharge{\sqrt{\frac{1}{\dext-2}}} & \dext & 1,\infty & \dext+1 
 & 
 \mathbf{2} & 4& \multirow{2}{*}{$\protect\segScaled{1.6}$}\\\cline{1-7}
 \splitA1 \oplus \mathbb{R} &\left(1\right)\Ucharge{\sqrt{\tfrac{1}{\dext-2}+\frac{1}{2}}} & \dext & 1,2 & \dext+2 
 & 
 \mathbf{2} & 2 & \\\hline
 \splitAkminus1 \oplus \mathbb{R} 
 & \left( 1,0_{k-2}\right)
 \text{\tiny$\left( \sqrt{\frac{1}{\dext-2}+\frac{1}{k}}\right)$} &
 \dext & 1,...,k&\dext+k 
 & \mathbf{k} & 2 &  \Delta_{k-1} \\\hline
\end{tabular}
\caption{Algebras and irreducible representations realising the perturbative polytopes of Table~\ref{tab:perturb_polytopes}. $D_{\text{max}}=\dext+{\text{max}}\{n\}$ is the maximum number of dimensions that the theory can decompactify to and the last three columns show the smallest allowed  representation, largest length-squared of the long roots and the Weyl polytope. We also use the shorthand $\mathfrak{sl}_k=\mathfrak{sl}(k,\mathbb{R})$ and $\splitCk=\mathfrak{sp}(2k,\mathbb{R})$.}
\label{tab:1irrep_integern_scaled_perturbative}
\end{table}

For the global polytopes given in Tables~\ref{tab:global_polytopes} and~\ref{tab:extra_polytopes}, the possible pairs $(\mathfrak{g},\lambda)$ correspond to a $\lambda$-extended Dynkin diagram such that all the $\lambda$-extended Dynkin subdiagrams for the $(r-1)$-dimensional faces $F_I$ correspond to either a simplex as in~\eqref{eq:simplex-dynkin} or an orthoplex as in~\eqref{eq:orthoplex-dynkin}. For example consider the four-dimensional $\mathfrak{sp}(8,\mathbb{R})$ polytope
\begin{equation}
\begin{aligned}
    \tikzset{/Dynkin diagram,root radius=.125cm,edge length=.55cm}
    \begin{dynkinDiagram}{C}{oooo}
    \node[circle,inner sep=0pt,minimum size=2.5mm,fill={\colExt},draw] (A) at ($(root 2) + (0,0.55)$) {}; 
    \draw[\colExt] (A) -- (root 2);
    \end{dynkinDiagram} & : & && 
    F_{\{1,2,3\}} &= 
    \tikzset{/Dynkin diagram,root radius=.125cm,edge length=.55cm}
    \begin{dynkinDiagram}[edge/.style={white,thick}]{A}{***o}
    \node[circle,inner sep=0pt,minimum size=2.5mm,fill={\colExt},draw] (A) at ($(root 2) + (0,0.55)$) {}; 
    \draw[\colExt] (A) -- (root 2);
    \draw (root 1) -- (root 2) -- (root 3);
    \end{dynkinDiagram} &
    F_{\{2,3,4\}} &= 
    \tikzset{/Dynkin diagram,root radius=.125cm,edge length=.55cm}
    \begin{dynkinDiagram}{C}{o***}
    \node[circle,inner sep=0pt,minimum size=2.5mm,fill={\colExt},draw] (A) at ($(root 2) + (0,0.55)$) {}; 
    \draw[\colExt] (A) -- (root 2);
    \draw[white, thick] (root 1) -- (root 2);
    \end{dynkinDiagram}
\end{aligned}
\end{equation}
We see that both faces $F_{\{1,2,3\}}$ and $F_{\{2,3,4\}}$ are octahedra (3-orthoplexes),  and so this polytope is of the correct type. On the other hand 
\begin{equation}
\begin{aligned}
    \tikzset{/Dynkin diagram,root radius=.125cm,edge length=.55cm}
    \begin{dynkinDiagram}{C}{oooo}
    \node[circle,inner sep=0pt,minimum size=2.5mm,fill={\colExt},draw] (A) at ($(root 3) + (0,0.55)$) {}; 
    \draw[\colExt] (A) -- (root 3);
    \end{dynkinDiagram} & : & && 
    F_{\{1,2,3\}} &= 
    \tikzset{/Dynkin diagram,root radius=.125cm,edge length=.55cm}
    \begin{dynkinDiagram}[edge/.style={white,thick}]{A}{***o}
    \node[circle,inner sep=0pt,minimum size=2.5mm,fill={\colExt},draw] (A) at ($(root 3) + (0,0.55)$) {}; 
    \draw[\colExt] (A) -- (root 3);
    \draw (root 1) -- (root 2) -- (root 3);
    \end{dynkinDiagram} &
    F_{\{2,3,4\}} &= 
    \tikzset{/Dynkin diagram,root radius=.125cm,edge length=.55cm}
    \begin{dynkinDiagram}{C}{o***}
    \node[circle,inner sep=0pt,minimum size=2.5mm,fill={\colExt},draw] (A) at ($(root 3) + (0,0.55)$) {}; 
    \draw[\colExt] (A) -- (root 3);
    \draw[white, thick] (root 1) -- (root 2);
    \end{dynkinDiagram}
\end{aligned}
\end{equation}
fails because while $F_{\{1,2,3\}}$ is 3-simplex (tetrahedron), the face $F_{\{2,3,4\}}$ is neither a simplex nor an orthoplex. In these examples, $\lambda$ is connected to a single node and so is proportional to a fundamental weight. In general, it could be connected to more than one node, but since the only simplex or orthoplex of this type is the two-dimension simplex for $\splitD2$, this can only happen for $r\leq 3$. 

The final list of solutions is given in Table \ref{tab:1irrep_integern_scaled_global}, together with $\rho_{\rm min}$, giving, in particular, the theories that would correspond to the biggest possible normalization for $|r_{\text{long}}|$. (For the corresponding $\lambda$-extended Dynkin diagrams see Table~\ref{tab:lambda-Dynkin} in Appendix~\ref{app:regandcross}.) It is worth reemphasising that from Proposition~3 of \cite{Baines:2025upi}, this analysis constrains only the algebra $\mathfrak{g}$, and is independent of the particular arithmetic duality group $\Gamma \subset G$.

\begin{table}[ht!]\renewcommand{\arraystretch}{1.2}
\centering
\begin{tabular}
{|
@{\,}>{$}c<{$}@{\,}|
@{\,}>{$}c<{$}@{\,}|
>{$}c<{$}|
@{\,}>{$}c<{$}@{\,}|
@{\,}>{$}c<{$}@{\,}|
@{\,}>{$}c<{$}@{\,}|
@{\,}>{$}c<{$}@{\,}|
@{\,}>{$}c<{$}@{\,}|
} \hline
\mathfrak{g} 
& \left|\tfrac{r_{\text{long}}}{\sqrt{2}}\right| \times  \lambda & \dext & n 
    & D_{\text{max}}  
& \rho_{\text{min}} & |r_{\text{long}}|^2_{\text{max}} & \text{polytope} \\ \hline\hline
\splitA1 
&
\left(\sqrt{\tfrac{2}{\dext-2}}\right)
 & \dext & \infty & \dext
&
{\mathbf{2}} & {\frac{4}{\dext-2}} & \multirow{2}{*}{$\protect\segScaled{1.6}$}  \\
\cline{1-7}
\splitA1 
&
\left(\sqrt{2\tfrac{\dext-1}{\dext-2}}\right) & \dext & 1 & \dext+1 
&  
{\mathbf{2}} & 4\frac{\dext-1}{\dext-2} & \\\hline
\splitA2  
& (1,1) & \multirow{3}{*}{$3$} & \multirow{3}{*}{$1,2$} &\multirow{3}{*}{$5$} 
& {\mathbf{8}} & 2 & \multirow{3}{*}{\huge$\hexagon$} \\
\splitG2 
&
(0,1) & 
&& 
& {\mathbf{14}} & 2 
&\\
\splitG2  
&\hspace{-0.55cm} \sqrt{3}(1,0) & 
&& 
& {\mathbf{7}} & 6
&
\\\hline
\splitD2  
& \hspace{-0.55cm} \sqrt{2}(1;1) &\multirow{3}{*}{$3$} & \multirow{3}{*}{$1,\infty$}&\multirow{3}{*}{$4$}  
& 
{\mathbf{(2,2)}} & (4,4) & \multirow{3}{*}{\LARGE$\square$}  \\
\splitC2 & \hspace{-0.2cm} 2(1,0)&&&&  {\mathbf{4}} & 8 
& 
\\
\splitC2  
& \hspace{-0.55cm} \sqrt{2}(0,1) &&&
& 

{\mathbf{5}}& 4
&
\\\hline
\splitA2 
& \hspace{-0.55cm} 
\sqrt{2}(1,0) & 5 & 1,\infty & 6 
& 

{\mathbf{3}} & 4  & \text{\Large$\triangle$}
\\\hline
\mathfrak{sl}_2^{\oplus 2} &(1;\sqrt{2}) & 4 & 1,2,\infty & 6 & {(\mathbf{2},\mathbf{2})} & (2,4)&\text{\Large$\rectangle$}\\\hline
\mathfrak{sl}_2^{\oplus 3} 
&(1;1;1) & \multirow{5}{*}{$4$} & \multirow{5}{*}{$1,2,\infty$}&\multirow{5}{*}{$6$} 
&{(\mathbf{2},\mathbf{2},\mathbf{2})} & (2,2,2)&\multirow{5}{*}{$\cubeScaled{0.5}$}  \\

\mathfrak{sp}_4 \oplus \mathfrak{sl}_2
&(0,1;1) & 
&&
& {(\mathbf{5},\mathbf{2})}&
(2,2)&\\

\mathfrak{sp}_4 \oplus \mathfrak{sl}_2
&\hspace{-0.40cm} (\sqrt{2},0;1) &  
&& 
& {(\mathbf{4},\mathbf{2})} & 
(4,2)&\\
\splitB3  
& \hspace{-0.55cm} \sqrt{2} (0,0,1) & 
&&
&{\mathbf{8}} & 4
&\\
\splitC3
& (0,0,1) &  
&&
&{\mathbf{14'}} & 2
&\\\hline
\splitC3 
& \hspace{-0.55cm} \sqrt{2}(0,1,0) & \multirow{3}{*}{$3$} & \multirow{3}{*}{$1,2,3,\infty$}&\multirow{3}{*}{$6$} 
&{\mathbf{14}} & 4
&\multirow{3}{*}{$\cuboctahedronScaled{0.4}$} \\
\splitB3  
&(0,1,0) &  
&&  
& {\mathbf{21}} & 2
&\\
\splitA3
& (1,0,1) &  
&& 
& {\mathbf{15}} & 2&\\\hline
\splitE3 \cong \splitA1 \oplus \splitA2 & (1;0,1)& 8 & 1,2,3,\infty &11 &
{(\mathbf{2},\mathbf{3}^*)}
& (2,2) & \triprismScaled{0.4}\\\hline
\splitD4 
& (0,1,0,0) & \multirow{5}{*}{$3$} &  \multirow{5}{*}{$1,2,3,\infty$} &   \multirow{5}{*}{$6$} 
&{\mathbf{28}} & 2& \multirow{5}{*}{octacube}\\
\splitB4 
& (0,1,0,0) & && 
&{\mathbf{36}} & 2
& \\
\splitC4 
& \hspace{-0.55cm} \sqrt{2} (0,1,0,0) & 
&&
&{\mathbf{27}} & 4
&\\
\splitF4
& (1,0,0,0) &  
& 
& 
&{\mathbf{52}} & 2
&\\
\splitF4 &
 \hspace{-0.55cm} \sqrt{2}(0,0,0,1) & 
&&
&{\mathbf{26}} & 4
&\\\hline
\splitE4 \cong \splitA4 & (0,0,1,0) & 7 & 1,...,4,\infty & 11   & {\mathbf{10}^*} & 2 & \text{tetroctahedric}
\\\hline
\splitE5 \cong \splitD5 & (0,0,0,0,1)& 6 & 1,...,5,\infty&11   & {\mathbf{16}} & 2 & \text{demipenteract}\\\hline
\splitE6 & (1,0,0,0,0,0)& 5 & 1,...,6,\infty&11   & {\mathbf{27}} & 2 & V_{27}\\\hline
\splitE7 & (0,0,0,0,0,1,0)& 4 & 1,...,7,\infty&11   & {\mathbf{56}} & 2 & V_{56}\\\hline
\splitE8 & (0,0,0,0,0,0,1,0)& 3 & 1,...,8,\infty&11   & {\mathbf{248}} & 2 & \EeightAlt 
\\\hline
\end{tabular}
\caption{Algebras and irreducible representations realising the global polytopes of Table~\ref{tab:global_polytopes}. $D_{\text{max}}=\dext+{\text{max}}\{n\}$ is the maximum number of dimensions that the theory can decompactify to and the last three columns show the smallest allowed  representation, largest length-squared of the long roots for each simple factor, and the Weyl polytope. The ordering of the Dynkin labels follows the convention in Figure \ref{fig:DynkinDiagrams}.}
\label{tab:1irrep_integern_scaled_global}
\end{table}

As anticipated, all the $\lambda_{\rm min}$ representations are fundamental except in rank two and three, and even then, a consequence of the requirement mentioned in footnote~\ref{foot:so22}, only two possibilities appear: $\mathbf{15}$ of $\mathfrak{sl}(4,\mathbb{R})$ and the $\mathbf{8}$ of $\mathfrak{sl}(3,\mathbb{R})$. The Tables~\ref{tab:1irrep_integern_scaled_global} and~\ref{tab:1irrep_integern_scaled_perturbative} also naturally contain the toroidal compactifications of M- and string theory, respectively in the exceptional series $\mathfrak{e}_{k(k)}$ for $k=3,\dots,8$ and the orthogonal series $\mathfrak{so}(k,k) \oplus \mathbb{R}$. 

A few comments are due about the rank-1 polytopes in the tables. Clearly, if $\Gamma=SL(2,\mathbb{Z}$), the $\mathfrak{sl}(2,\mathbb{R})$ polytope ${}_\infty\protect\segScaled{1.6}_{\infty}$ can correspond to Type IIB in ten dimensions. In line with the discussion above, in this case there are no particles transforming under the duality group, but instead a doublet of strings (see in Table \ref{tab:G/K}), and so the $n=\infty$ ending points of the polytope do not correspond to particle weights\footnote{In fact, the physically-relevant analysis is of the string tensions, where the weights of the $\mathbf{2}$ representation would be twice the values here, corresponding to the tensions of the $F1$ and $D1$ strings that are exchanged under $SL(2)$.}. From now on, every time that we will mention this $n=\infty$ polytope as describing the $\mathbf{2}$ of $\mathfrak{sl}_2$, we are implicitly remembering that in type IIB this actually refers to emergent strings and not particles. Moreover, Type IIA in 10 dimensions and M-theory in 9 dimensions appear to be absent in the table\footnote{We will show how to obtain these polytopes `indirectly' in Section \ref{sec:tree}.}. The latter case is easily explained, as here the particles do not transform in a single irreducible representation but in the $\mathbf{2}\oplus\mathbf{1}$ of $\mathfrak{e}_{2(2)}$, so that they are not included in our classification. As for Type IIA, the polytope would be ${}_1\protect\segScaled{1.6}_{\infty}$, which can be seen as the non-perturbative completion of the trivial $k=0$ limit in the $\mathfrak{so}(k,k) \oplus \mathbb{R}$ series -- i.e. the sole $n=\infty$ point -- to the strong coupling region described by an M-theory frame, where the relevant tower would have $n=1$, again in a $\mathbb{R}$ representation. 

As listed in Table~\ref{tab:extra_polytopes}, allowing $\nvertex\geq 1$ gave two further polytopes -- another hexagon and an octagon.  The possible $( \mathfrak{g}, \lambda, \langle \cdot,\cdot\rangle)$ with minimal $\nvertex$ for these cases are given in Table~\ref{tab:1irrep_integern_extra}.

\begin{table}[ht!] \renewcommand{\arraystretch}{1.2}\centering\begin{tabular}
{|>{$}c<{$}|>{$}c<{$}|>{$}c<{$}|>{$}c<{$}|>{$}c<{$}|>{$}c<{$}|>{$}c<{$}|>{$}c<{$}|} \hline
\mathfrak{g} & \left|\tfrac{r_{\text{long}}}{\sqrt{2}}\right| \times  \lambda  & \dext & n & D_{\text{max}} & \rho_{\text{min}} & |r_{\text{long}}|^2 & \text{polytope} \\\hline\hline
\splitA2 &\sqrt{\tfrac23}(1,1)&\multirow{3}{*}{$3$} & \multirow{3}{*}{$3,\infty$}& \multirow{3}{*}{$6$}&
\mathbf{8} & \frac43 & \multirow{3}{*}{\huge$\hexagon$}\\
\splitG2  &\sqrt{\tfrac23}(0,1) & & & & \mathbf{14} & \frac43
&\\
\splitG2  &\sqrt{2}(1,0) & & & & \mathbf{7} & 4
&\\
\hline
\splitC2  & \tfrac{1}{\sqrt{2}}(1,1) & \multirow{2}{*}{$3$} &  \multirow{2}{*}{$4,8,\infty$} &  \multirow{2}{*}{$11$}&
\mathbf{16} & 1
&\multirow{2}{*}{\huge$\octagon$}\\
\splitC2  & \tfrac12(2,1) & & & &\mathbf{35} & \frac12 
&\\
\hline\end{tabular}\caption{Algebras and irreducible representations realising the global polytopes of Table~\ref{tab:extra_polytopes}. Theories that are grouped together give rise to the same Weyl polytope (see Table \ref{tab:polytopes_all1}). }
\label{tab:1irrep_integern_extra}
\end{table}

For each polytope we have seen that there are several realisations in terms of different choices of $\mathfrak{g}$, $\lambda_{\rm min}$ and $\langle\cdot,\cdot\rangle$. For completeness, we summarise in Table~\ref{tab:polytopes_all1} the possible Weyl polytopes and the corresponding duality groups and representations that give rise to them, allowing for arbitrary $\nvertex$. Here we are using the fact that if $( \mathfrak{g}, \lambda, \dext, \langle \cdot,\cdot\rangle)$ is a solution for a given polytope with the minimal allowed value of $\nvertex=\nvertex^\text{min}$, then $( \mathfrak{g}, \lambda, \dext', \langle \cdot,\cdot\rangle')$ is a solution for the same polytope with $\nvertex=z\nvertex^\text{min}$ (with $z\in\mathbb{N}$) where 
\begin{equation}
    \dext'=z(\dext-2)+2 \, , \qquad \langle \cdot,\cdot\rangle'=(1/z)\langle \cdot,\cdot\rangle \, , 
\end{equation}
as can be seen from~\eqref{eq:dn2} and~\eqref{eq:distance_zz}.  

\begin{table}[t]\renewcommand{\arraystretch}{1.2}\centering\small{\begin{tabular}{|>{$}c<{$}|>{$}c<{$}|>{}c<{}|>{$}c<{$}|>{$}c<{$}|} \hline r & \dext & $\text{polytope}_{\{n_I\}}$  &  \mathfrak{g} &  z_{\text{max}} \\ \hline\hline
\multirow{2}{*}{1} & \multirow{2}{*}{$\dext$}& $\protect\seg_z$ (line  segment) &  \splitA1  & 11-\dext \\ 
 &  & $\IIB_\infty$ (line segment)& \splitA1  & - \\ \hline 
\multirow{6}{*}{2} & 2+z &$\hexagon_{z,2z}$ (hexagon)  & \splitG2,\splitA2  & 3 \\
 & 2+z & $\square_{z,\infty}$ (square) & \splitC2,\splitD2  & 4 \\
 & 2+3z & $\triangle_{z,\infty}$ (triangle) & \splitA2 & 2 \\
 & 2+2z & $\rectangle_{z,2z,\infty}$ (rectangle)  & \splitA1^{\oplus 2} & 2 \\
 & 2+z & $\hexagon_{3z,\infty}$ (hexagon$'$) & \splitG2,\splitA2 & 2 \\
 & 2+z & $\octagon_{4z,8z,\infty}$ (octagon)  & \splitC2  & 1 \\ 
 \hline \multirow{3}{*}{3} & 2+2z & $\cube_{z,2z,\infty}$ (cube) & \splitC3,\splitB3,\splitC2\oplus \splitA1,\splitA1^{\oplus 3}  &  2 \\
 & 2+z & $\cuboctahedron_{z,2z,3z,\infty}$ (cuboctahedron) & \splitC3,\splitB3,\splitA3 & 2  \\
 & 2+6z & $\triprism_{z,2z,3z,\infty}$ (triangular prism) & \splitE3 \cong \splitA1 \oplus \splitA2 & 1 \\ 
 \hline \multirow{2}{*}{4} & 2+z & ${\Ffour}_{z,2z,3z,\infty}$ (octacube) & \splitF4,\splitC4,\splitB4,\splitD4 &  2  \\ 
 & 2+5z & ${\Efour}_{z,2z,3z,4z,\infty}$ (tetroctahedric) &  \splitE4  \cong \splitA4 & 1  \\ \hline
5 & 2+4z & ${\Efive}_{z,2z,\dots,5z,\infty}$ (demipenteract) & \splitE5 \cong \splitD5 & 1  \\ \hline
6 & 2+3z & ${\Esix}_{z,2z,\dots,6z,\infty}$ ($V_{27}$) & \splitE6 & 1   \\ \hline
7 & 2+2z & ${\Eseven}_{z,2z,\dots,7z,\infty}$ ($V_{56}$) & \splitE7 & 1 \\ \hline
8 & 2+z & ${\Eeight}_{z,2z,\dots,8z,\infty}$ ($\EeightAlt$) & \splitE8 & 1 \\ \hline\hline
\multirow{2}{*}{2}  & \multirow{2}{*}{$\dext$} & $\segconic_{z,2z}$ (line segment) &  \splitA1 \oplus \mathbb{R} & 
\frac{11-\dext}{2} \\
 & & $\segconic_{z,\infty}$ (line segment) & \splitA1 \oplus \mathbb{R} &  11-\dext  \\ \hline
\multirow{2}{*}{3}  & \multirow{2}{*}{$\dext$} & $\triangleconic_{z,2z,3z}$ (triangle) &  \splitA2 \oplus \mathbb{R} &  \frac{11-\dext}{3}
 \\
  & & $\squareconic_{z,2z,\infty}$ (square) & \left(\splitC2,\splitD2\right) \oplus \mathbb{R} & \frac{11-\dext}{2} \\ \hline
\multirow{2}{*}{$k+1$}  & \multirow{2}{*}{$\dext$} & $(\Delta_{k})_{z,2z,\dots,(k+1)z}$ ($k$-simplex) &  \splitAk \oplus  \mathbb{R}& \frac{11-\dext}{k+1}\\
  & & $(\beta_{k})_{z,2z,\dots,kz,\infty}$ ($k$-orthoplex) 
  & \left(\splitCk,\splitBk,\splitDk\right) \oplus \mathbb{R} &  \frac{11-\dext}{k} \\ \hline
\end{tabular}}
\caption{Allowed theories, their Weyl polytopes and their possible choices of algebra $\mathfrak{g}$, for $\nvertex=z$ and maximum value $z_{\text{max}}$ such that the theory decompactifies to at most $11$ dimensions. 
Their representations are listed in Tables \ref{tab:1irrep_integern_scaled_perturbative}, \ref{tab:1irrep_integern_scaled_global} and \ref{tab:1irrep_integern_extra}. The subscripts $z,2z,...$ indicate the number of directions decompactifying at the centre of 0-, 1-, 2-, etc dimensional faces of the polytope. For the perturbative polytopes this includes the centre of the whole polytope itself, which we denote by including a point at the centre of the polytope.}\label{tab:polytopes_all1}
\end{table}

\section{Families of polytopes and trees of embeddings}
\label{sec:tree}

In Section \ref{sec:SDC}, we provided a list of all the polytopes that encode the mass decay rates of KK towers and the Sharpened Distance Conjecture, assuming a single irreducible representation for particles, and then all their possible group-theoretic realisation. 

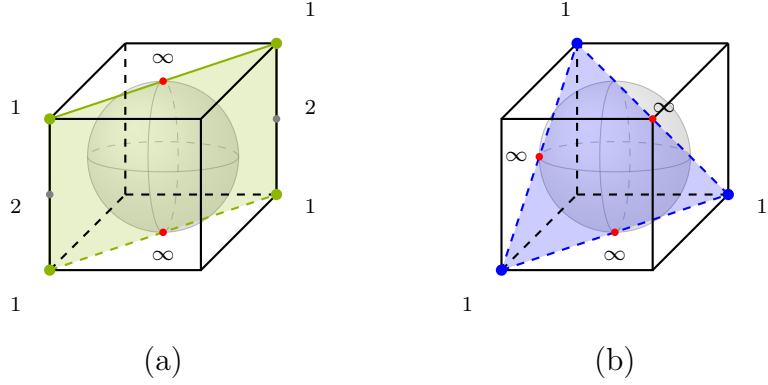
\begin{figure}[ht]
\def\colone{blue}
\def\coltwo{applegreen}
\def\colthree{cyan}
\def\colinf{Orchid}
    \centering
    \subfigure{\begin{tikzpicture}[scale=2]
\coordinatesCube
\coordinatesCubeExtra
\def\rad{0.5}\def\shiftx{0}\def\shifty{0}\drawSphere
\draw[thick, \colb] (B') -- (C') -- (D) -- (A);
\draw[thick, dashed, \colb] (A) -- (B') ;
\fill[\colb, opacity=0.2] (A) -- (B') -- (C') -- (D) -- cycle;
\drawCube
\foreach \i/\j in {A/A,D/D,B'/B',C'/C'}{\path ($0.5*(\i)+0.5*(\j)$) node[circle, fill=\colb, inner sep=1.5pt]{};}
\foreach \i/\j in {A/B',D/C'}{\path ($0.5*(\i)+0.5*(\j)$) node[circle, fill=red, inner sep=1.0pt]{};}
\foreach \i/\j in {A/D,B'/C'}{\path ($0.5*(\i)+0.5*(\j)$) node[circle, fill=gray, inner sep=1.0pt]{};}
\coordinate (Z) at (0,0);
\foreach \i/\j in {A/A,D/D,B'/B',C'/C'}{\path  ($1.3*0.5*(\i)+1.3*0.5*(\j)-0.3*(Z)$)  node[font=\scriptsize]{$1$};};
\foreach \i/\j in {A/B',D/C'}{\path  ($1.3*0.5*(\i)+1.3*0.5*(\j)-0.3*(Z)$)  node[font=\scriptsize]{$\infty$};};
\foreach \i/\j in {A/D,B'/C'}{\path  ($1.3*0.5*(\i)+1.3*0.5*(\j)-0.3*(Z)$)  node[font=\scriptsize]{$2$};};
\node[below=5pt of current bounding box.south, text width=2cm, align=center] 
            {(a)};
\end{tikzpicture}}\qquad \qquad 
\subfigure{\begin{tikzpicture}[scale=2]
\def\opafill{0.2}
\coordinatesCube
\def\rad{0.5}
\def\shiftx{0}
\def\shifty{0}
\drawSphere
\draw[\cola,dashed, thick] (A) -- (B') -- (D') -- cycle;
\fill[\cola,opacity=\opafill] (A) -- (B') -- (D') -- cycle;
\drawCube
\coordinate (Z) at (0,0);
\def\dlabels{0.3}
\def\mult{1.3}
\foreach \i/\j in {A/A,B'/B',D'/D'}{\path ($0.5*(\i)+0.5*(\j)$) node[circle, fill=\cola, inner sep=1.5pt]{};};
\foreach \i/\j in {A/B',B'/D',D'/A}{\path ($0.5*(\i)+0.5*(\j)$) node[circle, fill=red, inner sep=1.0pt]{};};
\foreach \i/\j  in {A/A,B'/B',D'/D'}{ \path  ($\mult*0.5*(\i)+\mult*0.5*(\j)-\dlabels*(Z)$) node[font=\scriptsize]{$1$};};
\foreach \i/\j  in {A/B',B'/D',D'/A}{ \path  ($\mult*0.5*(\i)+\mult*0.5*(\j)-\dlabels*(Z)$) node[font=\scriptsize]{$\infty$};};
\node[below=5pt of current bounding box.south, text width=2cm, align=center] 
            {(b)};
\end{tikzpicture}} 
    \caption{Examples of slices (a) passing through the centre of the original convex hull and (b) not passing through the centre, but orthogonal to a vertex.}
    \label{fig:exampleslices}
\end{figure}

Here we want to stress that -- as was already seen at the level of the polytopes in \cite{Etheredge:2024tok} -- these theories are actually related. A theory $(\mathfrak{g}',\lambda',d',\langle\cdot,\cdot\rangle)$ can be a descendent of the theory  $(\mathfrak{g},\lambda,d,\langle\cdot,\cdot\rangle)$ in two different ways:
\begin{itemize}
    \item the descendent theory arises from embedding the moduli space $\mathcal{M}'\subset\mathcal{M}$ as a totally geodesic submanifold, and the external dimension is unchanged ($d'=d$),
    \item the descendent theory arises as a decompactification, $\mathcal{M}'$ forms a ``boundary symmetric space'' of $\mathcal{M}$, and the external dimension increases ($d'>d)$.
\end{itemize}
In each case, the descendent polytope $C'$ lives on a slice of the original polytope $C$. (By a slice we mean an $\mathbb{R}^{r'}$ subspace of the $\mathbb{R}^r$ weight space in which $C$ lives.) For an embedded theory, the slice passes through the centre of $C$, and $C'$ is the convex hull of the projection of $C$ onto the slice, while for a decompactification the slice does not pass through the centre, but, as we will see, is orthogonal to one or more vertices, and $C'$ is the convex hull of the subset of the vertices of $C$ lying in the slice. A simple example of these two possibilities is given in Figure~\ref{fig:exampleslices}. We see that the cube $C$ admits a slice through its centre where $C'$ is a rectangle, and a slice orthogonal to back-left-lower vertex where $C'$ is a triangle.

We will discuss the details of how these two ways of obtaining descendent theories arise in the next two subsections, focussing on the global polytopes. For now, to summarise the result, we find that all the global polytopes of Table~\ref{tab:polytopes_all1} are in fact descendents of the $\splitE8$ polytope $V_{240}$, the theories arranging in a `tree' structure, in which the descendent $\mathfrak{g}'$ is a (regular or special) subalgebra of $\mathfrak{g}$. This is shown in Figure~\ref{fig:graph_poly_allranks}, where dashed arrows indicate embeddings and coloured arrows decompactifications. 

\def\cola{cyan}
\def\colb{applegreen}
\def\colc{blue}
\def\cold{teal}
\def\cole{orange}
\def\colf{RedViolet}
\def\colg{gray}
\begin{figure}[t!]\centering
\begin{tikzpicture}[scale=1.0,>=stealth]
\node (2) at (-2.0 , 15.0) {$\cubeScaled{0.4}$};
\node (4) at (1.0 , 15.0) {\Large$\rectangle$};
\node (13) at (-2.0 , 12.5) {\Large$\triangle$};
\node (24) at (0.5 , 14.0) {${\text{\Large{$\square$}}_{2}}$};
\node (5) at (1.0 , 17.0) {$\cuboctahedronScaled{0.3}$};
\node (1) at (3.0 , 17.0) {{\LARGE$\hexagon$}};
\node (14) at (2.0 , 16.5) {\Large{$\square$}};
\node (6) at (-2.0 , 17.0) {$\text{octa}$};
\node (18) at (-0.5 , 16.5) {$\text{\LARGE{$\hexagon$}}^{\infty}_{3}$};
\node (7) at (-3.5 , 7.0) {$\triprismScaled{0.45}$};
\node (23) at (-6.0 , 7.0) {$\text{\Large$\triangle$}_{2}$};
\node (8) at (-3.5 , 8.5) {$\text{tetro}$};
\node (9) at (-3.5 , 10.0) {$\text{demi}$};
\node (17) at (-6.0 , 10.0) {${\cubeScaled{0.4}_{2}}$};
\node (26) at (-8.0 , 9.5) {$\text{\Large{$\square$}}_{4}$};
\node (10) at (-3.5 , 12.0) {$\EsixAlt$};
\node (16) at (-2.0 , 11.5) {${\text{\LARGE{$\hexagon$}}_{3}}$};
\node (25) at (2.0 , 12.0) {${\text{\Large{$\square$}}_{3}}$};
\node (11) at (-3.5 , 15.0) {$V_{56}$};
\node (22) at (-6.0 , 15.0) {${\text{octa}_{2}}$};
\node (12) at (-3.5 , 17.0) {$\EeightAlt$};
\node (3) at (-5.0 , 17.0) {\LARGE$\octagon$};
\node (20) at (-8.5 , 10.5) {${\text{\Large$\rectangle$}_{2}}$};
\node (21) at (-8.5 , 14.0) {$\cuboctahedronScaled{0.3}_2$};
\node (15) at (-10.0 , 13.5) {$\text{\LARGE{$\hexagon$}}_{2}$};
\node (19) at (-7.0 , 13.5) {$\text{\LARGE{$\hexagon$}}^{\infty}_{6}$};
\node (27) at (-3.5 ,5.5) {${\isotriangleNoCircleScaled{0.3}}$};

\node (28) at (-7.0,4.0) {${}_{\infty}\protect\segScaled{1.4}_{\infty}$};
\node (29) at (-3.5,4.0) {${}_{1}\protect\segScaled{1.4}_{\infty}$};
\node (30) at (-8.5,7.0) {${}_{2}\protect\segScaled{1.4}_{\infty}$};
\node (31) at (-10.0,10.0) {${}_{2}\protect\segScaled{1.4}_{2}$};
\node (32) at (-5.0 , 8.5){${}_{4}\protect\segScaled{1.4}_{\infty}$};
\node (33) at (-2.0 , 7.0){${}_{3}\protect\segScaled{1.4}_{3}$};
\node (34) at (2.0 , 7.0){${}_{\infty}\protect\segScaled{1.4}_{\infty}$};
\node (35) at (2.0 , 13.5) {${}_{\infty}\protect\segScaled{1.4}_{\infty}$};
\node (36) at (3.0 , 15.0) {${}_{1}\protect\segScaled{1.4}_{1}$};
\node (37) at (1.0 , 12.5) {${}_{1}\protect\segScaled{1.4}_{\infty}$};
\node (38) at (-0.5 , 10.0) {${}_{\infty}\protect\segScaled{1.4}_{\infty}$};

\draw[->,dashed] (20) -- (38);
\draw[->,dashed] (26) -- (38);
\draw[->,color=\cold] (18) -- (38);
\draw[->,color=\colf] (24) -- (38);
\draw[->,color=\cola] (13) -- (38);

\draw[->,dashed] (4) -- (36);
\draw[->,color=\colb] (4) -- (37);

\draw[->,dashed] (4) -- (36);
\draw[->,color=\cole] (1) -- (36);

\draw[->,dashed] (4) -- (35);
\draw[->,dashed] (24) -- (35);
\draw[->,dashed] (19) -- (35);
\draw[->,color=\colf] (14) -- (35);

\draw[->,color=\colf] (25) -- (34);

\draw[->,color=\cole] (16) -- (33);
\draw[->,dashed] (7) -- (33);

\draw[->,color=\colg] (3) -- (32);
\draw[->,dashed] (8) -- (32);

\draw[->,color=\cole] (15) -- (31);
\draw[->,dashed] (20) -- (31);

\draw[->,dashed] (23) -- (30);
\draw[->,color=\colb] (20) -- (30);

\draw[->,color=\cola] (23) -- (28);
\draw[->,color=\colf] (26) -- (28);
\draw[->,color=\cold] (19) -- (28);
\draw[->,color=\colc] (27) -- (28);
\draw[->,color=\colc] (27) -- (29);

\draw[->,dashed] (2) -- (4);
\draw[->,color=\cola] (2) -- (13);
\draw[->,dashed] (2) -- (24);
\draw[->,dashed] (5) -- (1);
\draw[->,color=\colb] (5) -- (4);
\draw[->,dashed] (5) -- (14);
\draw[->,color=\cola] (6) -- (2);
\draw[->,dashed] (6) -- (5);
\draw[->,dashed] (6) -- (18);
\draw[->,dashed] (7) -- (23);
\draw[->,color=\colc] (8) -- (7);
\draw[->,color=\colc] (9) -- (8);
\draw[->,dashed] (9) -- (17);
\draw[->,color=\colc] (10) -- (9);
\draw[->,dashed] (10) -- (13);
\draw[->,dashed] (10) -- (16);
\draw[->,dashed] (10) -- (25);
\draw[->,dashed] (11) -- (2);
\draw[->,color=\colc] (11) -- (10);
\draw[->,dashed] (11) -- (22);
\draw[->,dashed] (12) -- (3);
\draw[->,dashed] (12) -- (6);
\draw[->,color=\colc] (12) -- (11);
\draw[->,dashed] (17) -- (20);
\draw[->,color=\cola] (17) -- (23);
\draw[->,dashed] (17) -- (26);
\draw[->,dashed] (21) -- (15);
\draw[->,color=\colb] (21) -- (20);
\draw[->,dashed] (21) -- (24);
\draw[->,color=\cola] (22) -- (17);
\draw[->,dashed] (22) -- (19);
\draw[->,dashed] (22) -- (21);
\draw[->,color=\colc] (7) -- (27);

 \node at (4,4.0) {$\dext =10$};
 \node at (4,5.5) {$\dext =9$};
 \node at (4,7.0) {$\dext =8$};
  \node at (4,8.5) {$\dext =7$};
  \node at (4,10.0) {$\dext =6$};
\node at (4,12.0) {$\dext =5$};
 \node at (4,14.5) {$\dext =4$};
\node at (4,16.5) {$\dext =3$};

\end{tikzpicture}\caption{Polytopes with no Abelian factors (except for $\protect\isotriangleNoCircleScaled{0.2}$ and ${}_{n}\protect\segScaled{1.4}_{\infty}$) and their descendents. Vertical arrows indicate decompactification limits, with each colour corresponding to a different polytope family. Dashed arrows indicate embeddings with the same value of $\dext$. Only the ones with $z\leq z_{\text{max}}$ are shown, such that the corresponding theories cannot be decompactified to more than $\dext=11$. For clarity, we show only those rank-1 polytopes which arise as decompactification limits.}
\label{fig:graph_poly_allranks} 
\end{figure}

In giving this tree we have focussed only on the descendent theories that correspond to polytopes of irreducible representations as listed in Table~\ref{tab:polytopes_all1} and so this is not an exhaustive list of all descendent polytopes. In particular, there are descendents that correspond to reducible representations. An example would be the isosceles triangle in $d=9$, obtained from the triangular prism as a decompactification slice as:
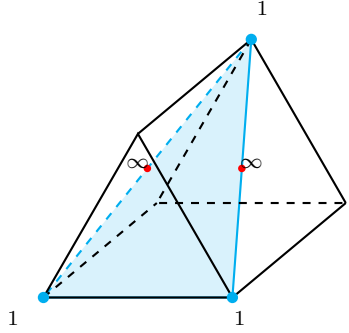
\begin{figure}[H]
    \centering
    \begin{tikzpicture}[scale=2.5]
\def\opaedges{1}
\def\opafill{0.15}
\coordinatesTriangularPrism
\coordinatesTriangularPrismExtra
\draw[\cola,thick,opacity=\opaedges](A) -- (B) -- (C');
\draw[\cola,dashed,thick,opacity=\opaedges] (C') -- (A);
\fill[\cola,opacity=\opafill] (A) -- (B) -- (C') -- cycle;
\drawTriangularPrism

\def\dlabels{0.2}
\def\mult{1.2}
\foreach \i/\j in {A/A,B/B,C'/C'}{\path ($0.5*(\i)+0.5*(\j)$) node[circle, fill=\cola, inner sep=1.5pt]{};};
\foreach \i/\j in {B/C',A/C'}{\path ($0.5*(\i)+0.5*(\j)$) node[circle, fill=red, inner sep=1.0pt]{};};
\foreach \i/\j  in {A/A,B/B,C'/C'}{ \path  ($\mult*0.5*(\i)+\mult*0.5*(\j)-\dlabels*(Z)$) node[font=\scriptsize]{$1$};};
\foreach \i/\j  in {B/C',A/C'}{ \path  ($\mult*0.5*(\i)+\mult*0.5*(\j)-\dlabels*(Z)$) node[font=\scriptsize]{$\infty$};};
\end{tikzpicture}
\end{figure}
\noindent (This is the one case of a reducible representation that, for convenience, we included in the tree in Figure~\ref{fig:graph_poly_allranks} since it corresponds to M-theory on $T^2$.)  An exhaustive list of reducible slices can be obtained, but we will leave it for future work. We also note that in principle we could include  the perturbative hulls in the tree of Figure~\ref{fig:graph_poly_allranks} provided they have a base with dimension\footnote{As we have seen eight is the maximum rank for a global polytope.} $r\leq7$. Indeed, these are trivially slices of appropriate global polytopes coincident with some face. Perturbative hulls whose base has dimension $r \geq 8$ clearly cannot be seen as such slices of allowed global hulls.

Finally, it is important to stress that the fact that the polytope $C'$ is a descendent of $C$ does not imply that every theory $(\mathfrak{g}',\lambda',d',\langle\cdot,\cdot\rangle)$ with polytope $C'$ is a descendent of every theory $(\mathfrak{g},\lambda,d,\langle\cdot,\cdot\rangle)$ with polytope $C$. First, it may be that $\mathfrak{g}'$ is not a subalgebra of $\mathfrak{g}$, and second the decomposition of the $\lambda$ representation under $\mathfrak{g}'$ may not contain the $\lambda'$ representation. In particular, even though the polytope $\EeightAlt$ associated to the $\mathbf{248}$ of $\splitE8$ contains all the others that appear in Table~\ref{tab:polytopes_all1}, it is not true that one can obtain all the theories in Table~\ref{tab:1irrep_integern_scaled_global} and~\ref{tab:1irrep_integern_extra} as descendents of the $\mathbf{248}$ of $\splitE8$. In other words, there are theories with a symmetric moduli space satisfying the Sharpened Distance Conjecture that do \emph{not} come from M-theory on tori. We will see explicit examples of this in the next two subsections. 

\subsection{Descendent polytopes: embeddings} 
\label{sec:embeddings}

The first step to seeing how a theory $(\mathfrak{g}',\lambda',d',\langle\cdot,\cdot\rangle)$ can descend from a theory $(\mathfrak{g},\lambda,d,\langle\cdot,\cdot\rangle)$ is to understand how the moduli spaces are related. The simplest possibility is that they embed $\mathcal{M}'\subset\mathcal{M}$ as a totally geodesic submanifold -- that is, geodesics on $\mathcal{M}'$ lift to geodesics on $\mathcal{M}$. We make this requirement because we want to be able to properly `separate' geodesic motion along $\mathcal{M}'$ from ones in the full space, and we can consistently `freeze' the moduli to live only on $\mathcal{M}'$. Since both spaces are locally symmetric and of non-compact type, the problem is to understand when one such space embeds as a totally geodesic space inside another such space. As mentioned above, in this case  $d=d'$ since we are simply freezing some of the moduli at finite values onto the submanifold $\mathcal{M}'\subset\mathcal{M}$, and the metric on $\mathcal{M}'$ is just the pull-back of the metric on $\mathcal{M}$ (and we hence continue to denote it as $\langle\cdot,\cdot\rangle$). 

For the corresponding globally symmetric spaces $\mathcal{S}' = G'/K'$ and $\mathcal{S} = G/K$, each assumed to be without a flat factor, the answer\footnote{More generally, totally geodesic submanifolds of globally symmetric spaces are in one-to-one correspondence with Lie triple systems, and have been studied fairly extensively in the mathematical literature (see for example \cite{berndt2020indexconjecturesymmetricspaces,kollross2023totallygeodesicsubmanifoldsexceptional}).\label{foot:Lie-triple}} is given by the \emph{Karpelevich--Mostow Theorem}~\cite{Karpelevich,mostow}: for any connected semi-simple subgroup $G' \subset G$, the orbit $G'\cdot p$ of any $p\in\mathcal{S}$ is a totally geodesic global symmetric space of the form $\mathcal{S}'=G'/K'$. For the corresponding locally symmetric spaces $\mathcal{M}'\subset\mathcal{M}$ we also require that $\Gamma'=G'\cap \Gamma$ is arithmetic. This will be true provided $G'$ is a subgroup of $G$ as a rational algebraic group, that is, $G'(\mathbb{Q})\subset G(\mathbb{Q})$. The boundaries then embed $\partial\mathcal{M}'\subset\partial\mathcal{M}$ such that a point $(P'(\mathbb{Q}),H')\in\partial\mathcal{M}'$ maps to the point $(P(\mathbb{Q}),H')\in\mathcal{M}$ with $P'(\mathbb{Q})\subset P(\mathbb{Q})$ and $H'\in \caa_{P'(\mathbb{Q}),1}^+\subset \caa_{P(\mathbb{Q}),1}^+$. This means embedded descendents  $(\mathfrak{g}',\lambda',d,\langle\cdot,\cdot\rangle)$ are such that $\mathfrak{g}'$ is semi-simple and the external dimension $d$ and metric $\langle\cdot,\cdot\rangle$ are unchanged. 

For the polytopes to match up we need a further condition. The Cartan subalgebra $\mathfrak{a}'$ of $\mathfrak{g}'$ defines a plane (or ``slice'') $\mathfrak{a}'\subset\mathfrak{a}$, passing through the centre of the polytope $C$. The polytope $C'$ of the embedded theory is the \emph{projection} of $C$ onto the $\caa'$ plane, since this is what contains the local extrema of restriction of the $\alpha$-hull to the $\mathfrak{a}'$ plane. For $C'$ to correspond to an EFT in the list of Table~\ref{tab:polytopes_all1}, we need that to further require that $C'$ is the Weyl polytope of some irreducible representation of $\cgg'$. In terms of representations, the irrep $\rho$ of $\cgg$, defined by the highest weight $\lambda$, decompose into irreps of $\cgg'$
\begin{equation}
\label{eq:irrep-decomp}
    \rho = \rho'_1 \oplus \rho'_2 \oplus \dots \oplus \rho'_s \, , 
\end{equation}
associated to highest weights $\lambda'_i\in\mathfrak{a}'$. Thus the condition is that $C'$ is the Weyl polytope of some irrep $\rho'\in\{\rho'_1,\dots,\rho'_s\}$. Note that when this happens $C'$ is also equal to the intersection of $C$ with the $\caa'$ plane.\footnote{Interestingly, when $C'$ is not equal to the intersection of $C$ with the $\caa'$ plane,  it can violate the geometrical constraints imposed by matching the ESC rates, specifically by having non-integer $n$. This arises because $C$ includes decompactification limits that are off the $\caa'$ plane. We leave the study of this  phenomenon for future work.} 

By way of an example, consider the $d=3$ cuboctahedron polytope $C$ of $\splitA3$ corresponding to the minimal $\mathbf{15}$ representation (see Table~\ref{tab:1irrep_integern_scaled_global}) shown in Figure~\ref{fig:rank2slicesC3}. The shaded sphere is the Sharpened Distance Conjecture bound $\alpha\geq1/\sqrt{d-2}=1$ in this case. 
\def\cola{teal}
\def\colb{violet}
\def\colg{magenta}

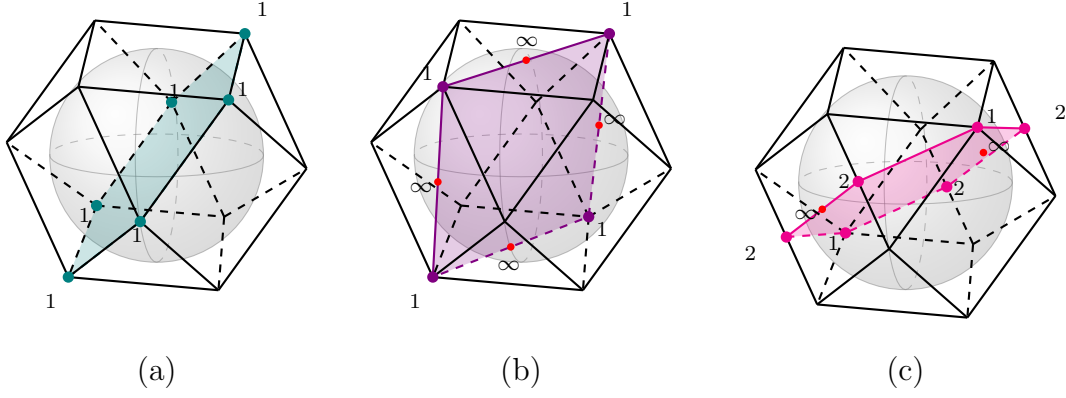
\begin{figure}[ht!]
\centering
\subfigure{\begin{tikzpicture}[scale=2]
\coordinatesCuboctahedronTilted
\def\rad{0.707}\def\shiftx{0}\def\shifty{0}\drawSphere;
\def\col{\cola}
\def\opafill{0.2}
\draw[\col,thick, dashed] (E) -- (G) -- (I) -- (B);
\fill[\col, opacity=\opafill]  (B) -- (J) -- (K) -- (E) -- (G) -- (I) -- cycle;
\drawCuboctahedron

\foreach \i/\j in {E/E,G/G,I/I,J/J,K/K,B/B}{\path ($0.5*(\i)+0.5*(\j)$) node[circle, fill=\col, inner sep=1.5pt]{};}
\foreach \i/\j in {E/E,G/G,I/I,J/J,K/K,B/B}{\path  ($1.2*0.5*(\i)+1.2*0.5*(\j)$)  node[font=\scriptsize]{$1$};}
\node[below=10pt of current bounding box.south, text width=2cm, align=center] 
            {(a)};
\end{tikzpicture}}\quad\quad
\subfigure{\begin{tikzpicture}[scale=2]
\coordinatesCuboctahedronTilted
\def\rad{0.707}\def\shiftx{0}\def\shifty{0}\drawSphere;
\def\opafill{0.2}
\def\col{\colb}
\draw[\col,thick, dashed] (B) -- (L) -- (E);
\draw[\col,thick] (E) -- (H) -- (B);
\fill[\col, opacity=\opafill] (B) -- (L) -- (E) -- (H) -- cycle;
\drawCuboctahedron
\foreach \i/\j in {B/B,L/L,E/E,H/H}{\path ($0.5*(\i)+0.5*(\j)$) node[circle, fill=\col, inner sep=1.5pt]{};}
\foreach \i/\j in {B/L,L/E,E/H,H/B}{\path ($0.5*(\i)+0.5*(\j)$) node[circle, fill=red, inner sep=1.0pt]{};}
\foreach \i/\j in  {B/B,L/L,E/E,H/H}{\path  ($1.2*0.5*(\i)+1.2*0.5*(\j)$)  node[font=\scriptsize]{$1$};}
\foreach \i/\j in  {B/L,L/E,E/H,H/B}{\path  ($1.2*0.5*(\i)+1.2*0.5*(\j)$)  node[font=\scriptsize]{$\infty$};}
\node[below=10pt of current bounding box.south, text width=2cm, align=center] 
            {(b)};
\end{tikzpicture}}\quad\quad\subfigure{\begin{tikzpicture}[scale=2]
\coordinatesCuboctahedronTilted
\def\rad{0.707}\def\shiftx{0}\def\shifty{0}\drawSphere;
\def\col{\colg}
\def\opafill{0.2}
\draw[\col,thick]($0.5*(A)+0.5*(B)$) -- (I) -- ($0.5*(G)+0.5*(H)$) -- ($0.5*(D)+0.5*(E)$);
\draw[\col,thick, dashed] ($0.5*(D)+0.5*(E)$) -- (K) -- ($0.5*(J)+0.5*(L)$) -- ($0.5*(A)+0.5*(B)$);
\fill[\col, opacity=\opafill] ($0.5*(A)+0.5*(B)$) -- (I) -- ($0.5*(G)+0.5*(H)$) -- ($0.5*(D)+0.5*(E)$) -- (K) -- ($0.5*(J)+0.5*(L)$) -- cycle;
\drawCuboctahedron

\foreach \i/\j in {A/B,I/I,G/H,D/E,K/K,J/L}{\path ($0.5*(\i)+0.5*(\j)$) node[circle, fill=\col, inner sep=1.5pt]{};}

\foreach \i/\j in {B/L,E/H}{\path ($0.5*(\i)+0.5*(\j)$) node[circle, fill=red, inner sep=1.0pt]{};}

\foreach \i/\j in  {I/I,K/K}{\path  ($1.2*0.5*(\i)+1.2*0.5*(\j)$)  node[font=\scriptsize]{$1$};}
\foreach \i/\j in  {A/B,G/H,D/E,J/L}{\path  ($1.3*0.5*(\i)+1.3*0.5*(\j)$)  node[font=\scriptsize]{$2$};}
\foreach \i/\j in  {B/L,E/H}{\path  ($1.2*0.5*(\i)+1.2*0.5*(\j)$)  node[font=\scriptsize]{$\infty$};}
\node[below=10pt of current bounding box.south, text width=2cm, align=center] 
            {(c)};
\end{tikzpicture}}
 \caption{Example of how we can get both single and multiple irrep slices from the same Weyl polytope $\protect\cuboctahedron_{1,2,3,\infty}$, all in $d=3$. (a) {\color{\cola}$\hexagon_{1,2}$} slice, (b) {\color{\colb}$\square_{1,\infty}$} slice and (c) hexagonal tower polytope with $n=\{1,2\}$ at the vertices.} \label{fig:rank2slicesC3} \end{figure}

Identifying rank-two subalgebras, we first have the regular subalgebra $\mathfrak{g}'=\splitA2\subset\splitA2\oplus\mathbb{R}\subset\splitA3$ under which we have the decomposition
\begin{equation}
    \mathbf{15} = \mathbf{8} \oplus \mathbf{3} \oplus \mathbf{3}^* \oplus \mathbf{1} \, ,
\end{equation}
corresponding to Figure~\ref{fig:rank2slicesC3}(a). The $\mathbf{8}$ and $\mathbf{1}$ representations lie in the green plane through the centre of $C$ and the $\mathbf{3}$ and $\mathbf{3}^*$ lie on the triangular faces above and below the plane. We see that $C'$ is simply the Weyl polytope of the $\mathbf{8}$ representation, that is the hexagon in Table~\ref{tab:1irrep_integern_scaled_global}. 

Under $\mathfrak{g}'=\splitD2\simeq\splitA1\oplus\splitA1$ we have two different embeddings, a special one and a regular one via $\splitA1\oplus\splitA1\subset\splitA1\oplus\splitA1\oplus \mathbb{R}\subset\splitA3$. Under the special embedding we have 
\begin{equation}
    \mathbf{15} = (\mathbf{3},\mathbf{3}) \oplus(\mathbf{3},\mathbf{1})\oplus(\mathbf{1},\mathbf{3}) \,, 
\end{equation}
and $C'$ is the violet square in Figure~\ref{fig:rank2slicesC3}(b),  formed by the Weyl polytope of the $(\mathbf{3},\mathbf{3})$ representation. This matches the $\splitA1\oplus\splitA1$ square with $\rho_\text{min}=(\mathbf{2},\mathbf{2})$ in Table~\ref{tab:1irrep_integern_scaled_global} but with $\lambda=2\lambda_{\text{min}}$ so that $\rho=(\mathbf{3},\mathbf{3})$ and $|r'_{\text{long}}|^2=\frac14\cdot4=1$. 

For the regular embedding we get the decomposition 
\begin{equation}
    \mathbf{15} = (\mathbf{3},\mathbf{1})  \oplus (\mathbf{1},\mathbf{3}) 
        \oplus 2\cdot (\mathbf{2},\mathbf{2}) 
        \oplus (\mathbf{1},\mathbf{1}) \,, 
\end{equation}
and again $C'$ is the violet square in Figure~\ref{fig:rank2slicesC3}(b). However now it is the convex hull of the $(\mathbf{3},\mathbf{1})$ and $(\mathbf{1},\mathbf{3})$ representations whose individual weights lie on the two diagonals. (The two $(\mathbf{2},\mathbf{2})$ representations are the squares above and below the violet plane, while $(\mathbf{1},\mathbf{1})$ lies at the centre of the plane.) In this case, we see that $C'$ is not defined by a single irrep and so the descendent theory is not one on our list. Even if we start with a different $\splitA3$ representation $\lambda=\gamma\lambda_{\text{min}}$, we find that $C'$ continues to be the convex hull of a reducible $\splitA1\oplus\splitA1$ representation. 

These last two examples stress an important point we already made above: the embedding of a polytope $C'$ as a slice is not enough to imply that that theory with polytope $C'$ is a descendent of the one with $C$. In this case the descent from cuboctahedron to square in the top row of Figure~\ref{fig:graph_poly_allranks} is possible with the minimal irrep of $\splitA3$ if we take $\mathfrak{g}'$ to be the special subgroup $\splitD2$ (and in the descendent theory we get a non-minimal representation as well), but if we take the regular $\splitD2$ subgroup, although the descendent theory still exists, it is not in our list because $C'$ is not the convex hull of an irrep. 

For completeness we also show in Figure~\ref{fig:rank2slicesC3}(c) the rank-two slice corresponding to $\mathfrak{g}'=\splitA1\oplus\mathbb{R}\subset\splitA1\oplus\splitA1\oplus \mathbb{R}\subset\splitA3$, under which we have 
\begin{equation}
    \mathbf{15} = \mathbf{3}_0 \oplus 2\cdot\mathbf{2}_2 \oplus 2\cdot\mathbf{2}_{-2} \oplus 4\cdot\mathbf{1}_0  \, .
\end{equation}
Since $\splitA1\oplus\mathbb{R}$ is not semi-simple, this is not guaranteed to define a totally geodesic submanifold. However, it is easy to show that this embedding defines a Lie triple system (see footnote~\ref{foot:Lie-triple}) and so $\mathcal{M}'$ is actually totally geodesic submanifold. Nevertheless, one can see that $C'$ is not the Weyl polytope of a single irrep. In fact it cannot be, since the only such examples for $\splitA1\oplus\mathbb{R}$ are the perturbative polytopes. 

In the examples of Figure~\ref{fig:rank2slicesC3}(a) and~\ref{fig:rank2slicesC3}(b) the vertices of $C'$ are also vertices of $C$. However this is not a requirement. As a simple counter example consider the $d=4$ cube of $\splitB3$ in the minimal $\mathbf{8}$ representation (see Table~\ref{tab:1irrep_integern_scaled_global}) shown in Figure~\ref{fig:embedCube}. Decomposing under $\cgg'=\splitD2\simeq\splitA1\oplus\splitA1$ we have 
\begin{equation}
    \mathbf{8} = (\mathbf{2},\mathbf{2}) \oplus (\mathbf{2},\mathbf{2}) \,,
\end{equation}
where the two $(\mathbf{2},\mathbf{2})$ are the front and back squares of the cube. Projecting onto the $\caa'$ slice we see that $C'$ is the green square passing through the centre of the cube. Note that none of the eight vertices of $C$ are vertices of $C'$. Instead the vertices are points on $C$ that correspond to decompactifying two dimensions. Thus $C'$ is the $d=4$ square of $\splitA1\oplus\splitA1$ in the $(\mathbf{2},\mathbf{2})$ representation with $z=2$ (see Table~\ref{tab:polytopes_all1}).  
\begin{figure}[h!]
\centering
\begin{tikzpicture}[scale=2]
\def\cola{blue}
\def\colb{applegreen}
\def\opafill{0.2}
\def\opaedges{1}
\def\sizebig{1.5pt}\def\sizesmall{1.0pt}
\coordinatesCube
\foreach \i/\j/\k in {A/A'/W,B/B'/X}{\coordinate[label=below:{\scriptsize$2$}] (\k) at ($0.5*(\i)+0.5*(\j)$);
\path ($0.5*(\i)+0.5*(\j)$) node[circle, fill=\colb, inner sep=\sizebig]{};};
\foreach \i/\j/\k in {C/C'/Y,D/D'/Z}{\coordinate[label={\scriptsize$2$}] (\k) at ($0.5*(\i)+0.5*(\j)$);
\path ($0.5*(\i)+0.5*(\j)$) node[circle, fill=\colb, inner sep=\sizebig]{};};
\draw[\colb, thick, opacity=\opaedges] (W) -- (X) -- (Y) -- (Z) -- cycle;
\fill[\colb, opacity=\opafill] (W) -- (X) -- (Y) -- (Z) -- cycle;
\drawCube
\end{tikzpicture}
\caption{\def\colb{applegreen}
  Embedding of {$C'=\color{\colb}\square_{2,\infty}$} in $C=\cube_{1,2,\infty}$. These polytopes can be realised by the EFTs $(\splitA1\oplus\splitA1,(\mathbf{2},\mathbf{2}),d=4)_{\nvertex=2}$ and $(\splitB3,{\bf 8},d=4)$ respectively.}
    \label{fig:embedCube}
\end{figure}
\noindent

\subsection{Descendent polytopes: decompactifications}
\label{sec:decompactifications}

For decompactification limits we again need a way of deriving a smaller, non-compact symmetric moduli space $\mathcal{M}'$ from the original moduli space $\mathcal{M}$. Rather than an embedding, this should be associated to a point on the boundary $(P(\mathbb{Q}),H)\in\partial\mathcal{M}$ that corresponds to a given decompactification limit. Intuitively $\mathcal{M}'$ should encode motion in the non-decompactified directions in $\mathcal{M}$, and can be thought of as ``blowing up'' the boundary point $(P(\mathbb{Q}),H)$ so as to keep track of these extra directions, moving within the equivalence class of geodesics associated to $(P(\mathbb{Q}),H)$.  

It turns out that such a notion already exists in the mathematical literature on compactifications of symmetric spaces~\cite{Borel} as we now describe. Consider a parabolic subgroup $P\subset G$. As reviewed in~\cite{Baines:2025upi}, the Lie algebra of $P$ has a Langlands decomposition
\begin{equation}
    \cpp \simeq \caa_P \oplus \cmm_P \oplus \cnn_P \, ,  
\end{equation}
where $\caa_P$ is the split component, $\cnn_P$ is nilpotent and $\cmm_P$ is reductive. The globally symmetric space $\mathcal{S}=G/K$ has a corresponding decomposition
\begin{equation}
    \mathcal{S} \simeq N_P A_P S_P \, , \qquad \text{with} \quad
    \mathcal{S}_P = M_P/K_P \, , 
\end{equation}
where $N_P=\ee^{\cnn_P}$, $A_P=\ee^{\caa_P}$, $A_P\times M_P$ is the centraliser of $A_P$ in $G$ and $K_P=M_P\cap K$ is its maximally compact subgroup. Although it can be viewed as a subspace of $\mathcal{S}$, the space $S_P$ is typically  referred to as the ``boundary symmetric space'' associated to $P$. We can then blow up the equivalence class of geodesics as we approach the boundary by keeping track of the geodesics position in $S_P$, that is the canonical geodesic is now
\begin{equation}
\label{eq:blowup-geodesic}
    \gamma_{H,s_P}(t) = \ee^{tH}\cdot s_P \,, \qquad \text{where $s_P\in \mathcal{S}_P\subset \mathcal{S}$} \text{ and } H\in \mathfrak{a}_{P,1}^+.
\end{equation}
Note that if $P$ is a minimal parabolic then $S_P$ is a point and so only non-minimal parabolics have such a blowup.\footnote{There is actually an alternative ``Satake compactification'' of $\mathcal{S}$ where one takes $\partial\mathcal{S}=\coprod_{P} \mathcal{S}_P$ rather than the union of $\caa^+_{P,1}$ as in~\eqref{eq:boundaryM}.} 

If the parabolic group is rational, denoted $P(\mathbb{Q})$, then $\Gamma_{P(\mathbb{Q})}=\Gamma\cap M_{P(\mathbb{Q})}$ is arithmetic and we can define a locally symmetric boundary space 
\begin{equation}
    \mathcal{M}' = \Gamma_{P(\mathbb{Q})}\backslash M_{P(\mathbb{Q})}/ K_{P(\mathbb{Q})} \, ,
\end{equation}
for each boundary point $(P(\mathbb{Q}),H)$, independent of $H\in\caa_{P(\mathbb{Q}),1}^+$. It is this space that we associate with the decompactification limit, so that $\cgg'=\cmm_{P(\mathbb{Q})}$. To see why this makes sense physically, we note that $\cmm_{P(\mathbb{Q})}$ is the part of the centraliser of $\caa_{P(\mathbb{Q})}$ that is orthogonal to $\caa_{P(\mathbb{Q})}$. (Since $\caa_{P(\mathbb{Q})}$ is Abelian, it is part of its own centraliser.) Thus the metric on $\mathcal{M}$ restricted to the space $\mathbb{R}\times\mathcal{M}'$ formed of geodesics of the form~\eqref{eq:blowup-geodesic} factorises
\begin{equation}
    \dd s^2 = \dd t^2 + \dd s^2(\mathcal{M}') \, ,
\end{equation}
where $\dd s^2(\mathcal{M}')$ is the symmetric space metric and, crucially, is independent of $t$ and also of the choice of $H\in\caa^+_{P(\mathbb{Q}),1}$. Thus $\mathcal{M}'$ is precisely the part of the moduli space that has a metric that is orthogonal to and independent of the geodesic motion towards the boundary. We also see that the metric on $\mathcal{M}'$ is just the pullback of the metric on $\mathcal{M}$ and hence continue to denote it as $\langle\cdot,\cdot\rangle$. 

For a Weyl polytope $C$ we have already seen that, for $k<r$, a $(k-1)$-dimensional face $F_I$ corresponds to decompactifying $kz$ dimensions where $z=n_\emptyset$ is the number of dimensions decompactified by each vertex (recall $\nvertex$ is the smallest number of dimensions one can decompactify). Furthermore, each face is associated to a particular parabolic subgroup $P(\mathbb{Q})$ as follows. The split component $\caa_{P(\mathbb{Q})}\subset\caa$ is the $k$-dimensional plane that contains the $k$-dimensional cone through the origin with base $F_I$, and the Cartan subalgebra $\caa'$ of $\cgg'=\cmm_{P(\mathbb{Q})}$ is the orthogonal $(r-k)$-dimensional plane. Equivalently $\caa'$ is orthogonal to all the vertices of $F_I$. It is worth stressing that since $\mathcal{M}'$ is independent of the choice of $H\in\caa_{P(\mathbb{Q}),1}^+$, every decompactification limit on the face $F_I$ gives the same decompactified moduli space $\mathcal{M}'$, that is 
\begin{equation}
    \text{faces $F_I$} \quad \stackrel{1:1}{\longleftrightarrow} \quad \text{decompactified moduli spaces $\mathcal{M}'\subset\mathcal{M}$. }
\end{equation}
In other words, we only keep track of how many different directions decompactify, not the rate at which they do so. As for the embedded moduli space, the highest weight representation $\rho$ of $G$ decomposes into irreducible highest weight representations $\rho'_i$ of $G'$ as in~\eqref{eq:irrep-decomp}. We can in addition grade the $\rho'_i$ by their weights under $\caa_{P(\mathbb{Q})}$, since this commutes with $\cgg'$. These live on different slices of $C$ parallel to the face $F_I$. The question is then whether there is any slice such that convex hull $C'$ formed by the intersection of $C$ with the slice is one in the list of~Table~\ref{tab:polytopes_all1} with the appropriate $\cgg'$ and external dimension $d'=d+k\nvertex$.\footnote{Notice that slices orthogonal to a weight but passing through the origin would still fall into the previous class. Indeed, the fact that they lie far from the origin is crucial to decompactification - because it forces to have a motion along the direction that we want to decompactify, contrary to freezing it when the slice passes through the origin.} 

To see how this works in an example, consider the $(\splitB3,{\bf 8},d=4)$ cube of Table~\ref{tab:1irrep_integern_scaled_global}, shown in Figure~\ref{fig:decompCube}. 
\begin{figure}[H]\centering
\subfigure{\begin{tikzpicture}[scale=2]
\def\cola{blue}
\def\colb{applegreen}
\def\opafill{0.2}
\def\opaedges{1}
\coordinatesCube
\draw[\cola,dashed, thick, opacity=\opaedges] (A) -- (B') -- (D') -- cycle;
\fill[\cola, opacity=\opafill] (A) -- (B') -- (D') -- cycle;
\draw[\colb, thick, opacity=\opaedges]    (D) -- (B) -- (C') --cycle;
\fill[\colb, opacity=\opafill] (D) -- (B) -- (C') -- cycle;
\drawCube
\path (C) node[circle, fill=black, inner sep=1.5pt]{};
\path (A') node[circle, fill=red, inner sep=1.5pt]{};
\end{tikzpicture}} \quad \quad
\subfigure{\begin{tikzpicture}[scale=2]
\def\cola{blue}
\def\opafill{0.2}
\coordinatesCube
\def\rad{0.5}
\def\shiftx{0}
\def\shifty{0}
\drawSphere
\draw[\cola,dashed, thick] (A) -- (B') -- (D') -- cycle;
\fill[\cola,opacity=\opafill] (A) -- (B') -- (D') -- cycle;
\drawCube
\coordinate (Z) at (0,0);
\def\dlabels{0.3}
\def\mult{1.3}
\foreach \i/\j in {A/A,B'/B',D'/D'}{\path ($0.5*(\i)+0.5*(\j)$) node[circle, fill=\cola, inner sep=1.5pt]{};};
\foreach \i/\j in {A/B',B'/D',D'/A}{\path ($0.5*(\i)+0.5*(\j)$) node[circle, fill=red, inner sep=1.0pt]{};};
\foreach \i/\j  in {A/A,B'/B',D'/D'}{ \path  ($\mult*0.5*(\i)+\mult*0.5*(\j)-\dlabels*(Z)$) node[font=\scriptsize]{$1$};};
\foreach \i/\j  in {A/B',B'/D',D'/A}{ \path  ($\mult*0.5*(\i)+\mult*0.5*(\j)-\dlabels*(Z)$) node[font=\scriptsize]{$\infty$};};
\end{tikzpicture}}\quad\quad
\subfigure{\begin{tikzpicture}[scale=2]
\def\cola{blue}
\def\colb{orange}
\def\opafill{0.2}
\def\opaedges{1}
\coordinatesCube
\def\rad{0.5}\def\shiftx{0}\def\shifty{0}\drawSphere
\draw[\cola,dashed, thick, opacity=\opaedges] (A) -- (B') -- (D') -- cycle;
\fill[\cola, opacity=\opafill] (A) -- (B') -- (D') -- cycle;
\draw[\colb,dashed, thick, opacity=\opaedges]    (D) -- (A') -- (C');
\draw[\colb,thick, opacity=\opaedges]  (C') -- (D);
\fill[\colb, opacity=\opafill] (D) -- (A') -- (C') -- cycle;
\draw[\colab,very thick]    ($0.25*(A)+0.25*(A')+0.25*(D)+0.25*(D')$) -- ($0.25*(A')+0.25*(B')+0.25*(C')+0.25*(D')$);
\drawCube

\coordinate (Z) at (0,0);
\def\dlabels{0.3}
\def\mult{1.3}
\foreach \i/\j in {A/A,B'/B',D'/D'}{\path ($0.5*(\i)+0.5*(\j)$) node[circle, fill=\cola, inner sep=1.5pt]{};};

\foreach \i/\j in {D/D,A'/A',C'/C'}{\path ($0.5*(\i)+0.5*(\j)$) node[circle, fill=\colb, inner sep=1.5pt]{};};

\foreach \i/\j in {A/D',B'/D'}{\path ($0.5*(\i)+0.5*(\j)$) node[circle, fill=\colab, inner sep=1.5pt]{};};

\foreach \i/\j  in {B'/D',D'/A}{ \path  ($\mult*0.5*(\i)+\mult*0.5*(\j)-\dlabels*(Z)$) node[font=\scriptsize]{$\infty$};};

\end{tikzpicture}}
  \caption{
  Decompactification limits of $\cube_{1,2,\infty}$ to {\color{\cola}$\triangle_{1,\infty}$} and {\color{\colab}${}_\infty\IIB_{\infty}$}. These can be realized respectively by $(\splitB3,{\bf 8},d=4)$, $(\splitA2,{\bf 3},d=5)$ and $(\splitA1,{\bf 2},d=6)$.}
    \label{fig:decompCube}
\end{figure}
\noindent
Moving along a geodesic corresponding to the $F_\emptyset$ face (or rather vertex) coloured in red in the first figure, decompactifies a single dimension. The corresponding parabolic group is maximal, with $\caa_{P(\mathbb{Q})}$ one-dimensional and given by the line $\mathbb{R}$ through the vertex. The decompactified Lie algebra is $\cgg'=\splitA2$ and the Cartan subalgebra $\caa'$ spans the plane orthogonal to the line through the vertex. Decomposing under $\splitA2\oplus\mathbb{R}$ we have 
\begin{equation}
    \mathbf{8} = \mathbf{1}_3 \oplus \mathbf{3}_1 \oplus \mathbf{3}^*_{-1} \oplus \mathbf{1}_{-3} \, ,
\end{equation}
corresponding to the red vertex, the blue triangle, the green triangle and black vertex respectively. One finds that the vertices of the blue triangle are at a distance $\sqrt{\frac13+1}$ from the centre of the triangle, corresponding to $d'=5$ and hence give the decompactified theory as shown in the second figure. Note that the green triangle also satisfies this condition. However, it has negative weight under $\mathbb{R}$ and as we discuss below in Section~\ref{sec:bdc} this means it cannot correspond to particles. One can then further decompactify by taking slices orthogonal to the upper-back-left vertex. This gives the rank-one polytope in cyan which is the intersection of the blue and orange triangles. It can equivalently be thought of as decompactifying relative to the one-dimensional face given by the edge connecting the upper-back-left and lower-back-left vertices. 

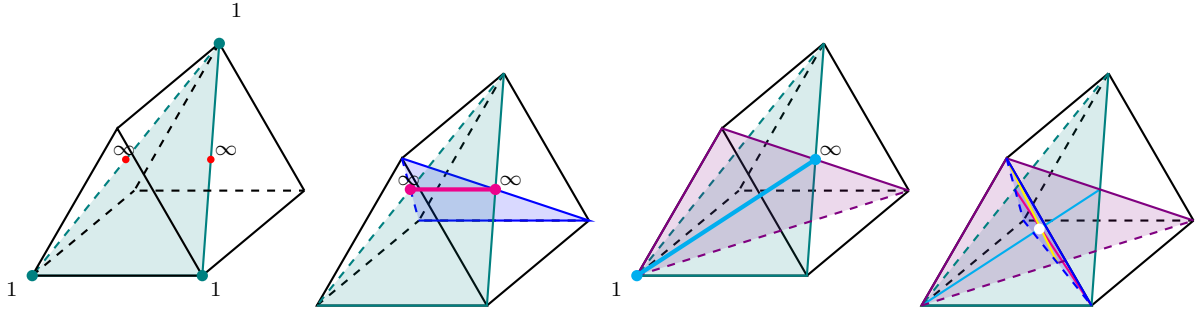
\begin{figure}[H]
\centering
\subfigure{
\begin{tikzpicture}[scale=2.25]
\def\opaedges{1}
\def\opafill{0.15}
\coordinatesTriangularPrism
\coordinatesTriangularPrismExtra
\draw[\cola,thick,opacity=\opaedges](A) -- (B) -- (C');
\draw[\cola,dashed,thick,opacity=\opaedges] (C') -- (A);
\fill[\cola,opacity=\opafill] (A) -- (B) -- (C') -- cycle;
\drawTriangularPrism

\def\dlabels{0.15}
\def\mult{1.2}
\foreach \i/\j in {A/A,B/B,C'/C'}{\path ($0.5*(\i)+0.5*(\j)$) node[circle, fill=\cola, inner sep=1.5pt]{};};
\foreach \i/\j in {B/C',A/C'}{\path ($0.5*(\i)+0.5*(\j)$) node[circle, fill=red, inner sep=1.0pt]{};};
\foreach \i/\j  in {A/A,B/B,C'/C'}{ \path  ($\mult*0.5*(\i)+\mult*0.5*(\j)-\dlabels*(Z)$) node[font=\scriptsize]{$1$};};
\foreach \i/\j  in {B/C',A/C'}{ \path  ($\mult*0.5*(\i)+\mult*0.5*(\j)-\dlabels*(Z)$) node[font=\scriptsize]{$\infty$};};
\end{tikzpicture}}
\subfigure{\begin{tikzpicture}[scale=2.25]
\coordinatesTriangularPrism
\coordinatesTriangularPrismExtra
\drawTriangularPrism
\def\opaedges{1}
\def\opafill{0.15}
\draw[\colc,thick,opacity=\opaedges] (A') -- (B') -- (C) ;
\draw[\colc,dashed, thick,opacity=\opaedges] (C) -- (A');
\fill[\colc, opacity=\opafill] (A') -- (B') -- (C) -- cycle;
\draw[\cola,thick,opacity=\opaedges] (A) -- (B) -- (C') ;
\draw[\cola,dashed, thick,opacity=\opaedges] (C') -- (A);
\fill[\cola, opacity=\opafill] (A) -- (B) -- (C') -- cycle;
\draw[\colac,ultra thick] ($0.25*(A)+0.25*(A')+0.25*(C)+0.25*(C')$) -- ($0.25*(B)+0.25*(B')+0.25*(C)+0.25*(C')$) ;

\def\dlabels{0.15}
\def\mult{1.2}
\foreach \i/\j in {B/C',A/C'}{\path ($0.5*(\i)+0.5*(\j)$) node[circle, fill=\colac, inner sep=1.5pt]{};};
\foreach \i/\j  in {B/C',A/C'}{ \path  ($\mult*0.5*(\i)+\mult*0.5*(\j)-\dlabels*(Z)$) node[font=\scriptsize]{$\infty$};};

\end{tikzpicture}}
\subfigure{\begin{tikzpicture}[scale=2.25]
\coordinatesTriangularPrism
\coordinatesTriangularPrismExtra
\drawTriangularPrism
\def\opaedges{1}
\def\opafill{0.15}
\draw[\cola,thick,opacity=\opaedges] (A) -- (B) -- (C') ;
\draw[\cola,dashed,thick,opacity=\opaedges] (C') -- (A);
\fill[\cola,opacity=\opafill] (A) -- (B) -- (C') -- cycle;
\draw[\colb,thick,opacity=\opaedges]  (B') -- (C) -- (A);
\draw[\colb,dashed,thick,opacity=\opaedges] (A) -- (B');
\fill[\colb,opacity=\opafill] (A) -- (B') -- (C) -- cycle;
\draw[\colab,ultra thick] (A) -- ($0.25*(B)+0.25*(B')+0.25*(C)+0.25*(C')$);

\def\dlabels{0.15}
\def\mult{1.2}
\foreach \i/\j in {A/A,B/C'}{\path ($0.5*(\i)+0.5*(\j)$) node[circle, fill=\colab, inner sep=1.5pt]{};};

\foreach \i/\j  in {B/C'}{ \path  ($\mult*0.5*(\i)+\mult*0.5*(\j)-\dlabels*(Z)$) node[font=\scriptsize]{$\infty$};};
\foreach \i/\j  in {A/A}{ \path  ($\mult*0.5*(\i)+\mult*0.5*(\j)-\dlabels*(Z)$) node[font=\scriptsize]{$1$};};
\end{tikzpicture}}
\subfigure{\begin{tikzpicture}[scale=2.25]
\def\opaedges{1}
\def\opafill{0.15}
\coordinatesTriangularPrism
\coordinatesTriangularPrismExtra
\drawTriangularPrism

\fill[\cola,opacity=\opafill] (A) -- (B) -- (C') -- cycle;
\fill[\colb,opacity=\opafill] (A) -- (B') -- (C) -- cycle;
\fill[\colc,opacity=\opafill] (B) -- (C) -- (A') -- cycle;

\draw[\colab, thick] (A) -- ($0.25*(B)+0.25*(B')+0.25*(C)+0.25*(C')$);
\draw[\colac, thick] (B) -- ($0.25*(A)+0.25*(A')+0.25*(C)+0.25*(C')$);
\draw[\colbc, thick] (C) -- ($0.25*(A)+0.25*(A')+0.25*(B)+0.25*(B')$);

\draw[\cola,thick,opacity=\opaedges] (A) -- (B) -- (C') ;
\draw[\cola,dashed,thick,opacity=\opaedges] (C') -- (A);
\draw[\colb,thick,opacity=\opaedges]  (B') -- (C) -- (A);
\draw[\colb,dashed, thick,opacity=\opaedges] (A) -- (B');

\draw[\colc,thick,opacity=\opaedges]  (B) -- (C);
\draw[\colc,dashed, thick,opacity=\opaedges] (C) -- (A') -- (B);

\def\tt{0.66}\def\ttt{1-\tt}
\coordinate (X) at ($\tt*0.25*(B)+\tt*0.25*(B')+\tt*0.25*(C)+\tt*0.25*(C')+\ttt*(A)$);
\foreach \pt in {X}{\fill[\colabc] (\pt) circle (0.9pt);};

\end{tikzpicture}}
    \caption{
    Decompactification limits of $\protect\triprism_{1,2,3,\infty}$ ($(\splitE3,{\mathbf{(2, 3)}},d=8)$) to {\color{\cola}$\protect\isotriangleNoCircleScaled{0.2}_{1,2,\infty} $}, {\color{\colac}${}_\infty\IIB_{\infty}^{10D}$},
{\color{\colab}${}_1\IIA_{\infty}^{10D}$} and 
    $\bullet^{11D}$ (white dot). The first two are realized by $(\splitE2,{\mathbf{2}}_{(-3/\sqrt{14})} \oplus {\mathbf{1}_{(4/\sqrt{14})}},d=9)$ and $(\splitA1,{\mathbf{2}},d=10)$ respectively. The isosceles triangle is not in our list in Tables~\ref{tab:1irrep_integern_scaled_global} and~\ref{tab:1irrep_integern_extra} as the particle content is not in a single irrep.}
    \label{fig:decompE3}
\end{figure}

Another illustrative example more familiar in string theory decompactification limits is that of the triangular prism in $d=8$, discussed also in \cite{Etheredge:2024tok}, corresponding to M-theory on $T^3$, with $\cgg=\splitE3 \simeq \splitA1 \oplus \splitA2$ and particles in the $(\mathbf{2},\mathbf{3})$ representation. Taking a slice orthogonal to a weight at any of the vertices, we get the isosceles triangle, as shown in Figure \ref{fig:decompE3}. This slice should be the polytope corresponding to the moduli space of compactifications of M-theory on $T^2$, with group $\splitE2\simeq\splitA1\oplus\mathbb{R}$. This polytope is however not in our list of two-dimensional polytopes, for the simple reason that the particles transform under more than one irreducible representation. Indeed, the $(\mathbf{2},\mathbf{3})$ representation of $\splitA1 \oplus \splitA2$ decomposes under the $\splitA1$ factor of $\splitE2$ as $\mathbf{1} \oplus \mathbf{1} \oplus  \mathbf{2} \oplus  \mathbf{2}$. The particles of the $\dext=9$ theory transform in a $\mathbf{1} \oplus \mathbf{2}$ representation, while the other representations correspond to higher-dimensional objects (M2-branes wrapping one cycle, and M2-branes not wrapping the $T^2$). We will come back to this in Section~\ref{sec:bdc}, and connect this to the $\mathbb{R}$ weight of the slice.

\begin{figure}[ht!]
	\centering
\def\cola{RedViolet}
\def\colb{blue}
\def\colc{applegreen}
\subfigure{
\begin{tikzpicture}[scale=0.9,>=stealth]
\coordinate (O) at (0,0);
\node[circle, fill=\cola, inner sep=2pt,label=90:\footnotesize $z$] (V) at (90:4cm) {};
\node[circle, fill=\colb, inner sep=2pt,label=45:{\footnotesize $z$}] (A) at (40:4cm) {};
\coordinate[label=above right:{\footnotesize $2z$}] (VA) at ($0.5*(V)+0.5*(A)$);
\coordinate (B) at (140:4cm);
\draw[->,\cola, very thick] (O) -- (V) node[midway, below left=-2pt, color=\cola] {$d_\emptyset$};
\draw[->,\colb, very thick] (O) -- (A) node[midway, below right=-2pt, color=\colb] {$d_\emptyset$};
\draw[\colc, very thick] (V) -- (A); 
\draw[->,\colc, very thick] (O) -- (VA) node[midway, right=1pt, color=\colc] {$d_{\{i\}}$};
\draw[->,very thick] ($0.5*(B)+0.5*(A)$) -- (A) node[midway, below right=0pt] {$d'_\emptyset$};
\end{tikzpicture}}
\qquad \qquad
\subfigure{
\begin{tikzpicture}[scale=0.9,>=stealth]
\coordinate (O) at (0,0);
\node[circle, fill=\cola, inner sep=2pt,label=90:\footnotesize $z$] (V) at (90:4cm) {};
\node[circle, fill=\colb, inner sep=2pt,label=45:{\footnotesize $z$}] (A) at (40:4cm) {};
\coordinate[label=above right:{\footnotesize $\infty$}] (VA) at ($0.5*(V)+0.5*(A)$);
\coordinate (B) at (140:4cm);
\coordinate (VB) at ($0.5*(V)+0.5*(B)$);
\draw[->,\cola, very thick] (O) -- (V) node[midway, below left=-2pt, color=\cola] {$d_\emptyset$};
\draw[->,\colb, very thick] (O) -- (A) node[midway, below right=-2pt, color=\colb] {$d_\emptyset$};
\draw[\colc, very thick] (V) -- (A); 
\draw[->,\colc, very thick] (O) -- (VA) node[midway, right=0pt, color=\colc] {$d_{\{i\}}$};
\draw[->,very thick] ($0.5*(VB)+0.5*(VA)$) -- (VA) node[midway, below=0pt] {$d'_\infty$};
\end{tikzpicture}}
\caption{Kaluza--Klein towers and emergent strings on the decompactified slice. }
\label{fig:d'}
\end{figure}
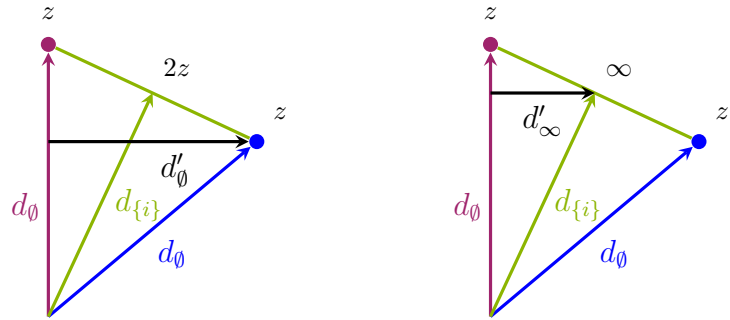

These two examples already capture the generic features of the decompactification limits. First that, at least for $r'>1$, the $C'$ polytope lies on a slice such that the vertices of $C'$ are also vertices of $C$. For $r=1$ the other possibility is that $C'$ is a segment ${}_{\infty}\protect\segScaled{1.4}_{\infty}$ or ${}_{n}\protect\segScaled{1.4}_{\infty}$ where one or both vertices give an emergent string limit. To see why this has to be the case, and also why there is always a slice with a suitable $d'=d+k\nvertex$, consider the left-hand diagram in Figure~\ref{fig:d'}. The face $F_{\{i\}}$ is an edge joining two vertices of $C$ and we assume that $r>2$, hence, as in~\eqref{norm_rho_n1} but allowing $z=\nvertex\geq1$,
\begin{equation}
   d_\emptyset^2 = \frac{1}{\dext-2} + \frac{1}{z} \, , \qquad 
   d^2_{\{i\}} = \frac{1}{\dext-2} + \frac{1}{2z} \, . 
\end{equation}
Suppose now that we decompactify $z$ dimensions in the direction along the violet vertex. The distance $d'_\emptyset$ of the blue vertex projected on the plane orthogonal to the violet vertex is given by the length of the black line. By similar triangles we have
\begin{equation}
    d'_\emptyset{}^2 = 4\bigl(d_\emptyset^2-d_{\{i\}}^2\bigr) \frac{d_{\{i\}}^2}{d_\emptyset^2} 
        = \frac{1}{\dext+z-2} + \frac{1}{z} \, , 
\end{equation}
which is precisely the value we need for decompactifying $z$ dimensions. Repeating this argument for $r=2$ and $d^2_{\{i\}}=1/(d-2)$ corresponding to an emergent string, we have the diagram on the right of Figure~\ref{fig:d'}, and find   
\begin{equation}
    d'^2_\infty = \bigl(d_\emptyset^2-d_{\{i\}}^2\bigr) \frac{d_{\{i\}}^2}{d_\emptyset^2} 
        = \frac{1}{\dext+z-2} \, , 
\end{equation}
again reproducing the value we need for decompactifying $z$ dimensions for a point representing an emergent string limit. We note that in both cases the slice containing the vertex or the emergent string point has a positive weight under the $\mathbb{R}$ subalgebra defined by the decompactifying the violet vertex. This is necessary in order to interpret the states as particles and strings, something we will discuss this in more detail in Section~\ref{sec:bdc}. 

Iterating this argument by decompactifying in the direction of the further vertices one finds that the particle hull $C'$ defined by decompactifying the face $F_I$ of dimension $k-1$ (with $k<r-1$) is given by 
\begin{equation}
    C' = \text{convex hull of vertices closest to $F_I$}
\end{equation}
where by ``closest'' we mean each vertex of $C'$ lies on a $k$-dimensional simplex whose other vertices all lie in $F_I$. If $k=r-1$ then $F_I$ is a $(r-2)$-dimensional simplex that is the common face of either two $(r-1)$-dimensional simplices or one simplex and one orthoplex or two orthoplexes. The one-dimensional $C'$ connects to the remaining simplex vertex or the centre of the orthoplex, giving ${}_{\nvertex}\protect\segScaled{1.4}_{\nvertex}$ or ${}_{\nvertex}\protect\segScaled{1.4}_{\infty}$ or ${}_{\infty}\protect\segScaled{1.4}_{\infty}$ respectively. 

As a further example, consider Figure~\ref{fig:decompCuboctahedron} representing the decompactification limits of the $d=3$ cuboctahedron in Table~\ref{tab:1irrep_integern_scaled_global}. Decompactifying with respect to the vertex facing the viewer gives the first diagram. The slice $C'$ closest to $F_\emptyset$ is the green rectangle, each vertex of which lies on an edge shared with the decompactification vertex. The second diagram shows a further decompactification via the vertex at the top left of the rectangle. The corresponding face $F_I$ is the edge joining the two decompactification vertices and $C'$ is the blue $\IIA_{1,\infty}$ interval closest to $F_I$. One sees that $F_I$ is the common face of a triangle (simplex) and a square (orthoplex) and the ``1'' vertex lies on the triangle and ``$\infty$'' vertex is at the centre of a square. The final diagram represents the full decompactification to six dimensions. 
\begin{figure}[H]\centering
\subfigure{\begin{tikzpicture}[scale=2]
\def\opafill{0.2}
\def\opaedges{1}
\coordinatesCuboctahedronTilted
\drawCuboctahedron
\def\rad{0.707}\def\shiftx{0}\def\shifty{0}\drawSphere;
\draw[\cola,thick,opacity=\opaedges](E) -- (F)  -- (I) -- (H) -- cycle;
\fill[\cola, opacity=\opafill](E) -- (F)  -- (I) -- (H) -- cycle;

\foreach \i/\j/\k in {E/E/E,F/F/F,I/I/I,H/H/H}{\path ($0.33*(\i)+0.33*(\j)+0.33*(\k)$) node[circle, fill=\cola, inner sep=1.7pt]{};}

\def\mult{1.2}

\foreach \i/\j/\k in {E/E/E,F/F/F,I/I/I,H/H/H}{\path ($\mult*0.33*(\i)+\mult*0.33*(\j)+\mult*0.33*(\k)$) node[font=\scriptsize]{$1$};}
\foreach \i/\j in {E/H,F/I}{\path ($0.5*(\i)+0.5*(\j)$) node[circle, fill=red, inner sep=1.3pt]{};}
\foreach \i/\j/\k in {E/H,F/I}{\path ($\mult*0.5*(\i)+\mult*0.5*(\j)$) node[font=\scriptsize]{$\infty$};}
\end{tikzpicture}}
\subfigure{\begin{tikzpicture}[scale=2]
\def\opafill{0.15}
\def\opaedges{1}
\coordinatesCuboctahedronTilted
\drawCuboctahedron
\def\rad{0.707}\def\shiftx{0}\def\shifty{0}\drawSphere;
\draw[\cola,thick,opacity=\opaedges](E) -- (F)  -- (I) -- (H) -- cycle;
\fill[\cola, opacity=\opafill](E) -- (F)  -- (I) -- (H) -- cycle;
\fill[\colb, opacity=\opafill] (C) -- (D)  -- (G) -- (I) -- cycle;
\draw[\colb,thick, opacity=\opaedges] (C) -- (D)  -- (G) -- (I) -- cycle;
\draw[\colab,ultra thick] ($0.5*(G)+0.5*(D)$) --  (I) ;

\foreach \i/\j in {G/D,I/I}{\path ($0.5*(\i)+0.5*(\j)$) node[circle, fill=\colab, inner sep=1.7pt]{};}
\foreach \i/\j in {G/D}{\path ($0.5*(\i)+0.5*(\j)$) node[circle, fill=red, inner sep=1.3pt]{};}

\def\mult{1.2}

\foreach \i/\j in {I/I}{\path ($\mult*0.5*(\i)+\mult*0.5*(\j)$) node[font=\scriptsize]{$1$};}
\foreach \i/\j in {G/D}{\path ($\mult*0.5*(\i)+\mult*0.5*(\j)$) node[font=\scriptsize]{$\infty$};}
\end{tikzpicture}}
\subfigure{\begin{tikzpicture}[scale=2]
\def\opafill{0.15}
\def\opaedges{1}
\coordinatesCuboctahedronTilted
\drawCuboctahedron
\def\rad{0.707}\def\shiftx{0}\def\shifty{0}\drawSphere;
\draw[\cola,thick,opacity=\opaedges](E) -- (F)  -- (I) -- (H) -- cycle;
\fill[\cola, opacity=\opafill](E) -- (F)  -- (I) -- (H) -- cycle;
\fill[\colb, opacity=\opafill] (C) -- (D)  -- (G) -- (I) -- cycle;
\draw[\colb,thick, opacity=\opaedges] (C) -- (D)  -- (G) -- (I) -- cycle;
\fill[\colc, opacity=\opafill] (A) -- (B)  -- (H) -- (G) -- cycle;
\draw[\colc,thick, opacity=\opaedges] (A) -- (B)  -- (H) -- (G) -- cycle;
\draw[\colab, thick]($0.5*(G)+0.5*(D)$) --  (I) ;
\draw[\colac, thick]($0.5*(G)+0.5*(A)$) --  (H) ;
\draw[\colbc, thick] ($0.5*(I)+0.5*(C)$) --  (G) ;
\fill[\colabc] ($0.333*(G)+0.083*(D)+0.25*(I)+0.083*(A)+0.167*(H)+0.083*(C)$) circle (1pt);
\end{tikzpicture}}
 \caption{
 Decompactification limits of $\protect\cuboctahedron_{1,2,3,\infty}$ to {\color{\cola}$\rectangle_{1,2,\infty}$}, {\color{\colab}$\IIA^{5D}_{1,\infty}$} and $\bullet^{6D}$.}
    \label{fig:decompCuboctahedron}
\end{figure}

\subsection{String theory interpretation}
\label{sec:stringinter}

In this section we make a few comments about the theories that we classified in light of string/M theory. 

First of all, as we have already mentioned, our list of theories includes, as it should, the standard known cases of $\mathfrak{e}_{d(d)}$ and $\splitDk \oplus \mathbb{R}$ with the correct string theoretic particle representation. Furthermore these have $|r_{\rm long}|^2=2$ and the representation is the minimal one. From these, although we cannot get examples of all the polytopes in Table~\ref{tab:polytopes_all1} as descendent theories, only a few cases are missing. This is done explicitly for the global polytopes in Appendix \ref{app:chains}. There we find all the possible embeddings of the theories in Table~\ref{tab:polytopes_all1} starting from a seed theory in the minimal representation and minimal $\nvertex$, for $d=3,4,5,6$ in Figures~\ref{fig:tree1}-\ref{fig:tree4} respectively. In particular, in Figure~\ref{fig:tree1}, all the theories connected by arrows to $\EeightAlt$ arise by restricting moduli to some totally geodesic submanifold of the moduli space of M-theory on $T^8$. However, it is also clear that there are theories that do not descend from the $T^8$ theory. Instead their `trees' are generated by three independent theories, namely 
\begin{equation*}
\begin{aligned}
   d=3\, : \quad  & (\splitG2, \mathbf{14})_{\nvertex=3} \, ,  \, (\splitC4, \mathbf{27})\, ,  \, (\splitF4, \mathbf{26}) \, . 
   \end{aligned}
\end{equation*}
It would clearly be interesting to see if these other seeds have realisations in string theory, or can be excluded by some other argument. It is notable that, while the first EFT has $|r_{\rm long}|^2=2$, the last two have $|r_{\rm long}|^2=4$. 

A further very striking fact is that although for the perturbative polytopes there was no limit on the space-time dimension, by contrast, for $r>1$
\begin{equation*}
    \text{for minimal $\nvertex$, all global polytope theories have $D_{max}\leq11$}
\end{equation*}
where for Table~\ref{tab:1irrep_integern_scaled_global} minimal means $\nvertex=1$ and for Table~\ref{tab:1irrep_integern_extra} it means $\nvertex=3$ for the hexagon and $\nvertex=4$ for the octagon. Thus all these theories could potentially be realised in M-theory. It is quite remarkable that the possible algebras $\cgg$ are bounded in this way: higher-dimensional polytopes satisfying the ESC rates are allowed but they cannot be realised as Weyl polytopes. 

Given these trees of descendent theories, it is natural to ask if the procedure of embedding moduli spaces can be realised as a generalised orbifolding of the seed $\mathfrak{e}_{r(r)}$ theory. Concretely, one can look for an orbifold action that lies in some discrete subgroup of the centraliser $\mathfrak{z}(\cgg)$ of $\cgg$ in $\mathfrak{e}_{r(r)}$. By construction this will leave the embedded moduli space invariant. (In principle it could leave a larger non-split algebra $\tilde{\cgg}\subset\mathfrak{e}_{r(r)}$ invariant, but with the same polytope.) One can also infer the number of preserved supercharges from the embedding $\cgg\subset\mathfrak{e}_{r(r)}$. Since in general the orbifolding would act by U-dualities, the resulting space would be a generalised `U-fold' and it would be a further check as to  whether these were actual string theory vacua (e.g.~if they were modular invariant).  

It is also interesting to look at full decompactification limits of the theories in the trees, that is, decompactifying all the non-compact moduli by progressively taking lower and lower rank slices until reaching rank 1 or 0. Each of the polytopes ending the decompactification chain corresponds to some $d$-dimensional theory. In particular, from Figure~\ref{fig:graph_poly_allranks}, we find two distinct types of such ending points:
\begin{itemize}
\item {\bf M-theory type}: these are rank-0 `trivial' polytopes corresponding to a moduli space without non-compact moduli.\footnote{Notice that `Type IIA' like segments, ${}_n\IIA_{\infty}$, can always be decompactified once more to an M-theory type theory.} They occur for $\dext = 11, 10, 8, 6, 5$ and are the decompactification limits of either non-regular polytopes with emergent string limits (rectangle, octagon) or regular polytopes without emergent string limit (hexagon).
\item {\bf Island type:} these are Type-IIB-like segments ${}_\infty\IIB_{\infty}$, with $n=\infty$ at both ends. They occur for $\dext = 10, 8, 6, 4$ and are the decompactification limits of all the regular polytopes presenting an emergent string limit, i.e. squares, triangles, hexagons. These theories only have one non-compact scalar which does does not lead to a decompactification limit but instead to an emergent string, and so behaves like the dilaton. We refer to the latter as `Island type' because they are the analogues of the list of `string island' theories found in~\cite{Dabholkar_1999}. Analysing the amount of preserved supersymmetry for these polytopes would allow us to see if some are in fact the same as those in~\cite{Dabholkar_1999}. 
\end{itemize}
It is also worth noticing that some of the polytopes allowed by the ESC rates do not feature an $n_I\rightarrow\infty$ limit. This corresponds to a moduli space with no emergent string limit, which we do not have any examples of standard string theory setups. These can nevertheless correspond to embedded slices of moduli space that simply do not include the dilaton (such as an S-fold~\cite{Garcia-Etxebarria:2015wns}) so that the full moduli space will have an emergent string limit. 

While we have seen that a large class of the allowed theories, at least with $\nvertex=1$ (or more generally with the minimal allowed $\nvertex$) fit within M-theory, one should emphasise that the full set of solutions is infinitely larger. For every theory in the lists in Table~\ref{tab:1irrep_integern_scaled_global} and~\ref{tab:1irrep_integern_extra}, we have infinitely many copies with bigger and bigger particle representations coming from the scaling invariance \eqref{eq:rescalingroots}. (We also have a further infinity of theories coming from allowing $\nvertex=z\nvertex^\text{min}$, unless one requires $D_{\text{max}}\leq 11$.) Though from a string-theoretic point of view the presence of these additional towers may look unusual, they are perfectly consistent mathematically. For trees of descendent theories given in Figures~\ref{fig:tree1}-\ref{fig:tree4}, we assumed we started with a seed in the minimal representation. Finding the corresponding trees for seeds in non-minimal representations is quite possible and, it is important to note, in general gives different results. However, the majority of these additional theories will not have a natural M-theory interpretation. 

One important point to note is that these additional theories all have smaller $|r_{\rm long}|^2=2$ than the $\lambda_{\text{min}}$ theories. Looking at Tables~\ref{tab:1irrep_integern_scaled_global} and~\ref{tab:1irrep_integern_extra} we see there is a universal bound
\begin{equation}
    |r_{\rm long}|^2 \leq 8 \, ,
\end{equation}
(and for all the perturbative polytopes and many of the global polytope theories we actually have $|r_{\rm long}|^2\leq 2$). This translates into a universal bound on the curvature of $\mathcal{M}$
\begin{equation}
    R_{ij} = - k g_{ij} \qquad k \geq \tfrac14 {h^\vee} \, ,
\end{equation}
where $h^\vee$ is the dual Coxeter number of the Lie algebra $\cgg$ (or of the semi-simple Lie algebra $\cgg'$ for the perturbative polytopes, for which the bound is $k \geq h^\vee$). As already noted in~\cite{Baines:2025upi}, in all supersymmetric theories with more than eight supercharges one has $|r_{\rm long}|^2=2$ (and so $k=h^\vee$), as do all symmetric space hypermultiplet theories, and vector multiplet theories have $|r_{\rm long}|^2\leq2$. Thus we immediately see that restricting to supersymmetric theories with at least eight supercharges already leads to a finite number of solutions (if in addition we require $D_{\text{max}}\leq11$). 

One might wonder if there is also a way to exclude the infinite families of theories by appealing to the  completeness of the spectrum~\cite{Polchinski_1996,Banks:2010zn,Harlow:2018tng,Heidenreich:2021xpr}. This conjecture states that, to be consistent with quantum gravity, a theory must contain states transforming in all finite-dimensional representations of the gauge group and is closely connected to the conjecture that quantum gravity admits no global symmetries. The latter implies, in this context, that  we should view the duality symmetries $\Gamma$ as spontaneously broken zero-form discrete gauge symmetries, and hence we must have a complete set of $\Gamma$ representations. Taking the particles in the SDC representation $\rho$ as being somehow fundamental, one can ask whether this representation can generate the complete set via tensor products. If not, this could restrict the allowed representations, for example to only fundamental representations (in the precise semi-simple Lie algebra sense).

As we now argue, completeness of the spectrum does not actually constrain the solutions, but rather defines which global form of the duality group we have. As we have already noted, the particle states entering the SDC transform in some representation $\rho$ which forms a faithful representation of (generically) some finite central extension $\tilde{\Gamma}$ of the moduli space duality group $\Gamma$. By~\cite{Margulis-rigidity,Corlette} representations of $\tilde{\Gamma}$ extend to representations of $\tilde{G}$ which is a central extension of the local isometry group $G$ of the symmetric space (which itself is centerless). In this sense, the representation $\rho$ defines the global form of the duality group $\tilde{\Gamma}$ and hence also $\tilde{G}$. The question then is whether $\rho$ generates all the other irreps of $\tilde{G}$ (and hence of $\tilde{\Gamma}$). However, by definition $\tilde{G}$ is algebraic and $\rho$ is faithful and so by standard theorems (see for example Prop.~2.20 of~\cite{Deligne1982}) generates all finite-dimensional representations via tensor products of the form $\rho^{\otimes p}\otimes (\rho^*)^{\otimes q}$. As a non-trivial example, consider M-theory on $T^7$. The particle representations is $\rho_{\text{p}}=\mathbf{56}$ of $\mathfrak{e}_{7(7)}$ while the string representation is $\rho_{\text{s}}=\mathbf{133}$. By definition, $E_{7(7)}$ is the group faithfully realised by $\rho_{\text{p}}$, while the group faithfully realised on $\rho_{\text{s}}=\mathbf{133}$ is $E_{7(7)}/\mathbb{Z}_2$ which is also the local isometry group of the locally symmetric space. Not all representations of $E_{7(7)}$ are then representations of $E_{7(7)}/\mathbb{Z}_2$, and indeed from tensor products of    $\mathbf{133}\simeq\mathbf{133}^*$ there is no way of obtaining the $\mathbf{56}$ representation. An important caveat to the discussion here is that all these representations are non-unitary since $G$ is non-compact. (Although they will form unitary representations of $K$.) For unitary representations of $G$ one would need to consider infinite-dimensional representations which takes us outside our original assumptions. 

\section{From particle polytopes to string and brane polytopes} \label{sec:bdc}

In Sections~\ref{sec:SDC} and~\ref{sec:tree}, we focused on the predictions that can be derived from the ESC in our setting, and by construction only analysed the particle content of the theory. Nevertheless, the fact that states should transform in representations of the duality group $\Gamma$ and hence also $G$, applies to string and brane states as well. Thus, using the same argument that led to the moduli-dependence of the mass formula for particles, the formula for the tension $\mathcal{T}$ of a set of $(p-1)$-brane states that become tensionless as one approaches the boundary must be associated to weights of some representation $\rho_{p}$ of $\cgg$ and there will be an associated weight polytope $C_{p}$. 

In this section, we will briefly analyse such higher-dimensional states and polytopes, and, in particular, the so-called  `$\pmax$-hull'~\cite{Etheredge:2024amg}. This is a (convex) polytope that is made of the leading charge-to-tension ratios of all the states in the theory with worldvolume of dimension up to $\pmax$, e.g.~$\pmax=2$ for particles and strings. In our language, it is the convex hull generated by the combined weights of the representations for the corresponding particle, string and brane states, that is 
\begin{equation}
    C^{\text{max}}_{p} = \conv( C_{1} \cup C_{2} \cup \dots \cup C_{p}) \, . 
\end{equation}
According to the Brane Distance Conjecture (BDC) the distances of the facets of such polytopes from the origin should satisfy~\eqref{eqn:BDC}, namely 
\begin{equation}
    \alpha_{\pmax} \geq \frac{1}{\sqrt{\dext-\pmax-1}} \, . 
\end{equation}

\subsection{String representations from emergent string limits}

The first and most straightforward check of the BDC is for the emergent string limits in the particle moduli space. As expected from the ESC rates, at these points the Sharpened Distance Conjecture bound is saturated, $\alpha_{\text{St}}=\sqrt{1/(d-2)}$, corresponding to towers of oscillators of weakly coupled and tensionless critical strings. Their mass scale is
\begin{equation} \label{mosc}
    m_{\text{osc}} \sim \mathcal{T}^{\frac{1}{2}} \, ,
\end{equation}
where $\mathcal{T}$ the tension of the string emerging along the specific limit expressed in $d$-dimensional Planck units. 

Let $\vec{\alpha}_{\text{St}}$ be the vector in weight space pointing to the emergent string point on the convex hull. As already noted, in general $\vec{\alpha}_{\text{St}}$ does not need to correspond to a weight of the particle representation, and indeed in our classification it never is for the minimal representation. However, the string states must arrange in representations of the duality group and hence of $G$. Thus, using the same argument that led to the moduli-dependence of the mass formula for particles,  the tension $\mathcal{T}$ of the string state must be associated to a weight of $\cgg$. In other words we must have 
\begin{equation}
    2\vec{\alpha}_{\text{St}} \in W \, , 
\end{equation}
where $W$ is the weight lattice of $\cgg$. For all our solutions we find this is indeed the case irrespective of the choice of representation $\rho$ and $z$. This is relatively easy to check explicitly, or alternatively one notes that each $\vec{\alpha}_{\text{St}}$ lies at the centre of an orthoplex. By definition (see Appendix~\ref{app:regandcross}) if $\omega$ is a vertex of an orthoplex, then $\sigma(\omega)$ is also a vertex, where $\sigma$ is the action of inversion through the centre of the orthoplex. Thus $\vec{\alpha}_{\text{St}}=\frac12(\omega+\sigma(\omega))$. But we also have by definition, for the particle polytope, $\omega,\sigma(\omega)\in W$ and hence $2\vec{\alpha}_{\text{St}} \in W$. 

By definition, the weight $2\vec{\alpha}_{\text{St}}$ (and its images under the action of the Weyl group) must be in the string convex hull $C_{2}$. Since these are the only emergent string limits on $C_{1}$ we also must have $\frac12C_{2}\subset C_{1}$, that is the rescaled string polytope must fit inside the particle polytope, touching precisely at the $\vec{\alpha}_{\text{St}}$ points, see Figure \ref{fig:Cmax}. In general these points will come in more than one Weyl orbit, each orbit corresponding to a particular orthoplex face $F_I$ defined by a subset of roots $I$ as described in section~\ref{sec:Weylpolytopes}. 
\def\colone{blue}
\def\coltwo{applegreen}
\def\colthree{cyan}
\def\colinf{Orchid}
\def\colhull{Dandelion}
\def\colhulledges{black}
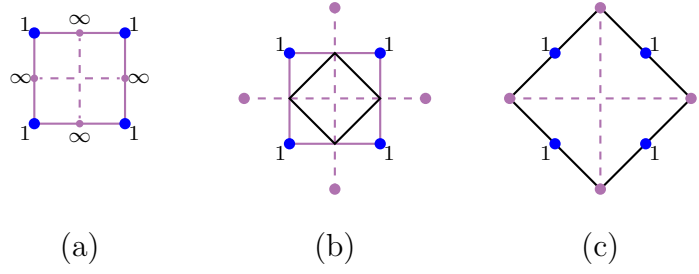
\begin{figure}[H]
    \centering
    \subfigure{
\begin{tikzpicture}[scale=0.6,>=stealth]
{\coordinatesSquare
\labelsSquareOneInf
\def\sizebig{1.5pt}\def\sizesmall{1.0pt}
\foreach \i/\j in {A/B,B/C,C/D,D/A}{
\draw[\colinf, thick] (\i) -- (\j);\path  ($0.5*(\i)+0.5*(\j)$) node[circle, fill=\colinf, inner sep=\sizesmall]{};};
\foreach \i in {A,B,C,D}{\path ($(\i)$) node[circle, fill=\colone, inner sep=\sizebig]{};};
\foreach \i/\j in {A/B,B/C,C/D,D/A}\draw[shorten >=2pt, shorten <=2pt, \colinf,thick,dashed] (0,0) -- ($0.5*(\i)+0.5*(\j)$) node[midway, below,color=\colinf]{};
} 
\node[below=26pt of current bounding box.south, text width=2cm, align=center] 
            {(a)};
\end{tikzpicture}} \quad \quad 
\subfigure{\begin{tikzpicture}[scale=0.6,>=stealth]
{\coordinate (A) at (1,1);
\coordinate (B) at (-1,1);
\coordinate (C) at (-1,-1);
\coordinate (D) at (1,-1);
\foreach \pt in {A,B,C,D}{ \path  ($1.2*(\pt)$)  node[font=\scriptsize]{1};};
\def\sizebig{1.5pt}\def\sizesmall{1.0pt}
\foreach \i/\j in {A/B,B/C,C/D,D/A}\draw[shorten >=2pt, shorten <=2pt, \colinf,thick,dashed] (0,0) -- ($(\i)+(\j)$) node[midway, below,color=\colinf]{};
\foreach \i/\j in {A/B,B/C,C/D,D/A}{
\draw[\colinf, thick] (\i) -- (\j);\path  ($(\i)+(\j)$) node[circle, fill=\colinf, inner sep=\sizebig]{};};
\foreach \i in {A,B,C,D}{\path ($(\i)$) node[circle, fill=\colone, inner sep=\sizebig]{};}
\draw[black, thick] (1,0) -- (0,1) -- (-1,0) -- (0,-1) -- cycle;

}
\node[below=10pt of current bounding box.south, text width=2cm, align=center] 
            {(b)};
\end{tikzpicture}} \quad \quad \subfigure{\begin{tikzpicture}[scale=0.6,>=stealth]
{\coordinate (A) at (1,1);
\coordinate (B) at (-1,1);
\coordinate (C) at (-1,-1);
\coordinate (D) at (1,-1);
\foreach \pt in {A,B,C,D}{ \path  ($1.2*(\pt)$)  node[font=\scriptsize]{1};};
\def\sizebig{1.5pt}\def\sizesmall{1.0pt}
\foreach \i/\j in {A/B,B/C,C/D,D/A}{
\draw[\colhulledges, thick] (\i) -- ($(\i)+(\j)$) -- (\j);\path  ($(\i)+(\j)$) node[circle, fill=\colinf, inner sep=\sizebig]{};};
\foreach \i in {A,B,C,D}{\path ($(\i)$) node[circle, fill=\colone, inner sep=\sizebig]{};};
\foreach \i/\j in {A/B,B/C,C/D,D/A}\draw[shorten >=2pt, shorten <=2pt, \colinf,thick,dashed] (0,0) -- ($(\i)+(\j)$) node[midway, below,color=\colinf]{};
}
\node[below=10pt of current bounding box.south, text width=2cm, align=center] 
            {(c)};
\end{tikzpicture}}
\caption{(a) $\color{\colinf}{C_{\text{1}}}$, (b) in black $\frac12 C_2$, and in violet the weights corresponding to the vertices of $C_2$ and (c) $C^{\rm max}_2$.} \label{fig:Cmax}
\end{figure}
If we take the minimal representation  $\lambda_{1}=\lambda_{\text{min}}$ then we can say something stronger about $C_{2}$. Let $\lambda_I=2\vec{\alpha}_{\text{St}}$ for each orthoplex face $F_I$.\footnote{It is easy to show that $\lambda_I$ is a multiple of $\omega_I$, where $\omega_I$ is the fundamental weight orthogonal to all the roots in $I$, or equivalently the one labelled by the deleted node on the $\lambda$-extended Dynkin diagram.} Then the only way to have  $\frac12C_{2}\subset C_{1}$ is if $\rho_{2}$ is the sum of the irrep defined by the highest weights $\lambda_I$, that is 
\begin{equation}
    \rho_{2} = \bigoplus_{I\in\text{orthoplex}} \rho_{\lambda_I} \, ,  
\end{equation}
and 
\begin{equation}
    C_{2} = \conv\left( \bigcup_{I\in\text{orthoplex}} C_{\lambda_I} \right) \, , 
\end{equation}
where $C_{\lambda_I}$ is the Weyl polytope of the highest weight $\lambda_I$. These representations are listed in Table~\ref{tab:facesparticlesNonAbelianNonMth}\footnote{In the particle polytopes with exceptional-type $\mathfrak{g}=\mathfrak{e}_{d(d)}$ the string representations we find agree with the M-theoretic ones in Table \ref{tab:G/K}, so we do not give them explicitly in Table \ref{tab:facesparticlesNonAbelianNonMth}. See Table \ref{tab:facesMthAllpNotParticles} for the relevant information about the respective $\pmax=2$ polytopes.}, together with the relevant data of the $\pmax=2$ polytope built from the union of the particle and the string polytopes. 

By way of example consider the faces of $\dext=3$, $\mathfrak{sp}(8,\mathbb{R})$ octacube
\begin{equation}
\begin{aligned}
    \tikzset{/Dynkin diagram,root radius=.125cm,edge length=.55cm}
    \begin{dynkinDiagram}{C}{oooo}
    \node[circle,inner sep=0pt,minimum size=2.5mm,fill={\colExt},draw] (A) at ($(root 2) + (0,0.55)$) {}; 
    \draw[\colExt] (A) -- (root 2);
    \end{dynkinDiagram} & : & && 
    F_{\{1,2,3\}} &= 
    \tikzset{/Dynkin diagram,root radius=.125cm,edge length=.55cm}
    \begin{dynkinDiagram}[edge/.style={white,thick}]{A}{***o}
    \node[circle,inner sep=0pt,minimum size=2.5mm,fill={\colExt},draw] (A) at ($(root 2) + (0,0.55)$) {}; 
    \draw[\colExt] (A) -- (root 2);
    \draw (root 1) -- (root 2) -- (root 3);
    \end{dynkinDiagram} &
    F_{\{2,3,4\}} &= 
    \tikzset{/Dynkin diagram,root radius=.125cm,edge length=.55cm}
    \begin{dynkinDiagram}{C}{o***}
    \node[circle,inner sep=0pt,minimum size=2.5mm,fill={\colExt},draw] (A) at ($(root 2) + (0,0.55)$) {}; 
    \draw[\colExt] (A) -- (root 2);
    \draw[white, thick] (root 1) -- (root 2);
    \end{dynkinDiagram}
\end{aligned}
\end{equation}
We see that there are two octahedral (3-orthoplex) faces, $F_{\{1,2,3\}}$ and $F_{\{2,3,4\}}$. Each defines an irrep contributing to an emergent string representation highest weights with $\lambda_{\{1,2,3\}}=\omega_4$ and $\lambda_{\{2,3,4\}}=2\omega_1$, corresponding to the deleted (white) nodes. These give the $\mathbf{42}$ and $\mathbf{36}$ representations respectively, as listed in Table~\ref{tab:facesparticlesNonAbelianNonMth}.

Already we notice  that we can  infer just from the values of $d$ and $\nvertex$ if a vertex of $C_{2}$, the string  polytope, is a vertex of $C^{\text{max}}_{2}$ . For this to happen, the associated KK tower must still be the leading one in that direction. This together with the convexity of the polytope, leads to two possibilities as follows 
\begin{equation}
    \nu \equiv \frac{d-2}{\nvertex} \, , \qquad 
    \begin{cases}
        \nu \leq 1 & C^{\text{max}}_{2} = C_{2} \, , \\
        \nu > 1 & C^{\text{max}}_{2} = \conv(C_{1}\cup C_{2}) \neq C_{2} \, . 
    \end{cases}
\end{equation}
Furthermore $C_2^{{\text{max}}}$ is in general semi-regular if $\nu \ne 3$. 

Remarkably, although we only imposed the ESC on the particle polytope, in all cases $C^{\text{max}}_{2}$ satisfies the BDC~\eqref{eqn:BDC}. We also stress that in general the dimension of the string representation $\rho_{2}$ is bigger than the number of emergent string limits, but the additional string-like objects do not have a leading mass decay rate  for any direction in moduli space, i.e.~the vector for their mass scale  would always fall inside the particle convex hull. 

\begin{table}\renewcommand{\arraystretch}{1.2}
\centering\footnotesize\begin{tabular}{
|@{}>{$}c<{$}@{}
|@{}>{$}c<{$}@{}
|>{$}c<{$}
|>{$}c<{$}
|>{$}c<{$}
|@{}>{$}c<{$}@{}
|>{$}c<{$}
|>{$}c<{$}
|>{$}c<{$}
|>{$}c<{$}|} \hline 
C_1 & C_2^{\text{max}} & \mathfrak{g} & \rho_{1} & \rho_{2} & \dext & V & E & F & C \\ \hline
 \hline 
 \multirow{5}{*}{ octacube }&
 \multirow{5}{*}{octacube}&\mathfrak{f}_{4(4)} & \mathbf{26} & \mathbf{52}&&&&&\\
 && \mathfrak{f}_{4(4)} & \mathbf{52} & \mathbf{324} &&&&&\\
 && \splitC4^{(4)} & \mathbf{27} &  \mathbf{42}\oplus\mathbf{36} & 3 & \begin{array}{@{}c@{}c@{}} 24 & (4) \\\end{array}&\begin{array}{@{}c@{}c@{}}96 & (3) \\\end{array}& \begin{array}{@{}c@{}c@{}}96 & (\frac{8}{3})\\
\end{array}& \begin{array}{@{}c@{}c@{}}24 & (2) \\\end{array}\\
&& \splitD4 & \mathbf{28} & 
\mathbf{35_v}\oplus\mathbf{35_s}\oplus\mathbf{35_c} &&&&&\\
&& \splitB4 & \mathbf{36} & \mathbf{126} \oplus \mathbf{44} &&&&&\\
\cline{1-10}
\EfourAlt & \text{co}(\Delta_4,\EfourAlt) & \splitA4 & \mathbf{10}^* & \mathbf{5} & 7 & \begin{array}{@{}c@{}>{\left(}c<{\right)}@{}}  10 & \frac{6}{5} \\  5 & \frac{4}{5} \\ \end{array}  & \begin{array}{@{}c@{}>{\left(}c<{\right)}@{}}  30 & \frac{7}{10} \\  30 & \frac{2}{3} \\ \end{array}  & \begin{array}{@{}c@{}>{\left(}c<{\right)}@{}}  60 & \frac{4}{7} \\  30 & \frac{8}{15} \\ \end{array}  & \begin{array}{@{}c@{}>{\left(}c<{\right)}@{}}  40 & \frac{1}{2} \\  5 & \frac{9}{20} \\ \end{array}\\ 
\cline{1-10}
\multirow{3}{*}{\begin{tabular}{c}$\cuboctahedronScaled{0.35}$
\end{tabular}} 
&
\multirow{3}{*}{\begin{tabular}{c}$\octahedronScaled{0.5}$
\end{tabular}} &\splitA3 & \mathbf{15} &  \mathbf{20'} &&&&\\
&&\splitB3 & \mathbf{21} & \mathbf{27} & 3 & \begin{array}{@{}c@{}c@{}} 6 & (4) \\\end{array} & \begin{array}{@{}c@{}c@{}} 12 & (2) \\\end{array} & 
\begin{array}{@{}c@{}c@{}} 8 & \left(\frac{4}{3}\right) \\\end{array} \\
&& \splitC3^{(4)} & \mathbf{14} & \mathbf{21} &&&&\\
 \cline{1-9} 
 \multirow{5}{*}{\begin{tabular}{c}$\cubeScaled{0.4}$\end{tabular}}
 &
 \multirow{5}{*}{\begin{tabular}{c}$\rhombicdodecScaled{0.5}$\end{tabular}}
 & \splitC3 & \mathbf{14'} & \mathbf{21}  & \multirow{5}{*}{4} & \multirow{5}{*}{$\begin{array}{c@{}c@{}} 6 & \left(2\right) \\ 8 & \left(\frac{3}{2}\right) \\\end{array}$} & \multirow{5}{*}{$\begin{array}{c@{}>{}c<{}@{}} 24 & \left(\frac{4}{3}\right) \\\end{array}$} & \multirow{5}{*}{$\begin{array}{c@{}c@{}} \color{blue}{\mathbf{12}} & \left(\color{blue}{\mathbf{1}}\right)\end{array}$}   \\
&& \splitB3 & \mathbf{8} & \mathbf{7} 
 &&&&\\
 &&\splitC2 \oplus \splitA1 & \mathbf{(4,2)} & 
\mathbf{(1,3)}\oplus\mathbf{(5,1)} &&&&\\
&& \splitC2 \oplus \splitA1 &\mathbf{(5,2)} & 
 \mathbf{(1,3)}\oplus\mathbf{(10,1)} &&&&\\
&&\splitA1^{\oplus 3} & \mathbf{(2,2,2)} & \text{sym}\mathbf{\left(3,1,1\right)} &&&&\\
 \cline{1-9} 
\multirow{1}{*}{$\triprismScaled{0.4}$} & J_{51} & \splitA1 \oplus \splitA2 & (\mathbf{2},\mathbf{3}^*)  & (\mathbf{1},\mathbf{3}) &  8 &  \begin{array}{@{}c@{}>{}c<{}@{}} 6 & \left(\frac{7}{6}\right) \\ 3 & \left(\frac{2}{3}\right) \\\end{array} & \begin{array}{@{}c@{}>{}c<{}@{}}9 & \left(\frac{2}{3}\right) \\ 12 & \left(\frac{4}{7}\right) \\\end{array} & \begin{array}{@{}c@{}>{}c<{}@{}} 14 & \left(\frac{1}{2}\right) \\\end{array} \\ \cline{1-9}
\multirow{2}{*}{\Large$\rectangleScaled{0.25}$}
&\multirow{2}{*}{\Huge${\recstringScaled{0.25}}$}
& \multirow{2}{*}{$\splitA1^{\oplus 2}$}
& \multirow{2}{*}{$\mathbf{(2,2)}$} & 
 \multirow{2}{*}{$\left(\mathbf{1,3}\right)$} & \multirow{2}{*}{$4$} &  2\, (2)  & 4\,(\tfrac43) \\
&&&&&&4\,(\tfrac32)&\color{blue}{\mathbf{2}}\,\color{blue}{\mathbf{(1)}}  \\
 \cline{1-8} 
\begin{tabular}{c}\Large$\triangle$ \end{tabular}
&
\begin{tabular}{c}\Huge$\rotatebox{30}{\hexagon}$ \end{tabular}
& \splitA2 & \mathbf{3} & \mathbf{3}^* & 5 & 
\begin{array}{@{}c@{}c@{}} 6 & \left(\frac{4}{3}\right) \\\end{array} &
\begin{array}{@{}c@{}c@{}} 6 & (1) \\\end{array}   \\
 \cline{1-8}
 \multirow{3}{*}{\begin{tabular}{c}{\LARGE$\square$}\end{tabular}}
 & 
 \multirow{3}{*}{\begin{tabular}{c}\rotatebox{45}{\huge$\square$}\end{tabular}}
 &\splitC2 & \mathbf{4} & \mathbf{5} &&&\\ 
&& \splitC2 & \mathbf{5} & \mathbf{10}
 & 3 & \begin{array}{@{}c@{}c@{}} 4 & (4) \\\end{array} & \begin{array}{@{}c@{}c@{}} 4 & (2) \\\end{array}   \\
&& \mathfrak{sl}_2^{\oplus 2} & \mathbf{(2,2)} & \left(\mathbf{3,1}\right) \oplus \left(\mathbf{1,3}\right) &&&\\ \cline{1-8}
\multirow{2}{*}{\begin{tabular}{c}{\bf\Large$\octagon$}\end{tabular}}  &
\multirow{2}{*}{\begin{tabular}{c}\rotatebox{45}{\huge$\square$}\end{tabular}} & \splitC2& \mathbf{16}&  \mathbf{35'}  & \multirow{2}{*}{\begin{tabular}{c} 3 \end{tabular}} & \multirow{2}{*}{$4\,\left(4
\right)$} &  \multirow{2}{*}{$4\,\left(2\right)$} \\ 
&& \splitC2 & \mathbf{35} & \mathbf{55}  & & & \\
\cline{1-8}
\multirow{3}{*}{\begin{tabular}{c}{{\LARGE$\hexagon$}}\end{tabular}} &
\multirow{3}{*}{\begin{tabular}{c}{\rotatebox{30}{\Huge$\hexagon$}}\end{tabular}} 
&\splitG2 & \mathbf{7} & \mathbf{14}  & \multirow{3}{*}{\begin{tabular}{c} 3 \end{tabular}} & \multirow{3}{*}{6\,(4)}  & \multirow{3}{*}{6\,(3)} \\ 
&& \splitG2 & \mathbf{14} & \mathbf{77} & & & \\
&& \splitA2 & \mathbf{8} & \mathbf{10} & & & \\
\cline{1-8}
\multirow{1}{*}{$\isotriangleNoCircleScaled{0.30}$} &
\multirow{1}{*}{$\isotriangleStringsScaled{0.30}$} 
&  \splitE2 & \mathbf{2}
\oplus \mathbf{1}
& {\mathbf{2}}
&  9 & 
\begin{array}{@{}c@{}>{}c<{}@{}}  3 & \left(\frac{8}{7}\right) \\  2 & \left(\frac{4}{7}\right) \\ \end{array} &
 \begin{array}{@{}c@{}>{}c<{}@{}}  4 & \left(\frac{1}{2}\right) \\  1 & \left(\frac{9}{14}\right) \\ \end{array} \\
\cline{1-8}
\end{tabular}\caption{$\pmax=1$ and $\pmax=2$ polytopes, together with the representations of particles $(\rho_{1})$ and strings ($\rho_{2}$) for each $(\mathfrak{g},\lambda_{1},d,\langle.,.\rangle_{\text{max}})$ with emergent string limit. We display the number of vertices (V), edges (E), faces (F) and 4-dimensional cells (C), and in parenthesis the distance of the facets of the $\pmax=2$ polytope built as the $\rho_{\text{p}}$ and $\rho_{\text{s}}$ convex hull. 
All of them satisfy the BDC bound \eqref{eqn:BDC}, and we show in blue the faces saturating it. The notation $\text{sym}\mathbf{\left(3,1,1 \right)}$ stands for $(\mathbf{3},\mathbf{1},\mathbf{1}) \oplus (\mathbf{1},\mathbf{3},\mathbf{1}) \oplus (\mathbf{1},\mathbf{1},\mathbf{3})$.   $J_{51}$ is the triaugmented triangular prism (dual to the associahedron $K_5$), $\text{co}(\Delta_4,\EfourAlt)$ is the convex hull of the combination of a hypertetrahedron and a tetroctahedric.}\label{tab:facesparticlesNonAbelianNonMth}\end{table}

Finally we note that for the polytopes without emergent string limits (as the hexagon) it is not possible using this method to unambiguously determine the string representation with leading tension decay-rate, nor if strings should exist at all in the theory. The only thing that can be argued is that if strings exist, we must have that $\frac12C_{2}$ is strictly contained within $C_{1}$.

\subsection{String and brane representations from decompactification}

\begin{figure}[H]\centering
\begin{tikzpicture}[scale=2]
\def\cola{blue}
\def\colb{applegreen}
\def\opafill{0.2}
\def\opaedges{1}
\coordinatesCube
\draw[\cola,dashed, thick, opacity=\opaedges] (A) -- (B') -- (D') -- cycle;
\fill[\cola, opacity=\opafill] (A) -- (B') -- (D') -- cycle;
\draw[\colb, thick, opacity=\opaedges]    (D) -- (B) -- (C') --cycle;
\fill[\colb, opacity=\opafill] (D) -- (B) -- (C') -- cycle;
\drawCube
\path (C) node[circle, fill=black, inner sep=1.5pt]{};
\path (A') node[circle, fill=red, inner sep=1.5pt]{};
\end{tikzpicture}
  \caption{
  Kaluza-Klein, particle, string and Kaluza-Klein monopole states in the decompactification of the $\cube_{1,2,\infty}$.}
    \label{fig:decompCube2}
\end{figure}
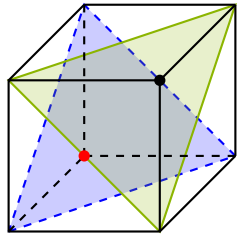

The second way to obtain string and higher brane representations is by considering decompactification limits of the particle polytope. For simplicity, let $\nvertex=1$ and consider decompactifying a single dimension. From the discussion in Section~\ref{sec:decompactifications} this corresponds to slicing the particle polytope on planes orthogonal to a vertex. For instance, returning to the example of the $(\splitB3,{\bf 8},d=4)$ cube of Table~\ref{tab:1irrep_integern_scaled_global}, the decompactified Lie algebra is $\cgg'=\splitA2$. Decomposing under $\splitA2\oplus\mathbb{R}$ we have 
\begin{equation}
    \mathbf{8} = \mathbf{1}_3 \oplus \mathbf{3}_1 \oplus \mathbf{3}^*_{-1} \oplus \mathbf{1}_{-3} \, ,
\end{equation}
corresponding to the red vertex, the blue triangle, the green triangle and black vertex respectively in Figure~\ref{fig:decompCube2} (which is just the repeat of the first diagram of Figure~\ref{fig:decompCube}).   

\noindent
From the orthogonal distance to the centre of each slice, we find the following exponential rates
\begin{align}
    \alpha_{\mathbf{1}_3} &= \sqrt{\tfrac32} \, , & 
    \alpha_{\mathbf{3}_1} &= \sqrt{\tfrac43} \, , & 
    \alpha_{\mathbf{3}^*_{-1}} &= -\sqrt{\tfrac43} \, , & 
    \alpha_{\mathbf{1}_{-3}} &= - \sqrt{\tfrac32} \, . 
\end{align}
Recall that when compactifying a single dimension to a $d$-dimensional theory we have~\cite{Etheredge:2022opl,Etheredge:2024amg}
\begin{equation}
\begin{aligned}
    \alpha_{\text{KK}} &= \sqrt{\frac{d-1}{d-2}}\, , \\
    \alpha_{\text{part}} &= \frac{1}{\sqrt{(d-1)(d-2)}} 
         = \alpha_{\text{KK}} - \sqrt{\frac{d-2}{d-1}} \, , \\
    \alpha_{\text{string}} &= -\frac{d-3}{\sqrt{(d-1)(d-2)}} 
         = \alpha_{\text{KK}} - 2\sqrt{\frac{d-2}{d-1}} \, , \\
    \alpha_{\text{mono}} &= -\sqrt{\frac{d-1}{d-2}}\,  , 
\end{aligned}
\end{equation}
for the rates of decay of the masses of Kaluza--Klein modes, conventional $(d+1)$-dimensional particles, wrapped strings\footnote{Note that $\alpha_{\text{string}}$, the decay rate of a wound string state, should not be confused with $\alpha_{St}$, the decay rate of an emergent, tensionless string.} and Kaluza--Klein monopoles respectively. (The Kaluza--Klein monopole rate only applies for $d=4$ and $d=3$.) In $d=4$ these match $\alpha_{\mathbf{1}_3}$, $\alpha_{\mathbf{3}_1}$,  $\alpha_{\mathbf{3}^*_{-1}}$ and $\alpha_{\mathbf{1}_{-3}}$ respectively. Thus we have a direct interpretation of the four slices. All the states on the cube are particles in $d=4$. In compactifying the 5d theory, they arise from KK modes, a triplet of particle states in 5d, a dual triplet of 5d string states wound on the circle, and KK monopoles. Thus we can directly read off the particle and string representations for the $(\splitA2,d=5)$ theory as
\begin{equation}
    \rho_{1} = \mathbf{3} \,,  \qquad \rho_{2} = \mathbf{3}^* \,,
\end{equation}
which agrees with our previous analysis of emergent string limits summarised in Table~\ref{tab:facesparticlesNonAbelianNonMth}. We also see that the 5d $C^{\text{max}}_{2}$ polytope is just the projection of the 4d particle polytope $C_{1}$ onto the plane orthogonal to the line through the decompactification vertex. 

One can repeat this analysis for all our global polytope examples listed in Tables~\ref{tab:1irrep_integern_scaled_global}, with $\nvertex=1$ and $\lambda_{\text{min}}$ and one finds 
\begin{equation}
\begin{aligned}
    \dext > 4: &\ \text{three slices -- KK, particles and strings} \\
    \dext = 4: &\ \text{four slices -- KK, particles, strings and KK monopoles} \\
    \dext = 3: &\ \text{five slices -- KK, particles, strings, gen.~KK monopoles and KK monopoles} 
\end{aligned}
\end{equation}
Here in $d=3$ we also have generalised KK monopole states, analogues of the supersymmetric generalised KK monopole states that appear as ``defect branes'', first introduced in~\cite{Bergshoeff:2011se,Kleinschmidt:2011vu}. In each case, the slices are evenly spaced in intervals of $\sqrt{(d-2)/(d-1)}$. Furthermore, just as in our discussion of decompactification descendents in Section~\ref{sec:decompactifications}, in some cases the slices are through emergent string points rather than vertices.\footnote{One can calculate the distance to the slices in Figure~\ref{fig:d'} and in both cases it is given by $\alpha_{\text{part}}$ and so indeed corresponds to the particle and emergent string states of the decompactified theory, justifying the claim made in that section.} One can then read off the string representation $\rho_{2}$ of the decompactified theory, and the corresponding convex hulls agree with the ones from the analysis of emergent string limits summarised  Table~\ref{tab:facesparticlesNonAbelianNonMth}, except that, since we are decompactifying we cannot get the string representations for the $\dext=3$ cases this way. 

In each case one finds that the convex hulls are related by 
\begin{equation}
\label{eq:Cpmax-projection}
    C^{{\rm max},d'}_p = \pi(C^{\dext}_{1}) \, , 
\end{equation}
where $C^{{\rm max},d'}_p$ is the $\pmax$-hull of the decompactified theory (here with $\pmax=2)$, $C^{\dext}_{1}$ is the particle polytope of the original undecompactified theory and $\pi$ is the projection onto the plane spanned by the Cartan subalgebra $\caa'$ of the decompactified theory.\footnote{The weights of the KK mode, KK monopole and generalised KK monopole states always lie inside or on the boundary of $C^{{\rm max},d'}_2$ and so are irrelevant in the projection.} The BDC (here with $\pmax=2$ and $\dext'=\dext+1$) is then a direct consequence of the ESC rates of the undecompactified theory, since 
\begin{equation}
    \frac{1}{\sqrt{d'-\pmax-1}} = \frac{1}{\sqrt{d-2}} \, , 
\end{equation}
in line with the original motivation for BDC as being required for consistency of the SDC under dimensional reduction~\cite{Etheredge:2024amg}. 

One can extend this general analysis to decompactifying more than one dimension by considering descendent decompactified theories defined by higher-dimension faces of the particle polytope, that is not just vertices. The split component of the parabolic subgroup $\caa_{P(\mathbb{Q})}$ is now higher dimensional, meaning that each slice is not defined by a single $\mathbb{R}$ charge but $k$ Abelian charges corresponding to the $k$ decompactifying dimensions. By keeping track of these charges one can identify a slice as corresponding to a particular $(p-1)$-brane (or generalised defect brane in some cases) and hence identify the representation of each set of $(p-1)$-branes. Just as for the case of a single decompactification, the corresponding $C^{{\rm max},d'}_p$ is just the projection of the particle polytope $C^{\dext}_{1}$ as in~\eqref{eq:Cpmax-projection}, and the BDC is a consequence of the ESC rates of the undecompactified theory. We leave detailed analysis of this case for later work. 

\subsubsection*{Examples}
    
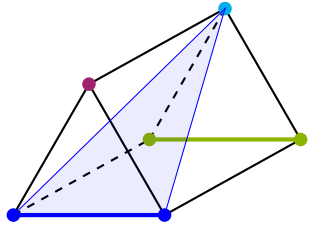
\begin{figure}[H]
\def\cola{blue}\def\colb{applegreen}\def\colc{RedViolet}\def\cold{cyan}
\centering
\begin{tikzpicture}[scale=2]
\def\opaedges{1}
\def\opafill{0.075}
\def\rad{1.8pt}
\def\thickness{ultra thick}
\coordinatesTriangularPrism
\drawTriangularPrism;
\draw[\cola,\thickness,opacity=\opaedges](A) -- (B);
\draw[\colb,\thickness,opacity=\opaedges](A') -- (B');
\path (C) node[circle, fill=\colc, inner sep=\rad]{};
\path (C') node[circle, fill=\cold, inner sep=\rad]{};
\foreach \i in {A,B}{\path (\i) node[circle, fill=\cola, inner sep=\rad]{};};
\foreach \i in {A',B'}{\path (\i) node[circle, fill=\colb, inner sep=\rad]{};};
\fill[\cola,opacity=\opafill] (A) -- (B) -- (C') -- cycle;
\draw[\cola,opacity=\opaedges] (A) -- (C') -- (B);
\end{tikzpicture}    
\caption{The slices of the $(\dext=8,\splitE3,(\mathbf{2},\mathbf{3}^*))$ polytope under the decompactification to $(\dext=9,\splitE2)$, showing KK mode, particle and string representations in purple, blue and green respectively.} \label{fig:E3decomposition}
\end{figure}

Let us briefly illustrate the $\pmax=2$ case with a few further examples. Consider first the familiar case of M-theory on $T^2$ as a decompactification of M-theory on $T^3$ that we also discussed in Section~\ref{sec:decompactifications}. The decompactified algebra is $\cgg'=\splitE2\simeq\splitA1\oplus\mathbb{R}\subset\splitE3\simeq\splitA2\oplus\splitA1$. The decompactification slices are shown in Figure~\ref{fig:E3decomposition} where we pick the decompactified direction to have KK tower associated to the weight in purple. 

Decomposing under $(\splitA1\oplus\mathbb{R})\oplus\mathbb{R}\subset\splitA2\oplus\splitA1$, where the second $\mathbb{R}$ factor is the Cartan $\caa_{P(\mathbb{Q})}$ of the parabolic group defining the decompactification and measures the distance of the slices from the centre of the polytope, we find 
\begin{equation} \label{eq:decomp}
\def\cola{blue}\def\colb{applegreen}\def\colc{RedViolet}\def\cold{cyan}
    (\mathbf{2},\mathbf{3}^*) =
        {\color{\colc}\mathbf{1}_{0,\sqrt{\frac76}}}
        \oplus \bigl({\color{\cola}\mathbf{2}_{-\frac{3}{\sqrt{14}},\frac{1}{\sqrt{42}}}}
        \oplus {\color{\cold}\mathbf{1}_{\frac{4}{\sqrt{14}},\frac{1}{\sqrt{42}}}} \bigr)
        \oplus {\color{\colb}\mathbf{2}_{\frac{1}{\sqrt{14}},-\frac{5}{\sqrt{42}}}} \, . 
\end{equation}
The second factor in the subscript is the distance of the slice from the centre of the polytope. This gives a single KK mode in purple, singlet and doublet particle representations on the blue triangular slices and a doublet of string states in green. These are exactly in agreement with the list in Table~\ref{tab:facesparticlesNonAbelianNonMth}. Projecting on the plane spanned by the Cartan of $\splitE2$ we get Figure~\ref{fig:Mth9D}. Restricting to just the particle slice we get the left-hand diagram, the convex hull matching that already discussed in Section~\ref{sec:decompactifications}. If instead we take the full set of weights, KK, particle and string, that is the projection of the $\splitE3$ particle polytope, we get the right-hand diagram. Again we see these agree with the $C_{1}$ and $C^{\text{max}}_{2}$ polytopes for $\splitE2$ listed in Table~\ref{tab:facesparticlesNonAbelianNonMth}. One can also see the string weights appearing at precisely $2\vec{\alpha}_\text{St}$ where $\vec{\alpha}_\text{St}$ is the position of the emergent string node on the particle polytope, just as discussed in the previous subsection. 
\begin{figure}[H]
\centering
\subfigure{
\begin{tikzpicture}[scale=2]{
\def\colinf{red}
\def\rad{2.0pt}
\coordinate (A) at (-0.707107,-0.801784);
\coordinate (B) at (0.707107,-0.801784);
\coordinate (C) at (0,0);
\coordinate (A') at (-0.707107,0.267261);
\coordinate (B') at (0.707107,0.267261);
\coordinate (C') at (0,1.06904);
\draw[thick] (A) -- (B) -- (C') -- cycle;

\foreach \x/\y/\z/\w/\r in {
A/\cola/1.06904/{}/1,
B/\cola/1.06904/$\mathbf{2}_{\text{\scalebox{.8}{$\left(\frac{-3}{\sqrt{14}}\right)$}}}$/1,
C'/\cold/1.06904/$\mathbf{1}_{\text{\scalebox{.8}{$\left(\frac{4}{\sqrt{14}}\right)$}}}$/1
}{
\draw[dotted,thick,color=\y] ($0.5*(\x)$) circle (\z*0.5);
\draw[dotted,thick,color=\y,fill=\y,opacity=0.05] ($0.5*(\x)$) circle (\z*0.5);
\path (\x) node[circle, fill=\y, inner sep=\r*\rad]{};
\node[above right, xshift=0.0cm, yshift=-0.1cm] at (\x) {\w};};

\def\sizebig{1.5pt}\def\sizesmall{1.0pt}
\foreach \i/\j in {A/C',B/C'}{
\path  ($0.5*(\i)+0.5*(\j)$) node[circle,fill=\colinf, inner sep=\sizesmall]{};};

\def\dlabels{0.0}
\def\mult{1.2}

\foreach \i/\j  in {A/C',B/C'}{ \path  ($\mult*0.5*(\i)+\mult*0.5*(\j)-\dlabels*(Z)$) node[font=\scriptsize]{$\infty$};};}
\end{tikzpicture} \qquad \qquad 
\subfigure{
\begin{tikzpicture}[scale=2]{
\def\rad{2.0pt}
\coordinate (A) at (-0.707107,-0.801784);
\coordinate (B) at (0.707107,-0.801784);
\coordinate (C) at (0,0);
\coordinate (A') at (-0.707107,0.267261);
\coordinate (B') at (0.707107,0.267261);
\coordinate (C') at (0,1.06904);

\draw[thick] (A) -- (B) -- (B') -- (C') -- (A') -- cycle;

\foreach \x/\y/\z/\w/\r in {
A/\cola/1.06904/{}/1,
B/\cola/1.06904/$\mathbf{2}_{\text{\scalebox{.8}{$\left(\frac{-3}{\sqrt{14}}\right)$}}}$/1,
C/\colc/0/{}/1,
A'/\colb/0.755929/{}/1,
B'/\colb/0.755929/$\mathbf{2}_{\text{\scalebox{.8}{$\left(\frac{1}{\sqrt{14}}\right)$}}}$/1,
C'/\cold/1.06904/$\mathbf{1}_{\text{\scalebox{.8}{$\left(\frac{4}{\sqrt{14}}\right)$}}}$/1
}{
\draw[dotted,thick,color=\y] ($0.5*(\x)$) circle (\z*0.5);
\draw[dotted,thick,color=\y,fill=\y,opacity=0.05] ($0.5*(\x)$) circle (\z*0.5);
\path (\x) node[circle, fill=\y, inner sep=\r*\rad]{};
\node[above right, xshift=0.0cm, yshift=-0.1cm] at (\x) {\w};};
}
\end{tikzpicture}}}
\caption{On the left the particle ($\pmax=1$), and on the right the $\pmax=2$ polytope for M-theory on a $T^2$.}\label{fig:Mth9D}
\end{figure}
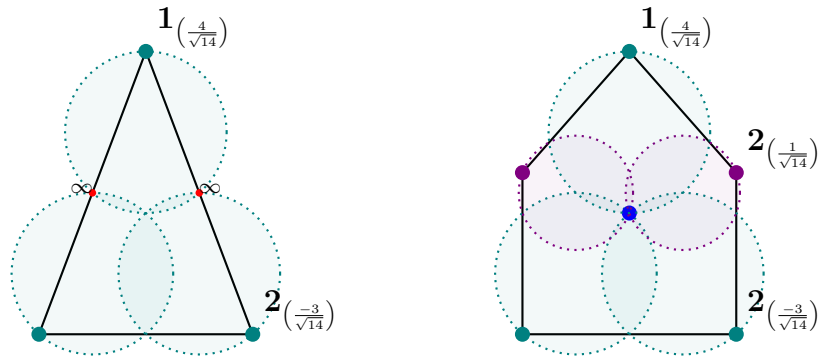

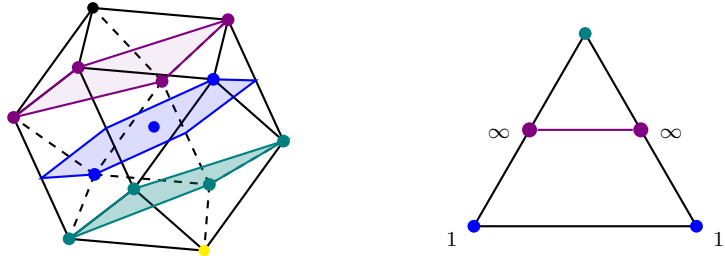
\begin{figure}[H]\centering
\subfigure{\begin{tikzpicture}[scale=1.8]
\def\opafill{0.2}
\def\opaedges{1}
\coordinatesCuboctahedronTilted
\drawCuboctahedron
\draw[\colb,thick,opacity=\opaedges] (B) -- (J)  -- (D) -- (H) -- cycle;
\fill[\colb, opacity=0.07](B) -- (J)  -- (D) -- (H) -- cycle;
\draw[\colc,thick,opacity=\opaedges] (I) -- ($0.5*(G)+0.5*(H)$)  -- ($0.5*(D)+0.5*(E)$) -- (K) -- ($0.5*(J)+0.5*(L)$)  -- ($0.5*(A)+0.5*(B)$) -- cycle;
\fill[\colc, opacity=0.14] (I) -- ($0.5*(G)+0.5*(H)$)  -- ($0.5*(D)+0.5*(E)$) -- (K) -- ($0.5*(J)+0.5*(L)$)  -- ($0.5*(A)+0.5*(B)$) -- cycle;
\draw[\cola,thick,opacity=\opaedges](A) -- (G)  -- (E) -- (L) -- cycle;
\fill[\cola, opacity=0.3](A) -- (G)  -- (E) -- (L) -- cycle;
\foreach \i in {B,J,D,H}{\path (\i) node[circle, fill=\colb, inner sep=1.7pt]{};};
\foreach \i in {A,G,E,L}{\path (\i) node[circle, fill=\cola, inner sep=1.7pt]{};};
\foreach \i in {I,K}{\path (\i) node[circle, fill=\colc, inner sep=1.7pt]{};};
\fill (C) circle (1.2pt);
\fill[\colbc] (F) circle (1.2pt);
\fill[\colc] ($0.5*(I)+0.5*(K)$) circle (1.2pt);
\end{tikzpicture}}
\qquad \qquad
\subfigure{
\begin{tikzpicture}[scale=1.7]
\coordinatesTriangle;
\drawTriangle;
\def\mult{1.2};
\node[circle, fill=\colb, inner sep=2pt,label=right:{\scriptsize $\infty$}] (U) at ($0.5*(A)+0.5*(B)$) {};
\node[circle, fill=\colb, inner sep=2pt,label=left:{\scriptsize $\infty$}] (V) at ($0.5*(A)+0.5*(C)$) {};
\draw[\colb,thick] (U) -- (V);
\foreach \i in {B,C}{\path (\i) node[circle, fill=\colc, inner sep=1.7pt]{};};
\foreach \i in {B,C}{\path ($\mult*(\i)$) node[font=\scriptsize]{$1$};};
\path (A) node[circle, fill=\cola, inner sep=1.7pt]{};
\end{tikzpicture}}
\caption{Decompactification slices of the cuboctahedron and triangle.}
\label{fig:decomp:d=3}
\end{figure}

As two final examples, consider the $\splitC3$ cubeoctahedron in $\dext=3$ and the $\splitA2$ triangle in $d=5$. Decompactifying one dimension for the cubeoctahedron we get five slices as shown in the first diagram of Figure~\ref{fig:decomp:d=3}. Decomposing $\splitC3\supset\splitA1\oplus\splitA1\oplus\mathbb{R}$ we have 
\begin{equation}
    \mathbf{14} = 
        (\mathbf{1},\mathbf{1})_{\sqrt2} \oplus (\mathbf{2},\mathbf{2})_{1/\sqrt{2}} 
        \oplus \bigl( (\mathbf{1},\mathbf{3}) \oplus (\mathbf{1},\mathbf{1}) \bigr)_0
        \oplus (\mathbf{2},\mathbf{2})_{-1/\sqrt{2}} \oplus (\mathbf{1},\mathbf{1})_{-\sqrt2} \, , 
\end{equation}
where we have normalised the $\mathbb{R}$ charge to be the distance to the slice, corresponding to KK, particle, string, generalised KK monopole and KK monopole states respectively. From the charge zero slice we see that the strings transform in the $(\mathbf{1},\mathbf{3})$ representation in agreement with Table~\ref{tab:facesparticlesNonAbelianNonMth}. But we actually see something more: there is also a string in the trivial $(\mathbf{1},\mathbf{1})$ representation. For the triangle under $\splitA2\supset\splitA1\oplus\mathbb{R}$ we get 
\begin{equation}
    \mathbf{3} = \mathbf{1}_{2/\sqrt{3}} \oplus \mathbf{2}_{-1/\sqrt{3}} \, , 
\end{equation}
where again we have normalised the $\mathbb{R}$ charge to be the distance to the slice. These give the KK and string states. But as we see from the diagram there is also a slice at a distance of $1/2\sqrt{3}$ that links emergent string points and gives the particle states, which in this case only arise from emergent strings. Again we see in the same sliced diagram, both the emergent strings on the particle slice and the string states themselves with weights at $2\vec{\alpha}_{\text{St}}$ on the string slice. 

\section{Conclusions}
\label{Conclusions}

In this paper we continued the study of the Swampland Distance Conjecture for locally symmetric  moduli spaces $\mathcal{M}=\Gamma\backslash G/K$ started in~\cite{Baines:2025upi}. In particular, we considered the constraints on the allowed moduli space and particle representations $\rho$ that came from imposing that the Emergent String Conjecture (ESC) holds. By this we mean that the rate of decay $\alpha$ of the leading massless towers are either that of a Kaluza--Klein particle or of the oscillator states of an emergent tensionless string~\eqref{eqn:alphakkstring}. For simplicity we restricted the simple factors in the Lie algebra $\cgg$ of $G$ to have split form and the particle representation $\rho$ to be irreducible. However neither of these were essential.  It is worth emphasising that nothing in the analysis of this paper or~\cite{Baines:2025upi} relied on supersymmetry and furthermore is independent of the duality group $\Gamma$, provided it is arithmetic. Thus in particular our classification also includes compactifications of non-supersymmetric strings that lead to symmetric moduli spaces, at least at the classical level, as for example in~\cite{Nair:1986zn,Ginsparg:1986wr,Itoyama:2021itj,Fraiman:2023cpa,DeFreitas:2024ztt,Fraiman:2025yrx}. 

As shown in~\cite{Baines:2025upi}, under very mild assumptions $\alpha$ is determined by the convex hull of the weights of the representation $\rho$. For an irreducible representation this is known as the Weyl polytope $C$ and the effect of the ESC is to constrain its shape. Irreducibility implies that $C$ is a ``spherical polytope'' meaning that all its vertices lie on a sphere, or equivalently, that each vertex corresponds to decompactifying the same number of dimensions. We then showed that this condition together with matching the ESC rates led to a complete classification of the allowed $C$ as a subset of what are known as semi-regular polytopes. The analysis split into two classes: ``global'' and ``perturbative'', depending on whether the reductive algebra $\cgg$ contains an Abelian factor or not. In each case the dimension $d$ of the EFT was fixed by the choice of $C$. 

The second step of the analysis was then to see which of the allowed $C$ corresponded to the
Weyl polytope of some symmetric space EFTs. The latter are specified by $(\mathfrak{g},\lambda,\dext,\langle \cdot,\cdot\rangle)$ where $\lambda$ is the highest weight defining the irrep $\rho$ and $\langle \cdot,\cdot\rangle$ is the $G$-invariant metric on $\mathcal{M}$ which is fixed by group theory only up to scale. Remarkably we showed that every allowed polytope had a symmetric space EFT realisation, and furthermore, several polytopes could be realised by more than one Lie algebra $\cgg$. In every case, there is a ``minimal'' theory where one takes the smallest representation $\lambda_\text{min}$ and also the minimum number of KK dimensions $\nvertex^{\text{min}}$ associated to a vertex of $C$. These are listed in Table~\ref{tab:1irrep_integern_scaled_perturbative} for the perturbative polytopes and Tables~\ref{tab:1irrep_integern_scaled_global} and~\ref{tab:1irrep_integern_extra} for the global polytopes. Essentially, the integrality of the ESC rate constraints -- specifically of the number of decompactifying dimensions and the EFT dimension $d$ -- means the polytope must be associated to a root algebra, and hence to a symmetric space theory, at least when the polytope is spherical. 

Furthermore, one immediately sees that the ESC rates are surprisingly constraining on all three of $\cgg$, $\rho$ and $d$. In particular for the global polytopes we find that, for the minimal theories,
\begin{itemize}
    \item there are only a finite number of possibilities,
    \item the maximum dimension of the decompactified theory $D_\text{max}\leq 11$, 
\end{itemize}
and they include the standard theories arising from M-theory on $T^n$. This strongly suggests that the minimal theories might all lie in the string landscape. Similarly for the perturbative polytopes, the minimal theories are essentially those of pure gravity compactified on $T^k$ ($\cgg=~\splitAkminus1\oplus\mathbb{R}$ and $C=\Delta_{k-1}$) or of a string on $T^k$ ($\cgg=\splitDk\oplus\mathbb{R}$ or $\cgg=\splitBk\oplus\mathbb{R}$ and $C=\beta_k$), though with unbounded $D_\text{max}$. However, the $\splitCk\oplus \mathbb{R}$ theories do not have such obvious physical interpretation.

It should be stressed however that this picture is somewhat misleading. The minimal theories are actually the base of two countably infinite series of allowed theories all with the same polytope: one where $\nvertex^\text{min}$ and $d$ are unchanged but one rescales $\lambda=\gamma\lambda_\text{min}$ and $\langle \cdot,\cdot\rangle=\langle \cdot,\cdot\rangle_\text{min}/\gamma^2$; and a second where $\lambda_\text{min}$ is unchanged but $\nvertex=z\nvertex^\text{min}$, $d= z(d_\text{min}-2)+2$ and $\langle \cdot,\cdot\rangle=\langle \cdot,\cdot\rangle_\text{min}/z$. As shown in Figure~\ref{fig:nface}, the first family includes many more states than the standard KK modes and emergent strings (although one or other of the latter two are always the leading tower as required by the ESC) and the second family allows unbounded $D_\text{max}$. From this perspective, unless one can restrict to the minimal (or perhaps close to minimal) theories, the naive landscape of allowed theories satisfying the ESC seems much larger than that of string models. Even within this enlarged class we can, however, still make some general statements. The first is that the number of allowed polytopes and Lie algebras $\cgg$ remains finite. The second is that the rescaling always shrinks the metric so that there is a universal bound on the curvature of $\mathcal{M}$
\begin{equation*}
\label{eq:curv-bound-concl}
    R_{ij} = - k g_{ij} \qquad k \geq \tfrac14 {h^\vee} \, ,
\end{equation*}
where $h^\vee$ is the dual Coxeter number of the Lie algebra $\cgg$ (or of the semi-simple Lie algebra $\cgg'$ for the perturbative polytopes, for which the bound is $k \geq h^\vee$).

Given this large class of solutions, one can ask whether these theories are really all independent, or whether a theory $(\mathfrak{g}',\lambda',\dext',\langle \cdot,\cdot\rangle)$ with moduli space $\mathcal{M}'$ is a special case of a theory $(\mathfrak{g},\lambda,\dext,\langle \cdot,\cdot\rangle)$ with moduli space $\mathcal{M}$. We showed that there are two natural ways one can get such descendent theories: either $\mathcal{M}'$ is embedded as a totally geodesic submanifold of $\mathcal{M}$, and corresponds to freezing some of the moduli, or it arises as a decompactification limit of $\mathcal{M}$ in which case it has a precise definition in terms of the parabolic subgroup associated to the decompactification geodesic. The polytope $C'$ of the descendent theory comes either from a projection of the polytope $C$ of the original theory, in the embedded case, or from a slice of $C$ at a fixed distance from a face of $C$, in the decompactification case. It is important to note that the existence of $C'$ as a suitable projection or slice of $C$ is not sufficient to imply that $(\mathfrak{g}',\lambda',\dext',\langle \cdot,\cdot\rangle)$ is a descendent of  $(\mathfrak{g},\lambda,\dext,\langle \cdot,\cdot\rangle)$, one always needs to check that the algebras and representations embed suitably. In particular, generically one gets a descendent theory with a reducible representation which is hence outside our classification. Starting with any given theory one then gets a complicated tree of descendent theories with the same or larger $d$. 

Returning to our previous suggestion,  we can then ask which of the minimal theories are descendents of the $d=3$, $\splitE8$  theory that corresponds physically to M-theory on $T^8$, and hence are indeed in the string landscape. We give the details of this analysis in Appendix~\ref{app:chains}. The result is that indeed nearly all of them are, with only three exceptional extra ``seed theories'' namely 
\begin{equation*}
\begin{aligned}
   d=3\, : \quad  & (\splitG2, \mathbf{14})_{\nvertex=3} \, ,  \, (\splitC4, \mathbf{27})\, ,  \, (\splitF4, \mathbf{26}) \, . 
   \end{aligned}
\end{equation*}
Whether these also have some string theory realisation is unclear to us. For the theories that are seeded by the $\splitE8$ theory, one should be able analyse how to freeze moduli to give an embedded descendent by forming generalised orbifolds, quotienting by elements of the duality group. (It is not clear however if such a quotient would yield a honest string theory with, for instance, a modular invariant partition function.) There are similar trees of theories starting with non-minimal seeds that one can derive. However since a descendent theory always inherits its metric $\langle \cdot,\cdot\rangle$ from its parent, one cannot relate minimal and non-minimal trees. 

The analysis of this paper is equally applicable to representations of $p$-brane states and the rates of decay of their tension as one approaches the boundary of moduli space. Furthermore, as we have shown, one can infer some of the string and brane content of a theory directly from the particle representation. This can be done in two ways: from the emergent string limits, or from decompactification limits, where different slices of the particle polytope correspond to weight polytopes for KK modes, particles, strings and (generalised) KK monopoles of the descendent theory, or if more than one dimension decompactifies, to higher dimensional branes. We gave a complete discussion of the string polytopes that arise from the minimal particle theories in the former case, and discussed several examples in the latter case, showing how they agreed with the emergent string analysis. We also showed that the combined ``$\pmax$-hull'' of all the brane states,  $C^\text{max}_p$, is just the projection of the parent particle polytope onto the Cartan algebra of the compactified descendent theory.

Following~\cite{Baines:2025upi} one might also wonder how much of our analysis extends to the case where $\mathcal{M}$ is only asymptotically locally symmetric, that is with regard to the analysis of the geometry of the asymptotic geodesics. The essential ingredients in our discussion are the emergence of an algebraic symmetry group defined over the rationals, and states forming a representation under the real form of the same group. In terms of the polytopes, if the group is parabolic $P(\mathbb{Q})$, a given cusp of $\mathcal{M}$ corresponds to a positive Weyl chamber $\caa_{P(\mathbb{Q})}^+$, and hence only one ``wedge'' of the polytope. However, the Weyl symmetry implies that the analysis of the allowed polytopes remains the same, even if other parts of the polytope do not correspond to asymptotic regions of $\mathcal{M}$. 

As we have already discussed, the majority of the minimal theories descend from the $\splitE8$ theory and so are part of the string landscape. One might then wonder if there is a way to exclude the infinite towers of allowed theories that are associated to each minimal one. As discussed in Section~\ref{sec:stringinter}, 
completeness of the spectrum does not by itself constrain the allowed representations, but only determines the global form of $G$ that acts faithfully on the particle states. Taking a rescaled $\lambda=\gamma\lambda_\text{min}$ highest-weight representation does introduce more particle states, and so should affect the calculation of the species scale. The analysis of how this could restrict the allowed theories is an interesting question that we leave for future work. However, this would still not exclude the tower of theories based on rescaling $\nvertex=z\nvertex^\text{min}$. 

By far the most natural way to exclude both towers is to put a lower bound on the metric $\langle \cdot,\cdot\rangle$ (or equivalently an upper bound on the curvature $k$ of the symmetric space in~\eqref{eq:curv-bound-concl}). Here supersymmetry is the most obvious candidate, since at least for theories with eight or more supercharges it fixes the length-squared $|r_\text{long}|^2$ of the long roots of $\cgg$, typically to $|r_\text{long}|^2=2$. However, as stands, many of our examples are not those of supersymmetric symmetric spaces in the given dimension $d$ (as listed for example in~\cite{Ferrara:2008de}). The crucial point here though is that, for simplicity, we have restricted our analysis to split Lie groups. In extending to non-split cases, one finds additional non-split minimal theories that can realise the polytopes in our lists in Tables~\ref{tab:1irrep_integern_scaled_global} and~\ref{tab:1irrep_integern_extra}. A preliminary analysis shows that in many of them there is a non~-~split form that is indeed supersymmetric although in some cases with less than half-maximal supersymmetry.  

Extending this analysis to the non-split case, and also to reducible representations, is consequently a very natural next step. The first aspect appears to be straightforward, and we expect will not introduce any additional polytopes to our list. Going to reducible representations is however more involved, leading to a more complicated relationship between polytope faces and algebras that will allow for new polytopes. The suggestion is nonetheless that the list of global polytopes should still be finite. In addition, there are indications that many and perhaps all polytopes admit a realisation with a supersymmetric EFT that lies in the string landscape. However, many also admit realisations in terms of symmetric spaces whose string interpretation is unclear.  We leave a full analysis of both these questions for future work.  

\vspace{0.5cm}

\noindent {\bf \large Acknowledgments\vspace{0.1in}}\\
We would like to thank Jos\'e Calderón Infante, Muldrow Etheredge, Carmine Montella, Héctor Parra de Freitas, Ignacio Ruiz Garc\'ia, Sanjay Raman, Cumrun Vafa and Irene Valenzuela for useful discussions.  V.C.~is supported by ERC Starting Grant QGuide101042568 - StG 2021. B.F. is supported by a Juan de la Cierva contract (JDC2023-050850-I) from Spain’s Ministry of Science, Innovation and Universities. The work of V.C.~and B.F.~is also supported through the grants CEX2020-001007-S, PID2021-123017NB-I00 and PID2024-156043NB-I00, funded by MCIN/AEI/10.13039/501100011033, and ERDF, EU. D.W.~is supported in part by the STFC Consolidated Grant ST/X000575/1. D.W.~also thanks the Galileo Galilei Institute for Theoretical Physics and the Leinweber Institute for Theoretical Physics at UC Berkeley for hospitality and support during the completion of this work.

\appendix

\section{Simplices, orthoplexes and Weyl polytopes} 
\label{app:regandcross}

We saw in section~\ref{sec:classification} that the Weyl polytopes that satisfy the ESC are built out of simplices and orthoplexes. In this appendix, for completeness, we define the latter and give some more details on the resulting allowed Weyl polytopes. 

A {\it regular $p$-dimensional simplex} $\Delta_p$ can be defined as the convex hull of its $p+1$ vertices embedded in $\mathbb{R}^{p+1}$ as $e_1=(1,0_{p})$ and $p+1$ permutations thereof. All of its lower dimensional faces are themselves regular simplices and the vertices of $\Delta_p$ all lie in the plane $x\cdot m=1$ where $m=(1.\dots,1)$. The centres of the $k$-dimensional faces lie at the points $(\frac1{k+1},\dots,\frac1{k+1},0_{p-k})$ and permutations thereof and the centre of the simplex lies at $(\frac{1}{p+1},\dots,\frac{1}{p+1})$. If we shift the origin so that the vertices lie at $e'_i=e_i+\gamma m$, we see that the distance to the centre of a $k$-dimensional face is given by 
\begin{equation}
    d_k^2 = \frac{1}{k+1} + \gamma + (p+1)\gamma^2 \, . 
\end{equation}
Thus setting $\gamma + (p+1)\gamma^2=\frac{1}{d-2}$ we see that the simplex satisfies the constraints of~\eqref{eq:dn2} with all the faces giving decompactification limits. 

We further note that $\Delta_p$ is the Weyl polytope of the vector representation of $\mathfrak{sl}(p+1,\mathbb{R})$ with the $\lambda$-extended Dynkin diagram
\begin{align}
\label{eq:ext-dyn-simplex}
    \mathbf{p+1} :     
    \tikzset{/Dynkin diagram,root radius=.125cm,edge length=.55cm}
    \begin{dynkinDiagram}{A}{oo.ooo}
    \node[circle,inner sep=0pt,minimum size=2.5mm,fill={\colExt},draw] (A) at ($(root 1) + (0,0.55)$) {}; 
    \draw[\colExt] (A) -- (root 1);
    \end{dynkinDiagram}  
\end{align}
A \textit{$p$-dimensional orthoplex}, labelled as $\beta_p$ by Coxeter, is a regular polytope which can be defined as the convex hull of the points $(\pm1,0_{p-1})$ in $\mathbb{R}^p$ and permutations thereof and by construction is centrally symmetric. All of its lower-dimensional faces are regular simplices. 

The centres of the $k$-dimensional faces lie at points of the form $(\pm\frac1{k+1},\dots,\pm\frac1{k+1},0_{p-k-1})$ and permutations thereof (where the signs are not correlated) so that their squared-distance from the centre of the orthoplex is given by $\frac{1}{k+1}$. Thus if we place the orthoplex at an orthogonal distance $\frac{1}{\sqrt{d-2}}$ from the origin in $\mathbb{R}^{p+1}$ we have the distance to the centre of each face is given by 
\begin{equation}
    d_k^2 = \begin{cases} 
        \frac{1}{k+1} + \frac{1}{d-2} & \text{if $k<p$} \, , \\ 
        \frac{1}{d-2} & \text{if $k=p$} \, ,
        \end{cases}
\end{equation}
and we see that the orthoplex satisfies the constraints of~\eqref{eq:dn2} with all the codimension two and higher faces giving decompactification limits while the codimension-one faces give emergent string limits. 

Finally, we also note that $\beta_p$ is the Weyl polytope of the vector representations of $\mathfrak{so}(p~+~1,p)$, $\mathfrak{sp}(2p,\mathbb{R})$ and $\mathfrak{so}(p,p)$ with the corresponding $\lambda$-extended Dynkin diagrams
\begin{align}
\label{eq:ext-dyn-ortho}
    \mathbf{2p+1} : 
    \tikzset{/Dynkin diagram,root radius=.125cm,edge length=.55cm}
    \begin{dynkinDiagram}{B}{oo.ooo}
    \node[circle,inner sep=0pt,minimum size=2.5mm,fill={\colExt},draw] (A) at ($(root 1) + (0,0.55)$) {}; 
    \draw[\colExt] (A) -- (root 1);
    \end{dynkinDiagram} \qquad
    \mathbf{2p} : 
    \tikzset{/Dynkin diagram,root radius=.125cm,edge length=.55cm}
    \begin{dynkinDiagram}{C}{oo.ooo}
    \node[circle,inner sep=0pt,minimum size=2.5mm,fill={\colExt},draw] (A) at ($(root 1) + (0,0.55)$) {}; 
    \draw[\colExt] (A) -- (root 1);
    \end{dynkinDiagram} \qquad
    \mathbf{2p} : 
    \tikzset{/Dynkin diagram,root radius=.125cm,edge length=.55cm}
    \begin{dynkinDiagram}{D}{oo.ooo}
    \node[circle,inner sep=0pt,minimum size=2.5mm,fill={\colExt},draw] (A) at ($(root 1) + (0,0.55)$) {}; 
    \draw[\colExt] (A) -- (root 1); 
    \end{dynkinDiagram} 
\end{align}
    
\begin{figure}[t]
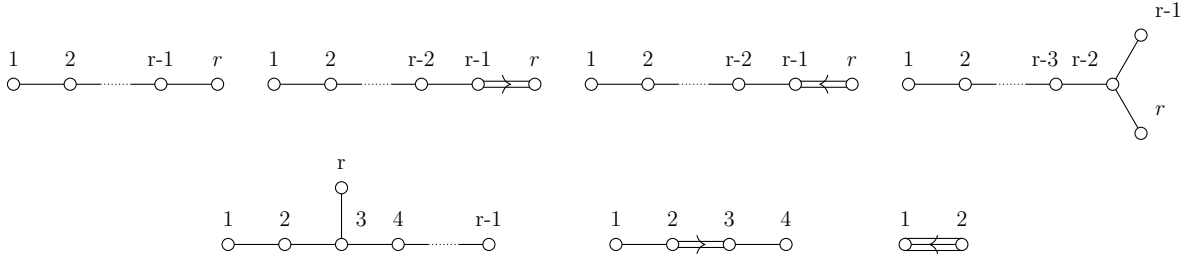

\centering

\begin{dynkinDiagram}[root
radius=.08cm,edge length=.75cm,labels*={1,2,$r{-}1$,r}]{A}{oo.oo}
\end{dynkinDiagram}\quad
\begin{dynkinDiagram}[root
radius=.08cm,edge length=.75cm,labels*={1,2,$r{-}2$,$r{-}1$,r}]{B}{oo.ooo}
\end{dynkinDiagram}\quad
\begin{dynkinDiagram}[root
radius=.08cm,edge length=.75cm,labels*={1,2,$r{-}2$,$r{-}1$,r}]{C}{oo.ooo}
\end{dynkinDiagram}\quad
\begin{dynkinDiagram}[root
radius=.08cm,edge length=.75cm,labels*={1,2,$r{-}3\quad$,$r{-}2$,$r{-}1$,r}]{D}{oo.oooo}
\end{dynkinDiagram}

\begin{dynkinDiagram}[root
radius=.08cm,edge length=.75cm,labels*={1,$r$,2,3,4,$r{-}1$}]{E}{ooooo.o}
\end{dynkinDiagram}\quad\quad\quad
\begin{dynkinDiagram}[root
radius=.08cm,edge length=.75cm,labels*={1,2,3,4}]{F}{oooo}
\end{dynkinDiagram}\quad\quad\quad
\begin{dynkinDiagram}[root
radius=.08cm,edge length=.75cm,labels*={1,2}]{G}{oo}
\end{dynkinDiagram}
\caption{Dynkin diagrams of simple Lie algebras and the ordering convention for the simple roots. On the upper row we show the classical series $A_r$, $B_r$, $C_r$, and $D_r$, while the exceptional types $E_r$, $F_4$, and $G_2$ are on the bottom.} \label{fig:DynkinDiagrams}
\end{figure}

In the next three tables we characterise more precisely the different polytopes in Table~\ref{tab:polytopes_all1}. For the global polytopes we list the Lie algebra and $\lambda$-extended Dynkin diagrams in Table~\ref{tab:lambda-Dynkin} and the number and types of faces in Table~\ref{tab:facesparticles}, grouped by polytope. For the perturbative polytopes the $\lambda$-extended Dynkin diagrams were given in~\eqref{eq:ext-dyn-simplex} and~\eqref{eq:ext-dyn-ortho} and the number and types of faces are shown in Table~\ref{tab:facesparticlesU1short}. 
We also give in Figure \ref{fig:DynkinDiagrams} the Dynkin labels for the simple Lie algebras used in Section \ref{sec:classification}.

\begin{table}[H]\footnotesize\centering
\begin{minipage}[t]{.48\linewidth}
\centering
\vspace{0pt}
\begin{tabular}{|>{$}c<{$}|>{$}c<{$}|>{$}c<{$}|>{$}c<{$}|>{$}c<{$}|} 
\hline
C & \mathfrak{g} & \rho & \dext & \text{$\lambda$-Dynkin} \\
\hline\hline
\EeightAlt & \splitE8 & \mathbf{248} & 3 & 
    \tikzset{/Dynkin diagram,root radius=.09cm,edge length=.35cm}
    \begin{dynkinDiagram}[backwards]{E}{oooooooo}
    \node[circle,inner sep=0pt,minimum size=1.8mm,fill={\colExt},draw] (A) at ($(root 8) + (0,0.35)$) {}; 
    \draw[\colExt, shorten >=-1.3pt] (A) -- (root 8);
    \end{dynkinDiagram} \\ 
\EsevenAlt &\splitE7 & \mathbf{56} & 4 & 
    \tikzset{/Dynkin diagram,root radius=.09cm,edge length=.35cm}
    \begin{dynkinDiagram}[backwards]{E}{ooooooo}
    \node[circle,inner sep=0pt,minimum size=1.8mm,fill={\colExt},draw] (A) at ($(root 7) + (0,0.35)$) {}; 
    \draw[\colExt, shorten >=-1.3pt] (A) -- (root 7);
    \end{dynkinDiagram}  \\
\EsixAlt & \splitE6 & \mathbf{27} & 5 & 
    \tikzset{/Dynkin diagram,root radius=.09cm,edge length=.35cm}
    \begin{dynkinDiagram}[backwards]{E}{oooooo}
    \node[circle,inner sep=0pt,minimum size=1.8mm,fill={\colExt},draw] (A) at ($(root 6) + (0,0.35)$) {}; 
    \draw[\colExt, shorten >=-1.3pt] (A) -- (root 6);
    \end{dynkinDiagram} \\
\text{demi} & \splitD5 & \mathbf{16} & 6 & 
    \tikzset{/Dynkin diagram,root radius=.09cm,edge length=.35cm}
    \begin{dynkinDiagram}[backwards]{D}{ooooo}
    \node[circle,inner sep=0pt,minimum size=1.8mm,fill={\colExt},draw] (A) at ($(root 4) + (0.35,0)$) {}; 
    \draw[\colExt, shorten >=-1.3pt] (A) -- (root 4);
    \end{dynkinDiagram} \\*[8pt] 
\text{tetro} & \splitA4 & \mathbf{10}^* & 7 & 
    \tikzset{/Dynkin diagram,root radius=.09cm,edge length=.35cm}
    \begin{dynkinDiagram}{A}{oooo}
    \node[circle,inner sep=0pt,minimum size=1.8mm,fill={\colExt},draw] (A) at ($(root 3) + (0,0.35)$) {}; 
    \draw[\colExt, shorten >=-1.3pt] (A) -- (root 3);
    \end{dynkinDiagram} \\ 
\triprism & \splitA2 \oplus \splitA1 & \mathbf{\left(3^*, 2\right)} & 8 & 
    \tikzset{/Dynkin diagram,root radius=.09cm,edge length=.35cm}
    \begin{dynkinDiagram}[name=left]{A}{oo}
    \node[circle,inner sep=0pt,minimum size=1.8mm,fill={\colExt},draw] (A) at ($(left root 2) + (0.25,0.35)$) {}; 
    \node (B) at ($(left root 2)+(0.5,0)$) {};
    \dynkin[at=(B),name=right,]A{o};
    \draw[\colExt, shorten >=-1.3pt] (A) -- (left root 2);
    \draw[\colExt, shorten >=-1.3pt] (A) -- (right root 1);
    \end{dynkinDiagram}  \\ 
\hline
\multirow{7}{*}{\text{oc}} & \splitF4 & \mathbf{26} & 3 & 
    \tikzset{/Dynkin diagram,root radius=.09cm,edge length=.35cm}
    \begin{dynkinDiagram}{F}{oooo}
    \node[circle,inner sep=0pt,minimum size=1.8mm,fill={\colExt},draw] (A) at ($(root 4) + (0,0.35)$) {}; 
    \draw[\colExt, shorten >=-1.3pt] (A) -- (root 4);
    \end{dynkinDiagram} \\ 
& \splitF4 & \mathbf{52} & 3 & 
    \tikzset{/Dynkin diagram,root radius=.09cm,edge length=.35cm}
    \begin{dynkinDiagram}{F}{oooo}
    \node[circle,inner sep=0pt,minimum size=1.8mm,fill={\colExt},draw] (A) at ($(root 1) + (0,0.35)$) {}; 
    \draw[\colExt, shorten >=-1.3pt] (A) -- (root 1);
    \end{dynkinDiagram}  \\
& \splitD4 & \mathbf{28} & 3 & 
    \tikzset{/Dynkin diagram,root radius=.09cm,edge length=.35cm}
    \begin{dynkinDiagram}[backwards]{D}{oooo}
    \node[circle,inner sep=0pt,minimum size=1.8mm,fill={\colExt},draw] (A) at ($(root 2) + (0.35,0)$) {}; 
    \draw[\colExt, shorten >=-1.3pt] (A) -- (root 2);
    \end{dynkinDiagram}  \\*[8pt]
& \splitC4 & \mathbf{27} & 3 & 
    \tikzset{/Dynkin diagram,root radius=.09cm,edge length=.35cm}
    \begin{dynkinDiagram}{C}{oooo}
    \node[circle,inner sep=0pt,minimum size=1.8mm,fill={\colExt},draw] (A) at ($(root 2) + (0,0.35)$) {}; 
    \draw[\colExt, shorten >=-1.3pt] (A) -- (root 2);
    \end{dynkinDiagram} \\
& \splitB4 & \mathbf{36} & 3 & 
    \tikzset{/Dynkin diagram,root radius=.09cm,edge length=.35cm}
    \begin{dynkinDiagram}{B}{oooo}
    \node[circle,inner sep=0pt,minimum size=1.8mm,fill={\colExt},draw] (A) at ($(root 2) + (0,0.35)$) {}; 
    \draw[\colExt, shorten >=-1.3pt] (A) -- (root 2);
    \end{dynkinDiagram} \\ 
\hline
\multirow{4}{*}{\cuboctahedron} & \splitA3 & \mathbf{15} & 3 & 
    \tikzset{/Dynkin diagram,root radius=.09cm,edge length=.35cm}
    \begin{dynkinDiagram}{A}{ooo}
    \node[circle,inner sep=0pt,minimum size=1.8mm,fill={\colExt},draw] (A) at ($(root 2) + (0,0.35)$) {}; 
    \draw[\colExt, shorten >=-2pt] (A) -- (root 1);
    \draw[\colExt, shorten >=-2pt] (A) -- (root 3);
    \end{dynkinDiagram} \\
& \splitB3 & \mathbf{21} & 3 & 
    \tikzset{/Dynkin diagram,root radius=.09cm,edge length=.35cm}
    \begin{dynkinDiagram}{B}{ooo}
    \node[circle,inner sep=0pt,minimum size=1.8mm,fill={\colExt},draw] (A) at ($(root 2) + (0,0.35)$) {}; 
    \draw[\colExt, shorten >=-1.3pt] (A) -- (root 2);
    \end{dynkinDiagram} \\
& \splitC3 & \mathbf{14} & 3 & 
    \tikzset{/Dynkin diagram,root radius=.09cm,edge length=.35cm}
    \begin{dynkinDiagram}{C}{ooo}
    \node[circle,inner sep=0pt,minimum size=1.8mm,fill={\colExt},draw] (A) at ($(root 2) + (0,0.35)$) {}; 
    \draw[\colExt, shorten >=-1.3pt] (A) -- (root 2);
    \end{dynkinDiagram} \\ 
\hline
\end{tabular} 
\end{minipage}
\begin{minipage}[t]{.48\linewidth}
\centering
\vspace{0pt}
\begin{tabular}{|>{$}c<{$}|>{$}c<{$}|>{$}c<{$}|>{$}c<{$}|>{$}c<{$}|} 
\hline
C & \mathfrak{g} & \rho & \dext & \text{$\lambda$-Dynkin} \\
\hline\hline
\multirow{7}{*}{\cube} & \splitC3 & \mathbf{14'} & 4 & 
    \tikzset{/Dynkin diagram,root radius=.09cm,edge length=.35cm}
    \begin{dynkinDiagram}{C}{ooo}
    \node[circle,inner sep=0pt,minimum size=1.8mm,fill={\colExt},draw] (A) at ($(root 3) + (0,0.35)$) {}; 
    \draw[\colExt, shorten >=-1.3pt] (A) -- (root 3);
    \end{dynkinDiagram} \\
& \splitB3 & \mathbf{8} & 4 & 
    \tikzset{/Dynkin diagram,root radius=.09cm,edge length=.35cm}
    \begin{dynkinDiagram}{B}{ooo}
    \node[circle,inner sep=0pt,minimum size=1.8mm,fill={\colExt},draw] (A) at ($(root 3) + (0,0.35)$) {}; 
    \draw[\colExt, shorten >=-1.3pt] (A) -- (root 3);
    \end{dynkinDiagram} \\
& \splitC2 \oplus \splitA1 &\mathbf{\left(4, 2\right)} & 4 & 
    \tikzset{/Dynkin diagram,root radius=.09cm,edge length=.35cm}
    \begin{dynkinDiagram}[name=left]{C}{oo}
    \node[circle,inner sep=0pt,minimum size=1.8mm,fill={\colExt},draw] (A) at ($(left root 2) + (0.25,0.35)$) {}; 
    \node (B) at ($(left root 2)+(0.5,0)$) {};
    \dynkin[at=(B),name=right,]A{o};
    \draw[\colExt, shorten >=-1.3pt] (A) -- (left root 2);
    \draw[\colExt, shorten >=-1.3pt] (A) -- (right root 1);
    \end{dynkinDiagram}  \\
& \splitC2 \oplus \splitA1 & \mathbf{\left(5, 2\right)} & 4 & 
    \tikzset{/Dynkin diagram,root radius=.09cm,edge length=.35cm}
    \begin{dynkinDiagram}[name=left]{B}{oo}
    \node[circle,inner sep=0pt,minimum size=1.8mm,fill={\colExt},draw] (A) at ($(left root 2) + (0.25,0.35)$) {}; 
    \node (B) at ($(left root 2)+(0.5,0)$) {};
    \dynkin[at=(B),name=right,]A{o};
    \draw[\colExt, shorten >=-1.3pt] (A) -- (left root 2);
    \draw[\colExt, shorten >=-1.3pt] (A) -- (right root 1);
    \end{dynkinDiagram}  \\
& \splitA1^{\oplus 3} & \mathbf{\left(2, 2, 2\right)} & 4 & 
    \tikzset{/Dynkin diagram,root radius=.09cm,edge length=.35cm,edge/.style={draw={\colExt}}}
    \begin{dynkinDiagram}{D}{oooo}
    \node[circle,inner sep=0pt,minimum size=1.8mm,fill={\colExt},draw] (A) at (root 2) {}; 
    \end{dynkinDiagram} \\*[8pt] 
\hline
\multirow{2}{*}{\octagon} & \splitC2 & \mathbf{16} & 3 & 
    \tikzset{/Dynkin diagram,root radius=.09cm,edge length=.35cm}
    \begin{dynkinDiagram}{C}{oo}
    \node[circle,inner sep=0pt,minimum size=1.8mm,fill={\colExt},draw] (A) at ($0.5*(root 1) + 0.5*(root 2) + (0,0.35)$) {}; 
    \draw[\colExt, shorten >=-2pt] (A) -- (root 1);
    \draw[\colExt, shorten >=-2pt] (A) -- (root 2);
    \end{dynkinDiagram} \\
& \splitC2 & \mathbf{35} & 3 & 
    \tikzset{/Dynkin diagram,root radius=.09cm,edge length=.35cm}
    \begin{dynkinDiagram}{C}{oo}
    \node[circle,inner sep=0pt,minimum size=1.8mm,fill={\colExt},draw] (A) at ($0.5*(root 1) + 0.5*(root 2) + (0,0.35)$) {}; 
    \draw[\colExt, shorten >=-2pt] (A) -- (root 1);
    \draw[\colExt, shorten >=-2pt] (A) -- (root 2);
    \node[label={[text={\colExt},label distance=-1.5mm]\tiny$2$}] at (root 1) {};
    \end{dynkinDiagram} \\
\hline
\multirow{3}{*}{\large\hexagon} & \splitG2 & \mathbf{14} & 3 & 
    \tikzset{/Dynkin diagram,root radius=.09cm,edge length=.35cm}
    \begin{dynkinDiagram}{G}{oo}
    \node[circle,inner sep=0pt,minimum size=1.8mm,fill={\colExt},draw] (A) at ($(root 2) + (0,0.35)$) {}; 
    \draw[\colExt, shorten >=-1.3pt] (A) -- (root 2);
    \end{dynkinDiagram} \\
& \splitG2 & \mathbf{7} & 3 & 
    \tikzset{/Dynkin diagram,root radius=.09cm,edge length=.35cm}
    \begin{dynkinDiagram}{G}{oo}
    \node[circle,inner sep=0pt,minimum size=1.8mm,fill={\colExt},draw] (A) at ($(root 1) + (0,0.35)$) {}; 
    \draw[\colExt, shorten >=-1.3pt] (A) -- (root 1);
    \end{dynkinDiagram} \\
& \splitA2 & \mathbf{8} & 3 & 
    \tikzset{/Dynkin diagram,root radius=.09cm,edge length=.35cm}
    \begin{dynkinDiagram}{A}{oo}
    \node[circle,inner sep=0pt,minimum size=1.8mm,fill={\colExt},draw] (A) at ($0.5*(root 1) + 0.5*(root 2) + (0,0.35)$) {}; 
    \draw[\colExt, shorten >=-2pt] (A) -- (root 1);
    \draw[\colExt, shorten >=-2pt] (A) -- (root 2);
    \end{dynkinDiagram} \\ 
\hline
\multirow{3}{*}{$\square$} & \splitC2 & \mathbf{4} & 3 & 
    \tikzset{/Dynkin diagram,root radius=.09cm,edge length=.35cm}
    \begin{dynkinDiagram}{C}{oo}
    \node[circle,inner sep=0pt,minimum size=1.8mm,fill={\colExt},draw] (A) at ($(root 1) + (0,0.35)$) {}; 
    \draw[\colExt, shorten >=-1.3pt] (A) -- (root 1);
    \end{dynkinDiagram} \\
& \splitC2 & \mathbf{5} & 3 & 
    \tikzset{/Dynkin diagram,root radius=.09cm,edge length=.35cm}
    \begin{dynkinDiagram}{C}{oo}
    \node[circle,inner sep=0pt,minimum size=1.8mm,fill={\colExt},draw] (A) at ($(root 2) + (0,0.35)$) {}; 
    \draw[\colExt, shorten >=-1.3pt] (A) -- (root 2);
    \end{dynkinDiagram} \\
& \splitA1^{\oplus 2} & \mathbf{\left(2, 2\right)} & 3 & 
    \tikzset{/Dynkin diagram,root radius=.09cm,edge length=.35cm,edge/.style={draw={\colExt}}}
    \begin{dynkinDiagram}{A}{ooo}
    \node[circle,inner sep=0pt,minimum size=1.8mm,fill={\colExt},draw] (A) at (root 2) {}; 
    \end{dynkinDiagram} \\  
\hline
\rectangle & \splitA1^{\oplus 2} & \mathbf{\left(2, 2\right)} & 4 &
    \tikzset{/Dynkin diagram,root radius=.09cm,edge length=.35cm,edge/.style={draw={\colExt}}}
    \begin{dynkinDiagram}{A}{ooo}
    \node[circle,inner sep=0pt,minimum size=1.8mm,fill={\colExt},draw] (A) at (root 2) {}; 
    \end{dynkinDiagram} \\  
\hline
\triangle & \splitA2 & \mathbf{3} & 5 & 
    \tikzset{/Dynkin diagram,root radius=.09cm,edge length=.35cm}
    \begin{dynkinDiagram}{A}{oo}
    \node[circle,inner sep=0pt,minimum size=1.8mm,fill={\colExt},draw] (A) at ($(root 1) + (0,0.35)$) {}; 
    \draw[\colExt, shorten >=-1.3pt] (A) -- (root 1);
    \end{dynkinDiagram} \\ 
\hline
\IIB & \splitA1 & \mathbf{2} & \dext & 
    \tikzset{/Dynkin diagram,root radius=.09cm,edge length=.35cm}
    \begin{dynkinDiagram}{A}{o}
    \node[circle,inner sep=0pt,minimum size=1.8mm,fill={\colExt},draw] (A) at ($(root 1) + (0,0.35)$) {}; 
    \draw[\colExt, shorten >=-1.3pt] (A) -- (root 1);
    \end{dynkinDiagram} \\ 
\hline
\end{tabular}
\end{minipage}
\caption{$\lambda$-extended Dynkin diagrams denoting the global polytopes in Tables~\ref{tab:1irrep_integern_scaled_global} and~\ref{tab:1irrep_integern_extra}.}
\label{tab:lambda-Dynkin}
\end{table}

\begin{table}[H]\centering\footnotesize
\setlength{\tabcolsep}{3pt}
\begin{tabular}{|>{$}c<{$}|>{$}c<{$}|>{$}c<{$}|>{$}c<{$}|>{$}c<{$}>{$}c<{$}>{$}c<{$}>{$}c<{$}>{$}c<{$}>{$}c<{$}>{$}c<{$}>{$}c<{$}|>{$}c<{$}|} \hline
 C & \mathfrak{g} & \rho & \dext & \Delta_{0} & \Delta_{1} & \Delta_{2} & \Delta_{3} & \Delta_{4} & \Delta_{5} & \Delta_{6} & \Delta_{7} & \beta_{k-1}
 \\ \hline\hline
 \EeightAlt & \splitE8 & \mathbf{248} & 3 & 240 & 6720 & 60480 & 241920 & 483840 & 483840 & 207360 & 17280  & 2160_{\splitD7} \\
 \EsevenAlt & \splitE7 & \mathbf{56} & 4 & 56 & 756 & 4032 & 10080 & 12096 & 6048 & 576 &  & 126_{\splitD6} \\
 \EsixAlt & \splitE6 & \mathbf{27} & 5 & 27 & 216 & 720 & 1080 & 648 & 72 &  &  & 27_{\splitD5} \\
 \EfiveAlt & \splitD5 & \mathbf{16} & 6 & 16 & 80 & 160 & 120 & 16 &  &  &  & 10_{\splitD4} \\
 \EfourAlt & \splitA4 & \mathbf{10}^* & 7 & 10 & 30 & 30  & 5 &  &  &  &  & 5_{\splitD3}\\
 \triprism & \splitA2 \oplus \splitA1 & \mathbf{\left(3^*, 2\right)} & 8 & 6 & 9 & 2 &  &  &  &  &  & 3_{\splitD2} \\ \hline
 \multirow{5}{*}{oc} & \splitF4 & \mathbf{26} & \multirow{5}{*}{3} & \multirow{5}{*}{24} & \multirow{5}{*}{96} & \multirow{5}{*}{96} & \multirow{5}{*}{0} & \multirow{5}{*}{} & \multirow{5}{*}{} & \multirow{5}{*}{} & \multirow{5}{*}{} & 24_{\splitC3} \\
 & \splitF4 & \mathbf{52} &  & & & &  &  &  &  &  & 24_{\splitB3} \\
 & \splitD4 & \mathbf{28} &  & & & &  &  &  &  &  & 24_{\splitD3}\\
 & \splitC4 & \mathbf{27} &  & & & &  &  &  &  &  & 16_{\splitD3}+8_{\splitC3}\\
 & \splitB4 & \mathbf{36} &  & & & &  &  &  &  &  & 16_{\splitD3}+8_{\splitB3}\\ \hline
 \multirow{3}{*}{\cuboctahedron} & \splitA3 & \mathbf{15} &  \multirow{3}{*}{3} & \multirow{3}{*}{12} & \multirow{3}{*}{24} & \multirow{3}{*}{8}  &  &  &  &  &  & 6_{\splitD2}\\
 & \splitB3 & \mathbf{21} &  &  &  &  &  &  &  &  &  & 6_{\splitB2}\\
 & \splitC3 & \mathbf{14} &  &  &  &  &  &  &  &  &  & 6_{\splitC2}\\ \hline
 \multirow{5}{*}{\cube} & \splitC3 & \mathbf{14'} & \multirow{5}{*}{4} & \multirow{5}{*}{8} & \multirow{5}{*}{12} & \multirow{5}{*}{0} &  &  &  &  &  & 6_{\splitB2}\\
 & \splitB3 & \mathbf{8} & & & &    &  &  &  &  &  & 6_{\splitC2}  \\
 & \splitC2 \oplus \splitA1 &\mathbf{\left(4, 2\right)} & & & &&&&&&& 4_{\splitD2}+2_{\splitC2}  \\
 & \splitC2 \oplus \splitA1 & \mathbf{\left(5, 2\right)} & & & &&&&&&& 4_{\splitD2}+2_{\splitB2}  \\
 & \splitA1^{\oplus 3} & \mathbf{\left(2, 2, 2\right)} & & & &&&&&&& 6_{\splitD2}\\ \hline
 \multirow{2}{*}{\octagon} & \splitC2 & \mathbf{16}& \multirow{2}{*}{3} & \multirow{2}{*}{8} & \multirow{2}{*}{4} &&&&&&& 4 \\
 & \splitC2 & \mathbf{35} & & & &&&&&&& 4 \\ \hline
  \multirow{3}{*}{\text{$\hexagon_\infty$}} & \splitG2 & \mathbf{14}& \multirow{3}{*}{3} & \multirow{3}{*}{6} & \multirow{3}{*}{0} &&&&&&& 6 \\
 & \splitG2 & \mathbf{7} & & & &&&&&&& 6 \\
 & \splitA2 & \mathbf{8} & & & &&&&&&& 6 \\ \hline
 \multirow{3}{*}{\hexagon} & \splitG2 & \mathbf{14}& \multirow{3}{*}{3} & \multirow{3}{*}{6} & \multirow{3}{*}{6} &&&&&&& 0 \\
 & \splitG2 & \mathbf{7} & & & &&&&&&& 0 \\
 & \splitA2 & \mathbf{8} & & & &&&&&&& 0 \\ \hline
  \multirow{3}{*}{\text{$\square$}} & \splitC2 & \mathbf{4} & \multirow{3}{*}{3} & \multirow{3}{*}{4} & \multirow{3}{*}{0} &  &  &  &  & & & 4 \\
 & \splitC2 & \mathbf{5} & & &  &  &  &  &  & & & 4 \\
 & \splitA1^{\oplus 2} &\mathbf{\left(2, 2\right)} & & &  &  &  &  &  & & & 4 \\  \hline
\rectangle & \splitA1^{\oplus 2} & \mathbf{\left(2, 2\right)} & 4 & 4 & 2 &&&&&&& 2\\  \hline
\triangle & \splitA2 & \mathbf{3} & 5 & 3 & 0 &  &  &  &  & & & 3  \\ \hline
\IIB_\infty & \splitA1 & \mathbf{2} & \dext & 0 &  &  &  &  &  &  &  & 2\\ \hline
\IIB_z & \splitA1 & \mathbf{2}& \dext & 2 &  &  &  &  &  &  &  & 0 \\
\hline
\end{tabular}\caption{Number of faces of a given type for the global polytopes of Tables~\ref{tab:1irrep_integern_scaled_global} and~\ref{tab:1irrep_integern_extra}. In the last column we indicate the subalgebra defining the orthoplex faces.}
\label{tab:facesparticles}\end{table}

\begin{table}[H]\centering\footnotesize
\begin{tabular}{|>{$}c<{$}|>{$}c<{$}|>{$}c<{$}|>{$}c<{$}|} 
\hline \mathfrak{g} & \rho & \Delta_{k'-1} & \beta_{k-1}  \\ \hline
 \splitAkminus1 \oplus \mathbb{R} & \mathbf{k}\left(\sqrt{\frac{1}{\dext-2}+\frac{1}{k}}\right) &  \binom{k}{k'} & 0  \\
\hline  \splitDk \oplus \mathbb{R} & \mathbf{2k}\left(\frac{1}{\sqrt{\dext-2}}\right) 
&\multirow{3}{*}{$2^{k'} \binom{k}{k'}$} & 1_{\splitDk}\\
   \splitCk \oplus \mathbb{R} & \mathbf{2k}\left(\frac{1}{\sqrt{\dext-2}}\right) && 1_{\splitCk} \\
   \splitBk \oplus \mathbb{R} & \mathbf{\left(2k+1\right)}\left(\frac{1}{\sqrt{\dext-2}}\right) && 1_{\splitBk}\\  \hline
\end{tabular}\caption{Number of faces of a given type  for the perturbative polytopes of Tables \ref{tab:1irrep_integern_scaled_perturbative}.}\label{tab:facesparticlesU1short}
\end{table}

\newpage

\section{\texorpdfstring{$E_{\dt(\dt)}$}{E\_\dt(\dt)} moduli spaces and \texorpdfstring{$\pmax$}{pmax}-polytopes}
\label{app:Edd}
Here we collect some extra information about the theories with $E_{\dt(\dt)}$ moduli spaces and their polytopes. The representations for particles, strings and branes are given in Table~\ref{tab:G/K}. In Table~\ref{tab:facesMthAllpNotParticles} we list the number of faces of different dimension and distances to the face of the corresponding $\pmax$ polytopes up to $d=8$ and $2\leq\pmax\leq d-2$. Together with Table~\ref{tab:facesparticlesNonAbelianNonMth} this in particular completes the list of $\pmax=2$ polytopes for the global polytope theories, 

\begin{table}[ht!]\renewcommand{\arraystretch}{1.3}
\centering \small
\setlength{\tabcolsep}{4pt}
\begin{tabular}{|>{$}c<{$}||>{$}c<{$}|>{$}c<{$}|>{$}c<{$}|>{$}c<{$}|>{$}c<{$}|>{$}c<{$}|>{$}c<{$}|>{$}c<{$}|>{$}c<{$}|>{$}c<{$}|} \hline
\dt & G=E_{\dt(\dt)} & K_{\dt} & p=1 & p=2 & p=3 & p=4 & p=5 & p=6 & p=7 & p=8 \\ \hline\hline
1 & \mathbb{R}^+ & 1 &  \mathbf{1} & \mathbf{1} & \mathbf{1} & {-} & \mathbf{1} & \mathbf{1} & \mathbf{1} & \mathbf{1}\\\hline
1 & SL(2) & SO(2) & {-} & \mathbf{2} & {-} & \mathbf{1} & {-} & \mathbf{2} & {-}&\mathbf{3} \\ \hline
2 & GL(2) & SO(2) & \mathbf{2}+\mathbf{1} & \mathbf{2} & \mathbf{1} & \mathbf{1} & \mathbf{2} & \mathbf{2}+\mathbf{1} & \mathbf{3}+\mathbf{1}  \\\cline{1-10}
3 & SL(2)\times SL(3) & SO(2) \times SO(3) & (\mathbf{2},\mathbf{3}^*)  & (\mathbf{1},\mathbf{3}) & (\mathbf{2},\mathbf{1}) & (\mathbf{1},\mathbf{3^*}) & (\mathbf{2},\mathbf{3}) & \begin{matrix} (\mathbf{1},\mathbf{8})\\ (\mathbf{3},\mathbf{1})\end{matrix} 
\\\cline{1-9}
 4 & SL(5) & SO(5) & \mathbf{10}^* & \mathbf{5} & \mathbf{5^*} & \mathbf{10} & \mathbf{24} 
 \\\cline{1-8}
 5 & SO(5,5) & SO(5)\times SO(5) & \mathbf{16} & \mathbf{10} & \mathbf{16}^* & \mathbf{45} 
\\\cline{1-7}
 6 & E_{6(6)} & USp(8) & \mathbf{27} & \mathbf{27}^* & \mathbf{78} 
 \\\cline{1-6}
7 & E_{7(7)} & SU(8) & \mathbf{56} & \mathbf{133} 
\\\cline{1-5}
 8 & E_{8(8)} & SO(16) &\mathbf{248}
 \\\cline{1-4}
\end{tabular}
\caption[caption]{Cosets $G/K$ of maximal supergravities in $11$-$\dt$ dimensions, and corresponding $E_{\dt(\dt)}$ representations of objects spanning a $p$-dimensional world-volume in external space-time (see for instance~\cite{Obers:1998fb}). }  
\label{tab:G/K}
\end{table}

\begin{table}[H]\centering
\footnotesize\begin{tabular}{|>{$}c<{$}|>{$}c<{$}|>{$}c<{$}|>{$}c<{$}|>{$}c<{$}|>{$}c<{$}|>{$}c<{$}|>{$}c<{$}|>{$}c<{$}|} \hline
\pmax & \dext &  V & E & F & C & f_4  & f_5  & f_6  \\ \hline\hline
 \multirow{11}{*}{2} & 4 & \begin{array}{@{}c@{}>{\left(}c<{\right)}@{}}  126 & 2 \\  56 & \frac{3}{2} \\ \end{array} & \begin{array}{@{}c@{}>{\left(}c<{\right)}@{}}  2016 & \frac{3}{2} \\  1512 & \frac{4}{3} \\ \end{array}  & \begin{array}{@{}c@{}>{\left(}c<{\right)}@{}}  10080 & \frac{4}{3} \\  12096 & \frac{5}{4} \\ \end{array}  & \begin{array}{@{}c@{}>{\left(}c<{\right)}@{}}  20160 & \frac{5}{4} \\  40320 & \frac{6}{5} \\ \end{array}  & \begin{array}{@{}c@{}>{\left(}c<{\right)}@{}}  16128 & \frac{6}{5} \\  60480 & \frac{7}{6} \\ \end{array}  & \begin{array}{@{}c@{}>{\left(}c<{\right)}@{}}  4032 & \frac{7}{6} \\  36288 & \frac{8}{7} \\ \end{array}  & \begin{array}{@{}c@{}>{\left(}c<{\right)}@{}}  576 & \frac{8}{7} \\  4032 & \frac{9}{8} \\ \color{blue}{\mathbf{756}} & \color{blue}{\mathbf{1}} \\\end{array} \\ \cline{2-9}
 & 5 & \begin{array}{@{}c@{}>{\left(}c<{\right)}@{}}  54 & \frac{4}{3} \\ \end{array}  & \begin{array}{@{}c@{}>{\left(}c<{\right)}@{}}  270 & 1 \\  432 & \frac{5}{6} \\ \end{array}  & \begin{array}{@{}c@{}>{\left(}c<{\right)}@{}}  2160 & \frac{4}{5} \\ \end{array}  & \begin{array}{@{}c@{}>{\left(}c<{\right)}@{}}  2160 & \frac{3}{4} \\ \end{array}  & \begin{array}{@{}c@{}>{\left(}c<{\right)}@{}}  720 & \frac{2}{3} \\ \end{array}  & \begin{array}{@{}c@{}>{\left(}c<{\right)}@{}}  \color{blue}{\mathbf{72}} & \color{blue}{\mathbf{\frac{1}{2}}} \\ \end{array} \\ \cline{2-8}
    &6 & \begin{array}{@{}c@{}>{\left(}c<{\right)}@{}}  16 & \frac{5}{4} \\  10 & 1 \\ \end{array}  & \begin{array}{@{}c@{}>{\left(}c<{\right)}@{}}  80 & \frac{4}{5} \\  80 & \frac{3}{4} \\ \end{array}  & \begin{array}{@{}c@{}>{\left(}c<{\right)}@{}}  240 & \frac{2}{3} \\  160 & \frac{7}{12} \\ \end{array}  & \begin{array}{@{}c@{}>{\left(}c<{\right)}@{}}  320 & \frac{4}{7} \\ \end{array}  & \begin{array}{@{}c@{}>{\left(}c<{\right)}@{}}  120 & \frac{1}{2} \\  16 & \frac{9}{20} \\ \end{array}\\ \cline{2-7}
&7 & \begin{array}{@{}c@{}>{\left(}c<{\right)}@{}}  10 & \frac{6}{5} \\  5 & \frac{4}{5} \\ \end{array}  & \begin{array}{@{}c@{}>{\left(}c<{\right)}@{}}  30 & \frac{7}{10} \\  30 & \frac{2}{3} \\ \end{array}  & \begin{array}{@{}c@{}>{\left(}c<{\right)}@{}}  60 & \frac{4}{7} \\  30 & \frac{8}{15} \\ \end{array}  & \begin{array}{@{}c@{}>{\left(}c<{\right)}@{}}  40 & \frac{1}{2} \\  5 & \frac{9}{20} \\ \end{array}\\ \cline{2-6}
  &8 &  \begin{array}{@{}c@{}>{\left(}c<{\right)}@{}} 6 & \frac{7}{6} \\ 3 & \frac{2}{3} \\\end{array} & \begin{array}{@{}c@{}>{\left(}c<{\right)}@{}}9 & \frac{2}{3} \\ 12 & \frac{4}{7} \\\end{array} & \begin{array}{@{}c@{}>{\left(}c<{\right)}@{}} 14 & \frac{1}{2} \\\end{array} \\ \cline{2-5}
   \cline{1-8}\multirow{9}{*}{3}&5 & \begin{array}{@{}c@{}>{\left(}c<{\right)}@{}}  72 & 2 \\  54 & \frac{4}{3} \\ \end{array}  & \begin{array}{@{}c@{}>{\left(}c<{\right)}@{}}  720 & \frac{3}{2} \\  864 & \frac{5}{4} \\ \end{array}  & \begin{array}{@{}c@{}>{\left(}c<{\right)}@{}}  2160 & \frac{4}{3} \\  4320 & \frac{6}{5} \\ \end{array}  & \begin{array}{@{}c@{}>{\left(}c<{\right)}@{}}  2160 & \frac{5}{4} \\  8640 & \frac{7}{6} \\ \end{array}  & \begin{array}{@{}c@{}>{\left(}c<{\right)}@{}}  432 & \frac{6}{5} \\  6480 & \frac{8}{7} \\ \end{array}  & \begin{array}{@{}c@{}>{\left(}c<{\right)}@{}}  864 & \frac{9}{8} \\  \color{blue}{\mathbf{270}} & \color{blue}{\mathbf{1}} \\ \end{array} \\\cline{2-8}
  &6 & \begin{array}{@{}c@{}>{\left(}c<{\right)}@{}}  32 & \frac{5}{4} \\  90 & 1 \\ \end{array}  & \begin{array}{@{}c@{}>{\left(}c<{\right)}@{}}  160 & \frac{4}{5} \\ \end{array}  & \begin{array}{@{}c@{}>{\left(}c<{\right)}@{}}  400 & \frac{3}{4} \\ \end{array}  & \begin{array}{@{}c@{}>{\left(}c<{\right)}@{}}  240 & \frac{2}{3} \\ \end{array}  & \begin{array}{@{}c@{}>{\left(}c<{\right)}@{}}  \color{blue}{\mathbf{40}} & \color{blue}{\mathbf{\frac{1}{2}}} \\ \end{array}  \\\cline{2-7} 
  &7 & \begin{array}{@{}c@{}>{\left(}c<{\right)}@{}}  10 & \frac{6}{5} \\  10 & \frac{4}{5} \\ \end{array}  & \begin{array}{@{}c@{}>{\left(}c<{\right)}@{}}  20 & \frac{3}{4} \\  30 & \frac{7}{10} \\  30 & \frac{2}{3} \\ \end{array}  & \begin{array}{@{}c@{}>{\left(}c<{\right)}@{}}  30 & \frac{2}{3} \\  60 & \frac{4}{7} \\  10 & \frac{8}{15} \\ \end{array}  & \begin{array}{@{}c@{}>{\left(}c<{\right)}@{}}  40 & \frac{1}{2} \\ \end{array}  \\\cline{2-6}
  &8 & \begin{array}{@{}c@{}>{\left(}c<{\right)}@{}} 6 & \frac{7}{6} \\ 3 & \frac{2}{3} \\\end{array} & \begin{array}{@{}c@{}>{\left(}c<{\right)}@{}} 9 & \frac{2}{3} \\ 12 & \frac{4}{7} \\\end{array} & \begin{array}{@{}c@{}>{\left(}c<{\right)}@{}} 14 & \frac{1}{2} \\\end{array} \\\cline{1-7}
  
\multirow{6}{*}{4}&6 & \begin{array}{@{}c@{}>{\left(}c<{\right)}@{}}  40 & 2 \\  32 & \frac{5}{4} \\ \end{array}  & \begin{array}{@{}c@{}>{\left(}c<{\right)}@{}}  240 & \frac{3}{2} \\  320 & \frac{6}{5} \\ \end{array}  & \begin{array}{@{}c@{}>{\left(}c<{\right)}@{}}  400 & \frac{4}{3} \\  960 & \frac{7}{6} \\ \end{array}  & \begin{array}{@{}c@{}>{\left(}c<{\right)}@{}}  160 & \frac{5}{4} \\  960 & \frac{8}{7} \\ \end{array}  & \begin{array}{@{}c@{}>{\left(}c<{\right)}@{}}  160 & \frac{9}{8} \\ \color{blue}{\mathbf{90}} & \color{blue}{\mathbf{1}} \\ \end{array}  \\\cline{2-7}
&7 & \begin{array}{@{}c@{}>{\left(}c<{\right)}@{}}  20 & \frac{6}{5} \\  10 & \frac{4}{5} \\ \end{array}  & \begin{array}{@{}c@{}>{\left(}c<{\right)}@{}}  30 & 1 \\  40 & \frac{3}{4} \\ \end{array}  & \begin{array}{@{}c@{}>{\left(}c<{\right)}@{}}  60 & \frac{2}{3} \\ \end{array}  & \begin{array}{@{}c@{}>{\left(}c<{\right)}@{}}  \color{blue}{\mathbf{20}} & \color{blue}{\mathbf{\frac{1}{2}}} \\ \end{array}  \\\cline{2-6}
 &8 & \begin{array}{@{}c@{}>{\left(}c<{\right)}@{}} 6 & \frac{7}{6} \\ 3 & \frac{2}{3} \\\end{array}& \begin{array}{@{}c@{}>{\left(}c<{\right)}@{}} 9 & \frac{2}{3} \\ 12 & \frac{4}{7} \\\end{array} & \begin{array}{@{}c@{}>{\left(}c<{\right)}@{}} 14 & \frac{1}{2} \\\end{array} \\\cline{1-6} 
 
 \multirow{4}{*}{5}&7 & \begin{array}{@{}c@{}>{\left(}c<{\right)}@{}}  20 & 2 \\  20 & \frac{6}{5} \\ \end{array}  & \begin{array}{@{}c@{}>{\left(}c<{\right)}@{}}  60 & \frac{3}{2} \\  120 & \frac{7}{6} \\ \end{array}  & \begin{array}{@{}c@{}>{\left(}c<{\right)}@{}}  40 & \frac{4}{3} \\  180 & \frac{8}{7} \\ \end{array}  & \begin{array}{@{}c@{}>{\left(}c<{\right)}@{}}  10 & \frac{5}{4} \\  40 & \frac{9}{8} \\  \color{blue}{\mathbf{30}} & \color{blue}{\mathbf{1}} \\ \end{array}  \\\cline{2-6}
&8 & \begin{array}{@{}c@{}>{\left(}c<{\right)}@{}}  12 & \frac{7}{6} \\ \end{array}  & \begin{array}{@{}c@{}>{\left(}c<{\right)}@{}}  12 & 1 \\  6 & \frac{2}{3} \\ \end{array}  & \begin{array}{@{}c@{}>{\left(}c<{\right)}@{}}  \color{blue}{\mathbf{8}} & \color{blue}{\mathbf{\frac{1}{2}}} \\ \end{array}   \\
   
\cline{1-5}   6&8 & \begin{array}{@{}c@{}>{\left(}c<{\right)}@{}}  8 & 2 \\  12 & \frac{7}{6} \\ \end{array}  & \begin{array}{@{}c@{}>{\left(}c<{\right)}@{}}  6 & \frac{3}{2} \\  36 & \frac{8}{7} \\ \end{array}  & \begin{array}{@{}c@{}>{\left(}c<{\right)}@{}}  12 & \frac{9}{8} \\  \color{blue}{\mathbf{12}} & \color{blue}{\mathbf{1}} \\ \end{array}  \\\cline{1-5}
\end{tabular}
\caption{
 Number of faces of each type, together with the distance to them in parenthesis for the M-theory on $T^{11-\dext}$ polytopes for $2\leq \pmax\leq \dext-2$. 
 All of them satisfy the BDC bound $\frac{1}{\dext-\pmax-1}$, and we show in blue the faces saturating it. The representations are given in Table \ref{tab:G/K}.}\label{tab:facesMthAllpNotParticles}\end{table}

\section{Chains of polytope descendents}
\label{app:chains}

In the following Figures \ref{fig:tree1}-\ref{fig:treeDecomp}, we show the various descendents and decompactifications related to the global polytope theories in Tables \ref{tab:1irrep_integern_scaled_global} and \ref{tab:1irrep_integern_extra}. Our notation for a theory $(\mathfrak{g},\rho,d)$ whose polytope has vertices characterised by $\nvertex$ will be of the kind
\begin{equation}
    \text{polytope}_{(\nvertex)}{}^{\rho}_{\mathfrak{g}} \, ,
\end{equation}
and we will specify $\nvertex$ only when it is bigger than 1. 
We start with a given `seed' polytope, and for simplicity consider only its possible realisations in terms of theories with minimal representation and minimal $\nvertex$, which means $\nvertex=1$ for those in Table~\ref{tab:1irrep_integern_scaled_global} and $\nvertex=3$ or $\nvertex=4$ for those in Table~\ref{tab:1irrep_integern_extra}. We then consider only embedded descendents, that is ones that preserve $\dext$, while also having an irreducible representation and so are in Table \ref{tab:polytopes_all1}. The `seed' theories of each graph, i.e.~the ones that do not come from such an embedding procedure, are shown in violet in Figures \ref{fig:tree1}-\ref{fig:tree4}. The decompactification slices, that connect such `seed' theories, are then shown in Figure \ref{fig:treeDecomp}.   

\begin{figure}[H]\centering
\begin{tikzpicture}[scale=0.8,>=stealth]
\node (2) at (1.0 , 11.5) {$\hexagon_{\splitG2}^{\mathbf{14}}$};
\node (1) at (1.0 , 9.5) {$\hexagon_{\splitA2}^{\mathbf{8}}$};
\node (5) at (-1.5 , 10.0) {$\cuboctahedron_{\splitA3}^{\mathbf{15}}$};
\node (16) at (5.0 , 10.0) {$\square_{\splitC2}^{\mathbf{10}}$};
\node (18) at (5.0 , 11.0) {  $\square_{\splitA1^{\oplus2}}^{\mathbf{(3,3)}}$};
\node (6) at (-1.5 , 11.5) {$\cuboctahedron_{\splitB3}^{\mathbf{21}}$};
\node (7) at (-1.5 , 9.0) {$\cuboctahedron_{\splitC3}^{\mathbf{14}}$};
\node (14) at (5.0 , 9.0) {$\square_{\splitC2}^{\mathbf{5}}$};
\node (8) at (-4.0 , 13.0) {$\FfourAlt_{\splitB4}^{\mathbf{36}}$};
\node (10) at (-4.0 , 11.5) {$\FfourAlt_{\splitD4}^{\mathbf{28}}$};
\node (9) at (-4.0 , 10.0) {\color{Orchid}{$\FfourAlt_{\splitC4}^{\mathbf{27}}$}};
\node (11) at (-4.0 , 8.0) {\color{Orchid}{$\FfourAlt_{\splitF4}^{\mathbf{26}}$}};
\node (19) at (3.0 , 9.5) {${\hexagon^{\infty}_{(3)}}_{\splitA2}^{\mathbf{8}}$};
\node (22) at (3.0 , 8.0) {${\hexagon^{\infty}_{(3)}}_{\splitG2}^{\mathbf{7}}$};
\node (12) at (-1.0 , 13.0) {$\FfourAlt_{\splitF4}^{\mathbf{52}}$};
\node (13) at (0.0 , 15.0) {\color{Orchid}{${\EeightAlt}_{\splitE8}^{\mathbf{248}}$}};
\node (3) at (-3.0 , 15.0) {$\octagon_{(4)\splitC2}^{\mathbf{16}}$};
\node (4) at (-3.0 , 14.0) {$\octagon_{(4) \splitC2}^{\mathbf{35}}$};
\node (15) at (3.5 , 12.5) {  $\square_{\splitC2}^{\mathbf{14}}$};
\node (20) at (3.0 , 14.0) {  ${\hexagon^{\infty}_{(3)}}_{\splitA2}^{\mathbf{27}}$};
\node (23) at (3.0 , 15.0) { ${\hexagon^{\infty}_{(3)}}_{\splitG2}^{\mathbf{27}}$};
\node (17) at (5.0 , 7.5) {$\square_{\splitA1^{\oplus2}}^{\mathbf{(2,2)}}$};
\node (21) at (3.0 , 11.5) {\color{Orchid}{${\hexagon^{\infty}_{(3)}}_{\splitG2}^{\mathbf{14}}$}};
\draw[->] (2) -- (1);
\draw[->] (5) -- (1);
\draw[->] (5) -- (16);
\draw[->] (5) -- (18);
\draw[->] (6) -- (2);
\draw[->] (6) -- (5);
\draw[->] (7) -- (1);
\draw[->] (7) -- (14);
\draw[->] (8) -- (10);
\draw[->] (9) -- (5);
\draw[->] (9) -- (7);
\draw[->] (10) -- (6);
\draw[->] (11) -- (7);
\draw[->] (11) -- (19);
\draw[->] (11) -- (22);
\draw[->] (12) -- (8);
\draw[->] (13) -- (3);
\draw[->] (13) -- (4);
\draw[->] (13) -- (12);
\draw[->] (13) -- (15);
\draw[->] (13) -- (20);
\draw[->] (13) -- (23);
\draw[->] (14) -- (17);
\draw[->] (15) -- (18);
\draw[->] (21) -- (19);
\end{tikzpicture}

\caption{Chain of embeddings of $\dext=3$ single irrep polytopes of rank 2 or more, coming from ones with a fundamental representation. The seed polytope for each chain is shown in violet.}  
\label{fig:tree1}
\end{figure}

\begin{figure}[H]\centering
\begin{tikzpicture}[scale=0.8,>=stealth]
\node (1) at (2.0 , 7.0) {\color{Orchid}{$\cube_{\splitB3}^{\mathbf{8}}$}};
\node (6) at (2.0 , 5.0) {$\rectangle_{\splitA1^{\oplus2}}^{\mathbf{(2,2)}}$};
\node (22) at (4.0 , 5.0) {  ${\square_{(2)}}_{\splitC2}^{\mathbf{4}}$};
\node (24) at (6.0 , 5.0) {  $  {\square_{(2)}}_{\splitA1^{\oplus2}}^{\mathbf{(2,2)}}$};
\node (2) at (5.0 , 10.5) {$\cube_{\splitC3}^{\mathbf{14'}}$};
\node (5) at (5.0 , 7.0) {$\cube_{\splitA1^{\oplus3}}^{\mathbf{(2,2,2)}}$};
\node (21) at (7.0 , 7.0) {  ${\square_{(2)}}_{\splitC2}^{\mathbf{5}}$};
\node (3) at (5.0 , 9.0) {{$\cube_{\splitC2\splitA1}^{\mathbf{(5,2)}}$}};
\node (4) at (3.0 , 8.0) {\color{Orchid}{$\cube_{\splitC2\splitA1}^{\mathbf{(4,2)}}$}};
\node (7) at (7.0 , 12.5) {\color{Orchid}{${\EsevenAlt}_{\splitE7}^{\mathbf{56}}$}};
\node (17) at (7.0 , 10.5) {  ${\FfourAlt_{(2)}}_{\splitC4}^{\mathbf{27}}$};
\node (18) at (9.0 , 10.5) {  ${\FfourAlt_{(2)}}_{\splitD4}^{\mathbf{28}}$};
\node (19) at (4.0 , 12.5) {  ${\FfourAlt_{(2)}}_{\splitF4}^{\mathbf{26}}$};
\node (25) at (10.0 , 12.5) {  ${\square_{(2)}}_{\splitA1^{\oplus2}}^{\mathbf{(3,2)}}$};
\node (9) at (12.0 , 7.0) {${\hexagon_{(2)}}_{\splitG2}^{\mathbf{14}}$};
\node (8) at (12.0 , 5.0) {  ${\hexagon_{(2)}}_{\splitA2}^{\mathbf{8}}$};
\node (10) at (1.0 , 10.5) {  ${\hexagon^{\infty}_{(6)}}_{\splitA2}^{\mathbf{8}}$};
\node (13) at (9.0 , 9.0)  {  ${\cuboctahedron_{(2)}}_{\splitA3}^{\mathbf{15}}$};
\node (23) at (10.0 , 5.0) {  ${\square_{(2)}}_{\splitC2}^{\mathbf{10}}$};
\node (26) at (8.0 , 5.0) {  ${\square_{(2)}}_{\splitA1^{\oplus2}}^{\mathbf{(3,3)}}$};
\node (14) at (12.0 , 9.0) {  ${\cuboctahedron_{(2)}}_{\splitB3}^{\mathbf{21}}$};
\node (15) at (7.0 , 9.0) {  ${\cuboctahedron_{(2)}}_{\splitC3}^{\mathbf{14}}$};
\node (12) at (3.0 , 10.5) {  ${\hexagon^{\infty}_{(6)}}_{\splitG2}^{\mathbf{7}}$};
\node (27) at (2.0 , 9.0) {{ $\rectangle_{\splitA1^{\oplus2}}^{\mathbf{(3,2)}}$}};
\draw[->] (1) -- (6);
\draw[->] (3) -- (27);
\draw[->] (1) -- (22);
\draw[->] (1) -- (24);
\draw[->] (2) -- (3);
\draw[->] (3) -- (5);
\draw[->] (3) -- (21);
\draw[->] (4) -- (6);
\draw[->] (4) -- (22);
\draw[->] (4) -- (24);
\draw[->] (5) -- (24);
\draw[->] (7) -- (2);
\draw[->] (7) -- (17);
\draw[->] (7) -- (18);
\draw[->] (7) -- (19);
\draw[->] (7) -- (25);
\draw[->] (9) -- (8);
\draw[->] (13) -- (8);
\draw[->] (13) -- (23);
\draw[->] (13) -- (26);
\draw[->] (14) -- (9);
\draw[->] (14) -- (13);
\draw[->] (15) -- (8);
\draw[->] (15) -- (21);
\draw[->] (17) -- (13);
\draw[->] (17) -- (15);
\draw[->] (18) -- (14);
\draw[->] (19) -- (10);
\draw[->] (19) -- (12);
\draw[->] (19) -- (15);
\draw[->] (21) -- (24);
\end{tikzpicture}
\caption{Chain of embeddings $\dext=4$ single irrep polytopes of rank 2 or more coming from fundamental ones. The seed polytope for each chain is shown in violet.}
\label{fig:tree2}
\end{figure}

\begin{figure}[H]\centering
\begin{tikzpicture}[scale=0.65,>=stealth]
\node (1) at (0.0 , 9.0) {\color{Orchid}{${\EsixAlt}_{\splitE6}^{\mathbf{27}}$}};
\node (2) at (-4.0 , 6.5) {  $\triangle_{\splitA2}^{\mathbf{6}}$};
\node (3) at (-2.0 , 6.5) {  ${\hexagon_{(3)}}_{\splitA2}^{\mathbf{8}}$};
\node (5) at (0.0 , 6.5) {  ${\hexagon_{(3)}}_{\splitG2}^{\mathbf{7}}$};
\node (8) at (2.0 , 6.5) {  ${\square_{(3)}}_{\splitC2}^{\mathbf{10}}$};
\node (10) at (4.0 , 6.5) { ${\square_{(3)}}_{\splitA1^{\oplus2}}^{\mathbf{(3,3)}}$};
\node (11) at (3.0 , 9.0) {  ${\seg_{(2)}}_{\splitA1}^{\mathbf{11}}$};
\node (15) at (4.0 , 8.0) {  ${\seg_{(5)}}_{\splitA1}^{\mathbf{9}}$};
\node (18) at (-4.0 , 8.0) {  ${\seg_{(6)}}_{\splitA1}^{\mathbf{4}}$};
\node (19) at (-3.0 , 9.0) {  ${\seg_{(6)}}_{\splitA1}^{\mathbf{7}}$};
\node (13) at (0.0 , 4.5) {  ${\seg_{(3)}}_{\splitA1}^{\mathbf{3}}$};
\node (14) at (-4.0 , 4.5) {  ${\seg_{(3)}}_{\splitA1}^{\mathbf{5}}$};
\node (16) at (2.0 , 4.5) {  ${\seg_{(6)}}_{\splitA1}^{\mathbf{2}}$};
\node (17) at (-2.0 , 4.5) {  ${\seg_{(6)}}_{\splitA1}^{\mathbf{3}}$};
\node (21) at (4.0 , 4.5) { ${\seg_{(\infty)}}_{\splitA1}$};
\draw[->] (1) -- (2);
\draw[->] (1) -- (3);
\draw[->] (1) -- (5);
\draw[->] (1) -- (8);
\draw[->] (1) -- (10);
\draw[->] (1) -- (11);
\draw[->] (1) -- (15);
\draw[->] (1) -- (18);
\draw[->] (1) -- (19);
\draw[->] (3) -- (13);
\draw[->] (3) -- (14);
\draw[->] (5) -- (13);
\draw[->] (5) -- (16);
\draw[->] (5) -- (17);
\draw[->] (8) -- (13);
\draw[->] (8) -- (21);
\draw[->] (10) -- (21);
\end{tikzpicture}

\caption{Chain of embeddings for $\dext=5$ single irrep polytopes coming from fundamental ones. The seed polytope is shown in violet.}
\label{fig:tree3}
\end{figure}

\begin{figure}[H]\centering
\begin{tikzpicture}[scale=0.8,>=stealth]
\node (1) at (0.0 , 6.0) {\color{Orchid}{$\EfiveAlt_{\splitD5}^{\mathbf{16}}$}};
\node (2) at (-2.0 , 4.0) {  ${\cube_{(2)}}_{\splitB3}^{\mathbf{8}}$};
\node (6) at (0.0 , 4.0) {  ${\cube_{(2)}}_{\splitA1^{\oplus3}}^{\mathbf{(2,2,2)}}$};
\node (8) at (2.0 , 4.0) {  ${\square_{(4)}}_{\splitC2}^{\mathbf{5}}$};
\node (11) at (3.0 , 6.0) {  ${\seg_{(2)}}_{\splitA1}^{\mathbf{4}}$};
\node (15) at (-3.0 , 6.0) {  ${\seg_{(5)}}_{\splitA1}^{\mathbf{4}}$};
\node (7) at (-4.0 , 2.0) {  ${\rectangle_{(2)}}_{\splitA1^{\oplus2}}^{\mathbf{(2,2)}}$};
\node (9) at (-2.0 , 2.0) {  ${\square_{(4)}}_{\splitC2}^{\mathbf{4}}$};
\node (10) at (0.0 , 2.0) {  ${\square_{(4)}}_{\splitA1^{\oplus2}}^{\mathbf{(2,2)}}$};
\node (13) at (3.0 , 2.0) {  ${\seg_{(4)}}_{\splitA1}^{\mathbf{3}}$};
\node (17) at (4.0 , 4.0) {  ${\rectangle_{(2)}}_{\splitA1^{\oplus2}}^{\mathbf{(3,2)}}$};
\node (5) at (-4.0 , 4.0) {{  ${\cube_{(2)}}_{\splitC2\splitA1}^{\mathbf{(4,2)}}$}};
\node (12) at (-4.0 , 0.0) {  ${\seg_{(4)}}_{\splitA1}^{\mathbf{2}}$};
\node (16) at (-2.0 , 0.0) { ${\seg_{(\infty)}}_{\splitA1}$};
\draw[->] (1) -- (17);
\draw[->] (1) -- (2);
\draw[->] (1) -- (5);
\draw[->] (1) -- (6);
\draw[->] (1) -- (8);
\draw[->] (1) -- (11);
\draw[->] (1) -- (15);
\draw[->] (2) -- (7);
\draw[->] (2) -- (9);
\draw[->] (2) -- (10);
\draw[->] (2) -- (13);
\draw[->] (5) -- (7);
\draw[->] (5) -- (9);
\draw[->] (5) -- (10);
\draw[->] (6) -- (10);
\draw[->] (7) -- (12);
\draw[->] (7) -- (16);
\draw[->] (8) -- (10);
\draw[->] (8) -- (13);
\draw[->] (9) -- (12);
\draw[->] (9) -- (16);
\draw[->] (10) -- (16);
\draw[->] (17) -- (16);
\end{tikzpicture}
\caption{Chain of embeddings for $\dext=6$ single irrep polytopes coming from fundamental ones. The seed polytope is shown in violet.}
\label{fig:tree4}
\end{figure}

\begin{figure}[H]\centering
\begin{tikzpicture}[scale=0.4,>=stealth]
\node (33) at (10 , 0) {\color{Orchid}${\hexagon^{\infty}_{(3)}}_{\splitG2}^{\mathbf{14}}$};
\node (51) at (10 , -9) {${\seg_{(\infty)}}_{\splitA1}
$};

\node (8) at (7 , 0) {\color{Orchid} $\FfourAlt_{\splitF4}^{\mathbf{26}}$};
\node (11) at (7 , -3) {\color{Orchid} $\cube_{\splitB3}^{\mathbf{8}}$};
\node (31) at (7 , -6) {$\triangle_{\splitA2}^{\mathbf{3}}$};

\node (6) at (4 , 0) {\color{Orchid} $\FfourAlt_{\splitC4}^{\mathbf{27}}$};
\node (14) at (4 , -3) {\color{Orchid} $\cube_{\splitC2\splitA1}^{\mathbf{(4,2)}}$};

\node (10) at (1 , 0) {\color{Orchid} ${\EeightAlt}_{\splitE8}^{\mathbf{248}}$};
\node (18) at (1 , -3) {\color{Orchid} ${\EsevenAlt}_{\splitE7}^{\mathbf{56}}$};
\node (30) at (1 , -6) {\color{Orchid} ${\EsixAlt}_{\splitE6}^{\mathbf{27}}$};
\node (35) at (1 , -9) {\color{Orchid} ${\EfiveAlt}_{\splitD5}^{\mathbf{16}}$};
\node (46) at (1 , -12) {${\EfourAlt}_{\splitA4}^{\mathbf{10}^*}$};
\node (47) at (1 , -15) {${\triprism}_{\splitA1\splitA2}^{\mathbf{\left(2,3^*\right)}}$};

\draw[->] (6) -- (14);
\draw[->] (8) -- (11);
\draw[->] (10) -- (18);
\draw[->] (11) -- (31);
\draw[->] (18) -- (30);
\draw[->] (30) -- (35);
\draw[->] (33) -- (51);
\draw[->] (35) -- (46);
\draw[->] (46) -- (47);
\node at (-3,-15){\footnotesize$\dext=8$};
\node at (-3,-12){\footnotesize$\dext=7$};
\node at (-3,-9){\footnotesize$\dext=6$};
\node at (-3,-6){\footnotesize$\dext=5$};
\node at (-3,-3){\footnotesize$\dext=4$};
\node at (-3,0){\footnotesize$\dext=3$};
\end{tikzpicture}
\caption{Chains of decompactification slices for minimal representation and minimal $\nvertex$. Chains for non-minimal $\nvertex$ theories can be obtained by substituting $\nvertex \rightarrow z \nvertex$ and ${\dext \rightarrow z(\dext-2)+2}$.}
\label{fig:treeDecomp}

\end{figure}
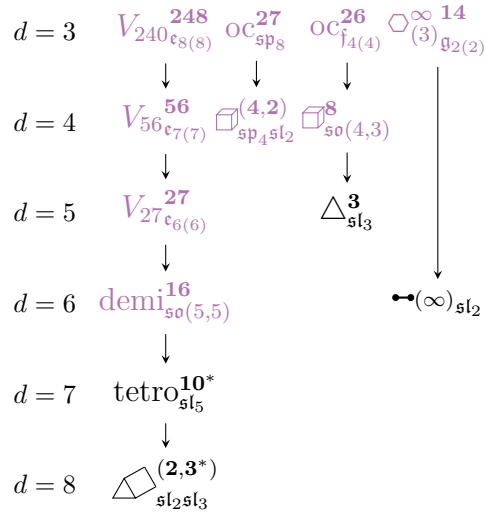
\newpage

\bibliographystyle{JHEP}
\bibliography{biblio}

\providecommand{\href}[2]{#2}\begingroup\raggedright\begin{thebibliography}{100}

\bibitem{Baines:2025upi}
S.~Baines, V.~Collazuol, B.~Fraiman, M.~Gra{\~n}a, and D.~Waldram, {\it {The Boundary of Symmetric Moduli Spaces and the Swampland Distance Conjecture}},  \href{http://arxiv.org/abs/2508.18401}{{\tt arXiv:2508.18401}}.

\bibitem{Vafa:2005ui}
C.~Vafa, {\it {The String landscape and the swampland}},  \href{http://arxiv.org/abs/hep-th/0509212}{{\tt hep-th/0509212}}.

\bibitem{Brennan:2017rbf}
T.~D. Brennan, F.~Carta, and C.~Vafa, {\it {The String Landscape, the Swampland, and the Missing Corner}},  {\em PoS} {\bf TASI2017} (2017) 015, [\href{http://arxiv.org/abs/1711.00864}{{\tt arXiv:1711.00864}}].

\bibitem{Palti:2019pca}
E.~Palti, {\it {The Swampland: Introduction and Review}},  {\em Fortsch. Phys.} {\bf 67} (2019), no.~6 1900037, [\href{http://arxiv.org/abs/1903.06239}{{\tt arXiv:1903.06239}}].

\bibitem{Agmon:2022thq}
N.~B. Agmon, A.~Bedroya, M.~J. Kang, and C.~Vafa, {\it {Lectures on the string landscape and the Swampland}},  \href{http://arxiv.org/abs/2212.06187}{{\tt arXiv:2212.06187}}.

\bibitem{vanBeest:2021lhn}
M.~van Beest, J.~Calder\'on-Infante, D.~Mirfendereski, and I.~Valenzuela, {\it {Lectures on the Swampland Program in String Compactifications}},  \href{http://arxiv.org/abs/2102.01111}{{\tt arXiv:2102.01111}}.

\bibitem{Grana:2021zvf}
M.~Gra\~na and A.~Herr\'aez, {\it {The Swampland Conjectures: A Bridge from Quantum Gravity to Particle Physics}},  {\em Universe} {\bf 7} (2021), no.~8 273, [\href{http://arxiv.org/abs/2107.00087}{{\tt arXiv:2107.00087}}].

\bibitem{Ooguri:2006in}
H.~Ooguri and C.~Vafa, {\it {On the Geometry of the String Landscape and the Swampland}},  {\em Nucl. Phys. B} {\bf 766} (2007) 21--33, [\href{http://arxiv.org/abs/hep-th/0605264}{{\tt hep-th/0605264}}].

\bibitem{Dvali:2007hz}
G.~Dvali, {\it {Black Holes and Large N Species Solution to the Hierarchy Problem}},  {\em Fortsch. Phys.} {\bf 58} (2010) 528--536, [\href{http://arxiv.org/abs/0706.2050}{{\tt arXiv:0706.2050}}].

\bibitem{Dvali:2007wp}
G.~Dvali and M.~Redi, {\it {Black Hole Bound on the Number of Species and Quantum Gravity at LHC}},  {\em Phys. Rev. D} {\bf 77} (2008) 045027, [\href{http://arxiv.org/abs/0710.4344}{{\tt arXiv:0710.4344}}].

\bibitem{Dvali:2009ks}
G.~Dvali and D.~Lust, {\it {Evaporation of Microscopic Black Holes in String Theory and the Bound on Species}},  {\em Fortsch. Phys.} {\bf 58} (2010) 505--527, [\href{http://arxiv.org/abs/0912.3167}{{\tt arXiv:0912.3167}}].

\bibitem{Dvali:2010vm}
G.~Dvali and C.~Gomez, {\it {Species and Strings}},  \href{http://arxiv.org/abs/1004.3744}{{\tt arXiv:1004.3744}}.

\bibitem{Hamada:2021yxy}
Y.~Hamada, M.~Montero, C.~Vafa, and I.~Valenzuela, {\it {Finiteness and the swampland}},  {\em J. Phys. A} {\bf 55} (2022), no.~22 224005, [\href{http://arxiv.org/abs/2111.00015}{{\tt arXiv:2111.00015}}].

\bibitem{Stout:2021ubb}
J.~Stout, {\it {Infinite Distance Limits and Information Theory}},  \href{http://arxiv.org/abs/2106.11313}{{\tt arXiv:2106.11313}}.

\bibitem{Stout:2022phm}
J.~Stout, {\it {Infinite Distances and Factorization}},  \href{http://arxiv.org/abs/2208.08444}{{\tt arXiv:2208.08444}}.

\bibitem{Calderon-Infante:2023ler}
J.~Calder\'on-Infante, A.~Castellano, A.~Herr\'aez, and L.~E. Ib\'a\~nez, {\it {Entropy bounds and the species scale distance conjecture}},  {\em JHEP} {\bf 01} (2024) 039, [\href{http://arxiv.org/abs/2306.16450}{{\tt arXiv:2306.16450}}].

\bibitem{Lee:2018urn}
S.-J. Lee, W.~Lerche, and T.~Weigand, {\it {Tensionless Strings and the Weak Gravity Conjecture}},  {\em JHEP} {\bf 10} (2018) 164, [\href{http://arxiv.org/abs/1808.05958}{{\tt arXiv:1808.05958}}].

\bibitem{Lee:2019wij}
S.-J. Lee, W.~Lerche, and T.~Weigand, {\it {Emergent strings from infinite distance limits}},  {\em JHEP} {\bf 02} (2022) 190, [\href{http://arxiv.org/abs/1910.01135}{{\tt arXiv:1910.01135}}].

\bibitem{Etheredge:2022opl}
M.~Etheredge, B.~Heidenreich, S.~Kaya, Y.~Qiu, and T.~Rudelius, {\it {Sharpening the Distance Conjecture in diverse dimensions}},  {\em JHEP} {\bf 12} (2022) 114, [\href{http://arxiv.org/abs/2206.04063}{{\tt arXiv:2206.04063}}].

\bibitem{Gendler:2020dfp}
N.~Gendler and I.~Valenzuela, {\it {Merging the weak gravity and distance conjectures using BPS extremal black holes}},  {\em JHEP} {\bf 01} (2021) 176, [\href{http://arxiv.org/abs/2004.10768}{{\tt arXiv:2004.10768}}].

\bibitem{Bedroya:2019snp}
A.~Bedroya and C.~Vafa, {\it {Trans-Planckian Censorship and the Swampland}},  {\em JHEP} {\bf 09} (2020) 123, [\href{http://arxiv.org/abs/1909.11063}{{\tt arXiv:1909.11063}}].

\bibitem{Andriot:2020lea}
D.~Andriot, N.~Cribiori, and D.~Erkinger, {\it {The web of swampland conjectures and the TCC bound}},  {\em JHEP} {\bf 07} (2020) 162, [\href{http://arxiv.org/abs/2004.00030}{{\tt arXiv:2004.00030}}].

\bibitem{Lanza:2020qmt}
S.~Lanza, F.~Marchesano, L.~Martucci, and I.~Valenzuela, {\it {Swampland Conjectures for Strings and Membranes}},  {\em JHEP} {\bf 02} (2021) 006, [\href{http://arxiv.org/abs/2006.15154}{{\tt arXiv:2006.15154}}].

\bibitem{Etheredge:2023odp}
M.~Etheredge, B.~Heidenreich, J.~McNamara, T.~Rudelius, I.~Ruiz, and I.~Valenzuela, {\it {Running decompactification, sliding towers, and the distance conjecture}},  {\em JHEP} {\bf 12} (2023) 182, [\href{http://arxiv.org/abs/2306.16440}{{\tt arXiv:2306.16440}}].

\bibitem{Raucci:2026fzp}
S.~Raucci, I.~Ruiz, and I.~Valenzuela, {\it {Alice in Warpland: KK modes, Warped Compactifications and the Swampland}},  \href{http://arxiv.org/abs/2603.11163}{{\tt arXiv:2603.11163}}.

\bibitem{Heidenreich:2015nta}
B.~Heidenreich, M.~Reece, and T.~Rudelius, {\it {Sharpening the Weak Gravity Conjecture with Dimensional Reduction}},  {\em JHEP} {\bf 02} (2016) 140, [\href{http://arxiv.org/abs/1509.06374}{{\tt arXiv:1509.06374}}].

\bibitem{Blumenhagen:2017cxt}
R.~Blumenhagen, I.~Valenzuela, and F.~Wolf, {\it {The Swampland Conjecture and F-term Axion Monodromy Inflation}},  {\em JHEP} {\bf 07} (2017) 145, [\href{http://arxiv.org/abs/1703.05776}{{\tt arXiv:1703.05776}}].

\bibitem{Grimm:2018ohb}
T.~W. Grimm, E.~Palti, and I.~Valenzuela, {\it {Infinite Distances in Field Space and Massless Towers of States}},  {\em JHEP} {\bf 08} (2018) 143, [\href{http://arxiv.org/abs/1802.08264}{{\tt arXiv:1802.08264}}].

\bibitem{Heidenreich:2018kpg}
B.~Heidenreich, M.~Reece, and T.~Rudelius, {\it {Emergence of Weak Coupling at Large Distance in Quantum Gravity}},  {\em Phys. Rev. Lett.} {\bf 121} (2018), no.~5 051601, [\href{http://arxiv.org/abs/1802.08698}{{\tt arXiv:1802.08698}}].

\bibitem{Blumenhagen:2018nts}
R.~Blumenhagen, D.~Kl\"awer, L.~Schlechter, and F.~Wolf, {\it {The Refined Swampland Distance Conjecture in Calabi-Yau Moduli Spaces}},  {\em JHEP} {\bf 06} (2018) 052, [\href{http://arxiv.org/abs/1803.04989}{{\tt arXiv:1803.04989}}].

\bibitem{Lee:2018spm}
S.-J. Lee, W.~Lerche, and T.~Weigand, {\it {A Stringy Test of the Scalar Weak Gravity Conjecture}},  {\em Nucl. Phys. B} {\bf 938} (2019) 321--350, [\href{http://arxiv.org/abs/1810.05169}{{\tt arXiv:1810.05169}}].

\bibitem{Ooguri:2018wrx}
H.~Ooguri, E.~Palti, G.~Shiu, and C.~Vafa, {\it {Distance and de Sitter Conjectures on the Swampland}},  {\em Phys. Lett. B} {\bf 788} (2019) 180--184, [\href{http://arxiv.org/abs/1810.05506}{{\tt arXiv:1810.05506}}].

\bibitem{Corvilain:2018lgw}
P.~Corvilain, T.~W. Grimm, and I.~Valenzuela, {\it {The Swampland Distance Conjecture for K\"ahler moduli}},  {\em JHEP} {\bf 08} (2019) 075, [\href{http://arxiv.org/abs/1812.07548}{{\tt arXiv:1812.07548}}].

\bibitem{Grimm:2018cpv}
T.~W. Grimm, C.~Li, and E.~Palti, {\it {Infinite Distance Networks in Field Space and Charge Orbits}},  {\em JHEP} {\bf 03} (2019) 016, [\href{http://arxiv.org/abs/1811.02571}{{\tt arXiv:1811.02571}}].

\bibitem{Buratti:2018xjt}
G.~Buratti, J.~Calder\'on, and A.~M. Uranga, {\it {Transplanckian axion monodromy!?}},  {\em JHEP} {\bf 05} (2019) 176, [\href{http://arxiv.org/abs/1812.05016}{{\tt arXiv:1812.05016}}].

\bibitem{Lust:2019zwm}
D.~L\"ust, E.~Palti, and C.~Vafa, {\it {AdS and the Swampland}},  {\em Phys. Lett. B} {\bf 797} (2019) 134867, [\href{http://arxiv.org/abs/1906.05225}{{\tt arXiv:1906.05225}}].

\bibitem{Joshi:2019nzi}
A.~Joshi and A.~Klemm, {\it {Swampland Distance Conjecture for One-Parameter Calabi-Yau Threefolds}},  {\em JHEP} {\bf 08} (2019) 086, [\href{http://arxiv.org/abs/1903.00596}{{\tt arXiv:1903.00596}}].

\bibitem{Erkinger:2019umg}
D.~Erkinger and J.~Knapp, {\it {Refined swampland distance conjecture and exotic hybrid Calabi-Yaus}},  {\em JHEP} {\bf 07} (2019) 029, [\href{http://arxiv.org/abs/1905.05225}{{\tt arXiv:1905.05225}}].

\bibitem{Marchesano:2019ifh}
F.~Marchesano and M.~Wiesner, {\it {Instantons and infinite distances}},  {\em JHEP} {\bf 08} (2019) 088, [\href{http://arxiv.org/abs/1904.04848}{{\tt arXiv:1904.04848}}].

\bibitem{Font:2019cxq}
A.~Font, A.~Herr\'aez, and L.~E. Ib\'a\~nez, {\it {The Swampland Distance Conjecture and Towers of Tensionless Branes}},  {\em JHEP} {\bf 08} (2019) 044, [\href{http://arxiv.org/abs/1904.05379}{{\tt arXiv:1904.05379}}].

\bibitem{Baume:2020dqd}
F.~Baume and J.~Calder\'on~Infante, {\it {Tackling the SDC in AdS with CFTs}},  {\em JHEP} {\bf 08} (2021) 057, [\href{http://arxiv.org/abs/2011.03583}{{\tt arXiv:2011.03583}}].

\bibitem{Perlmutter:2020buo}
E.~Perlmutter, L.~Rastelli, C.~Vafa, and I.~Valenzuela, {\it {A CFT distance conjecture}},  {\em JHEP} {\bf 10} (2021) 070, [\href{http://arxiv.org/abs/2011.10040}{{\tt arXiv:2011.10040}}].

\bibitem{Klaewer:2020lfg}
D.~Klaewer, S.-J. Lee, T.~Weigand, and M.~Wiesner, {\it {Quantum corrections in 4d $N$ = 1 infinite distance limits and the weak gravity conjecture}},  {\em JHEP} {\bf 03} (2021) 252, [\href{http://arxiv.org/abs/2011.00024}{{\tt arXiv:2011.00024}}].

\bibitem{Lanza:2021udy}
S.~Lanza, F.~Marchesano, L.~Martucci, and I.~Valenzuela, {\it {The EFT stringy viewpoint on large distances}},  {\em JHEP} {\bf 09} (2021) 197, [\href{http://arxiv.org/abs/2104.05726}{{\tt arXiv:2104.05726}}].

\bibitem{Lee:2021qkx}
S.-J. Lee and T.~Weigand, {\it {Elliptic K3 surfaces at infinite complex structure and their refined Kulikov models}},  {\em JHEP} {\bf 09} (2022) 143, [\href{http://arxiv.org/abs/2112.07682}{{\tt arXiv:2112.07682}}].

\bibitem{Lee:2021usk}
S.-J. Lee, W.~Lerche, and T.~Weigand, {\it {Physics of infinite complex structure limits in eight dimensions}},  {\em JHEP} {\bf 06} (2022) 042, [\href{http://arxiv.org/abs/2112.08385}{{\tt arXiv:2112.08385}}].

\bibitem{Rudelius:2022gbz}
T.~Rudelius, {\it {Asymptotic scalar field cosmology in string theory}},  {\em JHEP} {\bf 10} (2022) 018, [\href{http://arxiv.org/abs/2208.08989}{{\tt arXiv:2208.08989}}].

\bibitem{Baume:2023msm}
F.~Baume and J.~Calder\'on-Infante, {\it {On higher-spin points and infinite distances in conformal manifolds}},  {\em JHEP} {\bf 12} (2023) 163, [\href{http://arxiv.org/abs/2305.05693}{{\tt arXiv:2305.05693}}].

\bibitem{Rudelius:2023mjy}
T.~Rudelius, {\it {Revisiting the refined Distance Conjecture}},  {\em JHEP} {\bf 09} (2023) 130, [\href{http://arxiv.org/abs/2303.12103}{{\tt arXiv:2303.12103}}].

\bibitem{Alvarez-Garcia:2023gdd}
R.~\'Alvarez-Garc\'\i{}a, S.-J. Lee, and T.~Weigand, {\it {Non-minimal elliptic threefolds at infinite distance. Part I. Log Calabi-Yau resolutions}},  {\em JHEP} {\bf 08} (2024) 240, [\href{http://arxiv.org/abs/2310.07761}{{\tt arXiv:2310.07761}}].

\bibitem{Alvarez-Garcia:2023qqj}
R.~\'Alvarez-Garc\'\i{}a, S.-J. Lee, and T.~Weigand, {\it {Non-minimal elliptic threefolds at infinite distance II: asymptotic physics}},  {\em JHEP} {\bf 01} (2025) 058, [\href{http://arxiv.org/abs/2312.11611}{{\tt arXiv:2312.11611}}].

\bibitem{Castellano:2023jjt}
A.~Castellano, I.~Ruiz, and I.~Valenzuela, {\it {Stringy evidence for a universal pattern at infinite distance}},  {\em JHEP} {\bf 06} (2024) 037, [\href{http://arxiv.org/abs/2311.01536}{{\tt arXiv:2311.01536}}].

\bibitem{Castellano:2023stg}
A.~Castellano, I.~Ruiz, and I.~Valenzuela, {\it {Universal Pattern in Quantum Gravity at Infinite Distance}},  {\em Phys. Rev. Lett.} {\bf 132} (2024), no.~18 181601, [\href{http://arxiv.org/abs/2311.01501}{{\tt arXiv:2311.01501}}].

\bibitem{Ooguri:2024ofs}
H.~Ooguri and Y.~Wang, {\it {Universal Bounds on CFT Distance Conjecture}},  \href{http://arxiv.org/abs/2405.00674}{{\tt arXiv:2405.00674}}.

\bibitem{Calderon-Infante:2024oed}
J.~Calder\'on-Infante and I.~Valenzuela, {\it {Tensionless String Limits in 4d Conformal Manifolds}},  \href{http://arxiv.org/abs/2410.07309}{{\tt arXiv:2410.07309}}.

\bibitem{Ashmore:2015joa}
A.~Ashmore and D.~Waldram, {\it {Exceptional Calabi-Yau spaces: the geometry of $\mathcal{N}=2$ backgrounds with flux}},  {\em Fortsch. Phys.} {\bf 65} (2017), no.~1 1600109, [\href{http://arxiv.org/abs/1510.00022}{{\tt arXiv:1510.00022}}].

\bibitem{Aoufia:2024awo}
C.~Aoufia, I.~Basile, and G.~Leone, {\it {Species scale, worldsheet CFTs and emergent geometry}},  \href{http://arxiv.org/abs/2405.03683}{{\tt arXiv:2405.03683}}.

\bibitem{Friedrich:2025gvs}
B.~Friedrich, J.~Monnee, T.~Weigand, and M.~Wiesner, {\it {Emergent Strings in Type IIB Calabi--Yau Compactifications}},  \href{http://arxiv.org/abs/2504.01066}{{\tt arXiv:2504.01066}}.

\bibitem{Baume:2016psm}
F.~Baume and E.~Palti, {\it {Backreacted Axion Field Ranges in String Theory}},  {\em JHEP} {\bf 08} (2016) 043, [\href{http://arxiv.org/abs/1602.06517}{{\tt arXiv:1602.06517}}].

\bibitem{Klaewer:2016kiy}
D.~Klaewer and E.~Palti, {\it {Super-Planckian Spatial Field Variations and Quantum Gravity}},  {\em JHEP} {\bf 01} (2017) 088, [\href{http://arxiv.org/abs/1610.00010}{{\tt arXiv:1610.00010}}].

\bibitem{Etheredge:2023zjk}
M.~Etheredge and B.~Heidenreich, {\it {Geodesic Gradient Flows in Moduli Space}},  \href{http://arxiv.org/abs/2311.18693}{{\tt arXiv:2311.18693}}.

\bibitem{Etheredge:2024tok}
M.~Etheredge, B.~Heidenreich, T.~Rudelius, I.~Ruiz, and I.~Valenzuela, {\it {Taxonomy of Infinite Distance Limits}},  \href{http://arxiv.org/abs/2405.20332}{{\tt arXiv:2405.20332}}.

\bibitem{Etheredge:2023usk}
M.~Etheredge, {\it {Dense geodesics, tower alignment, and the Sharpened Distance Conjecture}},  {\em JHEP} {\bf 01} (2024) 122, [\href{http://arxiv.org/abs/2308.01331}{{\tt arXiv:2308.01331}}].

\bibitem{Grieco:2025bjy}
A.~Grieco, I.~Ruiz, and I.~Valenzuela, {\it {EFT strings and dualities in 4d $\mathcal{N}=1$}},  \href{http://arxiv.org/abs/2504.16984}{{\tt arXiv:2504.16984}}.

\bibitem{Etheredge:2024amg}
M.~Etheredge, B.~Heidenreich, and T.~Rudelius, {\it {A Distance Conjecture for Branes}},  \href{http://arxiv.org/abs/2407.20316}{{\tt arXiv:2407.20316}}.

\bibitem{Etheredge:2025ahf}
M.~Etheredge, {\it {Taxonomy of branes in infinite distance limits}},  \href{http://arxiv.org/abs/2505.10615}{{\tt arXiv:2505.10615}}.

\bibitem{Delgado:2024skw}
M.~Delgado, D.~van~de Heisteeg, S.~Raman, E.~Torres, C.~Vafa, and K.~Xu, {\it {Finiteness and the Emergence of Dualities}},  \href{http://arxiv.org/abs/2412.03640}{{\tt arXiv:2412.03640}}.

\bibitem{Margulis-rigidity}
G.~A. Margulis, {\em Discrete Subgroups of Semisimple Lie Groups}, vol.~17 of {\em Ergebnisse der Mathematik und ihrer Grenzgebiete (3) [Results in Mathematics and Related Areas (3)]}.
\newblock Springer-Verlag, Berlin, 1991.

\bibitem{Grimm:2025lip}
T.~W. Grimm, D.~Prieto, and M.~van Vliet, {\it {Tame embeddings, volume growth, and complexity of moduli spaces}},  {\em Phys. Rev. D} {\bf 112} (2025), no.~10 106015, [\href{http://arxiv.org/abs/2503.15601}{{\tt arXiv:2503.15601}}].

\bibitem{Cribiori:2024qsv}
N.~Cribiori and D.~Lust, {\it {String dualities and modular symmetries in supergravity: a review}},  11, 2024.

\bibitem{Calderon-Infante:2020dhm}
J.~Calder\'on-Infante, A.~M. Uranga, and I.~Valenzuela, {\it {The Convex Hull Swampland Distance Conjecture and Bounds on Non-geodesics}},  {\em JHEP} {\bf 03} (2021) 299, [\href{http://arxiv.org/abs/2012.00034}{{\tt arXiv:2012.00034}}].

\bibitem{hall2003lie}
B.~C. Hall, {\em Lie Groups, Lie Algebras, and Representations: An Elementary Introduction}, vol.~222 of {\em Graduate Texts in Mathematics}.
\newblock Springer, New York, 2nd~ed., 2003.

\bibitem{Satake60}
I.~Satake, {\it On representations and compactifications of symmetric riemannian spaces},  {\em Annals of Mathematics} {\bf 71} (1960), no.~1 77--110.

\bibitem{Dhillon17}
G.~Dhillon and A.~Khare, {\it Faces of highest weight modules and the universal weyl polyhedron},  {\em Advances in Mathematics} {\bf 319} (2017) 111--152.

\bibitem{Gosset1900}
T.~Gosset, {\it On the regular and semi-regular figures in space of n dimensions},  {\em Messenger of Mathematics} {\bf 29} (1900) 43--48.

\bibitem{Blind1991}
G.~Blind and R.~Blind, {\it The semiregular polytopes},  {\em Commentarii Mathematici Helvetici} {\bf 66} (1991), no.~1 150--154.

\bibitem{MR80123}
I.~Niven, {\em Irrational numbers}, vol.~No. 11 of {\em The Carus Mathematical Monographs}.
\newblock Mathematical Association of America, ; distributed by John Wiley \& Sons, Inc., New York, 1956.

\bibitem{berndt2020indexconjecturesymmetricspaces}
J.~Berndt and C.~Olmos, {\it The index conjecture for symmetric spaces},  2020.

\bibitem{kollross2023totallygeodesicsubmanifoldsexceptional}
A.~Kollross and A.~Rodríguez-Vázquez, {\it Totally geodesic submanifolds in exceptional symmetric spaces},  \href{http://arxiv.org/abs/2202.10775}{{\tt arXiv:2202.10775}}.

\bibitem{Karpelevich}
F.~I. Karpelevich, {\it Surfaces of transitivity of semisimple group of motions of a symmetric space},  {\em Dokl. Akad. Nauk} {\bf 93} (1953) 401–404.

\bibitem{mostow}
G.~Mostow, {\it Some new decomposition theorems for semi-simple groups},  {\em Mem. Amer. Math. Soc.} {\bf 14} (1955) 31–54.

\bibitem{Borel}
A.~Borel and L.~Ji, {\em Compactifications of Symmetric and Locally Symmetric Spaces}.
\newblock Birkhäuser Boston, MA, 2006.

\bibitem{Dabholkar_1999}
A.~Dabholkar and J.~A. Harvey, {\it String islands},  {\em Journal of High Energy Physics} {\bf 1999} (Feb., 1999) 006–006.

\bibitem{Garcia-Etxebarria:2015wns}
I.~Garc{\'\i}a-Etxebarria and D.~Regalado, {\it {$ \mathcal{N}=3 $ four dimensional field theories}},  {\em JHEP} {\bf 03} (2016) 083, [\href{http://arxiv.org/abs/1512.06434}{{\tt arXiv:1512.06434}}].

\bibitem{Polchinski_1996}
J.~Polchinski and E.~Witten, {\it Evidence for heterotic — type i string duality},  {\em Nuclear Physics B} {\bf 460} (Feb., 1996) 525–540.

\bibitem{Banks:2010zn}
T.~Banks and N.~Seiberg, {\it {Symmetries and Strings in Field Theory and Gravity}},  {\em Phys. Rev. D} {\bf 83} (2011) 084019, [\href{http://arxiv.org/abs/1011.5120}{{\tt arXiv:1011.5120}}].

\bibitem{Harlow:2018tng}
D.~Harlow and H.~Ooguri, {\it {Symmetries in quantum field theory and quantum gravity}},  {\em Commun. Math. Phys.} {\bf 383} (2021), no.~3 1669--1804, [\href{http://arxiv.org/abs/1810.05338}{{\tt arXiv:1810.05338}}].

\bibitem{Heidenreich:2021xpr}
B.~Heidenreich, J.~McNamara, M.~Montero, M.~Reece, T.~Rudelius, and I.~Valenzuela, {\it {Non-invertible global symmetries and completeness of the spectrum}},  {\em JHEP} {\bf 09} (2021) 203, [\href{http://arxiv.org/abs/2104.07036}{{\tt arXiv:2104.07036}}].

\bibitem{Corlette}
K.~Corlette, {\it Archimedean superrigidity and hyperbolic geometry},  {\em Annals of Mathematics} {\bf 135} (1992), no.~1 165--182.

\bibitem{Deligne1982}
P.~Deligne and J.~S. Milne, {\em Tannakian Categories}, pp.~101--228.
\newblock Springer Berlin Heidelberg, Berlin, Heidelberg, 1982.

\bibitem{Bergshoeff:2011se}
E.~A. Bergshoeff, T.~Ortin, and F.~Riccioni, {\it {Defect Branes}},  {\em Nucl. Phys. B} {\bf 856} (2012) 210--227, [\href{http://arxiv.org/abs/1109.4484}{{\tt arXiv:1109.4484}}].

\bibitem{Kleinschmidt:2011vu}
A.~Kleinschmidt, {\it {Counting supersymmetric branes}},  {\em JHEP} {\bf 10} (2011) 144, [\href{http://arxiv.org/abs/1109.2025}{{\tt arXiv:1109.2025}}].

\bibitem{Nair:1986zn}
V.~P. Nair, A.~D. Shapere, A.~Strominger, and F.~Wilczek, {\it {Compactification of the Twisted Heterotic String}},  {\em Nucl. Phys. B} {\bf 287} (1987) 402--418.

\bibitem{Ginsparg:1986wr}
P.~H. Ginsparg and C.~Vafa, {\it {Toroidal Compactification of Nonsupersymmetric Heterotic Strings}},  {\em Nucl. Phys. B} {\bf 289} (1987) 414.

\bibitem{Itoyama:2021itj}
H.~Itoyama, Y.~Koga, and S.~Nakajima, {\it {Target space duality of non-supersymmetric string theory}},  {\em Nucl. Phys. B} {\bf 975} (2022) 115667, [\href{http://arxiv.org/abs/2110.09762}{{\tt arXiv:2110.09762}}].

\bibitem{Fraiman:2023cpa}
B.~Fraiman, M.~Gra{\~n}a, H.~Parra De~Freitas, and S.~Sethi, {\it {Non-supersymmetric heterotic strings on a circle}},  {\em JHEP} {\bf 12} (2024) 082, [\href{http://arxiv.org/abs/2307.13745}{{\tt arXiv:2307.13745}}].

\bibitem{DeFreitas:2024ztt}
H.~P. De~Freitas, {\it {Non-supersymmetric heterotic strings and chiral CFTs}},  {\em JHEP} {\bf 11} (2024) 002, [\href{http://arxiv.org/abs/2402.15562}{{\tt arXiv:2402.15562}}].

\bibitem{Fraiman:2025yrx}
B.~Fraiman and H.~Parra~de Freitas, {\it {Symmetries and dualities in non-supersymmetric CHL strings}},  \href{http://arxiv.org/abs/2511.01674}{{\tt arXiv:2511.01674}}.

\bibitem{Ferrara:2008de}
S.~Ferrara and A.~Marrani, {\it {Symmetric Spaces in Supergravity}},  {\em Contemp. Math.} {\bf 490} (2009) 203--228, [\href{http://arxiv.org/abs/0808.3567}{{\tt arXiv:0808.3567}}].

\bibitem{Obers:1998fb}
N.~A. Obers and B.~Pioline, {\it {U duality and M theory}},  {\em Phys. Rept.} {\bf 318} (1999) 113--225, [\href{http://arxiv.org/abs/hep-th/9809039}{{\tt hep-th/9809039}}].

\end{thebibliography}\endgroup

\end{document}